\documentclass[epsf,here,12pt]{article}
\usepackage{mypennames}
\usepackage{epsfig}  
\usepackage{cite}  
\begin{document}

\newcommand{\rhobar}{\overline{\rho}}
\newcommand{\etabar}{\overline{\eta}}
\newcommand{\epsilonk}{\left | \epsilon_K \right |}
\newcommand{\vubovcb}{\left | \frac{V_{ub}}{V_{cb}} \right |}
\newcommand{\vtdovcb}{\left | \frac{V_{td}}{V_{cb}} \right |}
\newcommand{\vtdovts}{\left | \frac{V_{td}}{V_{ts}} \right |}
\newcommand{\BK}{B_K}
\newcommand{\Kos}{{\rm K^0_S}}
\newcommand{\pr}{{\rm P.R.}}
\newcommand{\Ds}{{\rm D}_s^+}
\newcommand{\Dsb}{{\rm D}_s^-}
\newcommand{\Dp}{{\rm D}^+}
\newcommand{\Dpb}{{\rm D}^-}
\newcommand{\Do}{{\rm D}^0}
\newcommand{\Dob}{\overline{{\rm D}}^0}
\newcommand{\piss}{\pi^{\ast \ast}}
\newcommand{\pis}{\pi^{\ast}}
\newcommand{\bbar}{\overline{b}}
\newcommand{\cbar}{\overline{c}}
\newcommand{\Dstar}{{\rm D}^{\ast}}
\newcommand{\Dstars}{{\rm D}^{\ast +}_s}
\newcommand{\Dstaro}{{\rm D}^{\ast 0}}
\newcommand{\Dstarp}{{\rm D}^{\ast +}}
\newcommand{\Dstarstar}{{\rm D}^{\ast \ast}}
\newcommand{\pistar}{\pi^{\ast}}
\newcommand{\pisstar}{\pi^{\ast \ast}}
\newcommand{\Dbar}{\overline{{\rm D}}}
\newcommand{\Bbar}{\overline{{\rm B}}}
\newcommand{\Bsbar}{\overline{{\rm B}^0_s}}
\newcommand{\Lcbar}{\overline{\Lambda^+_c}}
\newcommand{\Bstar}{{\rm B}^{*}}
\newcommand{\Bstarstar}{{\rm B}^{**}}
\newcommand{\Bstarstarb}{\overline{{\rm B}}^{**}}
\newcommand{\nubar}{\overline{\nu_{\ell}}}
\newcommand{\tautaubar}{\tau \overline{\tau}}
\newcommand{\Vcb}{\left | {\rm V}_{cb} \right |}
\newcommand{\Vub}{\left | {\rm V}_{ub} \right |}
\newcommand{\Vtd}{\left | {\rm V}_{td} \right |}
\newcommand{\Vts}{\left | {\rm V}_{ts} \right |}
\newcommand{\fleisher}{\frac{BR({\rm B}^0~(\overline{{\rm B}^0}) \rightarrow \pi^{\pm} {\rm K}^{\mp})}
{BR({\rm B}^{\pm} \rightarrow \pi^{\pm} {\rm K}^0)}}
\newcommand{\bptre}{\rm b^{+}_{3}}
\newcommand{\bp}{\rm b^{+}_{1}}
\newcommand{\bo}{\rm b^0}
\newcommand{\bos}{\rm b^0_s}
\newcommand{\bss}{\rm b^s_s}
\newcommand{\qq}{\rm q \overline{q}}
\newcommand{\cc}{\rm c \overline{c}}
\newcommand{\BsDmX}{{B_{s}^{0}} \rightarrow D \mu X}
\newcommand{\BsDsm}{{B_{s}^{0}} \rightarrow D_{s} \mu X}
\newcommand{\BsDsX}{{B_{s}^{0}} \rightarrow D_{s} X}
\newcommand{\BDsX}{B \rightarrow D_{s} X}
\newcommand{\BDomX}{B \rightarrow D^{0} \mu X}
\newcommand{\BDpmX}{B \rightarrow D^{+} \mu X}
\newcommand{\Dsfmn}{D_{s} \rightarrow \phi \mu \nu}
\newcommand{\Dsfipi}{D_{s} \rightarrow \phi \pi}
\newcommand{\DsfX}{D_{s} \rightarrow \phi X}
\newcommand{\DpfX}{D^{+} \rightarrow \phi X}
\newcommand{\DofX}{D^{0} \rightarrow \phi X}
\newcommand{\DfX}{D \rightarrow \phi X}
\newcommand{\DsD}{B \rightarrow D_{s} D}
\newcommand{\DsmX}{D_{s} \rightarrow \mu X}
\newcommand{\DmX}{D \rightarrow \mu X}
\newcommand{\Zbb}{Z^{0} \rightarrow \rm b \overline{b}}
\newcommand{\Zcc}{Z^{0} \rightarrow \rm c \overline{c}}
\newcommand{\Rbb}{\frac{\Gamma_{Z^0 \rightarrow \rm b \overline{b}}}
{\Gamma_{Z^0 \rightarrow Hadrons}}}
\newcommand{\Rcc}{\frac{\Gamma_{Z^0 \rightarrow \rm c \overline{c}}}
{\Gamma_{Z^0 \rightarrow Hadrons}}}
\newcommand{\bb}{\rm b \overline{b}}
\newcommand{\str}{\rm s \overline{s}}
\newcommand{\Bs}{\rm{B^0_s}}
\newcommand{\Bsb}{\overline{\rm{B^0_s}}}
\newcommand{\Bsh}{\rm{B^{heavy}_s}}
\newcommand{\Bsl}{\rm{B^{light}_s}}
\newcommand{\Bssh}{\rm{B^{short}_s}}
\newcommand{\Bslg}{\rm{B^{long}_s}}
\newcommand{\Gsh}{\Gamma^{{\rm heavy}}_{{\rm B^0_s}}}
\newcommand{\Gsl}{\Gamma^{{\rm light}}_{{\rm B^0_s}}}
\newcommand{\Gss}{\Gamma^{{\rm short}}_{{\rm B^0_s}}}
\newcommand{\Gslg}{\Gamma^{{\rm long}}_{{\rm B^0_s}}}
\newcommand{\Gs}{\Gamma_{{\rm B^0_s}}}
\newcommand{\Gd}{\Gamma_{{\rm B^0_d}}}
\newcommand{\Msh}{{\rm m}^{{\rm heavy}}_{{\rm B^0_s}}}
\newcommand{\Msl}{{\rm m}^{{\rm light}}_{{\rm B^0_s}}}
\newcommand{\Bp}{\rm{B^{+}}}
\newcommand{\Bm}{\rm{B^{-}}}
\newcommand{\Bo}{\rm{B^{0}}}
\newcommand{\Bd}{\rm{B^{0}_{d}}}
\newcommand{\Bdb}{\overline{\rm{B^{0}_{d}}}}
\newcommand{\Lb}{\Lambda^0_b}
\newcommand{\Lbb}{\overline{\Lambda^0_b}}
\newcommand{\Kstar}{\rm{K^{\star 0}}}
\newcommand{\phim}{\rm{\phi}}
\newcommand{\Dsp}{\mbox{D}_s^+}
\newcommand{\Dsm}{\mbox{D}_s^-}
\newcommand{\Dn}{\mbox{D}^0}
\newcommand{\Dm}{\mbox{D}^-}
\newcommand{\Dnb}{\overline{\mbox{D}}^0}
\newcommand{\Lc}{\Lambda_c^+}
\newcommand{\Lcb}{\overline{\Lambda}_c^-}
\newcommand{\Xic}{\Xi_c^{0,+}}
\newcommand{\Xicb}{\overline{\Xi}_c^{0,-}}
\newcommand{\Xib}{\Xi_b}
\newcommand{\Dstarm}{\mbox{D}^{\ast -}}
\newcommand{\Dsstarp}{\mbox{D}_s^{\ast +}}
\newcommand{\Dsstar}{\mbox{D}^{\ast \ast}}
\newcommand{\Km}{\mbox{K}^-}
\newcommand{\Pb}{P_{b-baryon}}
\newcommand{\KKpi}{\rm{ K K \pi }}
\newcommand{\GeV}{\rm{GeV}}
\newcommand{\MeV}{\rm{MeV}}
\newcommand{\nb}{\rm{nb}}
\newcommand{\Zzero}{{\rm Z}}
\newcommand{\MZ}{\rm{M_Z}}
\newcommand{\MW}{\rm{M_W}}
\newcommand{\GF}{\rm{G_F}}
\newcommand{\Gm}{\rm{G_{\mu}}}
\newcommand{\MH}{\rm{M_H}}
\newcommand{\MT}{\rm{m_{top}}}
\newcommand{\GZ}{\Gamma_{\rm Z}}
\newcommand{\Afb}{\rm{A_{FB}}}
\newcommand{\Afbs}{\rm{A_{FB}^{s}}}
\newcommand{\sigmaf}{\sigma_{\rm{F}}}
\newcommand{\sigmab}{\sigma_{\rm{B}}}
\newcommand{\NF}{\rm{N_{F}}}
\newcommand{\NB}{\rm{N_{B}}}
\newcommand{\Nnu}{\rm{N_{\nu}}}
\newcommand{\RZ}{\rm{R_Z}}
\newcommand{\rhob}{\rho_{eff}}
\newcommand{\Gammanz}{\rm{\Gamma_{Z}^{new}}}
\newcommand{\Gammani}{\rm{\Gamma_{inv}^{new}}}
\newcommand{\Gammasz}{\rm{\Gamma_{Z}^{SM}}}
\newcommand{\Gammasi}{\rm{\Gamma_{inv}^{SM}}}
\newcommand{\Gammaxz}{\rm{\Gamma_{Z}^{exp}}}
\newcommand{\Gammaxi}{\rm{\Gamma_{inv}^{exp}}}
\newcommand{\rhoZ}{\rho_{\rm Z}}
\newcommand{\thw}{\theta_{\rm W}}
\newcommand{\swsq}{\sin^2\!\thw}
\newcommand{\swsqmsb}{\sin^2\!\theta_{\rm W}^{\overline{\rm MS}}}
\newcommand{\swsqbar}{\sin^2\!\overline{\theta}_{\rm W}}
\newcommand{\cwsqbar}{\cos^2\!\overline{\theta}_{\rm W}}
\newcommand{\swsqb}{\sin^2\!\theta^{eff}_{\rm W}}
\newcommand{\ee}{{e^+e^-}}
\newcommand{\eeX}{{e^+e^-X}}
\newcommand{\gaga}{{\gamma\gamma}}
\newcommand{\mumu}{\ifmmode {\mu^+\mu^-} \else ${\mu^+\mu^-} $ \fi}
\newcommand{\eeg}{{e^+e^-\gamma}}
\newcommand{\mumug}{{\mu^+\mu^-\gamma}}
\newcommand{\tautau}{{\tau^+\tau^-}}
\newcommand{\qqb}{{q\overline{q}}}
\newcommand{\eegg}{e^+e^-\rightarrow \gamma\gamma}
\newcommand{\eeggg}{e^+e^-\rightarrow \gamma\gamma\gamma}
\newcommand{\eeee}{e^+e^-\rightarrow e^+e^-}
\newcommand{\eeeeee}{e^+e^-\rightarrow e^+e^-e^+e^-}
\newcommand{\eeeeg}{e^+e^-\rightarrow e^+e^-(\gamma)}
\newcommand{\eeeegg}{e^+e^-\rightarrow e^+e^-\gamma\gamma}
\newcommand{\eeeg}{e^+e^-\rightarrow (e^+)e^-\gamma}
\newcommand{\eemumu}{e^+e^-\rightarrow \mu^+\mu^-}
\newcommand{\eetautau}{e^+e^-\rightarrow \tau^+\tau^-}
\newcommand{\eehad}{e^+e^-\rightarrow {\rm hadrons}}
\newcommand{\eettg}{e^+e^-\rightarrow \tau^+\tau^-\gamma}
\newcommand{\eell}{e^+e^-\rightarrow l^+l^-}
\newcommand{\Ztopig}{{\rm Z}^0\rightarrow \pi^0\gamma}
\newcommand{\Ztogg}{{\rm Z}^0\rightarrow \gamma\gamma}
\newcommand{\Ztoee}{{\rm Z}^0\rightarrow e^+e^-}
\newcommand{\Ztoggg}{{\rm Z}^0\rightarrow \gamma\gamma\gamma}
\newcommand{\Ztomumu}{{\rm Z}^0\rightarrow \mu^+\mu^-}
\newcommand{\Ztotautau}{{\rm Z}^0\rightarrow \tau^+\tau^-}
\newcommand{\Ztoll}{{\rm Z}^0\rightarrow l^+l^-}
\newcommand{\Ztocc}{{\rm Z^0\rightarrow c \overline c}}
\newcommand{\Lamp}{\Lambda_{+}}
\newcommand{\Lamm}{\Lambda_{-}}
\newcommand{\Pt}{\rm P_{t}}
\newcommand{\Gee}{\Gamma_{ee}}
\newcommand{\Gpig}{\Gamma_{\pi^0\gamma}}
\newcommand{\Ggg}{\Gamma_{\gamma\gamma}}
\newcommand{\Gggg}{\Gamma_{\gamma\gamma\gamma}}
\newcommand{\Gmumu}{\Gamma_{\mu\mu}}
\newcommand{\Gtautau}{\Gamma_{\tau\tau}}
\newcommand{\Ginv}{\Gamma_{\rm inv}}
\newcommand{\Ghad}{\Gamma_{\rm had}}
\newcommand{\Gnu}{\Gamma_{\nu}}
\newcommand{\GnuSM}{\Gamma_{\nu}^{\rm SM}}
\newcommand{\Gll}{\Gamma_{l^+l^-}}
\newcommand{\Gff}{\Gamma_{f\overline{f}}}
\newcommand{\Gtot}{\Gamma_{\rm tot}}
\newcommand{\Rb}{\mbox{R}_b}
\newcommand{\Rc}{\mbox{R}_c}
\newcommand{\al}{a_l}
\newcommand{\vl}{v_l}
\newcommand{\af}{a_f}
\newcommand{\vf}{v_f}
\newcommand{\ael}{a_e}
\newcommand{\ve}{v_e}
\newcommand{\amu}{a_\mu}
\newcommand{\vmu}{v_\mu}
\newcommand{\atau}{a_\tau}
\newcommand{\vtau}{v_\tau}
\newcommand{\ahatl}{\hat{a}_l}
\newcommand{\vhatl}{\hat{v}_l}
\newcommand{\ahate}{\hat{a}_e}
\newcommand{\vhate}{\hat{v}_e}
\newcommand{\ahatmu}{\hat{a}_\mu}
\newcommand{\vhatmu}{\hat{v}_\mu}
\newcommand{\ahattau}{\hat{a}_\tau}
\newcommand{\vhattau}{\hat{v}_\tau}
\newcommand{\vtildel}{\tilde{\rm v}_l}
\newcommand{\avsq}{\ahatl^2\vhatl^2}
\newcommand{\Ahatl}{\hat{A}_l}
\newcommand{\Vhatl}{\hat{V}_l}
\newcommand{\Afer}{A_f}
\newcommand{\Ael}{A_e}
\newcommand{\Aferb}{\overline{A_f}}
\newcommand{\Aelb}{\overline{A_e}}
\newcommand{\AVsq}{\Ahatl^2\Vhatl^2}
\newcommand{\Iwk}{I_{3l}}
\newcommand{\Qch}{|Q_{l}|}
\newcommand{\roots}{\sqrt{s}}
\newcommand{\pT}{p_{\rm T}}
\newcommand{\mt}{m_t}
\newcommand{\Rechi}{{\rm Re} \left\{ \chi (s) \right\}}
\newcommand{\up}{^}
\newcommand{\abscosthe}{|cos\theta|}
\newcommand{\dsum}{\Sigma |d_\circ|}
\newcommand{\zsum}{\Sigma z_\circ}
\newcommand{\sint}{\mbox{$\sin\theta$}}
\newcommand{\cost}{\mbox{$\cos\theta$}}
\newcommand{\mcost}{|\cos\theta|}
\newcommand{\epair}{\mbox{$e^{+}e^{-}$}}
\newcommand{\mupair}{\mbox{$\mu^{+}\mu^{-}$}}
\newcommand{\taupair}{\mbox{$\tau^{+}\tau^{-}$}}
\newcommand{\gamgam}{\mbox{$e^{+}e^{-}\rightarrow e^{+}e^{-}\mu^{+}\mu^{-}$}}
\newcommand{\fullskip}{\vskip 16cm}
\newcommand{\halfskip}{\vskip  8cm}
\newcommand{\quarskip}{\vskip  6cm}
\newcommand{\abitskip}{\vskip 0.5cm}
\newcommand{\ba}{\begin{array}}
\newcommand{\ea}{\end{array}}
\newcommand{\bc}{\begin{center}}
\newcommand{\ec}{\end{center}}
\newcommand{\be}{\begin{eqnarray}}
\newcommand{\eeq}{\end{eqnarray}}
\newcommand{\bes}{\begin{eqnarray*}}
\newcommand{\ees}{\end{eqnarray*}}
\def\etal{{\it et al.}}
\newcommand{\Kz}{\ifmmode {\rm K^0_s} \else ${\rm K^0_s} $ \fi}
\newcommand{\Zz}{\ifmmode {\rm Z} \else ${\rm Z } $ \fi}
\newcommand{\qqbar}{\ifmmode {\rm q\overline{q}} \else ${\rm q\overline{q}} $ \fi}
\newcommand{\ccbar}{\ifmmode {\rm c\overline{c}} \else ${\rm c\overline{c}} $ \fi}
\newcommand{\bbbar}{\ifmmode {\rm b\overline{b}} \else ${\rm b\overline{b}} $ \fi}
\newcommand{\xxbar}{\ifmmode {\rm x\overline{x}} \else ${\rm x\overline{x}} $ \fi}
\newcommand{\rphi}{\ifmmode {\rm R\phi} \else ${\rm R\phi} $ \fi}
%\renewcommand {\pt}         {\rm p_t}
%========================================================================% 
%  definitions pour Vcb
% \def\vcb{\mbox{$|V_{cb}|$}}
\def\vcb{$\left | {\rm V}_{cb} \right |$}
\def\fw{${\cal F}(w)$}
\def\fone{${\cal F}_{D^{*}}(1)$}
\def\fvcb{${\cal F}_{D^{*}}(1)|{\rm}V_{cb}|$}
\def\btods{$ \Bdb \rightarrow {\rm D}^{*+}\ell^-{\overline \nu_{\ell}}$}
\def\Btau{$ \Bdb \rightarrow {\rm D}^{*+}\tau^-{\overline \nu_{\tau}}$}
\def\Bxc{$ \Bdb\rightarrow {\rm D}^{*+} {\rm X}_{\overline{c}}$}
\def\btodss{$ \Bdb\rightarrow {\rm D}^{**+}\ell^-{\overline \nu_{\ell}}$}
%***
\newcommand {\bl}    {{\rm BR}({b \rightarrow \ell})}
\newcommand {\cl}    {{\rm BR}({c \rightarrow \overline{\ell}})}
\newcommand {\bcbl}   {{\rm BR}({b \rightarrow \overline{c} \rightarrow \ell})}
\newcommand {\bcl}   {{\rm BR}({b \rightarrow c \rightarrow \overline{\ell}})}
\newcommand {\btaul}  {{\rm BR}({b \rightarrow \tau \rightarrow \ell})}
\newcommand {\bpsill} {{\rm BR}({b \rightarrow {\rm J}/\psi\rightarrow \ell^+\ell^-})}
\newcommand {\glcc}   {\rm{g \rightarrow c \overline c}}
\newcommand {\glbb}   {\rm{g \rightarrow b \overline b}}
\def\dsp{${\rm D}^{*+}$}
\def\bbar{$\overline{{\rm B}^0_d}$}
%***
% Definitions for Oscillations
%********
%%%%%%%%%%%%%%%%%%%%%%%%%%%%%%%%%%%%%%%%%%%%%%%%%%%%%%%%%%%%%%%%%%%%%
%
% current results
% ===============
%
%%%%%%%%%%%%%%%%%%%%%%%%%%%%%%%%%%%%%%%%%%%%%%%%%%%%%%%
%    Lifetimes
%%%%%%%%%%%%%%%%%%%%%%%%%%%%%%%%%%%%%%%%%%%%%%%%%%%%%%%
%   Inclusive b-hadron lifetime
%\newcommand{\taubav}{1.564 \pm 0.014} %% OK
\newcommand{\taubav}{1.561 \pm 0.014} %% OK
%   Bd lifetime
\newcommand{\taubd}{1.546 \pm 0.021} %% OK 2001-2
%   B+ lifetime
\newcommand{\taubp}{1.647 \pm 0.021} %% OK
%   Bs lifetime
\newcommand{\taubs}{1.464 \pm 0.057} %% OK
%   Lb lifetime
\newcommand{\taulb}{1.229 ^{+0.081}_{-0.079}} %% OK
%   Xib lifetime
\newcommand{\tauxib}{1.39 ^{+0.34}_{-0.28}} %% OK
%   b-baryon lifetime
\newcommand{\taubbar}{1.208^{+0.051}_{-0.050}} %% OK
%   tau(B+)/Tau(Bd)
\newcommand{\taubpovertaubd}{1.063 \pm 0.020} %% OK 2001-2
%%%%%%%%%%%%%%%%%%%%%%%%%%%%%%%%%%%%%%%%%%%%%%%%%
%   Rates et Oscillations
%%%%%%%%%%%%%%%%%%%%%%%%%%%%%%%%%%%%%%%%%%%%%%%%%
%\newcommand{\dmdl}{0.468\pm 0.019} %% dmd average from LEP
%\newcommand{\dmdlerr}{0.468\pm 0.012\ {\rm (stat)}\pm 0.014\ {\rm (syst)}}  %% dmd average from LEP
\newcommand{\dmdw}{0.470\pm 0.018} %% dmd average from LEP+SLD+CDF (world average from time-dep. analyses)
\newcommand{\dmdwerr}{0.486\pm 0.015}  %% dmd average from LEP+SLD+CDF (world average from time-dep. analyses)
\newcommand{\dmdx}{0.487\pm 0.014} %% dmd average from LEP+SLD+CDF+Upsilon4S (world average)
\newcommand{\dmdxev}{3.21 \pm 0.09} %% dmd average from LEP+SLD+CDF+Upsilon4S in eV (world average)
\newcommand{\chix}{0.181\pm 0.007} %% chid average from LEP+SLD+CDF+Upsilon4S (world average)
\newcommand{\chicls}{0.181\pm 0.008} %% chid average from LEP+SLD+CDF
%
%\newcommand{\fbsbr}{11.2^{+3.2}_{-2.7}}
%\newcommand{\fbbary}{11.0^{+2.3}_{-1.9}}  %% OK
% WARNING: the Bs and b-baryon fractions from BR measurements 
%          should also be updated in Table 1 !!!!
%          (look for string "WARNING" in this file)
%          Note that in fact all numbers from Table 1 and 2 should 
%          be updated by hand if they change ...
%%%%%%%%
% rates from direct measurements
%%%%%%%%
%    -Bs rate from Ds-l evts
\newcommand{\fbsdir}{(12.2^{+4.5}_{-3.1})\%} %% OK
%    -Bu rate from charged sec. vertices
\newcommand{\fbudir}{(41.4 \pm 1.6)\%} %% OK
%    -Lb rate from Lc-l evts
\newcommand{\flbdir}{(11.6^{+4.6}_{-3.1})\%} %% OK 2001-2
%    -Xib rate from Xi-l evts
\newcommand{\fxidir}{(1.1^{+0.6}_{-0.4})\%} %% OK
%    -b-baryon rate from direct measurements
\newcommand{\fbardir}{(13.7^{+4.8}_{-3.2})\%} %% OK 2001-2
%    -b-baryon rate from spectator protons
\newcommand{\fbarspec}{(10.2 \pm 2.8)\%} %% OK
%    -b-baryon rate from direct + spectator protons
\newcommand{\fbaravg}{(11.7 \pm 2.1)\%} %% OK 2001-2
%%%%%%%%%%%%%%%%%%%%%%%%
% combined rates from direct measurements, including CDF and
% imposing fu=fd and 1=fu+fd+fs+fb (STEP 1)
%%%%%%%%%%%%%%%%%%%%%%%%
%    -Bs rate 
\newcommand{\fsa}{(9.2 \pm 2.4)\%} %% OK 2001-2
%    -b-baryon rate
\newcommand{\fbara}{(10.5 \pm 2.0)\%} %% OK 2001-2
%    -Bu=Bd rate
\newcommand{\fua}{(40.1 \pm 1.3)\%} %% OK 2001-2
%    -correlation between Bs and b-baryon rates
\newcommand{\rhosbara}{-0.36} %% OK 2001-2
%    -correlation between Bu and Bs rates
\newcommand{\rhosua}{-0.68} %% OK 2001-2
%    -correlation between Bu and b-baryon rates
\newcommand{\rhobarua}{-0.45} %% OK 2001-2
%%%%%%%%%%%%
%   Bs fraction from mixing
\newcommand{\fbsmix}{(10.1\pm 1.4)\%} %% OK 2001
%%%%%%%%%%%%%
%   Final (STEP 3) rates and correlations
\newcommand{\fbs}{(9.8 \pm 1.2)\%} %% OK 2001-2
\newcommand{\fbar}{(10.3 \pm 1.8)\%} %% OK 2001-2
\newcommand{\fbd}{(39.9\pm 1.1)\%} %% OK 2001-2
\newcommand{\rhosbar}{0.01} %% OK 2001-2
\newcommand{\rhosd}{-0.55} %% OK 2001-2
\newcommand{\rhobard}{-0.84} %% OK 2001-2
%%%%%%%%%%%%%
% world combined dms limits
\newcommand{\dmslim}{15.0}  %%% World dms limit (method A) 2001-2
\newcommand{\dmsb}{12.1}    %%% World dms limit (method B)
\newcommand{\dmssen}{18.1}  %%% World dms sensitivity (method A) 2001-2
% LEP only dms limits
\newcommand{\ldmslim}{11.5} %%% LEP dms limit (method A)
\newcommand{\ldmsb}{10.8}   %%% LEP dms limit (method B)
\newcommand{\ldmssen}{12.9} %%% LEP dms sensitivity (method A)
%
%%%%%%%%%%%%%%%%%%%%%%%%%%%%%%%%%%%%%%%%%%%%%%%%%%%%%%%%%%%%%%%%%%%%%
%  Vcb results
%%%%%%%%%%%%%%%%%%%%
%%% Inclusive average
\newcommand{\vcbinc}{(40.7 \pm 0.5({\rm exp.}) \pm 2.4({\rm theo.})) \times 10^{-3}} %% OK
%%% Exclusive average
\newcommand{\favcb}{(35.6 \pm 0.8({\rm stat.}) \pm 1.5({\rm syst.})) \times 10^{-3}} %% OK
\newcommand{\rhobb}{1.38 \pm  0.08 \pm 0.26} %% OK
\newcommand{\vcbexc}{(40.5 \pm 1.9({\rm exp.}) \pm 2.3({\rm theo.}))\times 10^{-3}} %% OK
%%% Global average
\newcommand{\vcbavg}{(40.6 \pm 1.9) \times 10^{-3}} %% OK
%%%%%%%%%%%%%
%    Final (STEP 3) rates and correlations

%%%%%%%%%%%%%%%%%%%%%%%%%%%%%%%%%%%%%%%%%%%%%%%%%%%%%%%%%%%%%%%%%%%%%
%

\setlength{\textheight}{24.0cm}
\setlength{\topmargin}{-0.5cm}
\setlength{\textwidth}{15.0cm}
% Some Definitions
\newcommand{\dmd}{\ensuremath{\Delta m_d}}
\newcommand{\dms}{\ensuremath{\Delta m_s}}
\newcommand{\Bb}{\mbox{$b$-baryon}}
\newcommand{\chib}{\ensuremath{\overline \chi}}
\newcommand{\chid}{\ensuremath{\chi_d}}
\newcommand{\chis}{\ensuremath{\chi_s}}
\newcommand{\Yfs}{\ensuremath{{\Upsilon({\rm 4S})}}}
\newcommand{\fd}{f_{\PdB}}
\newcommand{\fs}{f_{\PsB}}
\newcommand{\fu}{f_{\PBp}}
\newcommand{\fb}{f_{\mbox{\scriptsize \Bb}}}
\newcommand{\gs}{g_{\PsB}}
\newcommand{\gd}{g_{\PdB}}
\newcommand{\gu}{g_{\PBp}}
\newcommand{\gb}{g_{\mbox{\scriptsize \Bb}}}
\newcommand{\ips}{${\mathrm{ps}}^{-1}$}
\newcommand{\ipsm}{{\mathrm{ps}}^{-1}}
\newcommand{\blankvalue}{\multicolumn{2}{c|}{ }}
\newcommand{\Brr}[2]{\ensuremath{\mathrm{BR}(#1 \rightarrow #2)}}
\newcommand{\mffbs}{\Brr{\overline b}{\PsBz}}
\newcommand{\mbsdslX}{\Brr{\PsBz}{\PsDm \ell^+\nu_{\ell} X}}
\newcommand{\prodbs}{$\mffbs \cdot \mbsdslX$}
\newcommand{\bsdslX}{\Brr{\PsBz}{\PsDm \ell^+ \nu_{\ell} \mathrm{X}}}
\newcommand{\prodlb}{\Brr{b}{\PbgLz}~
  \Brr{\PbgLz}{\PcgLp \ell^-\overline {\nu_{\ell}} \mathrm{X}}}
\newcommand{\prodxb}{\Brr{b}{\PbgX}~
  \Brr{\PbgXm}{\PgXm \ell^-\overline {\nu_{\ell}} \mathrm{X}}}
\newcommand{\brpkpi}{\Brr{\PcgLp}{\Pp \PKm \Pgpp}}
\newcommand{\mbXlnu}{\Brr{\overline b}{\ell^+\nu_{\ell}{\mathrm{X}}}}
\newcommand{\mBXlnu}{\Brr{\PB}{\ell^+ \nu_{\ell}{\mathrm{X}}}}
\newcommand{\mups}{\Upsilon(4\mathrm{S})}
\newcommand{\bzdlnu}{\Brr{\PdBz}{\PDm \ell^+ \nu_{\ell}}}
\newcommand{\bzdstlnu}{\Brr{\PdBz}{\PDstm\ell^+ \nu_{\ell}}}
\newcommand{\bpdlnu}{\Brr{\PBp}{\PaDz \ell^+ \nu_{\ell}}}
\newcommand{\bpdstlnu}{\Brr{\PBp}{\PDstz \ell^+ \nu_{\ell}}}
\newcommand{\mbdstpilnu}{\Brr{\PB}{\PDst \Pgp \ell^+ \nu_{\ell}}}
\newcommand{\mbdpilnu}{\Brr{\PB}{\PaD \Pgp \ell^+ \nu_{\ell}}}
\newcommand{\brphipi}{\Brr{\PsDm}{\phi \Pgpm}}
%%% New definitions Olivier
\newcommand{\mysection}[1]{\section{\boldmath #1}}
\newcommand{\mysubsection}[1]{\subsection[#1]{\boldmath #1}}
\newcommand{\mysubsubsection}[1]{\subsubsection[#1]{\boldmath #1}}
%%% Old definitions
%\newcommand{\mysection}[1]{\section[#1]{\boldmath #1}}
%\newcommand{\mysubsection}[1]{\subsection[#1]{\boldmath #1}}
%\newcommand{\mysubsubsection}[1]{\subsubsection[#1]{\boldmath #1}}
% definitions for dgamma/gamma
\newcommand{\dgs}{\Delta \Gamma_{\Bs}}
\newcommand{\dgbs}{\Delta \Gamma_{\rm \Bs}/\Gamma_{\rm \Bs}}
\newcommand{\dgd}{\Delta \Gamma_{\Bd}}
\newcommand{\dgbd}{\Delta \Gamma_{\rm \Bd}/\Gamma_{\rm \Bd}}
\newcommand{\tbs}{\tau_{\Bs}}
\newcommand{\tbd}{\tau_{\Bd}}
\newcommand{\tbssemi}{\tau_{\rm B^{semi.}_s}}
\newcommand{\tbsshort}{\tau_{\Bssh}}
\newcommand{\tbsdh}{\tau_{\rm B^{D_s-had.}_s}}
\newcommand{\tbspsi}{\tau_{\rm B^{{\rm J}/\psi \phi}_s}}
\newcommand{\tbsinc}{\tau_{\rm B^{incl.}_s}}
%################################################## titlepage declaration

\begin{titlepage}

\pagenumbering{arabic}
\vspace*{-0.9cm}
\begin{center}
{\large EUROPEAN ORGANIZATION FOR NUCLEAR RESEARCH}
\end{center}
\vspace*{1.1cm}
\begin{tabular*}{15.cm}{l@{\extracolsep{\fill}}r}
{  } & 
%===================> DELPHI note number       =====> To be filled <=====%
CERN-EP/2001-050
%========================================================================%
\\
& 
%===================> DELPHI note date         =====> To be filled <=====%
June 26, 2001\\
%\today \\
%Draft 2
%========================================================================%
\\
&\\
% \hline
\end{tabular*}
\vspace*{1.cm}
\begin{center}  
\Large 
{\bf Combined results on {\boldmath $b$}-hadron production rates and
decay properties} 
\vspace*{2.cm}
\\
\normalsize {    {\bf 
 ALEPH, CDF, DELPHI, L3, OPAL, SLD }}\\
\vskip 0.5truecm
\footnotesize{Prepared 
\footnote{The members of the working groups involved in 
this activity are:
D. Abbaneo, 
J. Alcaraz, V. Andreev, E. Barberio, M. Battaglia,
S. Blyth, G. Boix,
 M. Calvi, P. Checchia, P. Coyle,
L. Di Ciaccio,  P. Gagnon, R. Hawkings, O. Hayes, P. Henrard, T. Hessing, 
I.J. Kroll, O. Leroy, D. Lucchesi, M. Margoni, S. Mele, 
H.G. Moser, F. Muheim, F. Palla, D. Pallin, F. Parodi, M. Paulini, E. Piotto,
P. Privitera, Ph. Rosnet, 
P. Roudeau, D. Rousseau, O. Schneider,
C. Shepherd-Themistocleous,
 F. Simonetto, P. Spagnolo, A. Stocchi, D. Su, T. Usher, C. Weiser,
B. Wicklund and S. Willocq.}
from Contributions to the 2000 Summer conferences.} \\
\end{center}

\vskip 2.5truecm

\begin{abstract}
\noindent
 Combined results on $b$-hadron lifetimes,
$b$-hadron production rates,
$\Bd-\Bdb$ and $\Bs-\Bsb$ oscillations,
the decay width difference 
between 
the mass eigenstates of the $\Bs-\Bsb$ system,
the average number of $c$ and $\overline{c}$ quarks
in $b$-hadron decays, 
%the properties of the CKM unitarity triangle, 
and searches for CP violation 
in the $\Bd-~\Bdb$ system
% \rightarrow J/\psi \Kos$ decay channel of neutral $b$-mesons
are presented. They have been
obtained from published and preliminary measurements
available in Summer 2000 from the ALEPH, CDF, DELPHI, L3, OPAL and SLD
Collaborations. These results have been used to determine the parameters of
the CKM unitarity triangle.

\end{abstract}

\vskip 2.0truecm
\vspace{\fill}
\begin{center}
%To be submitted to  CERN-OPEN
\end{center}
\noindent

\vspace{\fill}
\end{titlepage}

\setcounter{page}{1}    
\tableofcontents

\setcounter{footnote}{0}

\newpage
%\linenumbers
\section {Introduction}
\label{sec:intro}
%{\bf Try to explain the differences as compared to Summer99 and why
%we have not included results from B-factories.}

This paper contains an update of similar results \cite{ref:summer99}
which were
based on measurements made available in Summer 1999.
Experimental results, made available in Summer 2000,
have been included and new sections have been prepared
on charm counting in $b$-hadron decays,
on direct $\sin{(2 \beta)}$
measurements and on other searches for CP violation
in the $\Bd-\Bdb$ system.
A large fraction of these measurements 
have been used, in addition, to determine the CKM 
unitarity triangle
parameters. These studies have been limited to combining
measurements
obtained by experiments which were running before the start of the 
asymmetric B factories.
As compared with the previous report \cite{ref:summer99}, 
apart from the new sections, the main improvements
have been obtained on $b$-lifetime and oscillation measurements.

%{\bf End modif.}

Accurate determinations of $b$-hadron decay properties provide constraints
on the values of the elements of the Cabibbo-Kobayashi-Maskawa (CKM) matrix
\cite{ckm}. The $\Vcb$ and $\Vub$ elements can be obtained from semileptonic
decay rates into charmed and non-charmed hadrons, and measurements
of the oscillation frequencies in $\Bd-\Bdb$ and $\Bs-\Bsb$ systems give
access to $\Vtd$ and $\Vts$.

Elements of the CKM matrix govern weak transitions between quarks.
Experimental results are obtained from processes involving $b$-hadrons. Effects
from strong interactions have thus to be controlled and
 $b$-hadrons are also
a good laboratory in this respect. 
Lifetime differences between the different
weakly decaying hadrons 
can be related to interactions between
the heavy quark and the light quark system inside the hadron. The polarization
of $\Lb$ baryons produced by $b$-quarks of known polarization,
 emitted from $\Zz$ decays at LEP or SLC,
indicates how polarization is transmitted from the 
heavy quark to the baryon(s) in the hadronization process.
%{\it Sentence on Delta(Gammas) ..............}
Rates and decay properties of $\Dstarstar$ 
states\footnote{The notation $\Dstarstar$ includes all 
charm mesons and non-resonant charmed final states
which are not simply D or $\Dstar$ mesons.} produced in
$b$-hadron semileptonic decays are needed to obtain accurate
determinations of $\Vcb$ as they provide constraints on experimental
and theoretical related uncertainties. 
Decays of $\Dstarstar$ states are also an important source
of background in other channels, and their properties have to be measured.
Finally, it is necessary to measure
%, as precisely as possible, 
the production rates
of the different weakly decaying $b$-hadrons emitted during
the hadronization of $b$-quarks created in high energy collisions,
because all these states have different properties so the study
of any one of them requires control of the background from the others.

Results obtained on $b$-hadron lifetimes,
%(ALEPH, CDF, DELPHI, L3, OPAL
%and SLD)
$b$-hadron production rates,
% (ALEPH, CDF, DELPHI, L3, OPAL),
$\Bd-\Bdb$ and $\Bs-\Bsb$ oscillations,
% (ALEPH, ARGUS, CDF, CLEO, DELPHI, L3, OPAL and SLD),
%$\Bs-\Bsb$ oscillations 
%(ALEPH, CDF, DELPHI, OPAL and SLD) 
$b$-hadron semileptonic decays, on the mean number of
$c$ and $\overline{c}$ quarks in $b$-hadron decays,
on the decay width difference 
between the $\Bs-\Bsb$ system mass eigenstates
and on CP violation in the $\Bd-\Bdb$ system
% observed
%in studies of the $J/\psi {\rm K}^0_s$ decay channel
%the values of the CKM matrix elements, $\Vcb$ 
%(ALEPH, DELPHI and OPAL) 
%and $\Vub$ 
made available 
%(ALEPH, DELPHI and L3) 
during Summer 2000, are presented here. 
%from
%measurements of the ALEPH, CDF, DELPHI and L3 Collaborations.
These quantities have been obtained by averaging
    published and preliminary measurements released publicly
    by the ALEPH, CDF, DELPHI, L3, OPAL and SLD experiments.
In addition, these results have been used to provide averaged
values for $\Vcb$, $\Vub$ and to obtain the parameters defining the
unitarity triangle, within the Standard Model framework.

Whenever possible, the input
    parameters used in the various analyses have been adjusted
    to common values, and all known correlations have been
    taken into account.
% These quantities have been obtained by adjusting all measurements,
%published as well as preliminary results contained in reports released 
%publicly by the 
%experiments ALEPH, CDF, DELPHI, L3, OPAL and SLD.
%Common values for external
%parameters have been used and all known correlations taken into account.
%and averaging after taking into account correlations. 
Close contacts have been 
established between representatives from the experiments and members of 
the different working
groups in charge of the averages, to ensure that the data are prepared in a form 
suitable for combinations.
Working group activities are coordinated by a steering 
group\footnote{The present members of the Heavy Flavour Steering Group are: 
D. Abbaneo, 
J. Alcaraz, E. Barberio, M. Battaglia, S. Blyth, 
D. Su, P. Gagnon, R. Hawkings, S. Mele, 
F. Palla, M. Paulini, P. Roudeau, O. Schneider,
A. Stocchi, Ch. Weiser, B. Wicklund and S. Willocq.}. 

Section \ref{sec:systgen} presents the values of the common input parameters 
that contribute to the systematic
uncertainties presented in this note.
%Some of these values have been taken 
%from averages obtained by the LEP ElectroWeak Working Group (LEPEWWG) 
%\cite{HFLEPEW,EWWG} or the 
%PDG \cite{PDG00}. 
Studies on the production rates of weakly decaying 
$b$-hadrons, on 
some characteristics of $\Dstarstar$
mesons
in semileptonic $b$-decays, and on the $\Lb$ polarization are 
%the results
%of the present work.
also presented.
As some measured quantities are needed
for the evaluation
of others, an iterative procedure
has been adopted to obtain stable results.

%Most of them can be considered as basic quantities in B physics.
%These are the semileptonic branching fractions of 
%$b$-hadrons, the production rates, some characteristics of $\Dstarstar$
%mesons\footnote{The notation $\Dstarstar$ includes all 
%charm-meson and non-resonant final states
%which are not D or $\Dstar$ mesons.}
%in semileptonic B decays, and the $\Lb$ polarization.
%Quantities not derived in this note are taken
%from averages obtained by the LEP ElectroWeak Working Group (LEPEWWG) \cite{HFLEPEW,EWWG} or the 
%PDG \cite{PDG00}.

Section \ref{sec:Atau} describes the averaging of the  $b$-hadron lifetime
 measurements.
The combined values
are compared with expectations from theory. The inclusive  
$b$-hadron and $\Bdb$ meson\footnote{Throughout the paper charge
 conjugate states are implicitly included unless stated otherwise.}
 lifetimes are needed in the 
determination of $\Vub$ and $\Vcb$ in order to convert
measured branching fractions into partial widths that can then be
compared with theory.

In Section \ref{sec:boscill}, oscillations of neutral B mesons are studied.
Also the production rates of the different $b$-hadrons in jets induced by a 
$b$-quark are determined using direct measurements 
together with the constraints provided by B mixing.
%and also constraints
%from the oscillation rate. 
The $\Bdb$ production rate is an important
input for the $\Vcb$ measurement using 
$\Bdb \rightarrow \Dstarp \ell^- \overline{\nu_{\ell}}$
decays, and the sensitivity
to $\Bs$-$\Bsb$ oscillations depends on the $\Bsb$ production rate.

In Section \ref{sec:deltag}, a limit on the decay width difference
between mass eigenstates of the  $\Bs$-$\Bsb$ system is given.
%As this average is obtained for the first time, more details
%on the adopted procedure have been given than for lifetimes
%or oscillations measurements.

The average number of $c$ and $\overline{c}$ quarks in $b$-hadron decays
is studied in Section \ref{sec:secccbar}. Combined measurements of
the inclusive semileptonic branching fraction and of charm production in
$b$-hadron decays have been compared with theoretical expectations.

%All measurements mentioned above correspond to world averages whereas for 
The determination of $\Vcb$ and $\Vub$ presented in Sections \ref{sec:vcb}
and \ref{sec:vub} includes only LEP results.
% have been combined so far.
%These averages are also released for the first time. 
The determination of $\Vub$ is based on a technique which uses
the lepton momentum and the mass of the hadronic system.
The accuracy of these results (especially that of $\Vcb$) depends mostly
on theoretical uncertainties. Appendix \ref{appendixC} gives details on the
theoretical inputs used.

Measurements of the time dependence of the $J/\psi {\rm K}^0_s$ 
decay channel for initially produced $\Bd$ and $\Bdb$ mesons
have been used, in Section \ref{sec:sintwobeta}, to evaluate the 
phase angle $\beta$ of CP violation. Other searches for CP violation 
involving $b$-quarks have been summarized also in this Section.

In Section \ref{sec:triangle}, the values of the CKM matrix parameters, 
$\rhobar$ and $\etabar$, obtained from the measurements of $b$-decay and 
oscillation
properties are given. These values are compared with those deduced
from measurements of 
CP violation in the K system. As the two approaches
give compatible results, a global average is obtained 
and parameters of the CKM unitarity triangle are determined.

A summary of all results obtained
by the different working groups is given  in Section \ref{sec:conclusion}.
In addition, Appendices \ref{appendixA}-\ref{appendixAe} contain,
respectively, the individual measurements of the production rates
of narrow $\Dstarstar$ states, $\Lb$ polarization, $b$-hadron
lifetimes, direct measurements of $b$-meson and $b$-baryon
production rates, and measurements
on $c$-hadron production in $b$ decays, that have been used in the
present averages.

\mysection{Common input parameters}
\label{sec:systgen}
The $b$-hadron properties used as common input parameters in these averages
%in the derivation 
%of the quantities studied in this paper,
% to obtain LEP or world averages,
are given in Table \ref{tab:gensys}. 
%The measurements 
Most of the quantities 
%that have not
%been the subject of dedicated studies
%from the working groups 
have been taken from results obtained
by the LEPEWWG \cite{HFLEPEW, EWWG} or quoted by the PDG \cite{PDG00}.
The others, which concern the production rates and decay properties
of $\Dstarstar$ mesons in $b$-hadron semileptonic decays and
the value of the $\Lb$ polarization, are explained later in this section.

\begin{table}[th!]
  \begin{center}
% Insert minipage to keep footnote in plot
    \begin{minipage}{\linewidth}
% Use footnote symbols instead of numbers and increment to get dagger.
    \renewcommand{\thempfootnote}{\fnsymbol{mpfootnote}}
    \addtocounter{mpfootnote}{1}
    \begin{tabular}{|l| l | l @{\,$\pm$\,}l | c |}
      \hline
      Quantity & Symbol &  \multicolumn{2}{c|}{Value}    
      & Reference \\
      \hline
      Fraction of $b$ events  & ${\rm R}_b$      & 0.21652  & 0.00069 
      & \cite{HFLEPEW} \\
      Fraction of $c$ events  & ${\rm R}_c$      & 0.1702  & 0.0034 
      & \cite{HFLEPEW} \\
      Beam energy fraction  & $<x_E>$            & 0.702  & 0.008 
      & \cite{EWWG} \\
     $b$-hadron sl. BR  & BR$(b \rightarrow \ell^- \overline{\nu_{\ell}}X)$ & 0.1056  & 0.0021       & \cite{HFLEPEW} \\
      Cascade $b$ sl. decay (r.s.) & BR$(b \rightarrow \overline{c}\rightarrow \ell^-\overline{\nu_{\ell}}X)$ & 0.0162  & $^{0.0044}_{0.0036}$       &  \cite{EWWG} \\
      Cascade $b$ sl. decay (w.s.) & BR$(b \rightarrow c \rightarrow \ell^+\nu_{\ell}X)$ & 0.0801  & 0.0026       &  \cite{HFLEPEW} \\
      $c$-hadron sl. BR  & BR$(c \rightarrow \ell^+\nu_{\ell}X)$ & 0.0984  & 0.0032       & \cite{HFLEPEW} \\
      $b$ quarks from gluons & P$(g \rightarrow b\overline{b})$ &  0.00254  & 0.00050       & \cite{HFLEPEW}  \\
       $c$ quarks from gluons & P$(g \rightarrow c\overline{c})$ & 0.0299  & 0.0039       &  \cite{HFLEPEW} \\
      $b$ decay charged mult.  & $n_{ch}^b$      & 4.955  & 0.062 
      & \cite{EWWG} \\
      $b$-hadron mixing & \chib\         & 0.1194   & 0.0043
      & \cite{HFLEPEW}      \\
      $\chid$ at the  \Yfs
                         & $\chid(\Yfs)$        & 0.182   & 0.015 
      & \cite{argchi,cleochi} \\
\hline
$\Dstarstar$ in sl. $b$ decays & ${\rm BR}(\Bdb \rightarrow \rm D^{\ast \ast +} \ell^- \overline{\nu_{\ell}})$ &  0.0304  & 0.0038 &  Sect. \ref{sec:dssa} \\ 
      $\Dstarp$ in $\Dstarstar$ sl. $b$ decays & BR(B$^- \rightarrow \Dstarp \pi^- \ell^-\overline{\nu_{\ell}}$)  & 0.0129  & 0.0016
      & Sect. \ref{sec:dssb}  \\
      $\Dstarp$ in $\tau$ sl. $b$ decays  & BR$(\Bdb \rightarrow \Dstarp
\tau^- \overline{\nu_{\tau}}X)$ & 0.0127  & 0.0021 
      &  Sect. \ref{sec:ajous1}  \\
      $\Dstarp$ in double charm & BR$(b \rightarrow \Dstarp
X_{\overline{c}}( \rightarrow \ell^-X))$ & 0.008  & 0.003 
      &  Sect. \ref{sec:ajous1}  \\
      $\Lb$ polarization  & ${\cal P}(\Lb)$ &-0.45   & $^{0.19}_{0.17}$
      &  Sect. \ref{sec:dssd}  \\
      \hline
      $\tau$ in sl. $b$ decays & BR$(b \rightarrow \tau^- \overline{\nu_{\tau}}X)$ & 0.026  & 0.004       &\cite{PDG00}\\
      J/$\psi$ in $b$ decays& BR$(b \rightarrow {\rm J}/\psi X $ & 0.0116  & 0.0010       &\cite{PDG00}\\
      $\Do$ branching fraction
   & BR($\Do \rightarrow {\rm K}^- \pi^+$) & 0.0383   & 0.0009 
      & \cite{PDG00} \\
      $\Dp$ branching fraction
   & BR($\Dp \rightarrow {\rm K}^- \pi^+ \pi^+$) & 0.090   & 0.006 
      & \cite{PDG00} \\
      $\Dsp$ branching fraction
   & BR($\Dsp \rightarrow \phi \pi^+$) & 0.036   & 0.009 
      & \cite{PDG00} \\
      $\Lc$ branching fraction
   & BR($\Lc \rightarrow {\rm p} {\rm K}^- \pi^+$) & 0.050   & 0.013 
      & \cite{PDG00} \\
      $\Dstarp$ branching fraction
   & BR($\Dstarp \rightarrow \Do \pi^+$) & 0.677   & 0.005
      & \cite{PDG00} \\
\hline
    \end{tabular}
%    $^\dag$ Average of measurements using inclusive vertexing techniques only.- \\
%    $^\ddag$ These parameters are used only in the \dmd\ fit.
  \end{minipage}
  \end{center}
    \caption{{\it Common set of input parameters used for the 
derivation of the various measurements presented in this paper.
The first set of results has been taken from those obtained
by the LEPEWWG \cite{HFLEPEW, EWWG} or quoted by the PDG \cite{PDG00},
the second set corresponds to averages obtained in the present report and,
for the sake of completeness, in the last set,
branching fractions in charm decays needed for
the production rates of these particles are listed as well.
The abbreviated notations sl., r.s. and w.s. correspond, respectively,
to semileptonic, right sign and wrong sign.}
    \label{tab:gensys}}
\end{table}

Note the following:
\begin{itemize}
\item {\it ${\rm R}_b,~{\rm R}_c$ and $<x_E>$}

These values apply only to $b$-hadrons produced in $\Zz$ decays.
${\rm R}_b$ and ${\rm R}_c$ are the respective branching fractions
of the $\Zz$ boson into $b \overline{b}$ and $c \overline{c}$ pairs
in hadronic events, $<x_E>$ is the mean fraction of the beam energy  
taken by a weakly decaying $b$-hadron.

\item {\it shape of the $b$-quark fragmentation function}

%{\bf To be updated .........}
%{\bf For this version I propose to continue with the value
%of $<XE>$ we were using already. I still hope to organize better discussions
%on this subject in the coming months so that we can have an update for
%the final version (corresponding to Summer 01).}

%The mean fraction of the beam energy taken by a weakly decaying $b$-hadron
%at LEP or SLC,

The value of $<x_E>$
%, the mean beam energy fraction of $b$-quarks,
is given
in Table \ref{tab:gensys}. To evaluate the corresponding systematic 
uncertainty, parameter(s) governing $b$-quark fragmentation functions
 have been varied in accordance with the uncertainty
quoted for $<x_E>$. Fragmentation functions taken from two
models \cite{ref:colsop,ref:kart} have been chosen to estimate
the systematic uncertainties coming from the shape of the function.
These models typically yield results on either side of those obtained
using the Peterson function \cite{ref:peters},
which is commonly used by the experiments.
%The present determination of the $b$-quark fragmentation function
%is based on rather old results and a better measurement of this
%distribution is needed using available data and
%more powerful algorithms.
For analyses which are rather insensitive to the fragmentation uncertainty,
it is considered adequate to use only the Peterson model and to inflate
the uncertainty on $<x_E>$ to $\pm 0.02$.

\item {\it the inclusive semileptonic branching fraction of $b$-hadrons}

The average LEP value for 
${\rm BR} (b \rightarrow \ell^- \overline{\nu_{\ell}} X) = (10.56 \pm 0.11({\rm stat.}) 
\pm 0.18({\rm syst.}))\% $ is taken from a LEPEWWG fit
which combines the heavy flavour measurements performed 
 at the $\Zz$ without including forward-backward
asymmetry measurements.
%but removing the
%forward-backward asymmetry measurements (see Section \ref{subs:vcb}). 
%Those measurements were all obtained 
%using a combination of lepton and lifetime tags.

 The largest contribution to the systematic error
% from the LEPEWWG fit
comes from the uncertainty on the semileptonic decay model. 
%This
%error is reduced from $\pm 0.084 \times 10^{-2}$ to 
%$\pm 0.065 \times 10^{-2}$
%when the fit is performed including the forward-backward asymmetry measurements.
%%in addition to all heavy flavour measurements performed at the $\Zz$.
%%Extensive studies have been done to understand the origin of this reduction.
%This happens because
%% from including various asymmetry measurements which
%%were obtained using different methods. 
%%The reason is that 
%only asymmetry measurements
%obtained using leptons depend on the semileptonic
%decay model, measurements using a lifetime tag
%combined with jet-charge or D-meson reconstruction do not.
%To achieve consistency between these measurements, the fit
%effectively constrains the size of the error attributed to
%the semileptonic decay model, thus reducing the 
%corresponding systematic error 
%%from this
%%source assigned to 
%on the semileptonic branching fraction. 
Including asymmetry measurements gives an
inclusive $b$-hadron semileptonic branching fraction  of
$(10.56 \pm 0.19)\%$ \cite{HFLEPEW}.
%The central
%values obtained from the fit are consistent with each other as mentioned in
%Section \ref{sec:systgen}. 
In the following the former value has been used.
%we have not
%included the asymmetry measurements in the fit to extract
%the semileptonic branching fraction used in the present averages.      

%{\bf I had always difficulties to buy this argument because everything seems 
%OK we go from 10.56+-0.21 to 10.57+-0.19 and we seem to make a big
%argument on the fact that we are conservative in using 0.21??? I would
%suppress the previous sentence.}

%The average value was obtained 
%using only the measurements of the semileptonic 
%branching fractions, 
%with the average $\rm B^0-\overline{B^0}$ mixing parameter 
%and  $ {\rm R}_b$ values
%%=\Gamma_{bb}/\Gamma_{had}$ 
%given in Table \ref{tab:gensys}.  
%The asymmetries values have been fixed to their standard model values. 
%The measurements of $BR (b \rightarrow X l \nu)$ used in this fit 
%{\it quote references}
%were all 
%obtained  using a combination of lepton and lifetime tags.
%If asymmetry measurements are included, a compatible value
%for the inclusive semileptonic branching fraction of $b$-hadrons of
%$(10.62 \pm 0.17)\%$ is obtained.

%In the following analyses,
%semileptonic widths of $b$-hadrons are assumed to be equal and, as a result,
%their respective semileptonic branching fractions are expected to
%be proportional to their measured lifetimes.

In the absence of direct measurements of the semileptonic branching
fractions for the different B meson states, it has been assumed, when 
needed in the following analyses, that all $b$-hadron semileptonic widths  
are equal. This hypothesis is strictly valid
for $\Bm$ and $\Bdb \rightarrow c \ell^- \overline{\nu_{\ell}}$ decays
because of isospin invariance originating from the $b \rightarrow c$ transition,
which is $\Delta {\rm I}=0$ (similar considerations apply also
to ${\rm D}^{0,+} \rightarrow s \ell^+ \nu_{\ell}$ decays). 
It is not valid in
$\Bm~{\rm and}~\Bdb \rightarrow u \ell^- \overline{\nu_{\ell}}$ decays, but
the induced difference between $\Bm$ and $\Bdb$ total semileptonic
decay rates can be neglected.
It has been assumed valid also
for $\Bsb$ mesons and $b$-baryons, but for $b$-baryons an uncertainty of 15$\%$ has been added,
estimated by comparing the lifetime ratios and semileptonic branching 
fraction 
ratios for $b$-mesons and $b$-baryons \cite{ref:pauline}.
%An uncertainty of
%%, respectively, 10$\%$ and 
%15$\%$ has been added 
%on the validity
%of this hypothesis for $b$-baryons as measured by comparing the 
%lifetime and semileptonic branching fractions ratios for $b$-mesons
%and $b$-baryons \cite{ref:pauline}.
Exclusive semileptonic branching fraction averages,
given in the following for the $\Bdb$ meson, have been obtained 
using this hypothesis.
% and quoted
%for the $\Bdb$ meson. 
Results for other $b$-hadron flavours
can be obtained using, in addition, the corresponding lifetime ratios.
The latter are obtained from $b$-hadron lifetimes given in 
Section \ref{sec:Atau}.
%have been used.

\item {\it gluon splitting to heavy quarks}

The quantities P$(g \rightarrow c\overline{c})$ and 
P$(g \rightarrow b\overline{b})$ are defined as the ratios
$\frac{{\rm BR}(\Zz \rightarrow q \overline{q} g,~g \rightarrow Q \overline{Q})}{{\rm BR}(\Zz \rightarrow {\rm hadrons})}$ in which, respectively, $Q$
is a $c$ or a $b$ quark.
% The value for P$(g \rightarrow c\overline{c})$
%corresponds to the recent result from OPAL \cite{ref:opalgcc} whereas
%the value for P$(g \rightarrow b\overline{b})$ is an average of all LEP 
%and SLD
%results \cite{ref:allgbb} obtained by assuming
%a common systematic uncertainty equal to $0.55~10^{-3}$, dominated
%by the modelling uncertainty.

%{\bf I did this average. Looking at the papers it is not clear what 
%systematics dominate even when different collaborations use similar procedures
%to measure gbb. The quoted systematic is conservative and based
%mainly on the new OPAL analysis.}

\item {\it $b$-hadron decay multiplicity}

The value given in Table \ref{tab:gensys} is an average of 
DELPHI \cite{multDELPHI} and OPAL
\cite{multOPAL} measurements which does not
include charged decay products from the long lived particles
$\Kos$ and $\Lambda^0$.
%of strange particles.
%, including hyperons.

\mysubsection{Inclusive $\Dstarstar$ production rate in $b$-hadron semileptonic decays}
\label{sec:dssa}
All results are quoted for the $\Bdb$ taken as reference but include
all available information. Corresponding values of the branching fractions
for the other $b$-hadron
states, ${\rm B}_i$, can be obtained by multiplying $\Bdb$ results
by the lifetime ratio $\tau({\rm B}_i) / \tau(\Bdb)$.

The inclusive $b$-hadron semileptonic branching fraction into 
$\Dstarstar$  mesons has been 
measured in three different ways:
\begin{itemize}
%\begin{enumerate}
\item[-]by subtracting the contributions of 
$\Bdb \rightarrow (\Dp + \Dstarp) \ell^- \overline{\nu_{\ell}}$ from the total
semileptonic branching fraction of $\Bdb$ mesons, yielding:
\begin{eqnarray}
{\rm BR}(\Bdb \rightarrow \rm D^{\ast \ast +} \ell^- \overline{\nu_{\ell}})& = &
(3.66 \pm 0.37 )\%
\label{eq:dsstarinc}
\end{eqnarray}
using the inclusive semileptonic branching 
fraction given in Table \ref{tab:gensys} and published values for the exclusive rates 
\cite{PDG00}.
\item[-]from a measurement at the $\Upsilon(4S)$ of the rate of final states
with a $\Dstarp$ and using models to account for the other channels
\cite{DssARGUS}:
\begin{eqnarray}
{\rm BR}(\Bdb \rightarrow \rm D^{\ast \ast +} \ell^- \overline{\nu_{\ell}}) &= &
(2.7 \pm 0.7 )\% 
\end{eqnarray}
\item[-]from an inclusive measurement of final states in which a D or a 
$\Dstarp$ meson is accompanied by a charged hadron and assuming
that non-strange $\Dstarstar$ decay channels involve only ${\rm D} \pi$ and
$\Dstar \pi$ final states:
\begin{eqnarray}
\nonumber {\rm BR}(\Bdb \rightarrow \rm D^{\ast \ast +} \ell^- \overline{\nu_{\ell}}) & = &(2.16 \pm 0.30 \pm 0.30) \%~ [15]\\
%\cite{DssALEPH}
     & =& (3.40 \pm 0.52 \pm 0.32) \%~  [16]
%\cite{DssDELPHI}
\end{eqnarray}
%\end{enumerate}
\end{itemize}
As the $\chi^2$ of these four measurements
is equal to 2.56 per degree of freedom,
and considering that systematic uncertainties may have been underestimated,
the uncertainty on the weighted
average has been multiplied by 1.6,
giving:
\begin{equation}
{\rm BR}(\Bdb \rightarrow \rm D^{\ast \ast +} \ell^- \overline{\nu_{\ell}}) ~=~ (3.04 \pm 0.38 ) \% 
\label{eq:ajous5}
\end{equation}

\mysubsection{$\Dstarstar$ decays to $\Dstar$ mesons in semileptonic $b$-decays}
\label{sec:dssb}

Semileptonic decays to excited charm states which subsequently decay 
to a $\Dstarp$\ are a source of correlated (physics) background in studies
of $\Bdb$ meson properties.
It is appropriate to express the different measurements in terms of the
parameter ${\rm b}^{\ast\ast}$, defined as the branching fraction 
of $\Bdb$
semileptonic decays involving $\Dstarstar$ final states in which a 
$\Dstar$, charged or neutral, is produced:
\begin{equation}
{\rm b}^{\ast\ast}={\rm BR}(\Bdb \rightarrow \rm D^{\ast \ast +} \ell^- \overline{\nu_{\ell}})
\times BR(D^{\ast \ast +} \rightarrow \Dstar X) 
\end{equation}
 This is to differentiate
such $\Dstarstar$ decays from those into a D meson directly.
%, not coming
%from a $\Dstar$, is emitted.
Throughout this section 
%and as there is, at present,
%no information on multi-pion final states, 
it is assumed that decays of 
$\Dstarstar$
mesons involve at most one pion (or one kaon for strange states).

The following measurements have been 
interpreted in terms of the quantity ${\rm b}^{\ast\ast}$ and the
production fractions ($f_{{\rm B}_i}$) and lifetimes ($\tau({\rm B}_i)$)
of the different types of weakly decaying $b$-hadrons:
%These measurements
%have then been combined to obtain the value for ${\rm b}^{\ast\ast}$:
%combined:
\begin{itemize}
%\begin{enumerate}
\item[-] semi-inclusive measurements of semileptonic decays in which
a $\Dstarp$ and a charged pion have been isolated:
\begin{eqnarray}
\nonumber {\rm BR}(b \rightarrow \Dstarp \pi^- {\rm X} \ell^- \overline{\nu_{\ell}}) & = &
(4.73 \pm 0.77 \pm 0.55)\times 10^{-3}~ [15]\\
%\cite{DssALEPH} \\
\nonumber     & =& (4.8 \pm 0.9 \pm 0.5) \times 10^{-3}~  [16]\\
%\cite{DssDELPHI}\\
     & =& \fu\frac{2}{3}{\rm b}^{\ast\ast}\frac{\tau(\Bm)}{\tau(\Bdb)}
\end{eqnarray}
\item[-] the ARGUS measurement \cite{DssARGUS}:
\begin{eqnarray}
\nonumber {\rm BR}(\Bdb \rightarrow \rm D^{\ast \ast +} \ell^- \overline{\nu_{\ell}}) &= &
(2.7 \pm 0.7 )\% \\
     & = &\frac{{\rm b}^{\ast\ast}}{0.77}
\end{eqnarray}
where the value of 0.77 corresponds to the modelling used for 
$\Dstarstar$ decays
in that analysis;
\item[-] the inclusive production rate of charged $\Dstar$ mesons
in semileptonic $b$ decays:
\begin{eqnarray}
\nonumber {\rm BR}(b \rightarrow \Dstarp {\rm X} \ell^- \overline{\nu_{\ell}}) & = &
(2.75 \pm 0.17 \pm 0.16)\%~  [16]\\
%\cite{DssDELPHI} \\
\nonumber     & =& (2.86 \pm 0.18 \pm 0.23) \%~  [17]\\
%\cite{DssOPAL}\\
& =& \fd {\rm b}^{\ast} + \fd \frac{1}{3}{\rm b}^{\ast\ast}+
\fu\frac{2}{3}{\rm b}^{\ast\ast}\frac{\tau(\Bm)}{\tau(\Bdb)}
\label{ajoute}\\
\nonumber & & +\fs\frac{\alpha}{2}{\rm b}^{\ast\ast}\frac{\tau(\Bsb)}{\tau(\Bdb)}
\end{eqnarray}
where the quantity ${\rm b}^{\ast}$ is the exclusive 
semileptonic branching fraction:
\begin{equation}
{\rm b}^{\ast}={\rm BR}(\Bdb \rightarrow \rm \Dstarp \ell^- \overline{\nu_{\ell}})  = (4.67 \pm 0.26) \% 
\end{equation}
which has been obtained by averaging results quoted by the PDG
\cite{PDG00} for corresponding decays of $\Bdb$ and $\Bm$ mesons, and
the scaling factor $\alpha=0.75\pm0.25$ has been
introduced in (\ref{ajoute}) to account for a possible SU(3) flavour violation
when comparing $\rm D^{**+}_s \rightarrow \Dstarp K^0$ and
$\rm D^{**0} \rightarrow \Dstarp \pi^-$ decays.
%\end{enumerate}
\end{itemize}
In the above, the
lifetimes ($\tau({\rm B}_i)$) and production fractions ($f_{{\rm B}_i}$) 
are taken from Sections \ref{sec:Atau} and \ref{sec:boscill} respectively.

These measurements form a coherent set of results yielding:
%and the average of
%${\rm b}^{\ast\ast}$
%${\rm b}^{\ast\ast}$, which is equal to the branching fraction of $\Bdb$
%semileptonic decays involving $\Dstarstar$ final states in which a 
%$\Dstar$ is produced, 
%amounts to:
%\footnote { This result is obtained assuming that $\Dstarstar$ mesons decay
%only into single pion (or kaon) final states.}:
\begin{equation}
{\rm b}^{\ast\ast}={\rm BR}(\Bdb \rightarrow \rm D^{\ast \ast +} \ell^- \overline{\nu_{\ell}})
\times BR(D^{\ast \ast +} \rightarrow \Dstar X)  = (1.82 \pm 0.21 \pm 0.08 )
 \% 
\label{eq:ajoute1}
\end{equation}

From this result, and using the same hypotheses, it is possible to derive 
other quantities of interest
in several analyses presented in this paper such as:
\begin{eqnarray}
 {\rm BR}(\rm B^- \rightarrow \rm D^{\ast \ast 0} (\rightarrow \Dstarp \pi^-) \ell^-\overline{\nu_{\ell}})~ = ~\frac{2}{3}b^{\ast\ast}\frac{\tau(\Bm)}{\tau(\Bdb)} ~=~
(1.29 \pm 0.16) \% \label{eq:ajoute2}\\
 {\rm BR}(\Bdb ~\rightarrow~ \Dstarp \pi^0 \ell^-\overline{\nu_{\ell}})~ = ~
\frac{1}{3}b^{\ast\ast}~=~
(0.61 \pm 0.08) \% \\
 {\rm BR}(\Bsb \rightarrow \Dstarp {\rm K}^0 \ell^-\overline{\nu_{\ell}})~ = ~
\frac{1}{2}b^{\ast\ast}\frac{\tau(\Bsb)}{\tau(\Bdb)}\alpha~=~
(0.65  \pm 0.23 ) \%. 
\end{eqnarray}

%$$\mathrm{B}^- \rightarrow \ell^- \overline{\nu_{\ell}} \Dstarstar^0 (\rightarrow \Dstarp \pi_{**}^-) 
%= 1.24 \pm 0.19 \pm 0.04$$
%Assuming isospin symmetry, charged current 
%conservation and SU(3) flavour symmetry, the corresponding values for $\Bdb$ and $\rm B_s$ mesons are obtained:
%In the last branching fraction a scaling factor $\alpha=0.75\pm0.25$ is 
%introduced to account for a possible SU(3) flavour violation,
%when comparing $\rm D^{**+}_s \rightarrow \Dstarp K^0$ and
%$\rm D^{**0} \rightarrow \Dstarp \pi^-$ decays. 
%There is no common systematic error between the two 
%measurements which are therefore
%averaged to provide the numbers reported in table \ref{tab:summary} and the 
%$\rm R^{**}$ used in the averaging procedure is $0.300+-0.036$.

It is also possible to determine the fraction of $\Bdb$ semileptonic 
decays which contain a $\Dstarp$:

\begin{equation}
\frac{{\rm BR}(\Bdb \rightarrow \Dstarp \ell^- \overline{\nu_{\ell}})+
{\rm BR}(\Bdb \rightarrow \Dstarp \pi \ell^- \overline{\nu_{\ell}})}
{{\rm BR}(\Bdb\rightarrow  \ell^- \overline{\nu_{\ell}} {\rm X})}
  = 0.50 \pm 0.03 
\label{eq:fracdstar}
\end{equation}

and the fraction of $b$-quark jets in which the $\Dstarp$ comes from a 
$\Dstarstar$ decay:
\begin{equation}
\frac{{\rm BR}(b \rightarrow \Dstarp \pi \ell^- \overline{\nu_{\ell}})}
{{\rm BR}(b\rightarrow \Dstarp \ell^- \overline{\nu_{\ell}})
+{\rm BR}(b \rightarrow \Dstarp \pi \ell^- \overline{\nu_{\ell}})}
  = 0.31 \pm 0.04
\label{eq:ajoute3}
\end{equation}

As these results (\ref{eq:ajoute1}) to (\ref{eq:ajoute3}) are highly correlated
they are represented in Table \ref{tab:gensys} by a single entry, 
chosen to be result
(\ref{eq:ajoute2}).

\mysubsection{Other semileptonic decays to $\Dstarp$ mesons}
\label{sec:ajous1}
Other mechanisms giving a $\Dstarp$ accompanied by a lepton
are important because they are further sources of background
for the exclusive channel $\Bdb \rightarrow \Dstarp \ell^- \overline{\nu_{\ell}}$.

\mysubsubsection{Charged $\Dstar$ production in semileptonic decays of 
$b$-hadrons involving $\tau$ leptons}

The value of  BR$(\Bdb \rightarrow \Dstarp \tau^- \overline{\nu_{\tau}} {\rm X})$ 
quoted in Table \ref{tab:gensys} is 
obtained from the average measured value of 
\mbox{BR($b \rightarrow \tau \overline{\nu_{\tau}} {\rm X}$)} taken from 
the PDG \cite{PDG00} and
%computed by the LEP Electroweak Working Group 
%\cite{EWWG}, 
assuming that a $\Dstarp$\ is produced in $(50 \pm 10) \%$ of the cases,
as measured in semileptonic decays involving light leptons. 
%(Equation (\ref{eq:fracdstar})). 
%This value is then multiplied by the PDG value of the 
%BR($\tau \rightarrow \ell^-\nu \nu $).\par
The uncertainty on this last number has been increased, as compared to 
the value
obtained in Equation (\ref{eq:fracdstar}), to account for a possible
different behaviour of decays when a $\tau$ is produced
instead of a light lepton because of differences in the masses
involved in the final state \cite{ref:ng}.

%{\bf I plan to have a look at a few predictions from models to see if,
%in case of tau semileptonic decays, one expects differences in the relative 
%production rates for D, D*, D** }

\mysubsubsection{Charged $\Dstar$ production in double charm semileptonic
$( b \rightarrow \overline{c} \rightarrow \ell)$ decays}

The double charm production rate in $b$-decays was measured by the ALEPH 
\cite{twocharma} and 
CLEO \cite{twocharmc} Collaborations, isolating the contributions of different
D meson species. From these measurements the inclusive semileptonic
branching fraction, involving a wrong sign charmed meson, has been obtained~\cite{EWWG}:
\begin{equation}
{\rm BR}(b \rightarrow \overline{c} \rightarrow {\ell})~=~
0.0162^{+0.0044}_{-0.0036}.
\end{equation}

%{\it Why not using Delphi result ?}

 The product  
\mbox{$\rm BR( b \rightarrow \Dstarp X_{\overline{c}} Y) \times 
BR( X_{\overline{c}} \rightarrow \ell^-X$)}
quoted in Table \ref{tab:gensys} 
is then determined assuming again that a $\Dstarp$\ is produced in 
$(50 \pm 10) \%$ of the cases.

\mysubsection {Production of narrow $\Dstarstar$ states in $b$-hadron
semileptonic decays}
\label{sec:dssc}

A modelling of $b$-hadron semileptonic decays requires detailed 
rate measurements of the different produced states which can be resonant 
or non-resonant 
${\rm D}^{(*)}n\pi$ systems\footnote{In the following the study has been 
limited to $n$=1.},
 each having characteristic 
decay properties. As an example, there are four orbitally excited states
with L=1. They can be grouped in two pairs according to the
value of the spin of the light system, $j={\rm L} \pm 1/2$ (L=1).

States with $j=3/2$ can have ${\rm J}^{\rm P}=1^+$ and $2^+$.
The $1^+$ state decays only through ${\rm D}^*\pi$, and the $2^+$
through ${\rm D}\pi$ or ${\rm D}^*\pi$.
% always within the hypothesis
%of decays involving a single pion. 
%Because of parity and angular momentum conservation, for the $2^+$, there is
%a D wave between the pion and the charmed particle.
%For the $1^+$, S and D waves are possible but, if one assumes that the heavy 
%quark spin is decoupled, the conservation of $j$(~=~3/2) implies that
%S waves are not possible. 
Parity and angular momentum conservation imply that in the $2^+$ the
${\rm D}^*$ and $\pi$ are in a D wave but allow both S and D waves
in the $1^+$. However, if the heavy quark spin is assumed to decouple, conservation of $j$(~=~3/2) forbids S waves even in the $1^+$.
An important D wave component, and the fact that the
masses of these states are not far from threshold, imply that the
$j=3/2$ states are narrow. These states have been observed with a typical
width of 20 $\MeV/c^2$, in accordance with the expectation.

On the contrary, $j=1/2$ states can have ${\rm J}^{\rm P}=0^+$ and $1^+$, so
they are expected to decay mainly through an S wave and to be broad resonances
with typical widths of several hundred $\MeV/c^2$. The first experimental
evidence for broad $1^+$ states has been obtained by CLEO \cite{ref:broaddss}.

At present, information on the composition 
%of this
%fraction 
of $b$-hadron semileptonic decays
involving $\Dstarstar$ is rather scarce.
% which corresponds to 30$\%$ of 
%the total rate (Section \ref{sec:dssa}).
ALEPH \cite{DssALEPH}, CLEO \cite{ref:cleodss} and DELPHI \cite{Dss2DELPHI}
have reported evidence for the 
production of narrow resonant states (${\rm D}_1$ and ${\rm D}_2^*$), 
which can be summarised as follows:
\begin{eqnarray}
{\rm BR}(\overline{{\rm B}} \rightarrow {\rm D}_1 \ell^-\overline{\nu_{\ell}} ) &=& (0.63 \pm 0.10) \% \label{eq:darate}\\
{\rm BR}(\overline{{\rm B}} \rightarrow {\rm D}_2^* \ell^-\overline{\nu_{\ell}} ) &=& (0.23 \pm 0.08)\%~{\rm or}~<0.4\% ~{\rm at~the~95\%~CL} 
\label{eq:dbrate} \\
{\rm R}^{**} = \frac{{\rm BR}(\overline{{\rm B}} \rightarrow {\rm D}^*_2 \ell^-\overline{\nu_{\ell}})}
{{\rm BR}(\overline{{\rm B}} \rightarrow {\rm D}_1 \ell^-\overline{\nu_{\ell}})} &=& 0.37 \pm 0.14 ~{\rm or}~<0.6 ~{\rm at~the~95\%~CL} \label{eq:Rss}
\end{eqnarray}
where ${\rm R}^{**}$ is the ratio between the production rates 
of ${\rm D}^*_2$ and
${\rm D}_1$ in $b$-meson semileptonic decays.
The original measurements are listed in Appendix \ref{appendixA}.
Absolute rates for the ${\rm D}_1$ and ${\rm D}_2^*$ mesons 
have been obtained
assuming that the former decays always into $\Dstar \pi$ only and the latter
decays into ${\rm D} \pi$ in $(71 \pm 7)\%$ of the cases, as measured
by ARGUS \cite{argusn} and CLEO \cite{cleon}.

It is of interest to note that these results are very different from
naive expectations from HQET
%. If the heavy quark mass goes to infinity
%it is expected that 
in which ${\rm R}^{**} \simeq 1.6$ \cite{Pene}.
This implies large $1/m_c$ corrections in these models, as was noted
for instance in \cite{ref:ligeti}. 
These corrections not only reduce the expected value
for ${\rm R}^{**}$, they also enhance the expected 
production rate of $j=1/2$ states which can become of similar importance
as the $j=3/2$ rate. 

Comparing Equations (\ref{eq:darate}-\ref{eq:dbrate}) and (\ref{eq:ajous5}),
it can be deduced that
narrow $\Dstarstar$ states account for less than one third
of the total $\Dstarstar$ rate.
%; the main fraction of these decays
%are thus through broad or non-resonant states.
If, in addition, model expectations for the production rates of
$j=1/2$ states are used, it can be concluded that non resonant
production of ${\rm D}^{(*)}\pi$ states is as important as, or
even larger than, the production of resonant states. 
As a consequence,
models like ISGW \cite{isgw,isgw2}, which contain only resonant charmed states,
do not provide a complete description of $b$-hadron semileptonic decays.

% which is in disagreement
%with theoretical expectations.
%{\it It seems according to the original ALEPH paper that R** is much lower 
%than 1. The number quoted above has to be checked .. What about of an 
%OPAL published result on this subject in which they quote evidence 
%for the 2+? }

%Non resonant states have not been identified at present. Inclusive 
%measurements of the decays $B\rightarrow D^{(*)} \ell^-\overline{\nu_{\ell}} X$ have 
%however been 
%performed by the ARGUS \cite{DssARGUS}, ALEPH 
%\cite{DssALEPH} and DELPHI \cite{DssDELPHI} Collaborations using different 
%techniques.
% \footnote{preliminary 
%results are obtained by DELPHI using both the methods}.\par
%ARGUS exploits the decay kinematics to determine the ratio
%\ba
%\nonumber \frac{N(\Dss \ell)}{N(\Ds \ell)} = 0.27 \pm 0.08 \pm 0.03
%\ea
%from which the product of branching ratios :
%\ba
%\nonumber BR(\overline{B}\rightarrow \Dss \ell^-\nu)\cdot BR(\Dss \rightarrow \Ds X) = (1.66 \pm 0.51 \pm 0.15)\%
%\ea
%can be obtained.
%ALEPH tags the additional charged pion ($\pi_{**}$) in the decay chain
%\mbox{$\mathrm{B}^- \rightarrow \ell^- \overline{\nu_{\ell}} \Dss^0 (\rightarrow \Ds \pi_{**}^-)$}
%and determines the BR for that process as:
%Averaging the  ALEPH and DELPHI measurements and using 
%$\fu$- the fraction of ${\rm B}^+$ mesons produced
%in a $b$ jet-, from Section \ref{sec:results}, we get:

\mysubsection{$\Lb$ polarization}
\label{sec:dssd}

%The $b$-quarks, originating from $\Zz$ decays during collisions of unpolarized
%$e^{\pm}$ beams, possess, according to the Standard Model,
%a high longitudinal polarization: ${\cal P}_b=-0.94$.
Even for unpolarized ${\rm e}^{\pm}$ beams, the $b$-quarks produced
in $\Zz$ decays are highly polarized: ${\cal P}_b=-0.94$.
Hard gluon emission and quark mass effects 
are expected to change ${\cal P}_b$ by only $3\%$
\cite{ref:popol}. However, during the hadronization process of the 
heavy quark, part or all of the initial $b$-quark polarization may be lost 
by the final weakly decaying $b$-hadron state. The $b$-mesons always 
decay finally to spin zero  pseudoscalar states, which do not retain
any polarization information. In contrast, $\Lb$ baryons
are expected to carry the initial $b$-quark polarization since the
light quarks are arranged, according to the constituent quark model, 
in a spin-0 and isospin-0 singlet. 
However, $b$-quark
fragmentation
into intermediate $\Sigma_b^{(*)}$ states can lead to a 
depolarization of the heavy quark
\cite{ref:popolb}.

The $\Lambda \ell$ final state is used to select event samples enriched
with $\Lb$ semileptonic decays, and the polarization is measured using
the average values of the lepton and neutrino energies. Samples
of events enriched in $b$-mesons, which carry no polarization information,
are used to calibrate the measurements.

Measurements from ALEPH \cite{ref:alpol}, DELPHI \cite{ref:delpol} and
OPAL \cite{ref:oppol}, collected in Appendix \ref{appendixAb},
 have been averaged, assuming
systematic uncertainties to be correlated, yielding:
\begin{equation}
{\cal P}(\Lb)~=~-0.45^{+0.17}_{-0.15}\pm 0.08
\end{equation}
where the uncertainty is still dominated by the statistics.
The following sources of systematic uncertainties, common to all
measurements, have been considered:
\begin{itemize}
\item $b$-quark fragmentation,
\item $\Lambda$ production in semileptonic and inclusive decays of the $\Lc$,
\item $\Lc$ polarization,
\item theoretical uncertainties in the modelling of polarized $\Lb$
semileptonic decays.
\end{itemize}

%\item parameters taken from PDG \cite{PDG00}.
%Quantities which are not better determined than values published
\end{itemize}

\mysection{Averages of $b$-hadron lifetimes}
\label{sec:Atau}

%{\bf Some changes have been made:
%\begin{itemize}
%%\item the b-quark fragmentation function: use a more up-to-date value for 
%%$<X_E>$ and evaluate the influence of a different shape, closer to data.
%%Evaluate the possible difference of the fragmentation function for 
%%$b$-baryon production,
%\item determine the average using exclusive lifetime results and production
%rates of the different $b$-hadrons (to be checked by the lifetime group)
%\item more details on systematics for $\Bs$ and $b-baryon$ lifetimes
%(still missing)
%\item include $B \rightarrow J/\psi X$ decays in inclusive lifetime meast.
%\item expand the comparison with theoretical expectations.
%\end{itemize}
%}

Best estimates for the various $b$-hadron lifetimes, for the ratio of the 
$\Bp$ and $\Bd$ lifetimes, and for the average lifetime of a sample
of $b$-hadrons produced in $b$-jets have been obtained by the B lifetime
working group\footnote{The present members of the B lifetime working group are:
J. Alcaraz, L. Di Ciaccio, T. Hessing, I.J. Kroll, H.G. Moser and
C. Shepherd-Themistocleous.}.
Details on the procedure used to combine the different measurements
can be found in  
\cite{ref:lifetimenote}.

%New measurements (n) available since Summer 98 and still preliminary
%values (p) have been indicated in the different Tables collected in Appendix 
%\ref{appendixB},
%which contain all measurements used in present averages.

The measurements used in the averages presented here are  
listed in Appendix \ref{appendixB}.
Possible biases can originate from the
averaging procedure. They depend on the statistics and time resolution
of each measurement and in the way systematic uncertainties have been
included. These effects have been studied in detail using simulations
and have been measured to be at a level 
which can be neglected as compared to the statistical accuracy.
%of 1$\%$ at most and thus
%can be neglected in most of present results.

\mysubsection{Dominant sources of systematic uncertainties}
 The dominant sources of systematic uncertainties leading to correlations
between measurements are briefly reviewed below (more details can be 
found in \cite{ref:lifetimenote}). 
%They include the estimation of
These are the background estimation, the evaluation of the $b$-hadron momentum, 
and the decay length reconstruction.
Depending on the level of correlation of the various components, the obtained 
systematic uncertainty on the average can be smaller than the total
systematic uncertainty affecting individual measurements.

 The background can be due either to physics processes leading
to a final state similar to that used to tag the signal, or to accidental 
combinations of tracks which simulate the decay of interest.
When ``physics'' background is present, the experimental uncertainties on the
branching fractions of the background processes and on the lifetimes of the
background particles lead to a systematic error which is correlated between
different experiments. When the background is combinatorial, the amount 
and/or the lifetime of the background ``particles'' is normally extracted 
from the data using the sidebands of mass distributions
 or wrong sign combinations. 
In this case the related systematic uncertainty is usually not correlated 
between experiments. But, in the measurement of the $b$-baryon lifetime
using $\Lambda \ell$ correlations, the amount of accidental background is 
obtained from the wrong sign combinations and a correction, common to all 
analyses, has to be applied
to this number to take into account the production asymmetry of accidental
$\Lambda \ell$ pairs.

\begin{figure}[tb!]
\begin{center}
\begin{tabular}{cc}
\mbox{\epsfxsize8.0cm \epsfysize10.0cm\epsffile{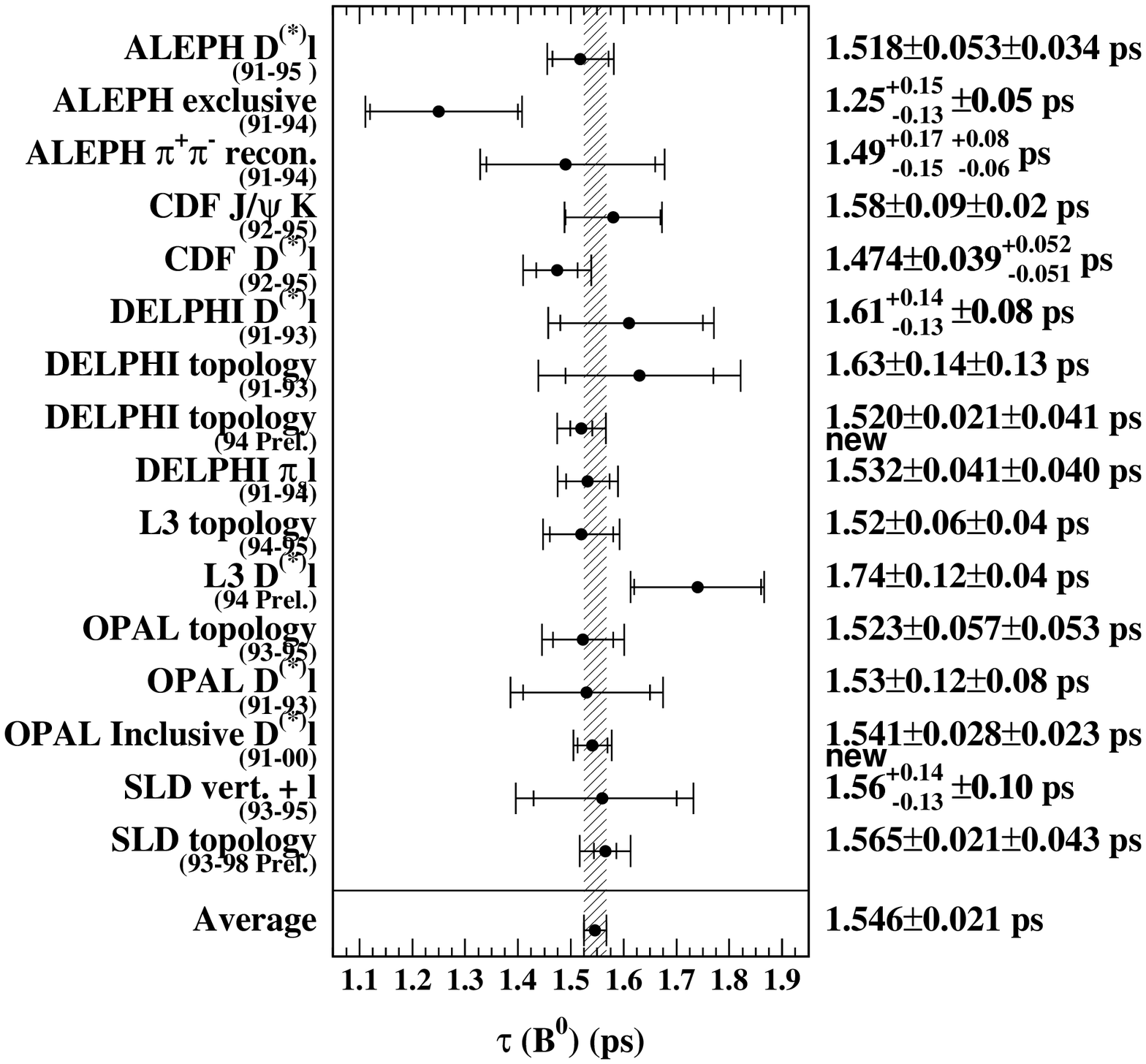}} &
\mbox{\epsfxsize8.0cm \epsfysize10.0cm\epsffile{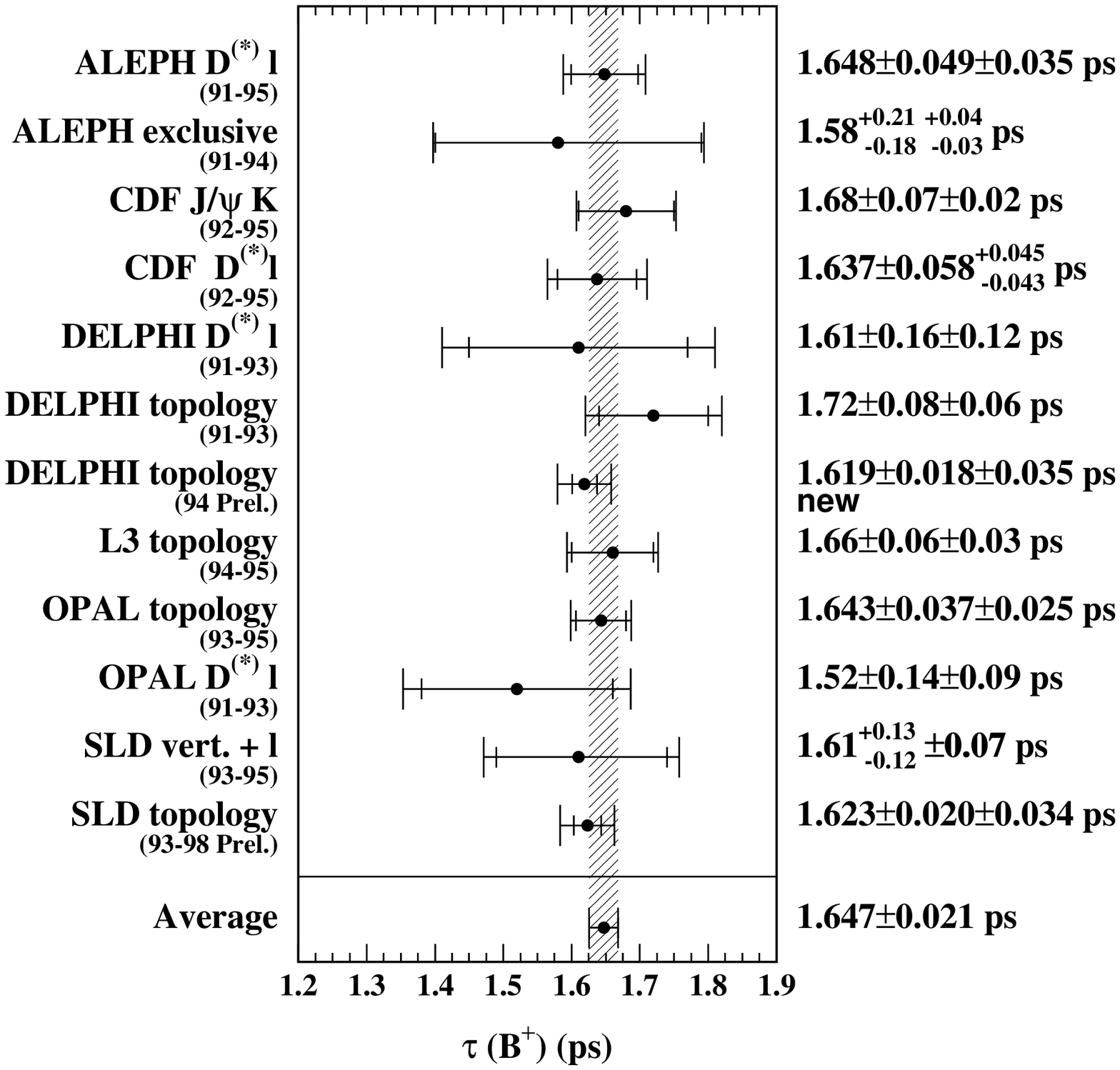}} \\
%\mbox{\epsfxsize8.0cm \epsfysize10.0cm\epsffile{/afs/cern.ch/user/r/roudeau/www/avb0d_bw.eps}} &
%\mbox{\epsfxsize8.0cm \epsfysize10.0cm\epsffile{/afs/cern.ch/user/r/roudeau/www/avbplus_bw.eps}} \\
\end{tabular}
\caption{{\it Left: $\Bd$ lifetime measurements.
 Right: $\Bp$ lifetime measurements.
Hatched areas correspond to present averages of the measurements given in the 
corresponding figures. Internal error bars correspond to statistical
uncertainties and full error bars include systematics.}
%{\bf Prepare b/w plots using same presentation as Summer 99}
\label{fig:tbdbp}}
\end{center}

\end{figure}

Most of the exclusive $b$-lifetime measurements are based on the reconstruction
of $b$-decay length and momentum. In most analyses the $b$-particles are 
only partially reconstructed, and their energies are estimated from 
the energies of the detected decay products. 
%No matter what estimator is used,
%however, 
In all cases, systematic uncertainties have been evaluated 
for the following effects:
\begin{itemize}
\item determination of the $b$-quark fragmentation function, 
\item branching fractions of $b$- and $c$-hadrons,
\item $b$-hadron masses,
\item $b$-baryon polarization and
\item modelling of neutral hadronic energy.
\end{itemize}
Finally there are uncertainties correlated within an experiment.
They are due to primary and secondary vertex
 reconstruction procedures, detector resolution, tracking errors, 
B flight direction reconstruction and detector alignment.

\begin{figure}[tb!]
\begin{center}
\begin{tabular}{cc}
\mbox{\epsfxsize8.0cm \epsfysize10.0cm\epsffile{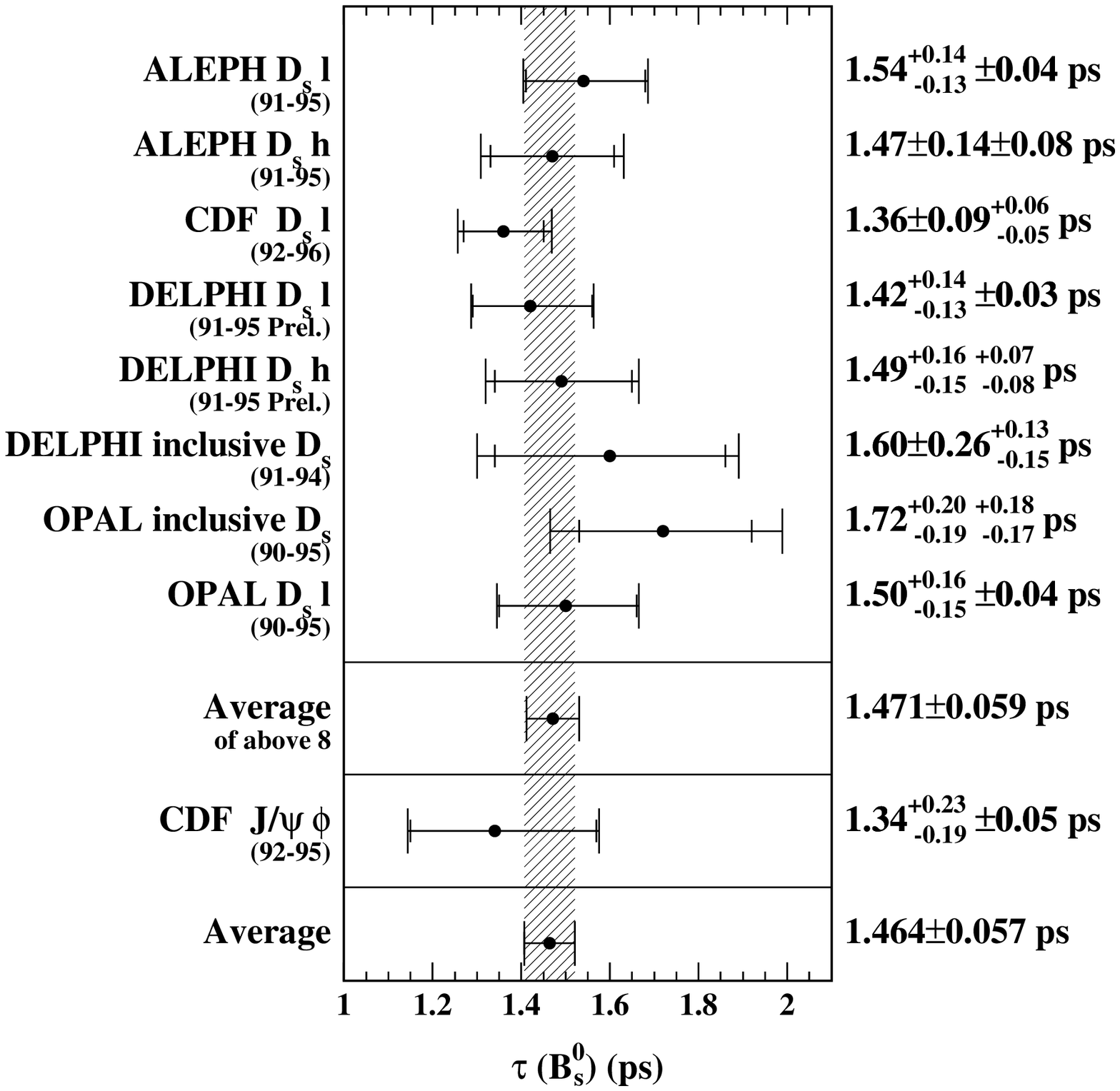}} &
\mbox{\epsfxsize8.0cm \epsfysize10.0cm\epsffile{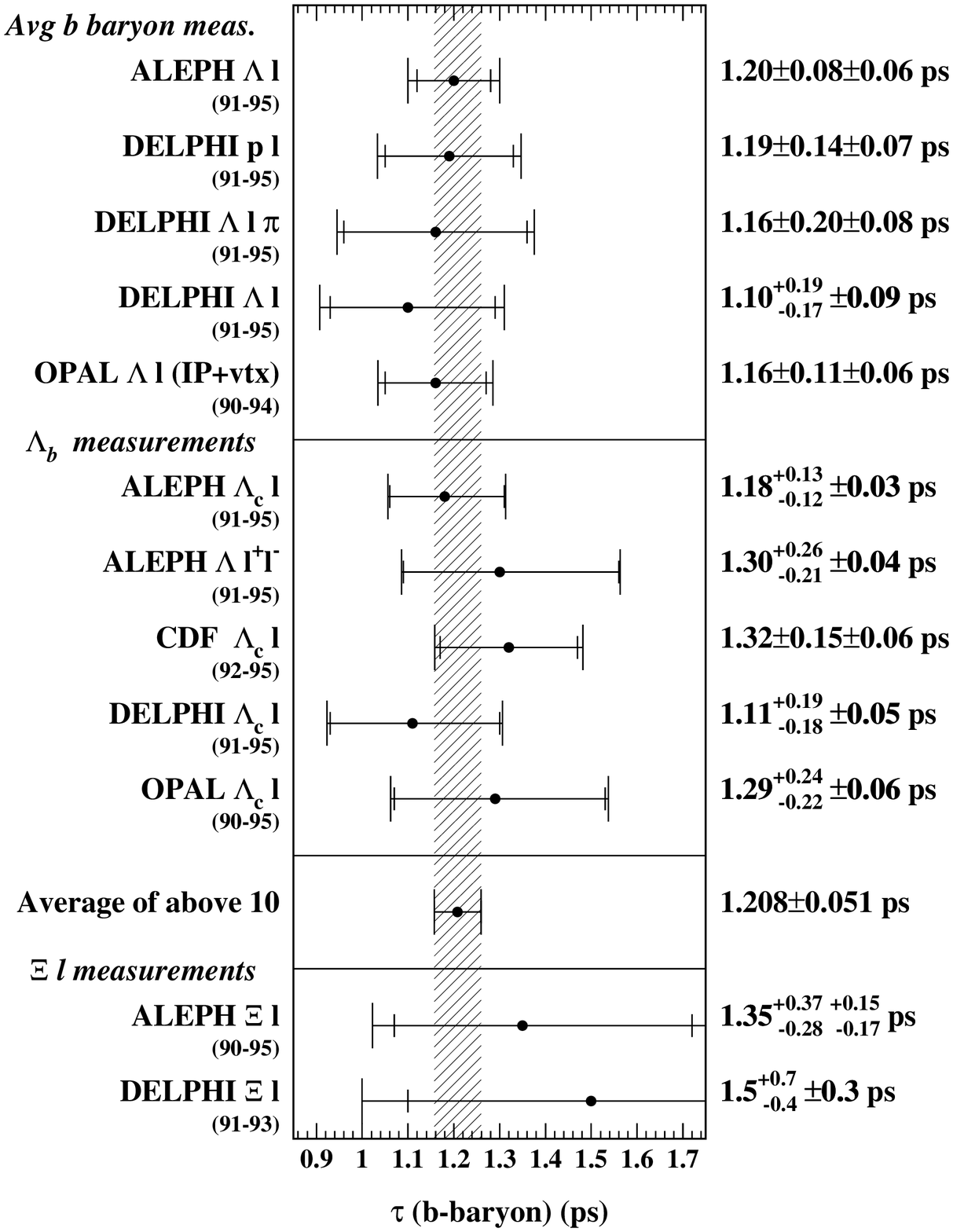}} \\
%\mbox{\epsfxsize8.0cm \epsfysize10.0cm\epsffile{/afs/cern.ch/user/c/claires/www/plots/summer99_paper/bw/avb0s.eps}} &
%\mbox{\epsfxsize8.0cm \epsfysize10.0cm\epsffile{/afs/cern.ch/user/c/claires/www/plots/summer99_paper/bw/bbarlife.eps}} \\
\end{tabular}
\caption{{\it Left: $\Bs$ lifetime measurements, 
Right: $b$-baryon lifetime measurements.
Hatched areas correspond to present averages of the measurements given in the 
corresponding figures. Internal error bars correspond to statistical
uncertainties and full error bars include 
systematics.} \label{fig:tbsbb}}
\end{center}

\end{figure}

\mysubsection{Measurements of $\Bd$ and $\Bp$ lifetimes}
\label{sec:taubd}
Apart from the measurements by CDF using large samples of exclusive 
$\Bd \rightarrow {\rm J}/\psi K^{(*)0}$ and $\Bp \rightarrow {\rm J}/\psi K^+$
decays, the most precise measurements of $\Bd$ and $\Bp$ lifetimes
originate from two classes of partially reconstructed decays.
In the first class the decay ${\rm B} \rightarrow \overline{{\rm D}^{(*)}} \ell^+ \nu_{\ell} {\rm X}$
is used in which the charge of the charmed meson distinguishes between
neutral and charged $b$-mesons. In the second class the charge attached to
the $b$-decay vertex is used to achieve this separation.

The following sources of correlated systematic uncertainties have been 
considered:
%\begin{itemize}
background composition (includes $\Dstarstar$ branching fraction uncertainties
as obtained in Section \ref{sec:systgen}),
momentum estimation,
lifetimes of $\Bs$ and $b$-baryons (as obtained in Sections \ref{sec:taubs}
and \ref{sec:taulb}),
and fractions of $\Bs$ and $b$-baryons
 produced in $\Zz$ decays (as measured in Section \ref{sec:results}).
%\end{itemize}

The world average lifetimes of $\Bd$ and $\Bp$ mesons are:
\begin{eqnarray}
 \tau(\Bd)& = & (\taubd)~{\rm ps} \\
 \tau(\Bp)& = & (\taubp)~{\rm ps}  .
\end{eqnarray}
The various measurements 
\cite{ALEB01,ALEB0,CDFB02,CDFB01,DELB01,DELB02,DELB03,DELB031,
L3B01,L3B02,OPAB0,OPAB1,OPAB11,SLDB01,SLDB02}
and the average values are shown
in Figure \ref{fig:tbdbp}. The respective values for the $\chi^2$/NDF
ratio of the fits are 9.1/16 and 2.2/12.

\mysubsection{$\Bs$ lifetime measurements}
\label{sec:taubs}

The most precise measurements of the $\Bs$ lifetime
originate from partially reconstructed decays in which
a ${\rm D}^-_s$ meson has been completely reconstructed
(see Figure \ref{fig:tbsbb}-left).

The following sources of correlated systematic uncertainties have been 
considered:
average $b$-hadron lifetime used for backgrounds,
$\Bs$ decay multiplicity,
and branching fractions of beauty and charmed hadrons.
%(i.e. BR($B \rightarrow D_s D$))

%{\it These systematices need to be explained and the text has to be expanded...}
 
The world average lifetime of $\Bs$ mesons is equal to:
\begin{eqnarray}
 \tau(\Bs)& = & (\taubs)~{\rm ps} 
\end{eqnarray}

This result has been obtained neglecting a possible difference, $\dgs$,
between the
decay widths of the two mass eigenstates of the $\Bs-\Bsb$ system. 
The measurement of $\dgs$ is explained in Section \ref{sec:deltag}. 
The various measurements 
\cite{CDFB01,ALEBS1,ALEBS2,CDFBS,DELBS0,DELBS1,DELBS2,OPABS1,OPABS2}
and the average value are shown
in Figure \ref{fig:tbsbb}-left.

\mysubsection{$b$-baryon lifetime measurements}
\label{sec:taulb}

The most precise measurements of the $b$-baryon lifetime
originate from two classes of partially reconstructed decays
(see Figure \ref{fig:tbsbb}-right).
In the first class, decays with an exclusively reconstructed $\Lc$ baryon
and a lepton of opposite charge are used.
In the second class, more inclusive final states with a baryon
(${\rm p},~\overline{{\rm p}},~\Lambda~{\rm or}~\overline{\Lambda}$) and a 
lepton have been used.

The following sources of correlated systematic uncertainties have been 
considered:
%\begin{itemize}
experimental time resolution within a given experiment, $b$-quark
fragmentation distribution into weakly decaying $b$-baryons,
$\Lb$ polarization,
decay model,
and evaluation of the $b$-baryon purity in the selected event samples.
%and Isgur Wise.
%{\it What means Isgur-Wise, expand the explanation of systematics}
%\end{itemize}
As the measured $b$-hadron lifetime is proportional to the assumed 
$b$-hadron mass,
%In computing the averages, 
%the errors due to the $\Lambda_b$ mass are scaled to 
%$\pm$ 100 MeV, except for $\Lambda_c \ell$, where $\pm$ 50 MeV is used.
the central values of the masses are scaled to m($\Lambda_b$) = ($5624 \pm 9$) 
$\MeV/c^2$ and
m($b$-baryon) = ($5670 \pm 100$) $\MeV/c^2$, before computing the averages.
%The meaning of uncertainties related to the decay model and 
%the correlations are not always clear.
%Mostly they 
Uncertainties related to the decay model are dominated by 
assumptions on the fraction of $n$-body decays.
 To be conservative it is 
assumed
that they are correlated whenever given as an error.
%DELPHI varies 4-body decays from 0 to 30$\%$ and, in computing the average,
% the DELPHI
%result is corrected using $(20 \pm 20)\%$.
Furthermore, in computing the average, results have been corrected 
for the effect of the measured value of the $\Lb$ polarization 
 and it has been assumed
that $b$-baryons have the same fragmentation distribution as all
$b$-hadrons (Section \ref{sec:systgen}).

%{\bf What does this mean exactly? same value for the parameter epsilon
%in the Peterson function or same <xE> for mesons and baryons ?}
  
%of -0.23$^{+0.25}_{-0.21}$ and a $\Lambda_b$ fragmentation parameter X$_E =0.70\pm 0.03$.
    
The world average lifetime of $b$-baryons is then:
\begin{eqnarray}
 \tau(b-{\rm baryon})& = & (\taubbar)~{\rm ps}
\end{eqnarray}
This value and the single measurements
\cite{ALELAM,DELLAM0,DELLAM1,OPALAM1,CDFLAM,OPALAM2}
  are given
in Figure \ref{fig:tbsbb}-right.
Keeping only $\Lambda^{\pm}_c \ell^{\mp}$ 
and $\Lambda \ell^- \ell^+$ final states, as representative of 
the $\Lb$ baryon, the following lifetime is obtained:
\begin{eqnarray}
 \tau(\Lb)& = & (\taulb)~{\rm ps}
\end{eqnarray}

Averaging the measurements based on the $\Xi^{\mp} \ell^{\mp}$
final states gives
\cite{ALELAM1,DELLAM2}
 a lifetime value for a sample of events 
containing $\Xi_b^0$ and $\Xi_b^-$ baryons:
\begin{eqnarray}
 \tau(\Xi_b)& = & (\tauxib)~{\rm ps}
\end{eqnarray}

\begin{figure}[tb!]
\begin{center}
\begin{tabular}{cc}
\mbox{\epsfxsize8.0cm \epsfysize10.0cm\epsffile{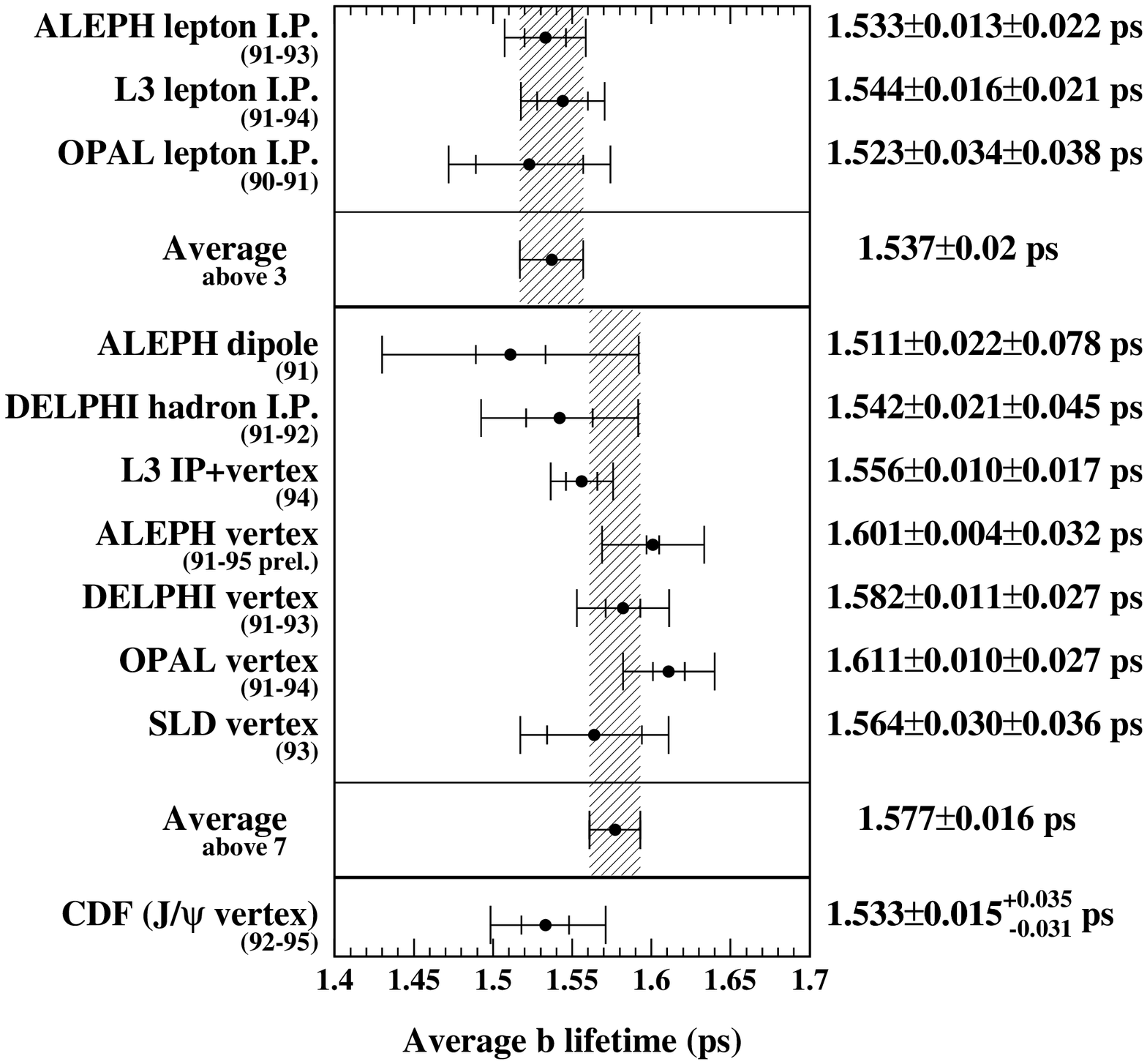}} &
\mbox{\epsfxsize8.0cm \epsfysize10.0cm\epsffile{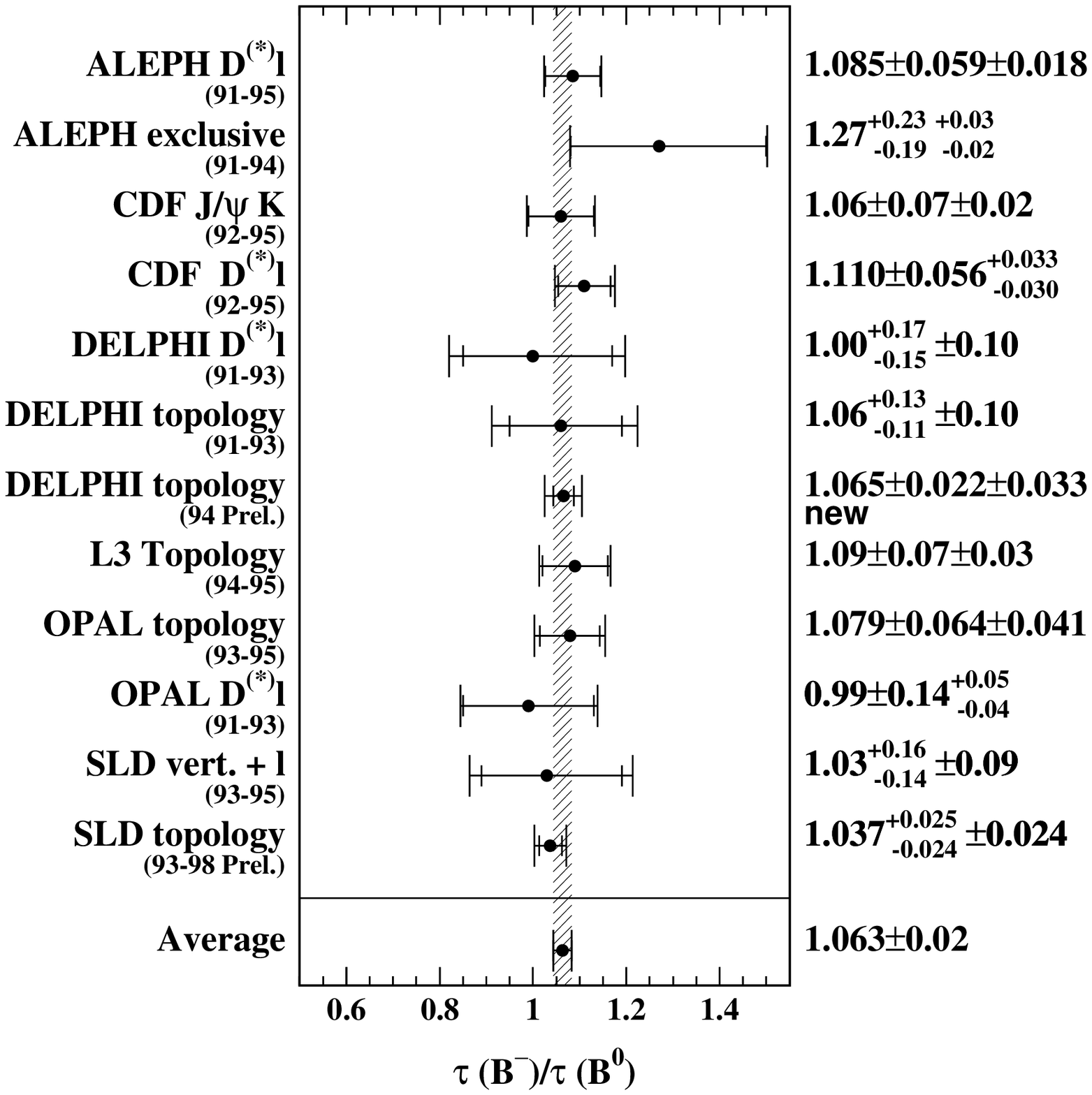}} \\
%\mbox{\epsfxsize8.0cm \epsfysize10.0cm\epsffile{/afs/cern.ch/user/c/claires/www/plots/summer99_paper/bw/avbhad2.eps}} &
%\mbox{\epsfxsize8.0cm \epsfysize10.0cm\epsffile{/afs/cern.ch/user/r/roudeau/www/avbratio_bw.eps}} \\
\end{tabular}
\caption{{\it Left: Measurements of the average $b$-hadron lifetime.
 Right: Ratio between the $\Bm$ and $\Bdb$ lifetimes.
Hatched areas correspond to present averages of the measurements given in the 
corresponding figures. Internal error bars correspond to statistical
uncertainties and full error bars include 
systematics.}  
%{\bf Prepare figures in b/w using same conventions as 
%Summer99.}
\label{fig:tbhad}}
\end{center}
\end{figure}

\mysubsection{Average $b$-hadron lifetime}
\label{sec:tauav}

%The average $b$-hadron lifetime is the average of the lifetimes of the 
%different
%types of $b$-hadrons weighted by their respective production rates:
Available results have been divided into three different sets and a 
separate average has been computed for each set
(see Figure \ref{fig:tbhad}-left):
\begin{itemize}
\item [(1)]measurements at LEP based on the identification of a lepton
      from a $b$-hadron decay:
\begin{equation}
\tau_b^{sl}~=~(1.537 \pm 0.020)~{\rm ps}
\end{equation}

\item [(2)]measurements at LEP and SLD which accept any $b$-hadron decay:
\begin{equation}
\tau_b^{incl.}~=~(1.577 \pm 0.016)~{\rm ps}
\end{equation}

%   2. the average of exclusive $b$-hadron lifetimes, as obtained 
%previously, weighted by the measured production rates of the different
%types of $b$-hadrons, given in Section \ref{sec:results}:
%%Using measured production rates (Section \ref{sec:results}) and lifetimes of the
%% different types
%%of $b$-hadrons the following average lifetime value is obtained:
%%which is compatible with the value of $(\tau_b^{incl.})$ and rather
%%accurate also.

\item [(3)]measurement at CDF based on the identification of a J/$\psi$ 
from a $b$-hadron decay: 
\begin{equation}
\tau_b^{{\rm J}/\psi}~=~(1.533^{+0.038}_{-0.034})~{\rm ps}
\end{equation}
\end{itemize}
%The three sets of measurements together with the corresponding averages
%are shown in Figure \ref{fig:tbhad}-left.

The reason for this division is that, since the lifetimes of the individual 
$b$-hadron species are different,
% (and it would appear the the b baryons are
%lower than b mesons)
a meaningful average lifetime 
%makes sense
can only be computed for samples in which the composition is the
same.

The following sources of correlated systematic uncertainties have been 
considered when evaluating the averages:
%\begin{itemize}
$b$- and $c$-quark fragmentation and decay models,
BR($b \rightarrow \ell$),
BR($b \rightarrow c \rightarrow \overline{\ell}$),
$\tau_c$,
BR($c \rightarrow \overline{\ell}$)
and the charged track multiplicity in $b$-hadron decays
(see Table \ref{tab:gensys}).
%\end{itemize}

The measurement in Set 2 is the average 
$b$-hadron lifetime for a sample of weakly decaying $b$-hadrons 
produced in $\Zz$ decays.
%\begin{equation}
%\tau_b^{excl.}~=~(1.544 \pm 0.019)~ps
%\end{equation}
\begin{equation}
\tau_b~=~\fd\tau(\Bd)+\fu\tau(\Bp)+\fs\tau(\Bs)+\fb\tau({\it b}-{\rm baryon})
\label{eq:tbexcl}
\end{equation}

%of the same composition as
%the production fractions at the Z0.  
%as obtained in a $b$-quark jet at high energy. 
%There is no reason
%to expect a sample composition different at LEP and at high energy 
%hadron colliders once jets are selected at high transverse momentum.

The measurements of Set 1 are based on samples 
%of
%hadrons depleted in $b$-baryons and 
containing more $\Bp$ and less $b$-baryons than in Set 2,
%with respect to other production
%fractions, 
and thus 
%bias the lifetime higher by about 0.01ps.
are expected to correspond to a higher average lifetime.
In practice this difference should be small.
%order\footnote{When considering that the relative differences 
%between $b$-hadron lifetimes 
%correspond to first order effects.}
% effect.
Assuming that all $b$-hadrons have the
same semileptonic partial width,
% and attributing an uncertainty of 
%15$\%$ on this hypothesis for $b$-baryons 
%(see Section \ref{sec:systgen}),
%Using these individual measurements,
the average lifetimes in the inclusive and semileptonic samples
can be related using the measured values of the production rates 
(Section \ref{sec:boscill})
and
lifetimes of the different types of $b$-hadrons, giving:
\begin{equation}
\frac{\tau_b^{sl}}{\tau_b}
~=~\frac{\sum_i{f_{{\rm B}_i} \tau ({\rm B}_i)^2}}
{\sum_i{f_{{\rm B}_i} \tau({\rm B}_i)}}
~=~1.006 \pm 0.003
\label{eq:taurapp}
\end{equation}
where the dominant uncertainty comes from the possible difference of
15$\%$ assumed between the semileptonic width of $b$-baryons as compared
to other $b$-hadrons.
%It can be also expected that the lifetimes obtained using the semileptonic
%and the ${\rm J}/\psi$ samples agree within the same level of uncertainties.
%It can be noticed that $\tau_b^{sl}$ is expected to be 
%slightly larger than $\tau_b$.
%at variance with present results.
%Finally the CDF ${\rm J}/\psi$ measurement probably corresponds
%to a different composition in $b$-hadron species 
%compared to any of
%the previous sets of measurements.
%
%As a consequence, the average $b$-lifetime, $\tau_b$, is obtained by combining
%the values of $\tau_b^{incl.}$
%%, $\tau_b^{excl.}$
%and $\tau_b^{sl}$ only, and neglecting the small correction expected in
%%after having corrected the latter using the result of 
%Equation (\ref{eq:taurapp}):
%Including this correction, 
%the difference between the values of $\tau_b^{incl.}$ and $\tau_b^{sl}$
%amounts to two standard deviations.

Combining the ten measurements corresponding to
$\tau_b^{incl.}$ and $\tau_b^{sl}$, with the correction of 
Equation (\ref{eq:taurapp}) included, the average $b$-hadron lifetime is:
\begin{equation}
\tau_b~=~(\taubav)~{\rm ps}.
\label{eq:taub}
\end{equation}
%$\tau_b = (1.560 \pm 0.014)~{\rm ps}$.
The global
$\chi^2$/NDF is equal to 10.1/10. It can be noted 
that there is a
2.1$\sigma$ difference between the partial averages obtained using semileptonic
decays or more inclusive final states which may indicate an underestimate of 
systematic uncertainties.
%the final quoted uncertainty has been increased 
%correspondingly giving:
%\begin{equation}
%\tau_b~=~(\taubav)~{\rm ps}.
%\label{eq:taub}
%\end{equation}
Measurements using J/$\psi$X final states, which may correspond to a different
composition in $b$-hadron species, as compared to any of the previous sets of
measurements, have not been included in the average.

%{\bf Below is a proposal which differs from 99 version .}
%
%As they involve similar topologies, processes corresponding to J/$\psi$
%production, have been combined in a similar way as semileptonic $b$-decays,
%with the value obtained using inclusive hadronic $b$-decays to evaluate
%the inclusive $b$-hadron lifetime, giving: 
%\begin{equation}
%\tau_b~=~(\taubav)~{\rm ps}.
%\label{eq:taub}
%\end{equation}

%{\bf The quoted uncertainty has been inflated so that the 
%$\chi^2$/ndf is unity. }
 
%The averaged value of $\tau_b^{incl.}$, $\tau_b^{sl}$ and $\tau_b^{{\rm J}/\psi}$,
%after having corrected these last two results by the ratio obtained
%in equation (\ref{eq:taurapp}) gives a compatible result: 
%($1.552 \pm 0.013)~{\rm ps}$ in which the quoted uncertainty has been inflated
%to account for the poor $\chi^2$ value of this fit.
%
%{\it These last 4 values are probably incorrect being obtained neglecting 
%common
%systematics and done naively ........}
%and, as the overall $\chi^2$
%of all ten corresponding measurements is equal to 8.2, 
%there is no reason
%to separate the two sets of results.

%This value of the inclusive $b$-lifetime can be compared with the value 
%obtained using the measured lifetimes and production rates of the various
%$b$-hadrons generated in a $b$-quark jet (see Equation (\ref{eq:tbexcl})):
%\begin{equation}
%\tau_b^{excl.}~=~(1.546 \pm 0.016)~{\rm ps}.
%\label{eq:taubexcl}
%\end{equation}
%The two values are compatible. 

%{\bf Can we quote a global average ? 
%End change}

\mysubsection{$b$-hadron lifetime ratios and expectations from theory}
\label{sec:tauth}

Ratios of other $b$-hadron lifetimes to the $\Bd$ lifetime
have also been obtained, using the averages of the 
individual lifetimes
and the direct measurements of the ratio between $\Bp$ and $\Bd$ lifetimes
shown in Figure \ref{fig:tbhad}-right. The results are given
in Table \ref{tab:lifetimefrac}.

\begin{table}[t]
  \begin{center}
    \begin{tabular}{|c| c |}
      \hline
      Lifetime ratio & Measured value \\
      \hline
     $\tau(\Bp)/\tau(\Bd)$ & $\taubpovertaubd$ \\
     $\tau(\Bs)/\tau(\Bd)$ & $0.946 \pm0.039$ \\
     $\tau(b-{\rm baryon})/\tau(\Bd)$ & $0.780 \pm0.035$ \\
\hline
    \end{tabular}
  \end{center}
    \caption{\it {Ratios of $b$-hadron lifetimes relative to the $\Bd$ 
lifetime.}    \label{tab:lifetimefrac}}
%and comparison with theoretical expectations

\end{table}
Inclusive properties of heavy hadron decays have been analyzed
using the Wilson formalism of Operator Product Expansion.
Total or partial decay widths are expressed as a series of local
operators with increasing dimension, where coefficients contain
inverse powers of the $b$-quark mass starting as $1/m_b^2$
\cite{ref:chay}. Average values of the local operators are also expanded
in inverse powers of the $b$-quark mass using the Heavy Quark Effective
Theory. Terms entering into this expansion receive contributions
from QCD operating in the non-perturbative regime and have been determined
using approximations or models.
% until
%lattice QCD evaluations become evailable. 
Within the factorization approximation, they obtain \cite{ref:bigilife}:
\begin{equation}
\frac{\tau(\Bp)}{\tau(\Bd)}~=~1 + 0.05 \times \frac{f_B^2}{(200~\MeV)^2},~
\frac{\tau(\Bs)}{\tau(\Bd)}~=~1 \pm {\cal O}(1\%),~
\frac{\tau(\Lb)}{\tau(\Bd)}~=~0.90-0.95.
\label{eq:predt}
\end{equation}
For $b$-meson lifetimes, these predictions 
%It can be seen that
%the predictions \cite{ref:bigilife} for the meson lifetimes
%based on factorization,
%given in Equation (\ref{eq:predt}),
%which 
had encountered theoretical criticism \cite{ref:sacrelife}
but the improvement in the experimental accuracy on $\Bd$ and $\Bp$
lifetime determinations shows that they were in agreement with
the measurements.
%are in agreement with the measurements. 
In addition, a recent lattice study \cite{ref:latlife}
finds a result quite consistent with \cite{ref:bigilife}.
% which was based on factorization.
%Initial expectations \cite{ref:bigilife}
%for the values of these ratios agree in case of mesons

However a significant discrepancy is observed for baryons.
It remains to be clarified if this problem is related only to the validity
of the quark models used to determine the parameters in the case of baryons, 
as was
suggested in reference \cite{ref:bigilife}, or to more basic
defects in the whole theoretical picture.
% as advocated for instance in 
%\cite{ref:sacrelife}. These last expectations have not been confirmed by 
%recent evaluations from lattice QCD 
%\cite{ref:latlife} which are, instead, in agreement with the initial 
%prediction for
%the ratio between the $\Bm$ and $\Bdb$ lifetimes. 
Lattice evaluations
for $b$-baryons are starting to be available \cite{ref:latlifeb} and
corrections appear to be larger for baryons than for mesons. 
Present results 
indicate that contributions involving spectator partons may account
for half of the discrepancy. The remaining difference can come 
from a violation of quark-hadron duality or, owing to present uncertainties,
to contributions of higher order terms. It should be noted that a possible
violation of duality in inclusive $b$-decays does not
impair the use of OPE to extract $\Vcb$ from measurements of the inclusive
$b$-hadron semileptonic decay width (Section \ref{sec:vcbinc})
 because the latter corresponds 
to a completely
different situation in which such violations are expected to be 
smaller \cite{ref:bigiosaka}. Possible evidence for duality violation
in $b$-semileptonic decays has to be searched for directly using the 
corresponding
final states.

\mysection{$\Bd$ and $\Bs$ oscillations and
  $b$-hadron production fractions}
\label{sec:boscill}
%{\bf References of measurements used in Dmd and Dms analyses 
%need to be updated .}

%\mysubsection{Introduction}
%\label{sec:intro}
The four LEP Collaborations and CDF and SLD have published or 
otherwise released
measurements of \dmd~\cite{A:dmd,C:dmd,D:dmd,L:dmd,O:dmd,S:dmd}
and lower limits for \dms~\cite{A:dms,C:dms,D:dms,O:dms,S:dms}.
%A combination of the experimental results is of interest for the
%determination of the CKM matrix elements with the greatest possible
%precision. In the first part of this note
Combined results, as well as estimates of $b$-hadron fractions, have been 
prepared by the B oscillation working 
group\footnote{The present members of the B oscillation working group are:
V. Andreev, E. Barberio, 
G. Boix, C. Bourdarios, P. Checchia, O. Hayes, R. Hawkings, O. Leroy, S. Mele,
 H-G. Moser, F. Parodi, M. Paulini,
P. Privitera, 
P. Roudeau, O. Schneider, A. Stocchi, T. Usher, C. Weiser
and S. Willocq.}.  
%The variation of the average oscillation rate of neutral $b$-hadrons in
%a $b$ jet is used, in addition to direct measurements, to obtain the
%production rates of $b$-hadrons.

The estimates of the $b$-fractions, described in Section \ref{sec:results},
are important inputs needed for the combination of the $\Bd$ and $\Bs$
oscillation results. 
The procedure used to combine \dmd\ results is explained in
Section \ref{sec:method}, followed by a description of the
common systematic uncertainties in the analyses, and the result for
the mean value of \dmd.
%Production rates of $b$-hadrons have been obtained in two ways. 
%Characteristic signatures allow to isolate samples enriched
%in $\Bs$, $b$-baryons or $\Bp$ and enable to do direct measurements
%of their corresponding rates. The other method is based on the comparison 
%between the measured and expected values of
%the average oscillation rate of $b$-hadrons emitted in a $b$-jet, the
%latter being dependent on the assumed production rates of these particles.
%The procedure used for
%combining \dmd\ results is explained, followed by a description of the
%common systematic uncertainties among the analyses, and the result for
%the mean value of \dmd.
%In the second part of the note the procedure
%used to combine limit results for \PsBz\ oscillations is
%explained. 
Then combined results on the \PsBz\ oscillation amplitude,
as well as an overall lower limit on \dms\, are presented
in Section \ref{sec:dmsosc}.
% are combined
% using a method
%similar to that used for \PdBz,
% with the additional step of a limit
%calculation on \dms\ after the combination is performed. 
%More details on these procedures can be found in reference \cite{ref:nimosc}.

\mysubsection{Measurements of $b$-hadron production rates in $b$-jets}
\label{sec:results}

In this analysis, the relative production rates
of the different types of weakly-decaying $b$-hadrons are
assumed to be similar in $b$-jets originating from $\Zz$ decays
and in high transverse momentum $b$-jets in ${\rm p}\overline{{\rm p}}$
collisions at 1.8 TeV.
%Informations on weakly decaying $b$-hadrons production
%rates in $b$-jets are obtained from direct measurements using channels
%providing characteristic signatures and from the time integrated 
%${\rm B}^0-\overline{{\rm B}^0}$ oscillation rate which is sensitive to
%the fractions of $\Bd$ and $\Bs$ mesons in a $b$-jet. It
%has been assumed that production rates of the different types
%of $b$-hadrons are similar in $b$-jets originating from $\Zz$
%decays or being emitted at high transverse momentum in 
%${\rm p}-\overline{{\rm p}}$ collisions at 1.8 TeV.
This hypothesis can be justified by considering the last steps of 
jet hadronization to be a non-perturbative QCD process occurring
at a scale of order $\Lambda_{{\rm QCD}}$.

Direct information on production rates is available
from measurements of branching fraction products using channels
with characteristic signatures. 
Dedicated analyses have also been developed which are sensitive
to the production of charged and neutral $b$-hadrons \cite{ref:delphibplus}
and of $b$-baryons \cite{A:flamdir} without requiring the 
knowledge of specific branching fractions.
All the measurements
used in the present analysis are listed in Appendix \ref{appendixAc}.
%All direct measurements, used in the present analysis, have been collected in 
%Appendix \ref{appendixAc}.

 At CDF and LEP, the $\Bsb$ production rate has been evaluated using
events with a $\Ds$ accompanied by a lepton of opposite sign in 
the final state (Table \ref{tab:sfrac}).
 Since the rate of these events is given by
$\fs \times \bsdslX$, it is necessary to evaluate \bsdslX.
%Assuming SU(3) symmetry, the \PdBz/\PBp\  system semileptonic branching 
%fractions,
%the total $b$-semileptonic decay rate and $b$-hadron lifetimes are used
%to estimate \bsdslX.
%To estimate \bsdslX, it 
This has been done by assuming, from SU(3) symmetry, that
the partial semileptonic decay widths into D, ${\rm D}^*$ 
and $\Dstarstar$ final states
are the same for all $b$-mesons.
A lower value for the $\Bs$ semileptonic branching fraction
with a $\Dsm$ in the final state is obtained by assuming
that all $\overline{{\rm D}^{**}_s}$ states decay into a non-strange D meson.
The branching fraction corresponding to
$\Bs \rightarrow {\rm D}_s^{**-} \ell^+ \nu_{\ell}$ is obtained
using Equation (\ref{eq:dsstarinc}) multiplied by the lifetime ratio
$\tau(\Bs)/\tau(\Bd)$. This corresponds to a lower limit
on $\Dsm$ production because the possibility that 
${\rm D}_s^{**-}$ states decay into $\Dsm \overline{{\rm K}}{\rm K}$
or $\Dsm \eta^{(')}$ has
been neglected.
In addition, assuming that only the measured ${\rm D}^{(*)} \pi$
final state in $\Bd$ or $\Bp$ semileptonic decays corresponds
to ${\rm D}^{(*)} {\rm K}$ transitions for the $\Bs$, an upper value
can be obtained for the $\Bs$ decay of interest using the relation:
\begin{equation}
\Gamma(\Bd \rightarrow \overline{{\rm D}^{(*)0}} \pi^- \ell^+ \nu_{\ell})
=\frac{4}{3} \Gamma(\Bs \rightarrow \overline{{\rm D}^{(*)0}} {\rm K}^- \ell^+ \nu_{\ell})
\end{equation}
based on isospin symmetry. In practice, the lower and upper values are 
compatible within uncertainties.
%In particular, the contribution from $\Dstarstar$ states,
% produced in $\Bsb$ decays, which is expected to generated only
%non-strange D mesons is obtained using expressions given in 
%Section \ref{sec:dssb}. The LEP measurements can be found in \cite{A:tbs,A:tbs2,A:tbs3}.
This allows $\fs=\fbsdir$ to be extracted from the LEP measurements.
%$\fs \times \bsdslX$ \cite{A:tbs,A:tbs2,A:tbs3}.
Using the same assumptions, the CDF Collaboration has measured the 
ratio $\fs/(\fu+\fd)=0.213\pm0.068$
from final states with an electron and a charm meson 
($\Do,~\Dp,~\Dstarp,~\Dsp$) 
\cite{A:flamcdf}. CDF has published a 
second measurement of the same ratio,
using double semileptonic B decays with $\phi \ell$ and ${\rm K}^* \ell$
final states as
characteristic signatures for $\Bs$ and non-strange B mesons 
\cite{ref:cdfrates}, giving $\fs/(\fu+\fd)=0.210\pm0.073$
\footnote{The uncertainties quoted for these two results from CDF
include the uncertainty on the evaluation of the branching fraction
for the $\Ds$ meson into $\phi \pi$; original measurements are given in
Table~\ref{tab:sfrac}.}.
%. At present,
%this result has not been included in the average  
%$\fs$ 
%for $b$-hadron production fractions because of
%difficulties in properly correlating certain assumptions in that analysis
%with the other measurements.}. 
%The input quantities are listed in 
%Table~\ref{tab:sfrac}.
\begin{table}[t]
\begin{center}
\begin{tabular}{|l|c|}
\hline
 $b$-hadron fractions  & correlation coefficients \\ \hline
 $\fs$ = $\fsa$ & \\
 $\fb$ = $\fbara$ &$\rho(\fs,\fb)=\rhosbara$ \\
$\fd = \fu$ = $\fua$ & $\rho(\fd,\fs)=\rhosua$,~$\rho(\fd,\fb)=\rhobarua$\\
\hline
\end{tabular}
\end{center}
\caption{
{\it Average values of $b$-hadron production rates
and their correlation coefficients obtained from direct measurements.}
\label{tab:ratesstepa}} 
\end{table}

%However, these rates are partially determined using

In a similar way, the fraction of $b$-baryons is estimated 
from the measured production rates
of $\Lc \ell^-$ \cite{AD:prodlamb, AD:prodlamb2} and $\Xi^- \ell^-$ 
\cite{AD:cascb, AD:cascb2} final states (Table \ref{tab:bfrac})
yielding, respectively,
$f_{\Lb}=\flbdir$ and $f_{\Xi_b^-}=\fxidir$. 
The value for BR($\Lb \rightarrow \Lc {\rm X} \ell^- \overline{\nu_{\ell}}$)
has been obtained considering that there could be one 
$\Lc$ produced in every decay or, at the other extreme, 
that all excited charmed
baryons, assumed to be produced with a rate similar to $\Dstarstar$ final 
states in B mesons, 
decay into a D meson and a non-charmed baryon.
Similar considerations
have been applied to $\Xi_b$ semileptonic decays.
The semileptonic decay width 
$\Gamma(\Xi_b \rightarrow \Xi_c {\rm X} \ell^- \overline{\nu_{\ell}}$)
has been taken to be equal to
$\Gamma(\Lb \rightarrow \Lc {\rm X} \ell^- \overline{\nu_{\ell}}$)
and it has been assumed that, within present uncertainties,
BR$(\Xi_c^0 \rightarrow \Xi^- {\rm X})$ = 
BR$(\Lc \rightarrow \Lambda^0 {\rm X})$.
The total $b$-baryon
production rate is then: $\fb=\fbardir$, assuming the same
production rates for $\Xi_b^0$ and $\Xi_b^-$ baryons.
%The input quantities are again taken from 
%Table~\ref{tab:sfrac}.
This is then averaged with a direct measurement of $\fb=\fbarspec$
from the number
of protons in $b$-events \cite{A:flamdir}
to give a result of $\fb = \fbaravg$ from LEP measurements alone.
%fbaryon=(11.0+2.3-1.9)%.
Finally, the CDF Collaboration
has measured the production rate of $b$-baryons relative
to non-strange B mesons,
$\fb/(\fu+\fd)=0.118\pm0.042$, using $\Lc e^-$ final states 
\cite{A:flamcdf}.

The fraction of $\Bp$ mesons has 
been estimated by subtracting $f_{\Xi_b^-}$ from the fraction of
all charged $b$-hadrons measured using the charge of
a large sample of inclusively reconstructed secondary
vertices
\cite{ref:delphibplus}:
% using the charge of a large
%sample of inclusively reconstructed secondary vertices: 
$\fu=\fbudir$ (Table \ref{tab:bpfrac}).

All these measurements have been combined to obtain average production rates
for $\Bs$, $b$-baryons, $\Bp$ and $\Bd$ \footnote{The production
of ${\rm B}_c$ mesons and of other weakly decaying
states made of several heavy quarks has been neglected.}, imposing that:
% the sum of all rates is equal to unity
\begin{equation}
\fu + \fd + \fs +\fb~=~1 
\label{eq:unitary}
\end{equation}
and 
\begin{equation}
\fu = \fd.
\label{eq:equal}
\end{equation}
%using the constraint that $\Bp$ and $\Bd$ production rates are the same.
The results obtained are given in Table \ref{tab:ratesstepa}.
Correlated systematics between the different measurements, 
coming mainly from
the poorly measured branching fractions of $\Ds$ and $\Lc$ charmed hadrons,
have been taken into account.
It may be noted that constraint (\ref{eq:equal}) is expected to be true for
$b$-mesons even though it is known to be wrong for $c$-mesons.
%This is because $\Bstar$ mesons decay electromagnetically, leaving
%the flavour of the light spectator quark unchanged. 
In both $b$- and $c$-jets, isospin invariance of strong 
interactions predicts
similar production rates of mesons
in which the heavy quark is associated to a $\overline{u}$ or
$\overline{d}$ antiquark. 
Strong and electromagnetic decays of these states result in different rates
for weakly decaying mesons with a $u$ or a $d$ flavour
because the $\Dstar \rightarrow {\rm D} \pi$ decays occur
very close to threshold, and the threshold prevents the transition
$\Dstarp \rightarrow \Dp \pi^-$. However
for $\Bd$ and 
%${\rm D} \pi$
$\Bp$ mesons, no asymmetry is expected: $\Bstar$ mesons decay 
electromagnetically, leaving the flavour of the light spectator quark
in the $b$-meson unchanged;
non-strange $\Bstarstar$ mesons decaying through strong interactions and having
masses away from threshold will not induce any asymmetry either;
${\rm B}^{**0}_s$ mesons decay also by strong interactions into
${\rm B}^{0(*)}_d \overline{{\rm K}^0}$ or 
${\rm B}^{+(*)} {\rm K}^-$ with equal probabilities,
although a possible tiny difference
between these two rates can be expected for decays of narrow states occurring
very close to threshold, because of the mass difference between ${\rm K}^0$
and ${\rm K}^+$ mesons\footnote{The mass difference between $\Bd$ and $\Bp$
mesons is compatible with zero within $\pm 0.3~\MeV/c^2$.}. 
%The largest part of the observed asymmetry between $\Do$ and $\Dp$ 
%production rates in $c$-jets
%comes from $\Dstar$ decays which occur, mainly, by strong interactions very
%close to threshold for ${\rm D} \pi$ production preventing the transition
%$\Dstaro \rightarrow \Dp \pi^-$.

Additional information on the production rates
can be obtained from measurements of the time-integrated
mixing probability of $b$-hadrons. For an unbiased sample of
semileptonic $b$-hadron decays in $b$-jets, with fractions
$\gd$ and $\gs$ of $\Bd$ and $\Bs$ mesons, this mixing
probability is equal to:
%Time integrated mixing measurements of neutral $b$-mesons depend also
%on $\Bd$ and $\Bs$ production rates:
\begin{equation}
\chib= \gs~\chis + \gd ~\chid
\label{eq:chibar}
\end{equation}
where $\chid$
%In this expression,
is the time-integrated mixing probability for $\Bd$ mesons 
(see Section \ref{sec:method})
and $\chis$=1/2 is the corresponding quantity for $\Bs$ 
mesons\footnote{The assumption $\chis =\frac{1}{2}$ can be justified
  by the existence of limits on $\dms$ obtained from $\rm D_s$-lepton 
analyses, which have negligible dependence on $\fs$.}.
% and $\chib= \gs \chis + \gd \chid$, where
%, $\chis
%\equiv \frac{1}{2}$, and 
As already mentioned in Section \ref{sec:systgen}, the semileptonic width
is assumed to be the same for all $b$-hadron species, implying
%$\gs$ and $\gd$ are the 
%product of the
%production fractions in a sample of semileptonic
%$b$-decays, corresponding to the type of events samples used
%in these measurements.
%The latter are assumed to be given by 
$g_i = f_i ~{\mbox R}_i$, where ${\mbox R}_i=\frac{\tau_i}{\tau_b}$
are the lifetime ratios. This leads to the relation:
\begin{equation}
%\fs~=~\frac{2 \tau_b \overline{\chi} - \tau_{B_d} \chi_d (1-\fb)}
%{2 \tau_{B_s} \chi_s-\tau_{B_d} \chi_d},
\fs~=~\frac{1}{{\mbox R}_s}~
\frac{(1+r)~\chib-(1-\fb ~{\mbox R}_{\mbox{\scriptsize \Bb}})~ \chid}
{(1+r)~ \chis - \chid}
\end{equation}
where $r={\mbox R}_u/{\mbox R}_d = \tau(\Bp)/\tau(\Bd)$.
This is used to extract another determination of $\fs$
from the $b$-baryon fraction of Table \ref{tab:ratesstepa},
the lifetime ratio averages of Table \ref{tab:lifetimefrac},
the world average value of $\chid$ from Equation (\ref{eq:dmdbb}) 
of Section \ref{sec:method}, and the $\chib$ average from the 
LEPEWWG (see Table \ref{tab:gensys}). This new estimate of
$\fs$,
%where $\tau_b$ is the average $b$-hadron lifetime. This is equivalent to 
%assuming that the semileptonic decay width is the same for all
%$b$-hadron species; see Section \ref{sec:systgen} for a discussion on the 
%validity of this hypothesis.
%%weighted by the semileptonic branching fractions
%The probability of $\Bs$ mixing, $\chis$, is assumed 
%to be equal to $\frac{1}{2}$.
%Ignoring the correlations between $\dmd$ and the $b$-hadron fractions 
%would introduce a circular dependence in
%the determination of \dmd\ for those (dilepton) measurements that use
%these parameters. This is avoided by expressing,
%the production fractions
%(of \PsB, \PBp\ and \PdBz) explicitly in terms of their
%dependence on \dmd, \chib, $\fb$, and the lifetimes.
%and
%\begin{equation}
%\fd~=~\frac{1}{2}(1-\fs-\fb),
%\end{equation}
%In addition it has been required that all $b$-hadron production rates
%add up to unity.
%{\it Notice that these expressions are different from those used at present.
%I probably have missed some point.}
%{\it The following lines have to be changed...}
%
%Assuming $\fd = \fu$ 
%and $\fd + \fu + \fs + \fb = 1$, the values of
%Table~\ref{tab:dmdsys} for \chib, \chis, $\fb$, and the $b$-lifetimes,
%the world average of \dmd\ 
%leads to $\fs = ( \fbsmix )$\%.
% Through the constraint coming from the expression of $\chib$, another
%determination of $\fs$ is obtained,
$\fs^{\rm mixing}= \fbsmix$, is then combined
with the $b$-hadron rates from direct measurements, 
taking into account correlations\footnote{
There is a small statistical correlation between 
$\chid$ and $\chib$, arising from the fact that a few $\dmd$ analyses at LEP 
are based on the same samples of dilepton events as the ones used to extract 
$\chib$. This correlation is ignored, with a negligible effect on the final result.} and imposing the conditions (\ref{eq:unitary}) and (\ref{eq:equal}).
The final $b$-hadron fractions are displayed in Table \ref{tab:rates}.

\begin{table}[t]
\begin{center}
\begin{tabular}{|l|c|}
\hline
 $b$-hadron fractions  & correlation coefficients \\ \hline
 $\fs$ = $\fbs$ & \\
 $\fb$ = $\fbar$ &$\rho(\fs,\fb)=\rhosbar$ \\
$\fd = \fu$ = $\fbd$ & $\rho(\fd,\fs)=\rhosd$,~$\rho(\fd,\fb)=\rhobard$\\
\hline
\end{tabular}
\end{center}
\caption{{\it Average values of $b$-hadron production rates
and their correlation coefficients obtained from direct measurements 
and including constraints from results on B mixing.} \label{tab:rates} }
\end{table}

%corresponding uncertainties,
%the following values have been obtained in the end:

%The tiny correlation between $\fs$ and $\fb$ results from an almost perfect 
%cancellation between two effects ...
%
%{\it  Christian can you add a few lines ..}

%Averaging this value of $\fs$
%from mixing with that from branching fraction measurements,
%the fraction of \PsBz\ mesons in a sample of weakly decaying $b$-hadrons is
%found to be $\fs = (\fbs )$\%. This further implies that
%$\fd = \fu = (\fbd)$\% and $\fb = (9.5 \pm 2.0 )\%$.
%These values have correlated errors; the correlation coefficients are:
%$\rho(\fd,\fs)=    ,~\rho(\fd,\fb)= $ and $\rho(\fs,\fb)=  $.

\begin{figure}[htp]
  \begin{center}
    \leavevmode
    \mbox{\epsfxsize16.0cm \epsfysize22.0cm
\epsffile{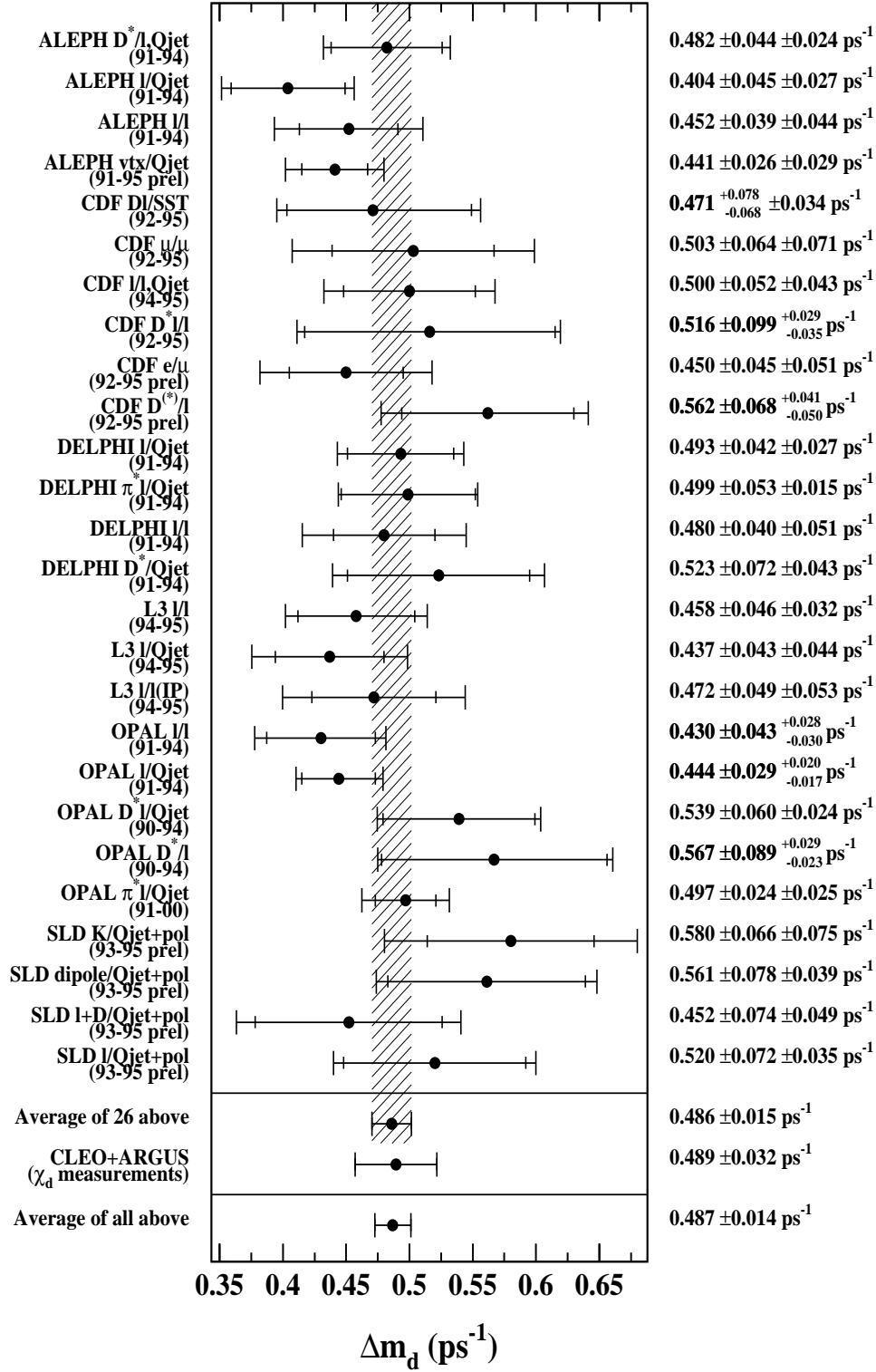}}
%\epsffile{/afs/cern.ch/phys/lep/lepbosc/combined_results/summer_2000_preprint/HF_preprint_plots/dmd.eps}}
    \caption{{\it Individual and combined measurements of $\dmd$ at LEP, CDF
      and SLD. 
Note that the
      individual measurements are as quoted in the original
      publications, but their average includes the effects of adjustments
      to a common set of input parameters.
The last average also includes $\chid$ measurements
\cite{PDG00} performed by CLEO and ARGUS experiments
      at the \Yfs\ resonance. }\label{fig:dmdvalues}}

  \end{center}
\end{figure}

The production rate for $b \overline{s}$ states 
is expected to be 
rather similar to the probability ($\gamma_s \sim 12\%$)
for creating an $s  \overline{s}$ pair
during the hadronization process. But, because of their masses,
non strange $\Bstarstar$ mesons are not expected to decay into
${\rm B}^{0(*)}_s \overline{{\rm K}}$, whereas strange $\Bstarstar$ 
mesons are expected to decay mainly into non-strange weakly 
decaying $b$-mesons.
As a result, the fraction of $\Bsb$ in $b$-jets is expected to be given by
$\fs \simeq \gamma_s~[1 -{\rm P}(b \rightarrow \Bstarstarb)] \simeq 10\%$,
as observed.

\mysubsection{Combination method for $\dmd$}
\label{sec:method}

The measurements of \dmd\ are now quite precise and it is
important that the correlated systematic uncertainties are correctly
handled. Furthermore, many results depend on physics parameters for which
different values were used in the original analyses. 
Before being combined, the measurements
of \dmd\ are therefore adjusted on the basis of a common consistent set of
input values.
% These tasks are performed in one program for which the
%experiments provide the following information for each of their
%analyses: the assumed values (and uncertainties) of the parameters,
%the measured value of \dmd\ and its statistical uncertainty, the
%systematic uncertainties from different sources, and the amount (if
%any) of statistical correlation with other analyses.  The program also
%requires the common parameter values and uncertainties to which the
%original results are to be adjusted.

For each input parameter the \dmd\ measurement and the corresponding
systematic uncertainty are linearly rescaled in accordance with
the difference between the originally used parameter value and error, and
the new common values.
%The procedure first linearly rescales each measurement of \dmd\ for
%each parameter, in accordance with the difference between the
%originally used parameter value and its new common value, and with the
%corresponding systematic uncertainty determined in the analysis.  This
%systematic uncertainty is also adjusted, if the uncertainty on the
%parameter is different from that originally used. 
%The statistical and
%systematic uncertainties are also symmetrized if necessary. An
%alternative approach using asymmetric errors produced negligible
%differences in the results.
This is done for the systematic uncertainties
originating from the $b$-hadron lifetimes, the
$b$-hadron fractions and the mixing parameters
$\chib$ and $\chid$. The
statistical and systematic uncertainties of each individual
measurement are symmetrized. An alternative approach using
asymmetric uncertainties (when quoted as such) would
produce negligible differences in the combined result.

The combination procedure makes a common fit of \dmd\ and the common input
parameters.
% related to the sources of the common systematic
%uncertainties. 
It is assumed that \dmd\ may be expressed as a function
$X(Y_1 \dots Y_{N_{sys}})$ with a weak dependence on the systematic
sources $Y_\alpha$. Expanding $X$ then gives:
\begin{equation}
  X(Y_1 \dots Y_{N_{sys}}) \approx
  X^0 + \sum_{\alpha=1}^{N_{sys}} Y_\alpha \sigma_\alpha^{syst}
\end{equation}
where the quantities $\sigma_\alpha^{syst}$ are the 
correlated errors on \dmd\
from systematic sources $\alpha$, and $Y_\alpha = (z_\alpha -
z^{fit}_\alpha)/\delta_\alpha$. In this last expression,
 $z_\alpha$ and $z^{fit}_\alpha$
are the input and fitted values of the systematic parameter $\alpha$, and
$\delta_\alpha$ is the variation used to calculate
$\sigma_\alpha^{syst}$.

The following $\chi^2$ is then constructed:
\begin{equation}
  \chi^2 = \sum_{i=1}^{N}
  \left(\frac{X_i - 
      \sum_{\alpha=1}^{N_{sys}} Y_\alpha \sigma_{i \alpha}^{syst} - X^0}
    {\sigma_i^{uncor}} \right)^2
      +\sum_{\alpha=1}^{N_{sys}} Y_\alpha^2,
\label{eq:compl}
\end{equation}
where $X_i$ are the measurements of \dmd\ and 
$\sigma_i^{uncor}$ the quadratic sum of the statistical and uncorrelated 
systematic uncertainties on $X_i$. This $\chi^2$ is minimized 
with respect to the parameters $X^0$ and $Y_\alpha$; the result for 
$\dmd$ is taken as the value of $X^0$ at the minimum (and the 
values of $Y_\alpha$ at the minimum are ignored).
%to solve for $X^0 =
%\dmd$, and the systematic parameters $Y_\alpha$. 
This method gives the
same results as a $\chi^2$ minimization with inversion of a global
correlation matrix. As several measurements also have a statistical
correlation, 
the first sum in Equation (\ref{eq:compl}) is then
    generalized to handle an $N \times N$ error matrix describing the
    statistical and uncorrelated systematic uncertainties.

The following sources of systematic uncertainties, common to analyses
from different experiments, have been considered:
\begin{itemize}
\item $b$-lifetime measurements (Section \ref{sec:Atau}).
Different measurements use the $b$-hadron lifetimes in different ways as
input: some use the actual lifetimes, and others use ratios of
lifetimes. As this leads to complicated correlations between the
analyses, the following procedure is adopted. For each measurement a
single combined value of all $b$-lifetime-related systematic
uncertainties is computed. This systematic is then treated as fully
correlated with all other such numbers from the other
measurements. Tests show that this procedure gives a 
negligible bias and results in a conservative evaluation
of the uncertainty.
%result that
%is different from a more rigorous treatment by an amount less than the
%resulting increase in the uncertainty.

\item $b$-quark fragmentation (Section \ref{sec:systgen}).
In some analyses the
$b$-hadron momentum dependence is treated through the variation of the
$\epsilon$ parameter in the Peterson fragmentation function, and in others
through the mean scaled $b$-hadron energy $<x_E>$. In such a case,
% the
%errors are treated as fully correlated without explicit dependence on
%$\epsilon$ or $<x_E>$.
the corresponding systematic uncertainties are treated as fully correlated, 
without attempting to adjust the individual results to a common value
of $\epsilon$ or $<x_E>$.

\item fraction of cascade decays (Section \ref{sec:systgen}),
related mainly to the value
of BR$(b \rightarrow c \rightarrow \ell^+\nu_{\ell}X)$.

\item fraction of $\Bp$ mesons remaining in $\Bd$ enriched samples
for analyses using $\Dstar$ mesons. This depends on decay branching
fractions and on the mistag probability.

\item production rate of $\Dstarstar$ mesons in $b$-hadron semileptonic decays
(Section \ref{sec:systgen}).

\item $b$-hadron fractions (Section \ref{sec:results}).
%obtained from direct measurements, without including
%information from mixing (Table \ref{tab:ratesstepa})

%\item uncertainties on $\chib$ and on $\chid$, the latter being measured
%at the $\Upsilon$(4S) (Section \ref{sec:systgen})

\item $\Do$ lifetime (\cite{PDG00}).

\end{itemize}

%\mysubsection{Common systematic uncertainties}
%\label{sec:params}
%Table~\ref{tab:dmdsys} lists the sources of common systematic
%uncertainty for the \dmd\ analyses from different experiments, and
%other parameters upon which the fit depends. These parameters may
%enter into the fit in different ways. The $b$-hadron lifetime values are
%used for the rescaling of \dmd. Production fractions are used as described
%in the previous section. Several sources of common systematic
%uncertainty are correlated, but not directly related to the same
%parameter in each of the analyses. For example,  
Common systematics due to purely experiment-dependent 
factors (i.e.\ common to different results in a particular
experiment) have not been included in the above list, 
but nonetheless are treated
as correlated in the fit.
Following this procedure, a
%discussed in Section~\ref{sec:method}, a
combined value:
% of $\dmd^{\rm LEP} = (\dmdlerr$ \ips $= \dmdl$) \ips\ is
%found.  Including results from the CDF and SLD experiments 
%gives 
\begin{eqnarray}
   \begin{array}{lll}
\dmd^{{\rm LEP+CDF+SLD}}& = &(\dmdwerr)~{\rm ps}^{-1}
   \end{array}
\label{eq:dmda}
\end{eqnarray}
is obtained. 
Using the 
relation\footnote{Equation (\ref{eq:chid}) assumes that there is no decay width
difference in the $\Bd-\Bdb$ system (current expectations are at the $10^{-3}$
level for $\dgbd$), and that effects from CP violation in the considered
final states can be neglected.}:
\begin{equation}
\chid = \frac{1}{2}~\frac{(\dmd~\tau_{B_d})^2}{(\dmd ~\tau_{B_d})^2 +1}
\label{eq:chid}
\end{equation}
and the $\Bd$ lifetime of Section \ref{sec:Atau},
the above $\dmd$ result can be converted to:
\begin{eqnarray}
   \begin{array}{lll}
\chid^{{\rm LEP+CDF+SLD}}& = & \chicls. 
   \end{array}
\label{eq:dmdb}
\end{eqnarray}
%$\chid$, using Equation (\ref{eq:chid}) and the $\Bd$ lifetime of
%    Section \ref{sec:Atau}, and then 
Averaging with the time-integrated mixing
results obtained by ARGUS and CLEO at the $\Upsilon$(4S)
(see Table \ref{tab:gensys}), yields finally
\footnote{In the following expressions, ALL refers to ARGUS, CDF, CLEO, LEP
and SLD measurements.}:
%    chi_d^Upsilon(4S) = 0.156 +- 0.024 [PDG 1998],
%to determine
%This value of \dmd, and the
%values of $\tau_\PdBz$ and $\chid^\Yfs$ listed in
%Table~\ref{tab:dmdsys}, may be used to determine 
\begin{eqnarray}
   \begin{array}{lll}
\chid^{{\rm ALL}}& = &\chix, 
   \end{array}
\label{eq:dmdbb}
\end{eqnarray}
or equivalently
\begin{eqnarray}
   \begin{array}{lllll}
\dmd^{{\rm ALL}}& = &(\dmdx)~{\rm ps}^{-1}& =& (\dmdxev)~10^{-4}~{\rm eV}/c^2.
   \end{array}
\label{eq:dmdc}
\end{eqnarray}
The individual measurements
    of $\dmd$ and their combined value are shown in 
Figure~\ref{fig:dmdvalues}.
%In Fig.~\ref{fig:dmdvalues} the individual
%measurements of \dmd\ and their combined value are shown.
The determination of $\dmd$ as described above and 
of the $b$-hadron fractions described in
    Section \ref{sec:results} cannot be performed sequentially: the
    values of the fractions are needed to perform the
    $\dmd$ fit, and the best estimates of the fractions can
    be obtained only once the final $\dmd$ average is known.
    This circular dependence has been handled by including
    the calculation of the fractions in the $\dmd$ fitting
    procedure, in such a way that the final results quoted for
    $\dmd$ and the $b$-hadron fractions form a consistent set.

\mysubsection{Combination method for $\Bs$ oscillation amplitudes
         and derived limits on $\dms$}
\label{sec:dmsosc}

\begin{figure}[htb]
  \begin{center}
    \leavevmode
    \mbox{\epsfxsize12.0cm \epsfysize12.0cm
\epsffile{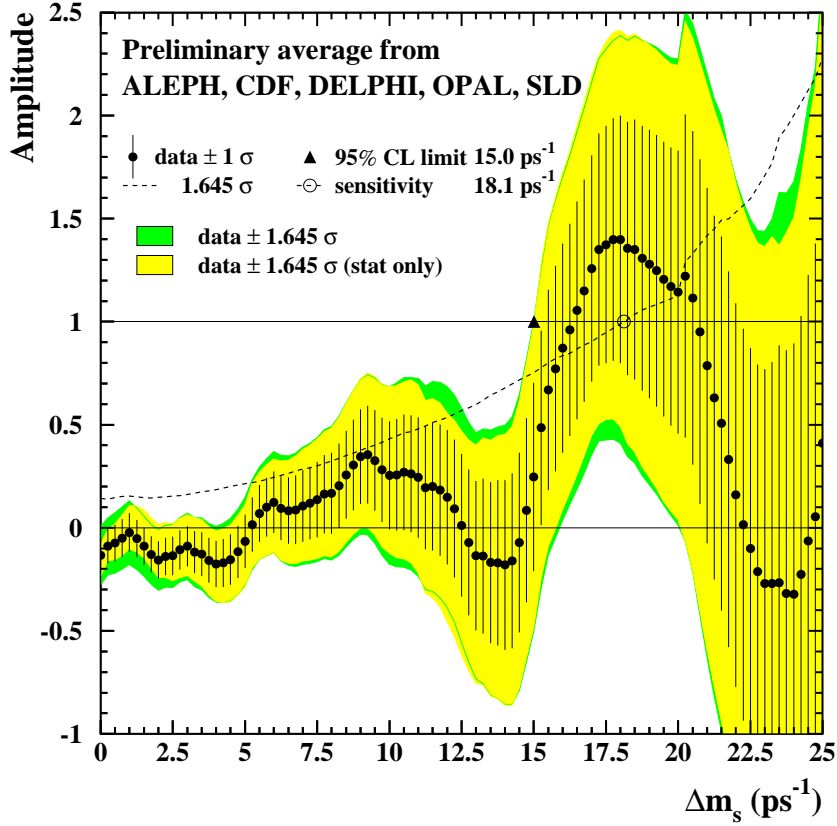}}
%\epsffile{/afs/cern.ch/phys/lep/lepbosc/combined_results/summer_2000_preprint/HF_preprint_plots/amplitude_vs_dms.eps}}
    \caption{{\it Combined \PsBz\ oscillation amplitude as a
      function of $\dms$.  A 95\% CL lower limit on $\dms$ of \dmslim\
      \ips\ is derived from this spectrum.
Points with error bars are the fitted amplitude values and 
corresponding uncertainties, including systematics. The dark shaded area
is obtained by multiplying these uncertainties
by 1.645 such that the integral of the probability distribution,
 assumed to be Gaussian, is equal to 5$\%$ above this range.
The light shaded area is obtained in a similar way but including only 
statistical uncertainties in the fit; as a consequence, central values
of the two regions do not necessarily coincide.
}\label{fig:dmsampl}}

  \end{center}
\end{figure}

No experiment has yet directly observed \PsBz\ oscillations, so the task
of a combination procedure is to calculate an overall limit 
or to quantify the evidence for a signal from the
information provided by each measurement. This is done using the
amplitude method of reference~\cite{AmpMeth}. At each value of $\dms$, in
the range of interest, an amplitude is measured in each analysis,
where the expected value of the amplitude is unity at the
true frequency.
%where an amplitude of unity indicates a successful observation of
%oscillation with that frequency. 
%A limit may be placed on regions of
%$\dms$ where amplitudes of unity are excluded.
An overall limit on $\dms$ is then inferred from the combined 
amplitude spectrum by excluding regions of $\dms$ where the amplitude
is incompatible with unity.

Studies with Monte Carlo simulation show that the measured amplitude
has a Gaussian distribution around its expected value, which is zero for
frequencies much lower than the true one, and unity at the true frequency.
In addition, the error on the amplitude depends on the statistical
power of the analysis but not on the true value of the oscillation
frequency. Therefore, if at a given frequency ($\Delta m_s$) a measured
amplitude ($\mu$) with error $\sigma$ is found, this value of the frequency
can be excluded at the~95\% C.L. if the probability of measuring an amplitude
equal or smaller than $\mu$, for a true amplitude of unity, is smaller than 5\%,
that is $ \int_{-\infty}^{\mu} G(A-1, \sigma) dA  < 0.05$, or equivalently
$\mu + 1.645 \sigma <1$ where $G$ is the Gaussian function.
%{\it Add a section on Duccio/ Gaelle approach ....}

Before combining the measured amplitudes, the individual central
       values and systematic uncertainties are adjusted, using the same
       procedure as for $\dmd$ measurements, to common values of the $b$-hadron
       fractions (Table \ref{tab:rates}), 
$b$-hadron lifetimes (Section \ref{sec:Atau}), and $\dmd$
       (Equation (\ref{eq:dmdc})).
       The adjustment to a common value of $\fs$ is performed first,
       and needs a special treatment for certain analyses,
       as explained below. Although the final value of $\fs$ is
       calculated under the assumption that $\Bs$ mixing is maximal, studies
       indicate that the effect on the combined $\dms$ result is negligible.

The statistical uncertainty on the measured amplitude is
    expected to be inversely proportional to the $\Bs$ purity
    of the analysed sample \cite{AmpMeth}. This purity is of the order
    of $\fs$ for inclusive analyses. For analyses where a
    full or partial $\Bs$ reconstruction is performed, the purity
    is much less dependent on the assumed value of $\fs$.
    Therefore, the statistical uncertainties on the amplitudes
    measured in inclusive analyses have been multiplied by
$\fs^{\rm used}/\fs^{\rm new}$, where
    $\fs^{\rm used}$ is the $\Bs$ fraction assumed in the
    corresponding analysis, and $\fs^{\rm new}$ is the $\Bs$
    fraction from Table \ref{tab:rates}. This correction is
    also applied on the
    central values of the amplitude measured in inclusive
    analyses, after having checked that the relative uncertainty
    on the amplitude is essentially independent of $\fs$ for
    large enough values of $\dms$.
%The central values and the statistical uncertainties of the measured
% amplitudes have been rescaled, before any other adjustment, with $1/\fs$
%for all inclusive analyses to account for differences between the values
%of $\fs$ used in the different measurements and the value determined
%in Table \ref{tab:rates}. This procedure has been adopted knowing that
%statistical uncertainties on the amplitude are expected to scale
%as $1/\fs$ (\cite{AmpMeth}) and after having verified
%that the relative error on the fitted amplitude is essentially constant with
%$\fs$ for large enough values of \dms.

The amplitudes (measured at each
value of $\dms$) are averaged using the same procedure as for 
the combination of $\dmd$ results.
%taking into account correlated
%systematics using the same procedure as for the combination of $\dmd$
%results. 
In addition to systematics which are correlated within the individual 
experiments, the following sources of correlated systematic uncertainties 
have been taken into account:
\begin{itemize}
\item $b$-hadron lifetime measurements,
\item $b$-quark fragmentation,
\item direct and cascade semileptonic branching fractions of $b$-hadrons,
\item $b$-hadron fractions,
\item $\dmd$ measurements,
\item $\Delta \Gs$ measurements.
\end{itemize}

\begin{figure}[t]
  \begin{center}
    \leavevmode
    \mbox{\epsfxsize12.0cm \epsfysize12.0cm
\epsffile{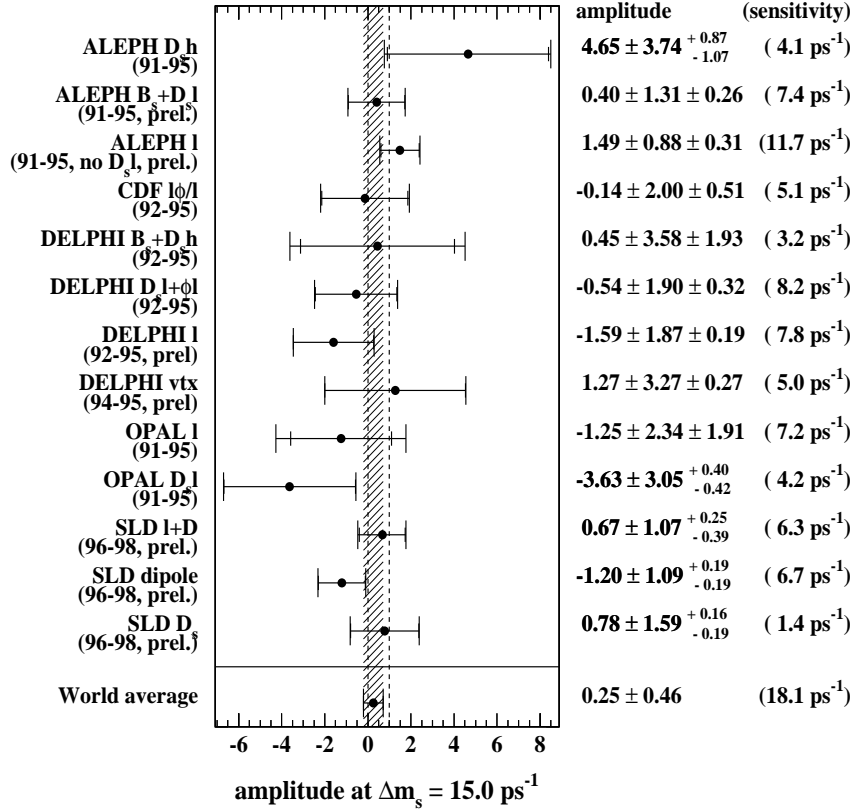}}
%\epsffile{/afs/cern.ch/phys/lep/lepbosc/combined_results/summer_2000_preprint/HF_preprint_plots/amplitude_at_dms15.eps}}
    \caption{{\it Measurements of the \PsBz\ oscillation
      amplitude. 
      The amplitudes are given at $\dms=15$ \ips, along with
      the relevant statistical and systematic errors. The exclusion
      sensitivities are given within parentheses in the column on the right.
      Note that the individual measurements are as quoted in the original
      publications, but the average, corresponding to the hatched area,
 includes the effects of adjustments
      to a common set of input parameters. The dashed lines correspond to 
amplitude values of 0 and 1.} \label{fig:dmsexpts}}

  \end{center}
\end{figure}

Data using the amplitude method have been provided by
ALEPH, DELPHI, OPAL
\footnote{The result released by the OPAL Collaboration
in September 2000, obtained from an analysis of the $\Ds$-lepton channel,
has been included in the combination.},
% (for their lepton-jet analysis 
%only\footnote{The OPAL Collaboration has other results on searches 
%for \PsBz\ oscillations 
%which are not included in this combination 
%(see reference~\cite{O:dms} for details).}),
CDF and SLD.
The
combined amplitude spectrum is shown in
Figure~\ref{fig:dmsampl}.  All values of \dms\ below
\dmslim\ \ips\ are excluded at the~95\% CL.
% while when using method b), the
%limit is lowered to \dmsb\ \ips. 
The exclusion sensitivity
to \dms, defined as the largest value of \dms\
that would have been excluded if the combined amplitudes were zero at
all values of \dms, is \dmssen\ \ips. 
%Using the LEP data alone, the
%combined limit is \ldmslim\ \ips\ using method a), \ldmsb\ \ips\ using 
%method b) and the sensitivity is \ldmssen\ \ips. 
In Figure~\ref{fig:dmsexpts} the values of the amplitude at 
$\dms =15$ \ips\,
from the various analyses, are shown. In addition, the
individual exclusion  sensitivities
are given.

%\mysubsection{Interpretation of the combined amplitude spectrum}

%%% WARNING: the following paragraph contains harcoded numbers !
%The combined amplitude in the region between 13 \ips\ and 17 \ips\ is
%greater than 1.645 standard deviations from zero. A scan in steps of
%0.25 \ips\ indicates that the most significant excursion occurs at 
%14.75 \ips, with a significance of 2.6 standard deviations. This is
%consistent with zero at a one-sided confidence level of 0.4\%. However,
%if the non-physical region corresponding to $A>1$ is excluded from the
%confidence level calculation the excursion remains consistent with $A=0$
%at a level of 3.1\%.

Positive measured amplitudes are found for frequency values near and
above the sensitivity limit. Some values deviate from zero by more than
twice the total estimated error.
The likelihood profile obtained from the amplitude spectrum, using the 
prescription of \cite{AmpMeth}, shows a minimum at 17.2 \ips. The likelihood
value at this minimum is 2.3 units below the asymptotic likelihood
value for $\dms \rightarrow \infty$. Because the measurements at different frequencies
are correlated, it is not possible to calculate analytically the probability 
that,
in the explored frequency range, a fluctuation as or more unlikely occurs in 
a sample where the true frequency is far beyond the sensitivity. Therefore, 
an estimation procedure
based
on fast Monte Carlo experiments has been developed \cite{ref:dega}: the above
probability is found to be around $3\%$.

Note that the combined sensitivity for a ``five-sigma discovery'' of
$\Bs-\Bsb$ oscillations, defined as the value of $\dms$ for which
the uncertainty on the combined amplitude is equal to 0.2, is 8 \ips.

\mysection{Limit on the decay width difference for mass eigenstates
in the $\Bs$-$\Bsb$ system}
\label{sec:deltag} 

%\section {Introduction}
The CKM picture of weak charge-changing transitions predicts that 
the $\Bs$ and its 
charge conjugate mix. This 
results in new states $\Bsh$ and $\Bsl$, with masses $\Msh$ and 
$\Msl$
and (probably) different widths $\Gsh$ and $\Gsl$. 

Neglecting CP violation, the mass eigenstates are also CP eigenstates, the 
$\Bsl$ being 
CP even and $\Bsh$ being CP odd. The decay of a $\Bsb$ meson via the quark 
subprocess $b (\overline{s})\rightarrow c \overline{c} s (\overline{s})$ gives rise to predominantly 
CP-even eigenstates, thus the CP-even eigenstate should have the greater decay rate 
and hence the shorter lifetime.  For convenience of notation, in the following
we therefore substitute 
$\Gsl \equiv \Gss$ and $\Gsh \equiv \Gslg$, and 
define $\Gs=1/\tbs=(\Gslg+\Gss)/2$ and 
$\Delta \Gs = \Gss-\Gslg$.

%{\bf The following lines have been changed as compared to the 
%previous version.}

Theoretical calculations \cite{Beneke2} of the ratio $\dgbs$ at 
next-to-leading order give:
\begin{equation}
\frac{\Delta \Gs}{\Gs} = \left({\frac{{\cal F}_{\Bs}}{230~\MeV}}\right)^2 [0.007\times {\cal B}_b(m_b)+(0.132^{+0.011}_{-0.027}) \overline{{\cal B}_s}(m_b)-(0.078 \pm 0.018)]
\label{eq:dgsth1}
\end{equation} 
where ${\cal F}_{\Bs}$ is the $\Bs$ decay constant,
${\cal B}_b(m_b)$ and 
$\overline{{\cal B}_s}(m_b)=\frac{m^2_{\Bs}{\cal B}_s(m_b) }
{(\overline{m_b}+\overline{m_s})^2}$ 
are bag parameters and where $\overline{m_q}$ are quark masses evaluated in the
$\overline{{\rm MS}}$ scheme at the scale $m_b=4.8~\GeV/c^2$.
The third term in square brackets in Equation (\ref{eq:dgsth1}) is an estimate
of the $1/m_b$ correction in the factorization approximation \cite{ref:bene3}.
Using the values from recent lattice calculations 
\cite{Beneke2,gimenez} of ${\cal B}_b(m_b)=0.9\pm0.1$,  
$\overline{{\cal B}_s}(m_b)=1.25\pm0.10$ and
${\cal F}_{\Bs}=(220\pm30)~\MeV$ (see Section \ref{sec:triangle}) yields 
$\dgbs =0.085^{+0.035}_{-0.046}$, where the dominant uncertainties are from 
${\cal F}_{\Bs}$, from the scale dependence of perturbative QCD corrections 
 and 
from the uncertainty on $1/m_b$ corrections.
% Care should be taken here as the values for 
%the bag constants are preliminary 
%and are correctly normalised at one loop only, 
%(they use renormalization constants in the static 
%theory rather than NRQCD) 
%so there may be some additional systematic 
%uncertainty to be included. 
Due to progress in lattice QCD evaluations, uncertainties on the ${\cal B}$
parameters are no longer dominant.

Assuming that the CKM matrix is unitary, 
%they have
the width difference of the $\Bs-\Bsb$ mass eigenstates can be related to
%related the width difference
%of $\Bs$ mass eigenstates to 
the mass difference of $\Bd-\Bdb$ mass eigenstates
in an expression in which non-perturbative QCD parameters enter through 
ratios, which are better determined than individual 
absolute values of contributing quantities \cite{becir,gimenez}. 
The dependence in the $\Bs$
decay constant also disappears. Assuming that there is no new physics
contributions in present determinations of the CKM parameters (see Section
\ref{sec:triangle}) they obtain:
\begin{equation}
\dgbs = (4.7 \pm 2.2)\%.
\label{eq:martidg}
\end{equation}
The width difference and the mass difference are correlated 
($\Delta \Gs/\dms =\frac{3}{2}\pi \frac{m_b^2}{m_t^2}$ to first approximation
\cite{ref:dgnaif}),
%and  
%detailed calculations yield \cite{Beneke2},
%\begin{equation}
%\frac{\Delta \Gs}{\dms} = 2.75\times 10^{-3}\left[2.13\frac{{\cal B}_s(m_b)}{{\cal B}_b(m_b)}-0.82 \right].
%\label{eq:dgs}
%\end{equation}
%This ratio is independent of the decay constant and depends only on the ratio of bag constants. 
%Using the bag constants from \cite{Hashimoto} yields 
%$\Delta \Gs/\dms = (6.5\pm 2.2) \times10^{-3}$, again with the caveat 
%on the values of the bag constants.
thus offering the 
possibility of measuring $\dms$ via the lifetime difference rather than the 
oscillation frequency. This could be particularly important if the oscillation 
frequency is too fast to be measured with the present experimental 
proper time resolution. In addition, if $\Delta \Gs$ does turn out to be 
sizable, the 
observation of CP violation and the measurement of CKM phases from untagged 
$\Bs$ samples 
can be imagined \cite{Dunietz}. 
Within the present level of uncertainties the two evaluations 
of $\dgbs$, given in Equations (\ref{eq:dgsth1}) and (\ref{eq:martidg}), 
are compatible
and progress is needed to reduce uncertainties related to the scale
dependence of perturbative QCD and to $1/m_b$ corrections before these
theoretical evaluations can be used to set a limit on $\dms$. 

%{\bf End modifs. }

The existing experimental constraints on the width difference and the 
combination of these constraints 
are reported here\footnote{The present members of the $\Delta \Gs$ working group are:
P. Coyle, D. Lucchesi, S. Mele, F. Parodi and P. Spagnolo.}.

%\newpage

\mysubsection{Experimental constraints on $\Delta \Gs/\Gs$}
Experimental information on $\Delta \Gs$ can be extracted by studying the 
proper time distribution of data samples enriched in $\Bs$ mesons. 
An alternative method based on  
measuring the branching fraction ${\rm B}_s \rightarrow{\rm D} _s^{(*)+}{\rm D} _s^{(*)-}$ has also
been proposed  \cite{ALEPH-phiphi}. 
The available results are summarised in 
Table \ref{tab:dgammat}. 
The values of the limit on $\dgs/\Gs$ quoted in the last column of this 
table have been obtained by the working group.
%~\ref{tab:dgammat}. 

Methods based on double exponential lifetime fits to samples containing a 
mixture of CP eigenstates have a quadratic sensitivity to $\Delta \Gs$ 
(inclusive, semileptonic, $\Ds$-hadron), whereas methods based on isolating
a single CP eigenstate have a linear dependence on $\Delta \Gs$ ($\phi\phi$,
${\rm J}/\psi\phi$). The latter are therefore, in principle, more sensitive to
$\Delta \Gs$; but they tend to suffer from reduced statistics.

In order to obtain an improved limit on $\Delta \Gs$,
the results based on fits to the proper time distributions 
are used to apply a constraint on the allowed 
range of $1/\Gs$. The world average $\Bs$ lifetime is not used, as its meaning 
is not clear if
%As the meaning of the ``average'' $\Bs$ lifetime is not well defined if 
$\Delta \Gs$ is non-zero.
% this constraint is not used. 
Instead, it is chosen to constrain $1/\Gs$
 to the world average $\tbd$ lifetime 
($1/\Gs \equiv 1/\Gd=\tbd= (\taubd)$~ps).
This is well motivated theoretically, as 
the total widths of the $\Bs$ and $\Bd$ mesons
%$\tbd$ and $\tbs$ 
are expected to be 
equal within less than one percent~\cite{ref:bigilife}, \cite{Beneke}
and $\Delta \Gd$ is expected to be small. 
  
Further information on the various individual measurements is now given.

\begin{table}
\begin{center}
{\small
% Insert minipage to keep footnote in plot
    \begin{minipage}{\linewidth}
% Use footnote symbols instead of numbers and increment to get dagger.
    \renewcommand{\thempfootnote}{\fnsymbol{mpfootnote}}
    \addtocounter{mpfootnote}{1}
\begin{tabular}{|l|c|c|c|} 

\hline
Experiment & Selection        & Measurement            & $\Delta \Gs/\Gs$ \\ 
\hline

L3~\cite{L3B01}         & inclusive $b$-sample              &                               & $<0.67$         \\

DELPHI~\cite{DELBS0}     & $\Bsb\rightarrow \Dsp  \ell^- \overline{\nu_{\ell}} X$ & $\tbssemi=(1.42^{+0.14}_{-0.13}\pm0.03)$~ps  & $<0.46$ \\

OTHERS~\cite{ref:others}& $\Bsb \rightarrow \Dsp \ell^-  \overline{\nu_{\ell}} X$  & $\tbssemi=(1.46\pm{0.07})$~ps & $<0.30$ \\

ALEPH~\cite{ALEPH-phiphi}      & $\Bs \rightarrow\phi\phi X$      & 
${\rm BR}(\Bssh \rightarrow {\rm D}_s^{(*)+} {\rm D}_s^{(*)-}) =(23\pm10^{+19}_{-~9})\%$       & $0.26^{+0.30}_{-0.15}$ \\ 

ALEPH~\cite{ALEPH-phiphi}      & $\Bs \rightarrow\phi\phi X$      & $\tbsshort=(1.27\pm0.33\pm0.07)$~ps           & 
$0.45^{+0.80}_{-0.49}$ \\

%CDF~\cite{CDF-Dsl}        & $\Bsb \rightarrow \Dsp \ell^- \overline{\nu_{\ell}}  X$ & $\tbssemi=(1.36\pm0.09^{+0.06}_{-0.05})$~ps  & $<0.83$ \\ 

DELPHI~\cite{DELBS0}$^\dag$    & $\Bsb \rightarrow \Dsp$ hadron     &  $\tbsdh=(1.53^{+0.16}_{-0.15}\pm0.07)$~ps                          & $<0.69$         \\

CDF~\cite{CDFB01}        & $\Bs \rightarrow {\rm J}/\psi\phi$        & $\tbspsi=(1.34^{+0.23}_{-0.19}\pm0.05)$~ps & $0.33^{+0.45}_{-0.42}$ \\ 

\hline
\end{tabular}
 $^\dag$ The value quoted for the measured 
lifetime differs
slightly from the one quoted in Table \ref{tab:bs} because it corresponds
to the present status of the analysis in which the information
on $\dgs$ has been obtained.\\
  \end{minipage}
}
\caption{{\it Experimental constraints on $\Delta \Gs/ \Gs$. The upper limits,
which have been obtained by the working group, are quoted at the~95~\%C.L.}
\label{tab:dgammat}}
%\vspace{0.5cm}
\end{center}
\end{table}
\begin{itemize}

\item {\it L3 inclusive $b$-sample:}
in an unbiased inclusive B sample, all $\Bs $ decay modes are measured, 
including decays into CP eigenstates. An equal number of $\Bssh$ and $\Bslg$ 
mesons 
are therefore selected, and the proper time dependence of the $\Bs$ signal is 
given by:
\begin{equation} 
P_{incl}(t) = \frac{1}{2}(\Gslg \exp{(-\Gslg t)}+\Gss \exp{(-\Gss t)}).
\end{equation}
If the proper time dependence of this sample
is fitted assuming only a single exponential lifetime
then, using the definitions of $\Gs$ and $\Delta \Gs$
and assuming that $\Delta \Gs/\Gs$ is small,
the measured lifetime is
 given by:
\begin{equation} 
\tbsinc = \frac{1}{\Gs} \frac{1}{1- \left (\frac{\Delta \Gs}{2\Gs}\right )^2}.
\end{equation}
%i.e. the measured $\Bs $ lifetime is expected to be longer than the true lifetime  
%if $\Delta \Gs$ is non-zero.

L3 effectively incorporates $P_{incl}(t)$ into the proper time fit of an inclusive $b$-sample and 
applies the constraint $1/\Gs=(1.49\pm0.06)$~ps. 
%They obtain $\Delta \Gs/\Gs <0.67$ at 95\%~C.L., 
%after inclusion of systematic uncertainties.
 
\item{\it DELPHI $\Bsb \rightarrow {\rm D}^+_s \ell^- \overline{\nu_{\ell}} X$:}
in a semileptonic selection, the ratio of short and long $\Bs $'s 
selected is proportional to the ratio of the decay widths and the proper time dependence of the 
signal is:
\begin{equation}  
 P_{semi}(t) =   \frac{\Gss~ \Gslg}{(\Gss+\Gslg)} 
(\exp{(-\Gss t)}+\exp{(-\Gslg t)}).
\end{equation}
If this sample is fitted assuming only a single exponential lifetime for the $\Bs $, then the 
measured lifetime is, always in the limit that $\Delta \Gs/\Gs$ is small, 
given by: 
\begin{equation}
\tbssemi = \frac{1}{\Gs} \frac{{1 + \left (\frac{\Delta \Gs}{ 2\Gs}\right )^2}}
{{1- \left (\frac{\Delta \Gs}{2\Gs}\right )^2}}.                  
\label{eqsemi}
\end{equation}
The single lifetime fit is thus more sensitive to the effects of 
$\Delta \Gs$ in the semileptonic than in the fully inclusive case.
%DELPHI measures $\tbssemi=(1.42^{+0.14}_{-0.13}\pm0.03)$~ps. 
Information on 
$\Delta\Gs$ is obtained by scanning the likelihood as a function the two parameters
$1/\Gs$ and $\Delta \Gs/\Gs$ and applying the $\tbd$ constraint.
%, they 
%obtain the upper limit $\dgbs<0.46$ at 95\% C.L.  

\item {\it OTHERS:}
other analyses of the $\Bs $ semileptonic lifetime \cite{ref:others}
have not explicitly 
considered the possibility 
of a non-zero $\Delta \Gs$ value.
Nevertheless, the fact that the single exponential lifetime 
for this case 
(Equation~(\ref{eqsemi})) is sensitive to $\Delta \Gs$ allows information on 
$\Delta \Gs$ to be extracted. 
The average $\Bs $ semileptonic lifetime has been recalculated
excluding the DELPHI result just discussed, and information on 
$\Delta\Gs$ has been obtained by using Equation~(\ref{eqsemi}) and applying
the $\tbd$ constraint.
%%is $\tbssemi=(1.46 \pm 0.07)$~ps. Using Equation~(\ref{eqsemi}) and applying the $\tbd$ constraint 
%this implies $\Delta \Gs/\Gs <0.30$ at 95\%~C.L.
The validity of Equation~(\ref{eqsemi}) in the presence of background contributions has 
been verified using a toy Monte Carlo~\cite{Moser}.

\item {\it ALEPH $\Bs  \rightarrow \phi \phi X$ (counting method):}
only those $\Bs $ decays which are CP eigenstates can contribute to a width difference 
between the CP even and CP odd states. An analysis~\cite{Aleksan} of such decays 
shows that $\Bs  \rightarrow D_s^{(*)+} D_s^{(*)-}$ is by far the dominant
contribution and is almost 100\% CP even. Under this assumption,  
$\Delta \Gs = \Gss (\Bssh \rightarrow  {\rm D}_s^{(*)+} {\rm D}_s^{(*)-})$
%and can be related to the branching fraction of
%$\Bssh \rightarrow {\rm D}_s^{(*)+} {\rm D}_s^{(*)-}$ 
%through: 
where:
\begin{equation}
{\rm BR}(\Bssh \rightarrow {\rm D}_s^{(*)+} {\rm D}_s^{(*)-})= 
\frac{\Gss (\Bssh \rightarrow  {\rm D}_s^{(*)+} {\rm D}_s^{(*)-})} {\Gsh} = 
\frac{\Delta \Gs}{\Gs (1+\frac{\Delta \Gs}{2 \Gs}) }. 
\end{equation}
ALEPH~\cite{ALEPH-phiphi} has measured this branching fraction, and
%${\rm BR}(\Bssh \rightarrow {\rm D}_s^{(*)+} {\rm D}_s^{(*)-})= (20 \pm 7(stat.)^{+19}_{-10}(sys.)) \%$, 
%which corresponds to $\frac{\Delta \Gs} {\Gs} = (22^{+22}_{-13})\%$.
this is the only constraint on $\Delta \Gs/\Gs$ which does not rely on 
a measurement of the average $\Bs$($\Bd$) lifetime.
%of the lifetime of the other CP eigenstate.
%the 
%assumption $\tbs \equiv \tbd$. 

\item {\it ALEPH $\Bs  \rightarrow \phi \phi X$ (lifetime method):}
as the decay $\Bs  \rightarrow D_s^{(*)+} D_s^{(*)-} \rightarrow \phi \phi X$, 
is  predominantly CP even,
the proper time dependence of the $\Bs $ component of the 
$\phi \phi$ sample is 
therefore just a simple exponential with the appropriate lifetime: 
\begin{equation} 
P_{single}(t)=\Gss \exp(-\Gss t).
\end{equation}
This lifetime is related to $\dgs$ and $\Gs$ via the expression:
\begin{equation}  
\frac{\Delta \Gs}{\Gs}=2(\frac{1}{\Gs~\tbsshort}-1).  
\end{equation} 
%In order to be independent of the assumption $\tbs \equiv \tbd$, ALEPH uses equation~1 to 
%reexpress $\tau_{B_s}$ in terms of $\tau^{semi}_{B_s}$ and $\dgbs$. 
ALEPH~\cite{ALEPH-phiphi} has measured 
the lifetime of $\phi\phi X$ events and extracted information on 
$\Delta \Gamma_s/\Gamma_s$ with the help of the world average
$\Bs$ lifetime obtained from semileptonic $\Bs$ decays.
The result listed in Table \ref{tab:dgammat} has been obtained 
by the working group with the assumption
$1/\Gs \equiv \tbd$.
% $\tau_{B_s^{short}}=(1.42\pm0.23\pm0.16)$~ps which, when
%combined with the assumption $1/\Gs \equiv \tbd$, yields 
%$\Delta \Gamma_s/\Gamma_s =0.20^{+0.67}_{-0.42}$.

\item {\it DELPHI inclusive ${\rm D}_s^+$:}
a fully inclusive $\Dsp$ selection is expected to have 
an increased CP-even content, as the $\Bs  \rightarrow {\rm D}_s^{(*)+} {\rm D}_s^{(*)-}$ 
contribution is enhanced by the selection criteria. 
If $f_{DsDs}$ is the fraction of ${\rm D}_s^{(*)+} {\rm D}_s^{(*)-}$ in the sample, 
then the proper time dependence is expected to be: 
\begin{equation} 
P(t)_{D_s-had.}= f_{DsDs}P_{single}(t)+ (1-f_{DsDs}) P_{semi}(t). 
\end{equation}

For the DELPHI analysis
%a $\Bs $ lifetime of $\tbsdh=(1.53^{+0.16}_{-0.15} \pm 0.07)$~ps is 
%measured and 
a value $f_{DsDs}=(22\pm7)\%$ is estimated from simulation.  
Scanning the likelihood as a function $1/\Gs$ and $\Delta \Gs/\Gs$ and 
applying the $\tbd$ constraint yields an upper limit  
on $\Delta \Gs/\Gs$.
% < 0.69$ at 95\%~C.L.

\item{\it CDF $\Bs  \rightarrow {\rm J}/\psi \phi$: } 
the final state $\Bs  \rightarrow {\rm J}/\psi \phi$ is thought to be predominantly CP even 
(i.e. measures mainly $\tbsshort$) \cite{Aleksan}.
An update \cite{CDF-angles} of the CDF measurement of the polarization 
in $\Bs  \rightarrow {\rm J}/\psi \phi$ decays measures the fraction of CP even in the final state 
to be $f_{short}=(79\pm19)\%$ and supports this expectation. 
For this case, the proper time dependence of the $\Bs $ component of the sample is: 
\begin{equation} 
P_{{\rm J}/\psi \phi}(t)=f_{short}P_{single}(\Gss, t)+(1-f_{short})P_{single}(\Gslg,t).
\end{equation}
%and the corresponding lifetime expected from a single exponential fit is 
%$$
%\tau_{B^{\psi \phi}_s} = ??
%$$
CDF measures the lifetime of ${\rm J}/\psi \phi$ events and information on
$\dgbs$ is obtained after applying the $1/\Gs \equiv \tbd$ constraint
%$\tbspsi=(1.34^{+0.23}_{-0.19}\pm0.05)$~ps which gives 
%$\dgbs=0.33^{+0.45}_{-0.42}$ after applying the $1/\Gs \equiv \tbd$ constraint
and including the experimental uncertainty on $f_{short}$.

\end{itemize}

\mysubsection {Combined limit on $\dgs$}

\begin{figure}
\begin{center}
\resizebox{0.95\textwidth}{!}{\includegraphics{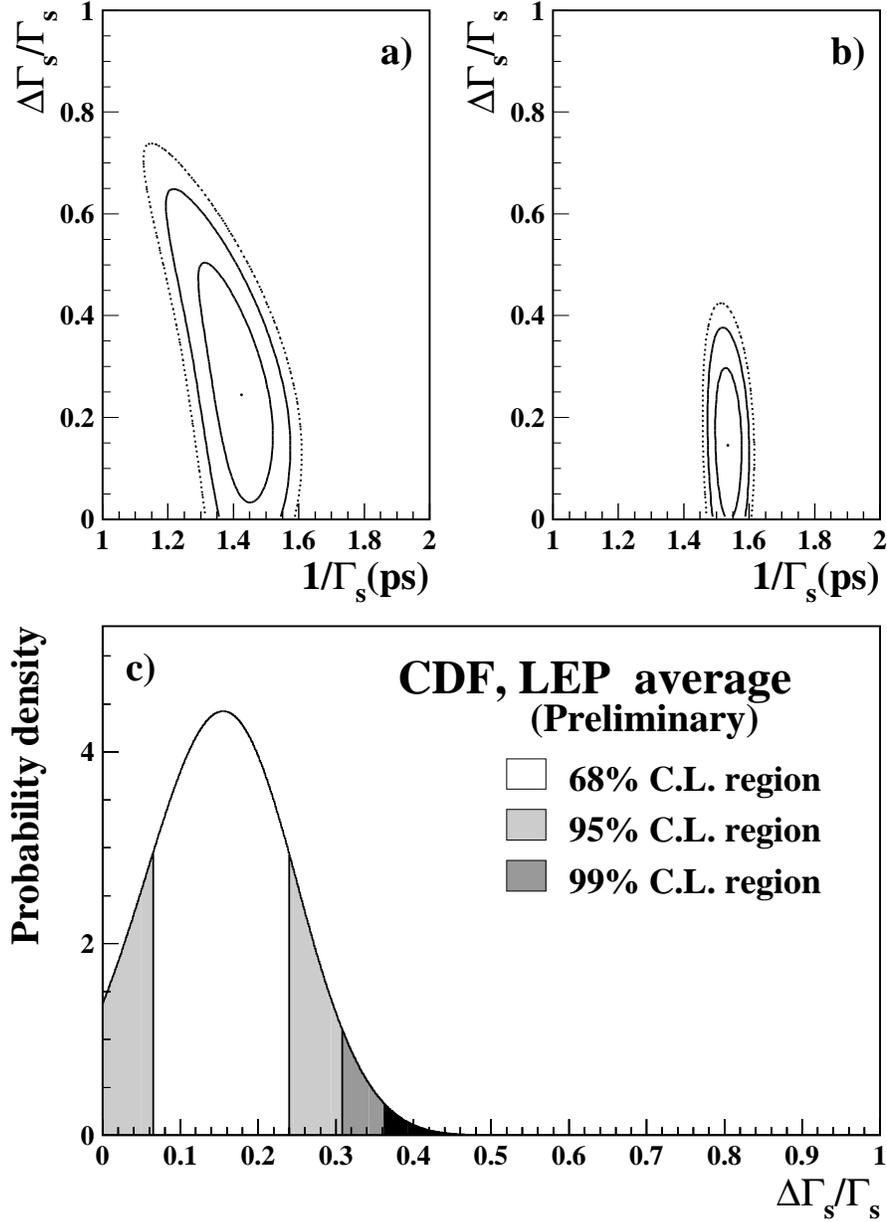}}
\end{center}
\caption[]{{ \it
          a) 68\%, 95\% and 99\% C.L. contours of the negative log-likelihood 
             distribution in the plane $1/\Gs$-$\dgbs$.
          b) Same as a) but with the constraint $1/\Gs \equiv\tau_{\Bd}$
supposed to be exact.
          c) Probability density distribution for $\dgbs$ after applying the constraint; 
             the three shaded regions show the limits at the 68\%, 95\% and 
99\% C.L. respectively.} \label{fig:dgplot}}

\end{figure}

In order to combine the analyses summarised in Table \ref{tab:dgammat},
%\ref{tab:dgammat}, 
the result of each analysis has been converted 
to a two-dimensional log-likelihood in the ($1/\Gs$,~$\dgbs$) plane. This log-likelihood has 
either been provided by each experiment or reconstructed from the measured lifetimes using 
the expected dependence of this quantity on $1/\Gs$ and $\dgbs$.
The latter procedure was necessary for the OTHERS and CDF entries
of Table \ref{tab:dgammat}.
%\ref{tab:dgammat}. 
The L3 analysis is not included in the
average as the two-dimensional likelihood was not provided and 
could not be reconstructed from
the available information. The $\tbd$ constraint is not applied
on $1/\Gs$ at this stage.
%No constraint is therefore applied on $1/\Gs$ at this stage. 
Systematic uncertainties are 
included in the individual log-likelihood distributions. 

The log-likelihood distributions  have been summed and 
the variation of the global negative log-likelihood function has been 
measured with respect to its minimum ($\Delta {\cal L}$). 
%The L3 analysis is not included as the 
%two-dimensional likelihood is not yet available.
The 68\%, 95\% and 99\% C.L. contours of the combined negative 
log-likelihood are shown in Figure~\ref{fig:dgplot}a.
%A significant correlation exists between $1/\Gs$ and $\dgbs$ indicating that 
%the current combination of the analyses doesn't have sufficient 
%sensitivity to 
%measure $1/\Gs$ and $\dgbs$ simultaneously.
The corresponding limit on $\dgbs$ is:
\begin{eqnarray}
\nonumber   \dgbs & = & 0.24^{+0.16}_{-0.12}  \\
   \dgbs & < & 0.53~{\rm at~the~} 95\%~{\rm C.L.} 
\end{eqnarray}

An improved limit on $\dgbs$
can be obtained by applying the $\tbd = (\taubd)~{\rm ps}$ constraint. 
When expressed as 
a probability density, this constraint is:
\begin{eqnarray}
\begin{array}{lll}
  f_C(1/\Gs) & = & 
   \frac{1}{\sqrt{2\pi} \sigma_{\tbd}} \exp \left(-\frac{(1/\Gs-\tbd)^2} {2\sigma^2_{\tbd}} \right). 
\end{array}
\end{eqnarray}
%The value of the uncertainty $\sigma_{\tbd}$ includes
%the theoretical uncertainty on the equality $1/\Gs=\tbd$,
%expected to be of the order of 1$\%$ or less.
Using Bayes theorem, the probability density for $\dgbs$, 
with the constraint 
\footnote{A flat, a priori probability density distribution has been assumed
for $\dgbs$.} applied, is obtained 
by convoluting $f_C(1/\Gs)$ with the 2-D probability density for $1/\Gs$ and $\dgbs$ 
(${\cal P}(\tbs,\dgbs)$), and normalizing the result to unity:
\begin{eqnarray}
   {\cal P}(\dgbs) = { {\int {\cal P}(1/\Gs,\dgbs) f_C(1/\Gs) d(1/\Gs) } \over 
    {\int {\cal P}(1/\Gs,\dgbs) f_C(1/\Gs) d(1/\Gs) d\dgbs } }  
\end{eqnarray}
where ${\cal P}(1/\Gs,\dgbs)$ is proportional to $\exp(-\Delta {\cal L})$. 
This probability has been normalized by taking only positive values of
$\dgbs$.
The two-dimensional log-likelihood obtained, after including the constraint,
supposed to be exact, is shown in 
Figure~\ref{fig:dgplot}b. The resulting probability density distribution for $\dgbs$ is 
shown in Figure~\ref{fig:dgplot}c. The corresponding limit on $\dgbs$ is:
\begin{eqnarray}
\nonumber   \dgbs & = & 0.16^{+0.08}_{-0.09}  \\
            \dgbs & < & 0.31~{\rm at~the}~ 95\%~{\rm C.L.} 
\end{eqnarray}

If an additional 2$\%$ uncertainty, assumed to be Gaussian distributed, 
is incorporated to account for the theory assumption $\tbd =1/\Gs$,
the effect on the result is small:
\begin{eqnarray}
\nonumber   \dgbs & = & 0.17^{+0.09}_{-0.10}  \\
  \dgbs & < & 0.32~{\rm at~ the}~95\%~{\rm C.L.} 
\end{eqnarray}

%Using the theoretical prediction $\Delta \Gamma_{\Bs}/\dms = (6.47^{+??}_{-??}) \times 10^{-3}$
%a constraint on $\dms$ can also be obtained:
%\begin{center}
%   $\dms=(??^{+??}_{-??})~ps^{-1}$ \\
%   \mbox{$??<   \dms < ??~ps^{-1}$~at the~95\% C.L.} 
%\end{center}
%As discussed previously, this indirect constraint on $\dms$ depends crucially on the assumed values 
%of the bag constants and relies on the assumption $\tbs \equiv \tbd$.
 
%At present, Equation (\ref{eq:dgs}) has not been used to set a limit on
%$\dms$ because of unstable results quoted for the ratio of the 
%bag parameters \cite{dgs2, hashimoto}.
%Given the current status of the theoretical estimates of the bag parameters, 
%it has been decided not to quote a corresponding constraint on the range 
%of $\dms$ values.

\mysection{Average number of $c$ and $\overline{c}$ quarks produced in
$b$-hadron decays}
\label{sec:secccbar}
%{\bf This is a new section}

The measured value of the $b$-hadron semileptonic decay branching fraction
is on the low side of theoretical expectations
\cite{neubsc1}. One way to
reconcile theory with experiment consists in assuming that the 
$c$-quark effective mass is lower than used in these evaluations.
Such considerations imply also that decays of the type
$b \rightarrow c \cbar s(d)$ correspond to a larger 
decay rate. The average number of $c$ and $\cbar$ quarks contributing
in $b$-hadron decays is thus negatively correlated with the expected value
for ${\rm BR}(b \rightarrow \ell {\rm X})$ and a
 simultaneous measurement of these two quantities may help
to clarify the theoretical picture.
Another way to lower the expected semileptonic branching fraction of 
$b$-hadrons consists in adding new physics contributions to the
$b \rightarrow s g$ transition \cite{kaganoc}.
This possibility has been constrained by the measurement of the decay
rate of $b$-hadrons with no-open charm in the final state.

Experimentally, the number of $c$ and $\cbar$ quarks contributing in
$b$-hadron decays can be obtained by measuring the production fractions
of charmed hadrons and charmonium states. 
Measurements originate from four sources:
\begin{itemize}
\item open-charm counting using exclusive signals of reconstructed charmed 
hadrons,
\item charmonium production,
\item inclusive measurements of the distribution of charged
track impact parameters and of secondary vertices allowing the determination of 
$b \rightarrow {\rm D}\overline{{\rm D}} {\rm X}$ and 
$b \rightarrow {\rm 0{\rm D}~charm}$ decay rates,

\item $b \rightarrow {\rm D}_i\overline{{\rm D}}_j {\rm X}$ branching fraction
measurements in which $\overline{{\rm D}}_j$, and sometimes ${\rm D}_i$,
are completely reconstructed.

\end{itemize}
Measurements done at LEP and at the $\Yfs$ have been analyzed 
using the same values for external parameters and the same hypotheses
to account for missing measurements. Consequently, present results
may differ from previous determinations \cite{ref:ncccount1,ref:ncccount3}
(see Section \ref{sec:countfive}).
%Rather similar values have been reported in \cite{ref:ncccount2}.

\mysubsection{Open-charm counting in $b$-hadron decays}

Exclusive signals of $\Dn$, $\Dp$, $\Ds$, $\Lc$ and $\Xic$ charmed hadrons
and of their anti-particles, produced in $b$-hadron decays, have been
obtained by the ALEPH \cite{ref:alinc}, CLEO \cite{ref:cleinc,ref:cleinc3},
 DELPHI \cite{ref:delinc} and OPAL \cite{ref:opainc} Collaborations.

%Tables \ref{tab:bxcbr1} and \ref{tab:bxcbr2} give the 
Values for the products 
of the production
rates for a given charmed or anti-charmed hadron by its decay branching
fraction into the reconstructed final state:
${\rm P}(b \rightarrow {\rm X}_{c}~{\rm or}~\overline{{\rm X}_{c}}) \times 
{\rm BR}({\rm X}_{c} \rightarrow {\rm X})$ are given in Appendix 
\ref{appendixAe}.
LEP results obtained by DELPHI and OPAL are quoted using the value
of ${\rm R}_b$ given in Table \ref{tab:gensys} whereas ALEPH measurements
are independent of ${\rm R}_b$.

To account for correlations between the different measurements
when evaluating averaged values,
systematic uncertainties have been split into three categories:
those which are typical of the considered measurement,
those which are common to several  channels within a given experiment
and those which are common to several experiments. 
%Uncertainties from the last category are included in the previous one.

 LEP and $\Yfs$ measurements have been averaged separately.
Using the values of the decay
branching fractions for charmed particles listed in Table \ref{tab:gensys},
%the measured open-charm production in $b$-hadron decays is the following:
%\begin{equation}
%n(\Dn+\Dp+\Ds+\Lc+~{\rm anti-particles})_{LEP}=1.093 \pm .
%\end{equation}
%\begin{equation}
%n(\Dn+\Dp+\Ds+\Lc+~{\rm anti-particles})_{CLEO}=1.062 \pm .
%\end{equation}
production rates of the different charmed hadrons are 
then obtained and given in Table
\ref{tab:ddbrates}. 
\begin{table}
\begin{center}
\begin{tabular}{|l|c|c|} \hline
Decay channel   & LEP & CLEO\\
 \hline
${\rm BR}(b \rightarrow \Dn,\Dnb {\rm X})$ & $(59.3 \pm 2.3 \pm1.4) \%$
&$(64.9 \pm 2.4 \pm 1.5)\%$ \\
${\rm BR}(b \rightarrow \Dp,\Dpb {\rm X})$ & $( 22.5\pm 1.0 \pm1.5) \%$
&$(24.0 \pm 1.3 \pm 1.6)\%$ \\
${\rm BR}(b \rightarrow \Ds,\Dsb {\rm X})$ & $( 17.3\pm 1.1 \pm 4.3) \%$
&$(11.8 \pm 0.9 \pm 2.9)\%$ \\
${\rm BR}(b \rightarrow \Lc,\Lcb {\rm X})$ & $(10.2 \pm 1.0 \pm2.7) \%$
&$(5.5 \pm 1.3 \pm 1.4)\%$ \\
\hline

\end{tabular}
\end{center}
\caption{{\it Measured production rates of charmed hadrons
in $b$-hadron decays. The second quoted uncertainty is due to
the error on charmed hadrons branching fractions.} \label{tab:ddbrates}}
\end{table}
Because of $\Bsb$ and $b$-baryon production in $b$-quark jets at high energy, 
it is expected that
branching fractions into $\Ds$ and $\Lc$ are             
larger at LEP than at the 
$\Yfs$, as observed.

To obtain the total number of $c$ and $\cbar$ quarks produced
 in $b$-hadron decays
these values have to be corrected for:
\begin{itemize}
\item unmeasured open-charm states such as strange-charm baryons and
\item charmonium production.
\end{itemize}

\mysubsubsection{Strange-charm baryon production in $b$-hadron decays}
There are currently no absolute 
measurement of any strange-charm baryon
branching fractions. As a result, the evaluation of the production rates 
of these baryons has 
to rely on models. Production of these states have been observed
by CLEO \cite{ref:xiccleo} in $b$-meson decays. 
They obtain:
\begin{equation}
\overline{\rm BR}(\overline{\rm B}\rightarrow \Xi_c^0,\overline{\Xi}_c^0 {\rm X})=(2.4 \pm 1.3)/f^0~\%,~
{\rm BR}({\rm B}\rightarrow \Xi_c^+,\overline{\Xi}_c^- {\rm X})=(1.5 \pm 0.7)/f^+~\%.
\end{equation}
The quoted uncertainties correspond only to statistical errors and the 
quantities $f^q$ are the fractions of semileptonic decays of $\Xi_c^q$
baryons containing a $\Xi$,
which are unknown at present and usually
taken to be equal to unity. 
Comparing the central value of the total $\Xi_c+\overline{\Xi}_c$ 
rate (3.9$\%$) with 
the corresponding measurement
of $\Lc+\Lcb$ production (5.5$\%$), using results quoted in Table
\ref{tab:ddbrates},
it appears that the production rate of strange-charm baryon can be much
larger than naively expected. This would imply that there is a mechanism
that enhances the production of $\Xic$ states but searches for
this process have not yet been successful as explained in the following.

Two production mechanisms can be considered for strange-charm baryons
in $b$-decays. 
In the first one, a strange diquark-antidiquark
%(s \overline{s})$ 
pair is created from the vacuum,
with the probability $f_s$, giving rise to a strange-charm baryon in
place of a $\Lc$.
As there are two possible weakly decaying strange-charm baryon states,
the production rate can be evaluated to be of the order of:
\begin{equation}
{\rm BR}_1(b \rightarrow \Xic,\Xicb X)= \frac{2 f_s}{1-2 f_s} \times 
{\rm BR}(b \rightarrow \Lc,\Lcb X).
\end{equation}
Using $f_s=0.10$ with an (arbitrary) uncertainty of 50$\%$, this gives:
\begin{equation}
{\rm BR}_1(b \rightarrow \Xic,\Xicb{\rm X})= (0.25 \pm 0.12)
{\rm BR}(b \rightarrow \Lc,\Lcb X).
\end{equation}

The second mechanism has been invoked for decays
in which the virtual W$^-$ couples to a $(\cbar s)$ pair. The $s$
quark can form a strange-charm baryon with the $c$-quark from the
$b \rightarrow c {\rm W}^-$ decay. A possible,
but indirect, signature for this mechanism
consists of the presence of $\Lcb$ in 
$\overline{{\rm B}}$ meson decays as observed by CLEO \cite{ref:xiccleo}.
They 
measure 
\footnote{The published value for this ratio ($0.19 \pm 0.14 \pm 0.04$)
has been corrected using the present value for the oscillation parameter of
$\Bd$ mesons.} the ratio:
\begin{equation}
{\rm R}_{\Lc}=\frac{
{\rm BR}(\overline{{\rm B}} \rightarrow \overline{\Lambda_c}^- {\rm X})}
{{\rm BR}(\overline{{\rm B}} \rightarrow \Lc {\rm X})}=0.17\pm0.13
\end{equation}
It is then possible to evaluate the fraction of $\Lcb$
baryons produced in this
mechanism, relative to the inclusive $\Lc$ and $\Lcb$
production rate:
\begin{equation}
{\rm BR}(\overline{{\rm B}} \rightarrow \Lcb {\rm X})=\frac{{\rm R}_{\Lc} \times
{\rm BR}(\overline{{\rm B}}\rightarrow \Lc,\Lcb {\rm X})}{1 +{\rm R}_{\Lc} }.
\label{eq:lcbar}
\end{equation}
 To evaluate, from this result, the fraction of events with a $\Xic$ baryon
is not direct because a $\Lcb$ can be 
produced without an accompanying strange-charm baryon and also
a strange-charm baryon, formed with the $cs$ quark pair, may not be
accompanied by a $\Lcb$. It has been assumed, 
in the following, that $\Xi_c$ production from the second mechanism
is similar to $\Lcb$ production giving:
%that every wrong-sign $\overline{\Lambda_c}^-$
%is accompanied by a $\Xi_c$ baryon; this gives:
\begin{equation}
{\rm BR}_2(\overline{{\rm B}}\rightarrow \Xic,\Xicb{\rm X})= 
(0.15\pm0.10)
{\rm BR}(\overline{{\rm B}} \rightarrow \Lc,\Lcb {\rm X}).
\end{equation}

The total production rate of strange-charm baryons, in $b$-meson decays,
 is then evaluated, summing the two contributions and inflating
the errors, to be:
\begin{equation}
{\rm BR}(\overline{{\rm B}}\rightarrow \Xic,\Xicb{\rm X})=(0.4 \pm 0.3)
{\rm BR}(\overline{{\rm B}} \rightarrow \Lc,\Lcb {\rm X})
\end{equation}
In the absence of accurate results on $\Xic$ production in $b$-meson decays
owing to the small statistical evidence for the signals and
the indetermination of the decay branching fractions of these states,
the quoted central value and uncertainty are supposed to accomodate
the various possible scenarios for $\Xic$ production.

For $b$-hadrons produced in the hadronization of $b$-quark jets, 
an additional source of strange-charm baryons can originate from 
$\Xib$ decays.
It will be assumed that:
\begin{equation}
{\rm BR}(\Xib \rightarrow \Xic) =
{\rm BR}(\Lb \rightarrow \Lc).
\end{equation}
and the value of ${\rm BR}(\Lb \rightarrow \Lc)$ is obtained by comparing the 
inclusive production rates of $\Lc$ measured at CLEO and LEP given in Table 
\ref{tab:ddbrates}:
\begin{equation}
{\rm BR}(\Lb \rightarrow \Lc) =\frac{1}{f_{\Lb}}
[{\rm BR}(b \rightarrow \Lc,\Lcb)_{LEP}-(1-f_{\Lb}){\rm BR}
(b \rightarrow \Lc,\Lcb)_{CLEO}]
\end{equation}
$f_{\Lb}$ is the fraction of $b$-baryons produced in a $b$-jet,
as measured in Section \ref{sec:results}.

\mysubsubsection{Average result for open charm production}
Adding $\Xic$ particles, open charm production measured at LEP and CLEO are
respectively:
\begin{equation}
n({\rm open-charm})_{LEP}=1.130 \pm 0.039 \pm 0.063 .
\end{equation}
\begin{equation}
n({\rm open-charm})_{CLEO}=1.084 \pm 0.041 \pm 0.045.
\end{equation}
where the last uncertainty comes from $c$-hadron 
branching fraction measurements.
Including correlated systematics between the two results which mainly
originate from $c$-hadron branching fractions, the average production
of open-charm in $b$-hadron decays is:
\begin{equation}
n({\rm open-charm})=1.095 \pm 0.059.
\label{eq:openlepcleo}
\end{equation}
The dominant uncertainties originate from the poorly known decay
branching fractions of the $\Ds$, $\Lc$ and $\Xic$ charmed hadrons.

\mysubsection{Charmonium production}
Bound $(c \cbar)$ states (charmonium) contribute also to
charm quark production in $b$-hadron decays.
J/$\psi$ production is measured to be similar at LEP \cite{PDG00} and at the 
$\Yfs$ \cite{ref:cleoccb}:
\begin{equation}
{\rm BR}({\rm B} \rightarrow {\rm J}/\psi {\rm X})=(1.16 \pm 0.10)\%(LEP),~
=(1.15 \pm 0.06)\% (CLEO)
\end{equation}
and in the following it has been assumed that charmonium production
is the same for all $b$-hadrons. 

%{\bf There one can let open the possibility that the charmonium production bR 
%is proportional to the lifetime of the considered b-hadron to be in-line with
% what is said in the lifetime section. This means that one can add
%an additional uncertainty to leave open this possibility. 
%In anycase this is a small effect. }
Combining LEP and CLEO results
for J/$\psi$ and $\psi^{\prime}$ production and using only 
CLEO \cite{ref:cleochi2}
measurements
for the $\chi_c^1$ and $\chi_c^2$ rates, measured branching fractions have 
been given in Table \ref{tab:charmvu}.
\begin{table}
\begin{center}
\begin{tabular}{|l|c|c|c|} \hline
Decay channel   & measured rate ($\%$)& direct production rate ($\%$)& reference\\
 \hline
${\rm BR}(b \rightarrow {\rm J}/\psi {\rm X})$ & $1.153 \pm 0.051$
&$0.812 \pm 0.064$ & CLEO/LEP average  \\

${\rm BR}(b \rightarrow \psi^{\prime} {\rm X})$ & $0.355 \pm 0.049$
&$0.355 \pm 0.049$ & CLEO/LEP average \\

$ {\rm BR}(b \rightarrow \chi_c^1 {\rm X})$ &$0.414 \pm 0.051 $
& $0.383 \pm 0.051$ & \cite{ref:cleochi2} \\

$ {\rm BR}(b \rightarrow \chi_c^2 {\rm X})$ &$0.10 \pm 0.05$
& $0.07 \pm 0.05$ & \cite{ref:cleochi2}\\
\hline
\end{tabular}
\end{center}
\caption{{\it Charmonium production
in $b$-hadron decays.} \label{tab:charmvu}}

\end{table}
These values have to be corrected for cascade decays to determine the direct 
production rates of charmonium states. This has been done
in the third column of Table \ref{tab:charmvu} using
corresponding branching fractions given in the PDG \cite{PDG00}.

The model of \cite{ref:ccbarth} has been used to account for unmeasured states 
as $\chi_c^0$, $\eta_c$ and $h_c$.
\begin{equation}
{\rm BR}(b \rightarrow \chi_c^0,~\eta_c,~h_c {\rm X})=(0.04\pm0.04)\%,~
(0.4\pm0.2)\%,~(0.2\pm0.2)\%
\end{equation}
Theoretical 
expectations for contributions of other $(c \cbar)$ states have been 
considered and found to be negligible \cite{ref:heavycharmo}.
% and,
%as mentioned in the following,
%there is not either experimental evidence for a larger contribution of
%$b \rightarrow {\rm 0{\rm D}~charm}$ decays than expected.
The total production rate for charmonium in $b$-hadron decays is then
evaluated to be:
\begin{equation}
{\rm BR}(b \rightarrow (c\cbar) {\rm X})=(2.3\pm0.3)\%.
\end{equation}
The quoted uncertainty depends mainly on the evaluation of $\eta_c$
and $h_c$ production rates. It may be noted that no model can account for 
the measured $\chi_c^1$ and $\chi_c^2$ rates. Colour singlet models predict
no $\chi_c^2$ production, as observed essentially, but they predict also 
a $\chi_c^1$ rate which is too low by an order of magnitude. Colour octet
models, in the contrary, can explain the measured $\chi_c^1$ rate but
predict that the $\chi_c^2$ rate  is larger by a factor 5/3 at variance with
the observations.

Using the average value for open-charm production given in Equation 
(\ref{eq:openlepcleo}),
the average number of charm quarks in $b$-hadron decays is then
evaluated to be:
\begin{equation}
n_c + n_{\overline{c}}= n({\rm open-charm})+2 \cdot
{\rm BR}(b \rightarrow (c\cbar) {\rm X})=
1.141 \pm 0.059
\end{equation}

\mysubsection{Measurements of $b$-hadron decays into no and two open-charm
particles}

DELPHI \cite{ref:incldel} and SLD \cite{ref:inclsld} have measured, 
using respectively the distributions of charged track impact
parameters relative to the beam interaction point position
and of secondary vertices, the fractions
of $b$-hadron decays in which zero or
two charmed hadrons are produced
(see Table \ref{tab:inclbr} in Appendix \ref{appendixAe}). These values have 
been averaged, giving:
\begin{eqnarray}
 {\rm BR}(b \rightarrow 0{\rm D} {\rm X}) & = &
(4.3 \pm 1.8) \% \label{eq:0d}\\
{\rm BR}(b \rightarrow {\rm D} \overline{{\rm D}} {\rm X}) & =&
 (22.3 \pm 5.6) \% \label{eq:2d}.
\end{eqnarray}
Quoted uncertainties for the measured ${\rm D} \overline{{\rm D}}$
rates by DELPHI \cite{ref:incldel}
and SLD \cite{ref:inclsld}
have been 
increased because the corresponding $\chi^2$ for the average of the two
results is equal to 10.7.
The difference between these two measurements may indicate an underestimate 
of systematic uncertainties in the present analyses.
Without this correction, the average
${\rm BR}(b \rightarrow {\rm D} \overline{{\rm D}} {\rm X})$ was equal to 
$(19.1 \pm 3.8)\%$.

The value obtained for no-open charm corresponds to the sum of 
$b \rightarrow u \overline{u} d$, $b \rightarrow s g$ and 
$b \rightarrow (c\overline{c}) {\rm X}$ decay processes.

From these values the average number of charm quarks produced
in $b$-hadron decays 
amounts to:
\begin{equation}
n_c + n_{\overline{c}}= 1 - {\rm BR}(b \rightarrow 0{\rm D} {\rm X})
+{\rm BR}(b \rightarrow {\rm D} \overline{{\rm D}} {\rm X})
+2 {\rm BR}(b \rightarrow (c\cbar) {\rm X})= 1.226 \pm 0.060.
\label{eq:ccinclu}
\end{equation}
%{\bf The last error needs to checked and has to be commented }

%This value can be averaged with inclusive results mentioned before taking 
%into account correlated systematics:
%\begin{equation}
%n_c + n_{\overline{c}}= 1.183 \pm 0.042.
%\end{equation}

\mysubsection{Measurements of $b$-hadron decays with a $\overline{c}$
quark in the final state}
%The approach used by the LEPEWWG, explained in \cite{ref:lepwwhf}, 
%has been used in the following. It consists in evaluating the
%branching fraction ${\rm BR}({\rm W}^- \rightarrow \overline{c} q)$
%in $b$-hadron decays. 
Branching fractions of $b$-hadrons into two charm hadrons can be obtained
by reconstructing completely these two hadrons or only the wrong-sign
one ($\overline{\rm B} \rightarrow \overline{\rm D}~{\rm X}$). These 
measurements can be expressed in terms of the wrong-sign charm branching
fraction corrected for the contribution of the tiny $b \rightarrow u$
transition:
\begin{equation}
{\rm BR}^{\prime}(b \rightarrow \overline{{\rm D}}_iX)=
{\rm BR}(b \rightarrow \overline{{\rm D}}_iX)(1- \vubovcb^2 \alpha_u).
\end{equation}
The quantity $\alpha_u= 2\pm1$ accounts for the larger phase space available
in a $b \rightarrow u$ as compared to a $b \rightarrow c$ transition.
The total double-charm rate can be obtained by summing over all types of
produced wrong-sign charm hadrons:
\begin{equation}
{\rm BR}(b \rightarrow {\rm D \overline{D}X})=\sum_i{
{\rm BR}^{\prime}(b \rightarrow \overline{{\rm D}}_iX)}.
\end{equation}

\mysubsubsection{Results at the $\Yfs$}
CLEO \cite{ref:cleoexcc} has measured the ratio of wrong sign relative
to right sign $\Dn$ and $\Dp$ mesons, called
D in the following
expressions, in $b$-meson decays
\footnote{The measured value $0.100 \pm 0.026 \pm 0.016$ has been
corrected using the new value of the $\Bd$ oscillation parameter
given in Table \ref{tab:gensys}.}. 
\begin{equation}
r_D= \frac{{\rm BR}(\overline{{\rm B}} 
\rightarrow \overline{{\rm D}} {\rm X})}{{\rm BR}(\overline{{\rm B}} 
\rightarrow {\rm D} {\rm X})}=
0.088 \pm 0.026 \pm 0.005
\end{equation}
The probability of having a $\Dnb$ or a $\Dpb$ from the virtual ${\rm W}^-$
is then:
\begin{equation}
{\rm BR}(\overline{{\rm B}} \rightarrow \overline{{\rm D}} {\rm X})=
\frac{r_D \times {\rm BR}(\overline{{\rm B}} \rightarrow {\rm D},~
\overline{{\rm D}} {\rm X})}
{1 + r_D}
\label{eq:wsdcleo}
\end{equation}

In non-strange $b$-meson decays the largest fraction of produced 
$\Dsb$ mesons is expected to originate from 
${\rm W}^- \rightarrow \overline{c} s$ final states. The measured
value by CLEO \cite{ref:cleinc2}
includes $\Ds$ and $\Dsb$ mesons and has to be corrected
for $\Ds$ production at the lower vertex when, after the $b \rightarrow c$
transition, the $c$ quark hadronises into a $\Ds$ meson by combining
with a 
$\overline{s}$ quark.
\begin{equation}
{\rm BR}(\overline{{\rm B}} \rightarrow \Dsb {\rm X})=
{\rm BR}(\overline{{\rm B}} \rightarrow \Ds,~\Dsb {\rm X})-
{\cal P}^{{\rm low.~V}}_{\overline{{\rm B}} \rightarrow \Ds}
\label{eq:corrds}
\end{equation}

CLEO \cite{ref:wcleods} has searched also for the production of right sign
$\Ds$ mesons, in $\overline{\rm B}$ decays and obtains 
a relative fraction equal to $0.16 \pm 0.08 \pm 0.03$.
 The lower vertex production rate of $\Ds$ mesons 
can then be obtained corresponding to $(2 \pm 1)\%$. The correction
which has to be applied in Equation (\ref{eq:corrds}) concerns the
production of strange $\Ds$ mesons at the lower vertex, in double-charm decays,
and its value may differ from this measured number because kinematical and 
dynamical situations are different. In the following it has been assumed
that:
\begin{equation}
{\cal P}^{{\rm low.~V}}_{\overline{{\rm B}} \rightarrow \Ds}= (2 \pm 2)\%
\end{equation}
and the same value has been used for charm baryon production, at the lower 
vertex. We introduce the notation:
\begin{equation}
p_{sdq}={\cal P}^{{\rm low.~V}}_{\overline{{\rm B}} \rightarrow \Ds}=
{\cal P}^{{\rm low.~V}}_{\overline{{\rm B}} \rightarrow \Lc}.
\end{equation}

The branching fraction into wrong-sign $\Lcb$ baryons 
${\rm BR}(\overline{{\rm B}} \rightarrow \Lcb {\rm X})$ has been 
given already in Equation (\ref{eq:lcbar}).

\mysubsubsection{ALEPH measurements}
\label{sec:alephm}
The ALEPH Collaboration \cite{ref:alephdoubled} has measured several
branching fractions of $b$-hadron decays into two charmed hadrons
by reconstructing completely their decay final states
(see Appendix \ref{appendixAe}, Table \ref{tab:alephcc})

 From decays with no strange $c$-meson, emitted in the final state,
it is possible to obtain:
\begin{equation}
{\rm BR^{\prime}(\overline{B} \rightarrow \overline{D}X)}=
\frac{{\rm BR}(b \rightarrow {\rm D \overline{D}X})}
{(\fu+\fd)(1-2p_{sdq})+(\fs+\fb)p_{nsdq}}.
\end{equation}
The quantity $p_{nsdq}$ corresponds to the probability,
supposed to be the same for $\Bsb$ and $b$-baryons, that
there is, respectively, no $\Ds$ and no charm baryon in their decay final state,
emitted at the lower vertex. The central value, measured in $b$-hadron
 semileptonic decays into $\Dstarstar$ states has been used as an estimate
but its attached
uncertainty has been increased
to account for the fact that this number is used in case of
double-charm decays:
\begin{equation}
p_{nsdq}={\cal P}^{{\rm low.~V}}_{\Bsb \rightarrow D}=
{\cal P}^{{\rm low.~V}}_{b-baryon \rightarrow D}=(30 \pm 10)\%.
\end{equation}

In a similar way, it is possible to relate the measured branching
fraction corresponding to ${\rm D}\Dsb {\rm X}$ and $\overline{{\rm D}}\Ds {\rm X}$
final states to the branching fraction into wrong sign D or $\Dsb$ mesons:
\begin{eqnarray}
{\tiny
{\rm BR}( b \rightarrow {\rm D}\Dsb {\rm X~+~\overline{{\rm D}}\Ds X} )}
       {\tiny =[(\fu+\fd+\fb)p_{sdq}+\fs p_{nsdq}]}
{\tiny {\rm BR^{\prime}(\overline{B}} \rightarrow {\rm \overline{D}X})} \nonumber \\
\nonumber
   {\tiny + [(\fu+\fd)(1-2p_{sdq})+(\fs+\fb)p_{nsdq}]
{\rm BR^{\prime}(\overline{B}} \rightarrow \Dsb {\rm X})  }
 \label{eq:alephdsd} \\
\end{eqnarray}

\mysubsubsection{DELPHI measurement}
DELPHI \cite{ref:delcbar} has measured the fractions, $r_{\Do}$ and $r_{\Dp}$, of wrong-sign
$\overline{\Do}$ and $\Dm$ in $b$-hadron decays:
\begin{equation} 
r_{\Do}=\frac{{\rm BR(b \rightarrow \Dob~X})}{{\rm BR(b \rightarrow \Do~X})}
=(12.9 \pm2.8)\%,~
r_{\Dp}=\frac{{\rm BR(b \rightarrow \Dm~X})}{{\rm BR(b \rightarrow \Dp~X})}
=(12.3 \pm6.7)\%
\end{equation}
Using the values of the inclusive
rates measured at LEP, given in Table \ref{tab:ddbrates}, and correcting
for the $b \rightarrow u$ contribution, a value for
${\rm BR}^{\prime}(\overline{\rm B} \rightarrow \overline{{\rm D}} {\rm X})$ 
has been obtained (Table \ref{tab:ccbexcl}).

\mysubsubsection{Average of exclusive measurements}

Production rates for wrong sign $c$-hadrons measured by CLEO, DELPHI and 
ALEPH have
been combined and their averages are given in Table \ref{tab:ccbexcl}.
\begin{table}
\begin{center}
\begin{tabular}{|l|c|c|c|} \hline
&${\rm BR}^{\prime}(\overline{\rm B} \rightarrow \overline{{\rm D}} {\rm X})$  &
 ${\rm BR}^{\prime}(\overline{\rm B} \rightarrow \Dsb {\rm X})$ & 
${\rm BR}^{\prime}(\overline{\rm B} \rightarrow \Lcb {\rm X})$ \\
 \hline
ALEPH & $0.094 \pm 0.032 \pm 0.006$
&$0.148 \pm 0.035 \pm 0.043$ & -- \\

DELPHI & $0.090 \pm 0.022 \pm 0.002$
& --   & --\\

CLEO & $0.070 \pm 0.020 \pm 0.002$
&$0.096 \pm 0.022 \pm 0.029$ & $0.008 \pm 0.006 $ \\
\hline
Average & $0.082 \pm 0.013 $
&$0.098 \pm 0.037 $ & $0.008 \pm 0.006 $ \\

\hline
\end{tabular}
\end{center}
\caption{{\it Probability of producing a charm hadron from the upper vertex
in $b$-decays estimated from ALEPH, DELPHI and CLEO data, 
together with the average.}
\label{tab:ccbexcl}}

\end{table}
From these values, and using Equation (\ref{eq:ccinclu}),
the number of $c$ and $\overline{c}$ quarks
produced in
$b$-meson decays has been obtained:
\begin{equation}
n_c + n_{\overline{c}}= 
%1 - {\rm BR}(b \rightarrow {\rm no}~{\rm D} {\rm X})
%+{\rm BR}(\overline{\rm B} \rightarrow \Dnb,~\Dsb,~\Lcb {\rm X}) 
%+2 {\rm BR}(b \rightarrow (c\cbar) {\rm X})= 
1.191 \pm 0.040.
\end{equation}

\begin{figure}[htb!]
\centerline{\epsfxsize 14.0truecm \epsfbox{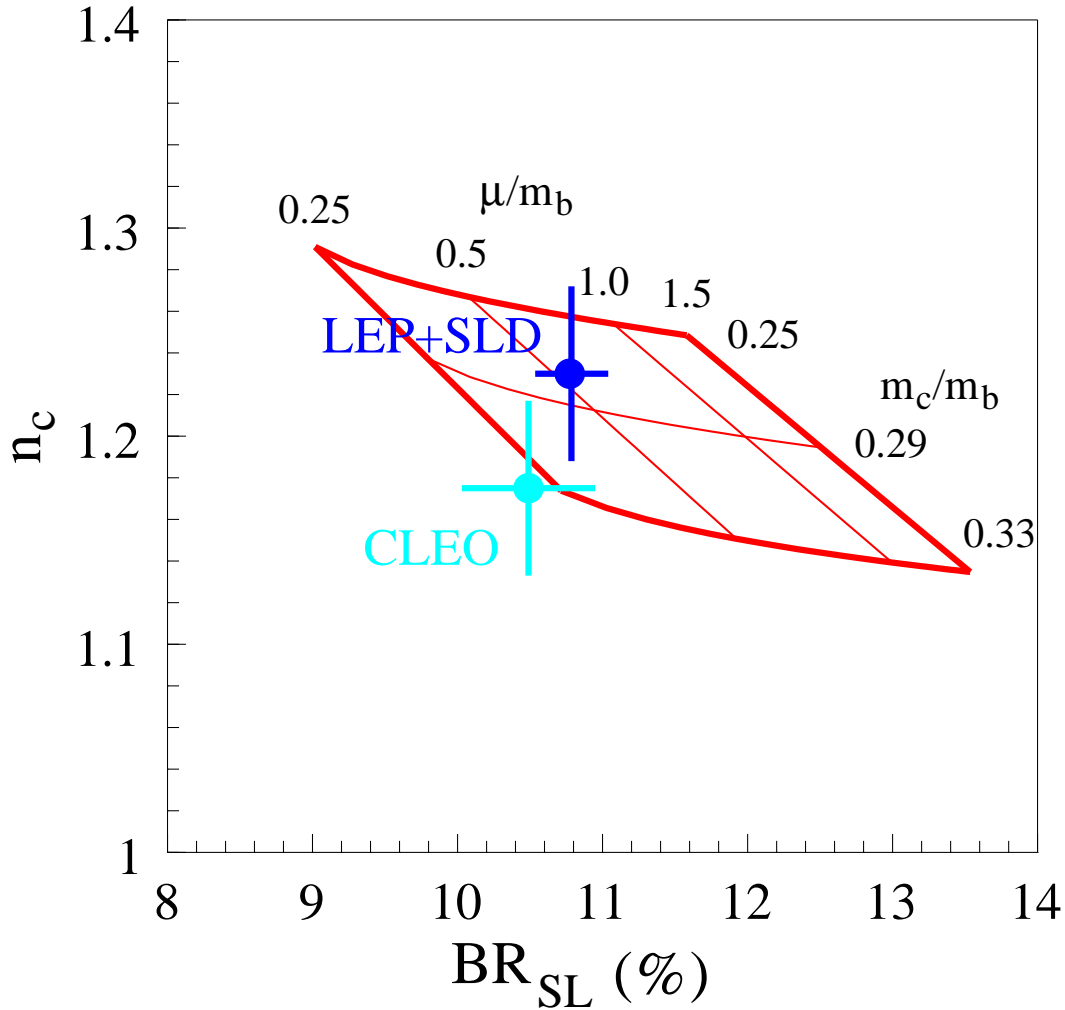}}
\caption{{\it Comparison between the measured number of 
$c$ and $\overline{c}$ quarks in $b$-hadron decays and of the
inclusive semileptonic branching fraction, with theoretical predictions.
The value of the semileptonic branching fraction reported
for CLEO is taken from \cite{ref:cleoslbr}.}
%{\bf The two points with errors have to be updated using the same
%values as given in the text. The LEP point is now at 1.23 because
%of inclusion of the inclusive SLD result. A slight displacement
%upwards of CLEO and LEP is also due to a different procedure
%to use double charm exclusive results.}
\label{fig:brslnc}}
\end{figure}

\mysubsection{Average number of $c$ and $\overline{c}$ quarks from all 
measurements}
\label{sec:countfive}
%All previous measurements have been combined taking into account
%correlations among the different systematic uncertainties. 
Measurements
done at the $\Yfs$ and at the $\Zz$ have been combined separately
\footnote{The value corresponding to $\Yfs$ measurements differs
from the one given in \cite{ref:ncccount1} $(1.10 \pm 0.05)$
mainly because of different rates taken for the production of charmed
baryons in $b$-meson decays and by the inclusion, in the present average,
of double charm rates measured by CLEO. The value obtained for 
$LEP+SLD$ measurements differs from the one given in \cite{ref:ncccount3}
$(1.171 \pm 0.040)$ because the inclusive double charm rate measured by SLD,
made available later in year 2000, has now been included.}; they
are found to be
compatible and an overall
average has been obtained.
\newpage
%1.154 \pm 0.036
\begin{eqnarray}
n_c + n_{\overline{c}}(\Yfs) & =& 1.175 \pm 0.042 \\
n_c + n_{\overline{c}}(LEP+SLD) & =& 1.230 \pm 0.042 \\
n_c + n_{\overline{c}}(\Yfs+LEP+SLD) & =& 1.206 \pm 0.033
\end{eqnarray}

These values and the corresponding measured semileptonic
branching fractions of $b$-hadrons have been indicated in 
Figure \ref{fig:brslnc} and compared with theoretical estimates
\cite{ref:sacrelife}.
The value of the semileptonic branching fraction measured at LEP
has been corrected so that it corresponds to a sample containing the 
same amount of $\Bdb$ and $\Bp$ mesons to be compared with the
corresponding measurement at the $\Yfs$:
\begin{equation}
{\rm BR}(b \rightarrow \ell^- \overline{\nu_{\ell}} X)({\rm corrected})=
\frac{\frac{1}{2}[\tau(\Bp)+\tau(\Bd)]}{\tau_b}
{\rm BR}(b \rightarrow \ell^- \overline{\nu_{\ell}} X).
\end{equation}
Using the lifetime values given in Section \ref{sec:Atau}, this correction 
amounts to $1.02 \pm 0.02$.

Present results favour a rather standard value for the charm quark
mass 
and a low scale, $\mu$, at which QCD corrections have to be evaluated:
\begin{equation}
\frac{m_c}{m_b} = 0.30 \pm 0.02, ~\frac{\mu}{m_b} \sim 0.35
\label{eq:mbmcmu}
\end{equation}
It should be noted that heavy quark masses entering into this
evaluation \cite{ref:sacrelife} have been defined using a different 
renormalization scheme
as compared to the one used in Sections \ref{sec:vcb} and \ref{sec:vub}
for the determination
of $\Vcb$ and $\Vub$. As a consequence, their values cannot be directly compared.
In Equation (\ref{eq:mbmcmu}), $m_Q$ are pole masses defined at one-loop in
perturbation theory.
%{\bf I dont know if this is a remark that we can make here: 
%if we use the value for mb=4.58 +- 0.06 GeV and use the ratio obtained
%for mc/mb we get: mc=1.37+-0.09 GeV whereas we use, in the inclusive Vcb 
%analysis mb-mc=3.50 which corresponds to mc=1.08 +- 0.04 GeV.....
%I have, since months submitted this remark to Bigi but no real answer. 
%}
Using recent determinations of $\overline{{\rm MS}}$ heavy quark masses
\cite{ref:mb,ref:mc}:
\begin{equation}
\overline{m_b}(\overline{m_b})=(4.25 \pm 0.08)~\GeV/c^2~{\rm and}~
\overline{m_c}(\overline{m_c})=(1.23 \pm 0.09)~\GeV/c^2
\end{equation}
and the expression, at one loop, relating these two definitions
of quark masses:
\begin{equation}
m_Q^{pole,1} =  \overline{m_Q}(\overline{m_Q}) 
\left ( 1 + \frac{4}{3} \frac{\alpha_S(\overline{m_Q})}{\pi}\right ),
\end{equation}
one obtains:
\begin{equation}
m_b^{pole,1} =(4.7 \pm 0.1)~\GeV/c^2~{\rm and}~m_c^{pole,1} =(1.4 \pm 0.1)~\GeV/c^2.
\end{equation}
The ratio of these two quantities agrees with Equation (\ref{eq:mbmcmu}).

The $0c$ rate can be obtained from the measured $0{\rm D}$
rate given in Equation 
 (\ref{eq:0d}) after having subtracted 
charmonium production:
\begin{equation}
{\rm BR(\overline{B} \rightarrow }0c) = (2.0 \pm 1.8)\%.
\end{equation}
This value is in agreement with the theoretical estimate of 
$(2.1 \pm 0.8)\%$ \cite{ref:zeroc} which includes $b \rightarrow u,~
b \rightarrow sg,~sgg$ transitions evaluated at NLO. There is no room
for a large additional $0c$ component as advocated for instance in 
\cite{ref:newbid}.

The $c \overline{c}$
rate can be obtained by combining results given in 
Equation (\ref{eq:2d}), Table \ref{tab:ccbexcl} and adding the 
charmonium component:
\begin{equation}
{\rm BR(\overline{B} \rightarrow }c\overline{c})= (22.3 \pm 3.2)\%.
\end{equation}
This value is in reasonable agreement with theoretical expectations:
$(21 \pm 6)\%$ as already illustrated in Figure \ref{fig:brslnc}.
These predictions depend mainly on the assumed value for the
$c$-quark mass.

The main conclusions from this study are that inclusive semileptonic
$b$-hadron decays can be explained
by theory once QCD corrections are evaluated at a rather low scale, of
the order of 0.35 $m_b$, and that the measured number of charmed quarks 
is in agreement with the value expected using independent determinations
of $b$ and $c$-quark masses.
%allows
%to extract a value for the $c$-quark mass which agrees with expectations.
The large variation of the estimates for the $b$-hadron semileptonic
branching fraction versus $\mu$ and the low value favoured for this
parameter indicate that computations of higher order QCD corrections are
needed. As already explained, it would be of interest to have a similar
analysis done using values for quark masses evaluated 
in the same 
renormalization scheme as in Sections \ref{sec:vcb} and \ref{sec:vub}.

\mysection{Average of LEP $\Vcb$ measurements}
\label{sec:vcb}

Within the framework of the Standard Model of electroweak interactions, the
elements of the Cabibbo-Kobayashi-Maskawa mixing matrix
are free parameters, constrained only by the requirement that the matrix
be unitary. 
The Operator Product Expansion (OPE) and Heavy Quark Effective
Theory (HQET) provide means to determine \vcb\
with relatively small theoretical uncertainties, 
by studying the decay rates of
inclusive and exclusive semileptonic $b$-decays respectively.
Relevant branching fractions have to be determined experimentally.
Inputs from theory are needed to obtain the values of the matrix 
elements.

There are two methods to measure \vcb: the inclusive
method, which 
uses the semileptonic decay width of $b$-decays and the OPE; and the  exclusive
method, where \vcb\ is extracted by studying the exclusive 
\btods\ decay process using HQET.   
%as a function of the recoil kinematics of the \dsp\ meson
%\cite{ref:neua}-\cite{ref:neud}
The  ${\rm B} \rightarrow {\rm D} \ell^- \overline{\nu_{\ell}}$ channel
has not been averaged to date.

%This note summarises the procedure to compute the LEP \Vcb\ average.
%It includes the correction applied to each measurement (central value 
%and uncertainties) to bring them to a common set of inputs and the averaging 
%procedure itself.

In this note, both methods are used to determine values for $\Vcb$, which 
are then combined to produce a single 
average\footnote{The present members of the $\Vcb$ working group are: 
D. Abbaneo, E. Barberio,
S. Blyth, M. Calvi, P. Gagnon, R. Hawkings, M. Margoni, S. Mele,
D. Rousseau and F. Simonetto.}.
 The semileptonic $b$-decay width,
determined by the LEP heavy flavour electroweak fit to ALEPH, DELPHI, L3
and OPAL data, is used to 
determine ${\rm BR}(b \rightarrow \ell^- \overline{\nu_{\ell}} {\rm X})$.
%needed to 
%extract \vcb\ from the inclusive method.
Results from ALEPH  \cite{ALEPH_vcb},
%\footnote{
%Updated to use the parametrisation of \cite{CLN}, the Ligeti 
%model \cite{ref:ligeti,ref:grinstein} for \btodss\ decays, and various updated inputs.},  
DELPHI \cite{DELPHI_vcb}
%\footnote{Updated to use 
%the Ligeti model  \cite{ref:ligeti} for \btodss~ decays.}
 and 
OPAL \cite{OPAL_vcb} are used to perform a LEP \vcb\ average in the
\btods\ decay channel. These measurements are combined using
a method similar to that used by the B oscillations working group.
OPAL has published two measurements, quoted in the following 
OPAL$_{{\rm inc}}$ and OPAL$_{{\rm exc}}$, using respectively a
partial or a complete reconstruction of the $\Dstarp$ decay products. 

Theoretical input parameters needed to 
extract $\Vcb$ from actual measurements
are detailed in Appendix \ref{appendixC}. 
In both methods, the extraction of $\Vcb$ is systematics limited and 
the dominant error is from theory. Thus it is important that a consistent
approach to theoretical uncertainties is adopted by the authors, even
though some arbitrariness in uncertainty definitions remain
(see Appendix \ref{appendixC}). Further work from theorists is required to 
provide uncertainties in a consistent framework and to identify methods to 
experimentally control these uncertainties.
In both analyses, the validity
of the quark/hadron duality hypothesis has been assumed.
This assumption implies that the cross sections and the decay rates
defined in the region of the time-like momenta are calculable in QCD 
after a ``smearing'' or ``averaging'' procedure is applied. 
In semileptonic decays, it is considered to be valid
\cite{ref:bigiosaka} as the integration over the lepton 
and neutrino phase space 
provides the ``smearing'' over the invariant hadronic 
mass (global duality). 
Global duality cannot be derived from first principles and 
it is an assumption in many QCD applications. 
Ultimately experimental 
data, for example the moment-type analyses of the lepton momentum
distribution in $b$-hadron semileptonic decays, will decide the real 
size of any deviations from global duality. 
As explained in \cite{ref:bigiosaka},
violations of duality will affect more the exclusive than the inclusive
determination of $\Vcb$.

As compared with previous averages \cite{ref:summer99}, a new analysis
from OPAL has been included and the same form factors parametrization,
as used in CLEO, has been adopted. 

\mysubsection{Inclusive $\Vcb$ determination}
\label{sec:vcbinc}

In the inclusive method, the partial width for semileptonic 
B meson decays to charmed mesons is related to \vcb\ 
using the following expression (Appendix \ref{appendixC}):
%by \cite{patricknote}:
\begin{eqnarray}
 \Vcb &=& 0.0411
~  \sqrt{\frac{1.55}{0.105} \Gamma(b \rightarrow \ell^-  \overline{\nu_{\ell}}{\rm X}_c )} 
\left(1-0.024 \left(\frac{\mu_\pi^2-0.5}{0.2}\right)\right)\times  \nonumber \\ 
 & & (1 \pm 0.030(pert.) \pm 0.020(m_b) \pm 0.024(1/m_b^3)).\label{eq:incl}
% \nonumber
\end{eqnarray}
where $\rm{X}_c$ represents all final states containing a charmed quark.
The definition adopted to fix the values of the running $b$- and $c$-quark
masses and the scale at which they are evaluated are given in
Appendix \ref{appendixC}.
The rest of the expression, within parentheses, represents the
correction to the muon decay formalism, depending on the $b$- and $c$-quark
masses and on the strong coupling constant; $\mu_\pi^2$ is the average of the 
square of the $b$-quark momentum inside the $b$-hadron.
%For the parameter
%$\mu_\pi^2$ the value $0.5 \pm 0.1$ is assumed.

%Experimentally, the semileptonic width of $b$-hadrons is determined from 
%its semileptonic branching fraction and lifetime.
%$$ \Gamma( \overline{{\rm B}} \rightarrow \ell^- \overline{\nu_{\ell}}{\rm X}_c ) = \frac{{\rm BR}(\overline{{\rm B}} \rightarrow \ell^- \overline{\nu_{\ell}}{\rm X}_c  )}{\tau_{\rm B}}.$$
In $\Zz$ decays, a mixture of 
$\Bdb,~ \Bm, ~\Bsb$ and $b$-baryons is produced, such that the 
inclusive semileptonic branching fraction measured at LEP is an average over
the different hadrons produced: 
%Assuming that the semileptonic widths of all b-hadrons are equal, the 
%following relation holds:
\begin{eqnarray}
{\rm BR}( b \rightarrow  \ell^- \overline{\nu_{\ell}}{\rm X}_c  )& =& 
       \fd \frac{\Gamma(\Bdb \rightarrow  \ell^- \overline{\nu_{\ell}}{\rm X}_c  )}{\Gamma(\Bdb)} +
       \fu \frac{\Gamma(\Bm \rightarrow \ell^- \overline{\nu_{\ell}} {\rm X}_c )}{\Gamma(\Bm)}  \nonumber \\
    & &  + \fs \frac{\Gamma(\Bsb \rightarrow  \ell^- \overline{\nu_{\ell}}{\rm X}_c  )}{\Gamma(\Bsb)} +
\fb \frac{\Gamma( b-{\rm baryon} \rightarrow  \ell^- \overline{\nu_{\ell}}{\rm X}_c  )}{\Gamma( b-{\rm baryon})}  
  \nonumber \\
   &  \simeq & \Gamma(b \rightarrow \ell^-  \overline{\nu_{\ell}}{\rm X}_c  ) ( \fd \tau_{B^0} + \fu \tau_{B^-} +
   \fs \tau_{B_s} + \fb\tau_{ b-{\rm baryon}})  \nonumber \\ 
  &  =& \Gamma(b \rightarrow  \ell^- \overline{\nu_{\ell}}{\rm X}_c ) \tau_b \label{eq:bri}
\end{eqnarray}
where 
%$\tau_b = \fd \tau_{B^0} + \fu \tau_{B^-} +
%    \fs \tau_{B_s} + \fb\tau_{ b-{\rm baryon}} $
$\tau_b $ is the average 
$b$-hadron lifetime.
Therefore the semileptonic width of $b$-hadrons can be obtained 
using the inclusive 
semileptonic branching fraction and the average  $b$-hadron lifetime.
 The two last equalities in Equation (\ref{eq:bri})
assume that all $b$-hadrons have the same semileptonic width.
This hypothesis may be incorrect for $b$-baryons.
%The b-baryon semileptonic width may differ slightly from that of the other
%b-hadrons, producing a correction to Equation~(\ref{eq:bri}).
Taking into account the present precision of LEP measurements of $b$-baryon 
semileptonic branching fractions and lifetimes, an estimate of
the correction 
to Equation (\ref{eq:bri}) is about 1.5\% (see Section \ref{sec:systgen}). 

The average LEP value for 
${\rm BR} (b \rightarrow \ell^- \overline{\nu_{\ell}}{\rm X}) = (10.56 \pm 0.11{\rm (stat.)} 
\pm 0.18{\rm (syst.)})\% $ is taken from the global fit
which combines the heavy flavour measurements performed 
at the $\Zz$ (see Section \ref{sec:systgen}).
%\cite{HFLEPEW}~\footnote{Hereafter, this fit is referred to as
% the HFLEPEW fit}.
%This value was obtained using only the measurements of the semileptonic 
%branching fractions, the average $\rm B^0-\overline{B^0}$ mixing parameter 
%and  $\rm R_b=\Gamma_{bb}/\Gamma_{had}$.  
%The asymmetries values have been fixed to their standard model values. 
%The measurements of $BR (b \rightarrow X l \nu)$ used in this fit were all 
%obtained  using a combination of lepton and lifetime tags.
The ${\rm BR}( b \rightarrow \ell^- \overline{\nu_{\ell}}{\rm X}_u )$ 
contribution is subtracted from  
${\rm BR} (b \rightarrow \ell^- \overline{\nu_{\ell}}{\rm X} )$,
 using the LEP average value 
%$(1.67 \pm 0.55 ) \times 10^{-3} $ \cite{vubwg} . 
given in Section \ref{sec:vub}.
For the average $b$-hadron lifetime, the world average value of
$\tau_b$
% = 1.564 \pm 0.014$~ps \cite{PDG00} 
is used, as obtained in Equation (\ref{eq:taub}).

\mysubsubsection{Sources of systematic errors}
\label{subs:vcb}
The systematic errors assumed at present in the determination of 
the inclusive semileptonic decay width
can be grouped into the following categories:
\begin{itemize}
\item {\it errors related to the efficiency and purity of the $b$-tagging algorithm}

 As a lifetime $b$-tag is involved, effects due to the uncertainties
in the sample composition in terms of different heavy hadrons, 
and uncertainties in the  heavy hadron lifetimes, are considered. 

\item 
{\it input parameters influencing the signal and background normalisation}

The values of the production fractions: $\fd$, $ \fu$, $\fs$, 
$ \fb$ were taken from Section \ref{sec:results}.
% $\rm R_c$ was fixed to its 
%Standard Model value {\it Which is ....}.
The branching fractions $\bcbl$, $\btaul$, $\bpsill$, and the rates of gluon 
splitting P($\glcc$) and P($\glbb$) have been fixed to the values of \cite{HFLEPEW}. 
$\Rb$,  $\rm R_c$, $\bcl$ and $\cl$ are parameters of the LEPEWWG fit. Values 
for all these quantities are 
listed in Table~\ref{tab:gensys}.
% The value of $BR( b \rightarrow X_u l)$:  
\item  
{\it the average fraction of the beam energy carried by the weakly decaying 
$b$-hadron} 

Different models have been considered for the shape of the 
fragmentation function and the free parameters of the models have
been determined 
from the data. 
\item 
{\it $\Lambda_b$ polarization}

These effects on the lepton spectra have been 
included.
\item 
{\it semileptonic decay models}

The average LEP value for 
${\rm BR} (b \rightarrow \ell^- \overline{\nu_{\ell}}{\rm X}) $ 
is taken from the global LEPEWWG fit
which combines the heavy flavour measurements performed 
 at the $\Zz$ \cite{HFLEPEW},
but removing the
forward-backward asymmetry measurements (see Section \ref{sec:systgen}). 
%Those measurements were all obtained 
%using a combination of lepton and lifetime tags.

%The dominant error on the branching fraction from the LEPEWWG fit is the 
%dependence on the semileptonic decay model.
%%In a fit which combines  only the measurements of the semileptonic 
%%branching fractions 
%%$\rm BR (b \rightarrow X \ell^-\nu)$,  $\rm BR (b \rightarrow c \rightarrow X \overline{\ell} \nu)$, 
%%$\rm BR (c \rightarrow X \ell^-\nu )$,
%%the average $\rm B^0-\overline{B^0}$ mixing parameter $\overline{\chi_b}$ and 
%% $\rm R_b=\Gamma_{bb}/\Gamma_{had}$ , this error amounts to 
%%$\rm \Delta \bl (\rm{model}) \, = \, 0.084 \times 10^{-2}$.
%In the LEPEWWG fit to all heavy flavour measurements, 
%a consistent result is obtained, with a 
%modelling error of $0.065 \times 10^{-2}$.
%Extensive studies have been made to understand the size of this error.
%The reduction of the modelling uncertainty is due to the inclusion of
%asymmetry measurements using different methods. Those using leptons depend
%on the semileptonic decay model while those using a lifetime tag and
%jet charge or D-mesons do not. The mutual consistency of the asymmetry
%measurements effectively constrains the semileptonic decay model, and
%reduces the uncertainty in the semileptonic branching fraction.
%The result of the fit which does not include the
%asymmetry measurements is considered here.  
\item 
{\it detector specific items}

These include: lepton efficiencies, misidentification 
probabilities, detector resolution effects, jet reconstruction,
etc.
\end{itemize}
The errors listed in the last item are uncorrelated among the different 
experiments. The others have been split into their uncorrelated and  
correlated parts.
The error on the average $b$-hadron lifetime is assumed to be uncorrelated
with the error on the semileptonic branching fraction.
The propagation of these errors to the error on \vcb\ is done assuming that 
they are Gaussian in the branching fraction and the lifetime, respectively.

\mysubsubsection{Inclusive $\Vcb$ average}

Using the  expression given in Equation (\ref{eq:incl}), the following value is 
obtained:
%$${\rm V_{cb}} = (41.11 ^{+0.38}_{-0.40}(exp) \pm 2.05 (th)) \times 10^{-3}.$$
\begin{equation}
\Vcb^{incl.} = \vcbinc
\end{equation}
where the first error is experimental and the second is from theory.
The experimental contributions due to the semileptonic
branching fraction and the lifetime are $\pm 0.37\times 10^{-3}$ and 
$\pm 0.18\times 10^{-3}$, respectively.
The dominant systematic uncertainty, of theoretical origin, comes from the
determination of the kinetic energy of the $b$-quark inside the $b$-hadron
as explained in Appendix \ref{appendixC}.

\mysubsection{Exclusive $\Vcb$ determination}
\label{sec:vcbexclu}

In the exclusive method, the value of \vcb\ is extracted by studying
the decay rate for the process \btods\ as a function of the recoil kinematics
of the \dsp\ meson. The decay rate is parameterized as a function
of the variable $w$, defined as the product of the four-velocities of the 
\dsp\ and the $\Bdb$ mesons.
This variable is related to the square of
the four-momentum transfer from the $\Bdb$ to the $\ell^-{\overline \nu}_\ell$
system, $q^2$, by:
\begin{equation}
w = \frac{m_{\rm D^{*+}}^2+m_{\rm B^0_d}^2-q^2}{2m_{\rm B^0_d}
%\cdot 
m_{\rm D^{*+}}},
\end{equation}
and its values range from $1.0$, when the \dsp\ is produced at rest in the \bbar\ rest 
frame, to about $1.50$.  Using HQET, the differential partial width for this
decay is given by: 
\begin{eqnarray}
\label{eq:decayw}
{\frac{{\rm d} \Gamma}{{\rm d} w}}&=
&{\cal K}(w){\cal F}_{D^*}^2(w) \Vcb^2
\label{eq:dgdw}
\end{eqnarray}
where ${\cal K}(w)$ contains kinematic factors
and ${\cal F}_{D^*}(w)$ is the hadronic form factor for
the decay. 
\begin{equation}
{\cal K}(w)=m_{D^*}^3 (m_B-m_{D^*})^2 \sqrt{w^2-1}(w+1)^2
\left( 1 + \frac{4w}{w+1} \frac{1-2wr+r^2}{(1-r)^2} \right)
\end{equation}
with $r=m_{D^*}/m_B$.

 Although the shape of the
form factor, ${\cal F}_{D^*}(w)$, is not known, its magnitude
at zero recoil, corresponding to $w=1$, can be estimated using HQET.  
It is found to be convenient to express ${\cal F}_{D^*}(w)$ in terms 
of the axial form factor
$h_{A_1}(w)$ and of the reduced helicity form factors $\tilde{H}_0$ 
and $\tilde{H}_{\pm}$:
\begin{equation}
{\cal F}_{D^*}(w)= h_{A_1}(w)\sqrt{\frac
{\tilde{H}_0^2+\tilde{H}_{+}^2+\tilde{H}_{-}^2}
{1 + \frac{4w}{w+1} \frac{1-2wr+r^2}{(1-r)^2}}}.
\end{equation}
The reduced helicity form factors are themselves expressed in terms of the 
ratios between the other HQET form factors 
($h_{V}(w),~h_{A_2}(w),~h_{A_3}(w)$) and $h_{A_1}(w)$:
\begin{eqnarray}
\tilde{H}_0(w)=1+\frac{w-1}{1-r}\left [ 1- R_2(w)\right] \\
\tilde{H}_{\pm}(w)=\frac{\sqrt{1-2wr+r^2}}{1-r}
\left[1 \mp \sqrt{\frac{w-1}{w+1} R_1(w)} \right]
\end{eqnarray}
with
\begin{equation}
 R_1(w)=\frac{h_V(w)}{h_{A_1}(w)}~{\rm and}~
 R_2(w)=\frac{h_{A_3}(w)+rh_{A_2}(w) }{h_{A_1}(w)}.
\label{eq:r12}
\end{equation}
Values for $R_1(w)$ and $R_2(w)$ have been obtained by CLEO 
\cite{ref:cleoform} using different models.

The  unknown function $h_{A_1}(w)$ is approximated with an expansion
around $w=1$ due to Caprini, Lellouch and Neubert (CLN) \cite{CLN}:
\begin{equation}
h_{A_1}(w)= h_{A_1}(1) \times 
\left[ 1 - 8 \rho_{A_1}^2 z + (53 \rho_{A_1}^2 - 15)z^2
- (231 \rho_{A_1}^2 -91) z^3 \right],
\label{eq:clnextr}
\end{equation}
where $\rho_{A_1}^2$ is the slope parameter at zero recoil and 
$\large z= \frac{\sqrt{w+1} - \sqrt{2}}{\sqrt{w+1} + \sqrt{2}}$.
An alternative parametrization, obtained earlier, can be found in
\cite{ref:grinstein}.
%The ratio between the axial and vector form factors is included in
%${\cal K}(w)$.
%Theoretical predictions restrict 
%values of $\rho^2$ to be in the range: $-0.14<\rho^2<1.54$.

In the heavy quark limit ($m_{\rm b}\rightarrow \infty$), 
${\cal F}_{D^*}(1)= h_{A_1}(1)$
coincides with the Isgur-Wise function \cite{neub,neuc} which is
normalised to unity at the point of zero recoil.
Corrections to \fone\ have been calculated to take into account
the effects of finite quark masses and QCD corrections \cite{luke}.
They yield \fone\ $=0.88 \pm 0.05$ 
(Appendix \ref{appendixC}).
%\cite{patricknote}. 

Experiments determine the product ${\cal F}_{D^*}^2(1) \Vcb^2$
by fitting this quantity and the slope $\rho_{A_1}^2$,
using the expression (\ref{eq:dgdw}), convoluted with the experimental
resolution on the $w$ variable.
%the measured $\frac{{\rm d} \Gamma}{{\rm d} w}$ distribution.
Since the phase space factor ${\cal K}(w)$ tends to zero
as $w\rightarrow 1$, the decay rate vanishes at
$w=1$ and the 
accuracy of the extrapolation relies on achieving a reasonably
constant reconstruction efficiency in the region close to $w=1$.  

\mysubsubsection{Sources of systematic uncertainties}

The systematic 
uncertainties in the determination of 
\vcb\ using the semileptonic decay \btods\ can be
%It is based on the papers published on the 
%subject by several Collaborations \cite{Argus,Cleo,Delphi,Aleph,Opal}, 
%on the work of the LEP \Vcb\ 
%Working Group, and on the conclusions of an informal workshop on 
%\Vcb, \Vub\ determination recently held at CERN \cite{WKSnote}. 
grouped into the following categories:
\begin{itemize}
\item {\it normalisation}: $\Bdb$ meson production rate, 
${\rm D}^{(*)}$ branching fraction to 
the tagged final states (including topological BR), 
$\Bdb$ lifetime (this is needed to obtain the 
%\BtoD\ 
$\Bdb \rightarrow \Dstarp   \ell^-\overline{\nu_{\ell}}$
decay partial 
width), and the $b \rightarrow {\rm B}$ fragmentation function
(which influences the reconstruction efficiency).
%, and also affects the efficiency in LEP measurements);

%\item {\it input parameters which influence the reconstruction efficiency:}
%(e.g. $b \rightarrow {\rm B}$ fragmentation function, 
%value of the $\Bdb$ lifetime);

\item {\it background from physical processes:} comprising \Btau, \Bxc\ 
(followed by the 
semileptonic decay $\tau/\mathrm{X}_{\overline{c}} \rightarrow \ell^-
\overline{\nu_{\ell}} {\rm X}$) 
and, particularly, the intermediate production of excited charm mesons
$\Dstarstar$
which subsequently decay to a $\Dstarp$;

\item {\it detector specific items}: selection efficiency 
(lepton identification, tracking, vertexing),
non physics background 
(combinatorial, hadron mis-identification), resolution, 
fitting, etc. 
This last set is treated as uncorrelated among experiments, and 
therefore will be ignored in the following discussion.
\end{itemize}
%Knowledge of many of the input parameters has improved considerably since
%the publication of some of the results. Similarly, HQET has
%provided more reliable descriptions of some important background processes. 
%The present situation is reviewed in the following sections.
%Some of the parameters are not likely to be updated in the near future,
%whilst others are still evolving and are subject to 
%change (most of them are determined by dedicated working groups following
%well established procedures). We will make reference to 
%\begin{itemize}
%\item published results (as reported in the PDG \cite{PDG00}) ;
%\item LEP+SLD+CDF Working Group approved notes ;
%\item contents of papers or conference contributions devoted to 
%\vcb\ determination.
%\end{itemize}

\mysubsubsection{Normalisation}
The  $\Bdb$  production rate at LEP is given by the product:
\begin{eqnarray}
\frac{\Gamma(\Zz \rightarrow b \overline{b})}{\Gamma(\Zz \rightarrow \mathrm{hadrons})} 
~ {\rm BR}(b \rightarrow \Bdb) ~=~ \Rb ~ \fd 
\end{eqnarray}
where $\Rb$\ is taken from
%with four digit precision by 
\cite{HFLEPEW} (Table \ref{tab:gensys}) and 
$\fd$\ 
%obtained by the B oscillation 
%working group as part of the $\Delta m_d$
%averaging procedure of 
from Section \ref{sec:results}. 
%This systematic error is specific to LEP experiments.
The values for the 
 charm meson decay branching fractions are taken from the PDG \cite{PDG00}. 
%Care has been 
%taken in properly accounting for 
Correlations among some of them (e.g. 
$\Do \rightarrow {\rm K}^-\pi^+$ with $\Do \rightarrow {\rm K} n\pi$, etc.)
are included. Analyses based on the inclusive 
reconstruction
of pions from $\Dstarp$ cascade decays
 may be affected by the knowledge of the topological 
$\Do$\ branching fractions; they are taken from MARKIII measurements
 \cite{ref:mark3}.
The $\Bdb$ lifetime determined in
Section \ref{sec:taubd} is used.
% This may change in the near
%future due to new precise measurements provided by the SLD, ALEPH, DELPHI 
%and OPAL Collaborations. 
%In this case, the updated value provided by the b lifetime working 
%group will be used.
Knowledge of the $\Bdb$\ fragmentation function is necessary  in order to 
compute the fraction of $\Bdb$\ which were not reconstructed
because they did not have enough energy to be detected.
$\Bdb$\ 
hadrons produced in $e^+ e^-$ annihilations carry on average 
%$<x_E>$ = ($0.702 \pm 0.008 $) 
a large fraction, $<x_E>$, 
%(Section \ref{sec:systgen})
 of the beam energy (Table \ref{tab:gensys}); 
consequently only a small fraction
 of them are outside the selection acceptance. 

\mysubsubsection{Physics background}
Semileptonic decays of $b$-hadrons to charm excited states ($\Dstarstar$)
which then decay to a $\Dstarp$ are the major source of correlated (physics)
background. The $\Dstarstar$ can be either a narrow resonant state,
${\rm D}_1$(narrow) or ${\rm D}_2^*$, or a broad and/or a non-resonant state,
${\rm D}_1$(broad) or ${\rm D}_0^*$.
The production rates of the different $\Dstarstar$
states (see Sections \ref{sec:dssb} and \ref{sec:dssc}) and the variation of 
their corresponding form factors as a function of $w$, have also
to be considered.
%is also required. 
\begin{figure}[htb!]
\centerline{\epsfxsize 14.0truecm \epsfbox{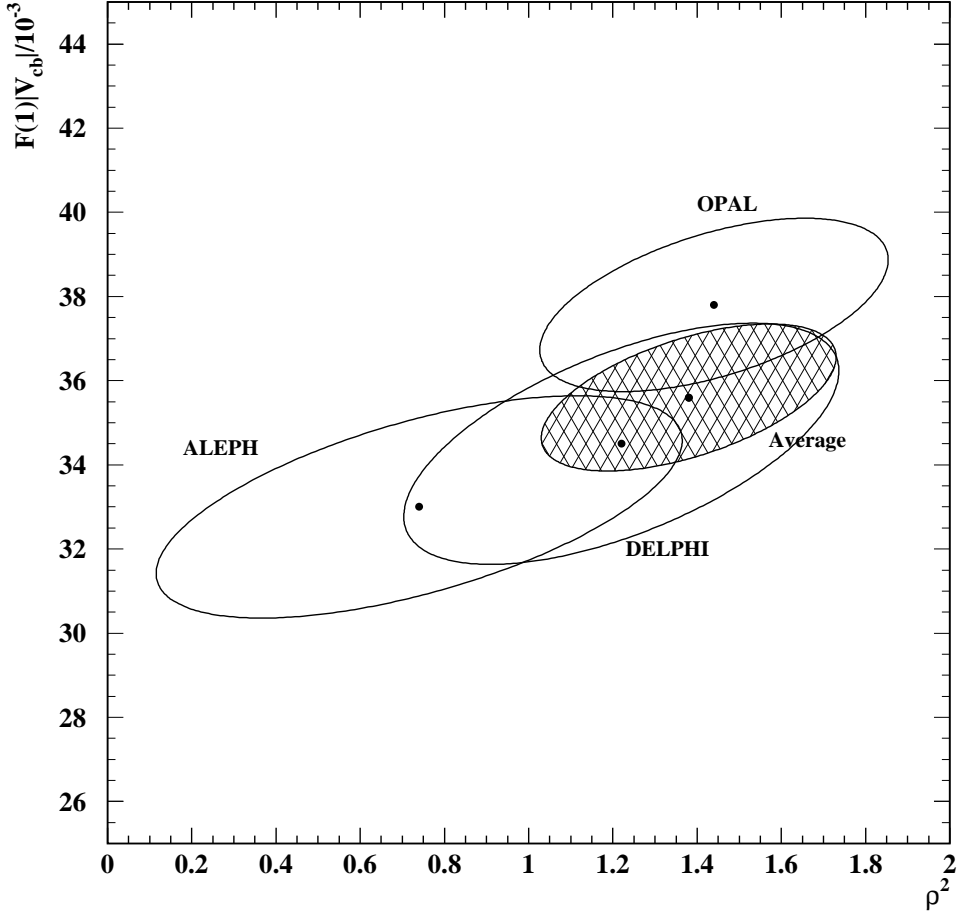}}
%\centerline{\epsfxsize 14.0truecm \epsfbox{/afs/cern.ch/phys/lep/lepvcb/pdg_rep/vcbell_win01_bw.eps}}
\caption{{\it Corrected measurements and LEP average of the
quantities \fvcb\ and $\rho^2$ using the
exclusive method.
The error ellipses, centred on the different measurements,
correspond to contours at the 68$\%$ C.L. level and include systematics.
The hatched ellipse corresponds to the average of the three experiments.
%{\it Give more details on the meaning of the different 
%ellipses}
}
%{\bf Needs to checked, updated and done in black/white.}
\label{f:vcbell}}
\end{figure}
%Published results on $\Vcb$
%are based on 
The Isgur-Wise model \cite{isgw}
predicts a sizeable $\Dstarstar$ 
background rate near the end-point spectrum.
As a consequence, in this model the error on the overall amount 
of $\Dstarstar$
has a large effect on $\Vcb$\ while having a negligible contribution
on the slope parameter $\rho_{A_1}^2$. However, HQET predicts that,
in the infinite charm mass limit, the rate near $w=1$ is suppressed by a 
further factor $(w^2-1)$ when compared with the signal \cite{Pene},
\cite{Babar}. 
In this case, the $\Dstarstar$ rate uncertainty would have a large 
effect on the 
slope with only a  small influence on $\Vcb$ \cite{DELPHI_vcb}.
However, models in this extreme case fail to predict the 
ratio ${\rm R}^{**}$ (Equation~(\ref{eq:Rss}))
between the production rates of the two narrow states.

A treatment which accounts for ${\cal O}(1/m_c)$ corrections 
is proposed in \cite{ref:ligeti}. 
%Here, the form factors are predicted in terms of a parameter,
%$\tau_1$ which can be tuned in order to reproduce the value of $R_{**}$. 
%Following the suggestion of \cite{WKSnote}, it is proposed to use this 
%model to describe the $\Dsstar$ form factors.  The systematic error due to the 
%model dependence will be computed by varying $\tau_1$ in such a way as to 
%accommodate values of $R_{**}$ between 0 and 1 \cite{franco_note}.
Several possible approximations of the form factors are provided,
depending on five different expansion schemes
and on three different input parameters.
% To be conservative, at the
%present stage of the analysis, each proposed scheme was
%tested in turn and the input parameters were varied over
%their full range. The quoted central value
%corresponds to the arithmetic average of the values 
%obtained with the two extreme models.
%The systematic error due to the
%modelling of the $\rm D^{**}$ background was computed as half the difference
%between the two extreme results.
%(the smallest
%\vcb\ value was obtained in the A scheme with $\eta = -0.75$, the largest in
%the B1 scheme with $z_{1h} = 0.2$). 
In the calculation of the systematic errors
each scheme proposed was tested and for each scheme, 
the input parameters were varied over the range consistent with the 
measured value of $\rm R^{**}$. 
The systematic error  
was then computed as half the difference
between the two most extreme results.
The branching ratios of the processes which affect the \vcb\ 
value are taken from Section \ref{sec:systgen}.
For the average, the ALEPH analysis
which was based on the Isgur-Wise model \cite{isgw} , has been 
updated to use the same model as the others.
\begin{table}[t]
\begin{center}
\begin{tabular}{|l|lc|} \hline
Experiment   & ${\cal F}_{D^{*}}(1)|V_{\rm cb}|~(\times 10^3)$& $ \rho_{A_1}^2$ \\ \hline
 ALEPH ~~\cite{ALEPH_vcb}      &  33.0 $\pm$  2.1 $\pm$ 1.6 & 0.74 $\pm$ 0.25 $\pm$ 0.41 \\
 DELPHI~\cite{DELPHI_vcb}      &  34.5 $\pm$  1.4 $\pm$ 2.5 & 1.22 $\pm$ 0.14 $\pm$ 0.36 \\
 OPAL$_{{\rm inc}}$~~\cite{OPAL_vcb}        &  37.9 $\pm$  1.3 $\pm$ 2.4 & 1.21 $\pm$ 0.15 $\pm$ 0.35 \\ 
 OPAL$_{{\rm exc}}$~~\cite{OPAL_vcb}        &  37.5 $\pm$  1.7 $\pm$ 1.8 & 1.42 $\pm$ 0.21 $\pm$ 0.24 \\ \hline
\end{tabular}
\end{center}
\caption{{\it Experimental results on \fvcb\ and $\rho_{A_1}^2$ corrected for common 
inputs.} \label{t:correxp} }

\end{table}

To account for the lack of knowledge of the non-resonant part of
the $\Dsstar$ background, the fraction of narrow resonances is varied
from 0 to 100\%. In the average, this error is 
added in quadrature with the one derived from the $\Dsstar$ shape.   
Since HQET models predict the variation of the 
$\rm D^{**}$ form factors as a function of $w$ only for resonant 
states 
%(whose existence is  established), in the LEP working group on
%heavy flavours (LEPHFWG)  
the present approach does not account for the shape variation 
from non-resonant $\rm D^{(*)} \pi$ background states 
if they have marked differences as compared to $\rm D^{**}$
resonant states.
\begin{table}[h]
\begin{center}
\begin{tabular}{|l|cccc|cccc|}\hline
 & \multicolumn{4}{|c|}{$\frac{\Delta {\cal F}(1)|V_{\rm cb}|}{{\cal F}(1)|V_{\rm cb}|}$} & \multicolumn{4}{|c|}{$
\Delta\rho_{A_1}^2$} \\ \hline
Source & A & D & $\rm O_{inc}$ & $\rm O_{exc}$ & A & D & $\rm O_{inc}$ 
&  $\rm O_{exc}$ \\ \hline
{\bf Correlated errors}  &&&&&&&& \\ \hline
$\Gamma _{{\mathrm b }{\overline{ \mathrm b}}}/\Gamma_{\mathrm{ had}}$
                        & 0.2& 0.3 & 0.2 & 0.2 & -   &   -  &  -  & -  \\
$\rm{BR}( b \to \Bdb)$       
                        & 1.1 & 1.8 & 1.3  & 1.3  & -   & 0.02 &  -  & -   \\
$\rm{BR}(D^{*+} \to D^0 \pi^+)$  
                        & 0.4 & 0.4 & 0.4  & 0.4  & -   & 0.01 &  -  & -  \\
$\rm{BR}( \Do\to K^+\pi^-)$
                        & 0.6 &  -  & -    & 0.3  & -   &   -  &  -  & 0.01\\
$\rm{BR}( \Do\to K3\pi )$
                        & 1.3 &   - & -    & -    & -   &   -  &  -  & - \\
$\rm{BR}( \Do \to K^02\pi)$
                        & 0.6 &  -  & -    &  -   & -   &   -  &  -  & -    \\
$\rm{BR}( \Do \to \ K^+\pi^-\pi^0)$
                        &  -  &  -  & -    & 2.6  & -   &   -  &  -  & 0.02 \\ 
$\rm{BR}(\mathrm D\to \mathrm K n \pi )$
                        & -   & 0.2 & -    & -    & -   &  -   & -   & - \\
$\rm D^{**}$  rate      & 0.9 & 1.5 & 0.4  & 0.7  & 0.02& 0.07 & 0.03& 0.05 \\
$\rm D^{**}$  shape
                        & 1.0 & 5.3  & 4.1 & 1.0  & 0.06& 0.19 & 0.15&0.13  \\
%$\rm D^{**}$  r-nr     & 0.9 & 0.5  &  -  & -    &-    & 0.05 & 0.01&0.10  \\
$\Bm \to D^* X_c$          
                        & 0.3 & 0.2  & 0.3 & 0.2  & -   & 0.01 & 0.01& -  \\ 
$\Bm \to D^* \tau \bar{\nu}$                
                         & 0.1 & 0.2 & 0.1  & 0.1 & -   &  -   & -   & - \\ 
Fragmentation            & 0.9 & 1.0 & 1.0  & 0.5 & 0.01&  -   & 0.11& -\\ 
$\tau_{\rm b}$ lifetime  & 0.9 & 0.9 & 0.7  & 0.6 & -   &  -   & -   & -\\
$\rm R_1$ and $\rm R_2$  & 2.4 & 1.1 & 1.0  & 1.0 & 0.4& 0.3 & 0.2& 0.2\\
\hline 
 {\bf Uncorrelated errors}  &&&&&&&& \\ \hline
Combinatorial and fake $\rm D^0$
                        & 1.1 &  0.5 & -   & 1.2 & -   &  0.07 &  -  & 0.01\\
Fake lepton             & 0.7 &  -   & 1.2 & 0.2 & -   &  -    &  -  & -\\ 
$\ell$ efficiency/modelling      
                         & 0.7 & 1.1 & -   & 1.2 & -   &  0.03 &  -  & -\\ 
Selection efficiency/modelling
                         & 1.2 & 2.6 & 2.9 & 3.6 & 0.02& 0.08  & 0.12& 0.01\\
MC statistics            & 1.6 & 0.2 &  -  &  -  & 0.05& -     & -   & -\\ 
$w$ resolution           & 1.5 & 2.1 & 2.2 & 1.4 & 0.07&0.07   & 0.12& 0.035\\  
\hline
{\bf Total Systematic}   & 4.0 & 7.2 & 6.0& 4.8 & 0.41& 0.36& 0.35& 0.24 \\ 
 \hline  
{\bf Statistical }      & 6.5 & 4.0 & 3.5& 4.5 &0.25 & 0.14 & 0.15&0.21 \\ 
 \hline  
\end{tabular}
\end{center}                 
\caption{{\it Dominant systematic uncertainties on \fvcb~ and $\rho_{A_1}^2$ expressed,
respectively, as relative and absolute values.} \label{t:sys} }
\end{table}

%\bc \bt
%\begin{table}
%\begin{center}
%\begin{tabular}{|l|c|c|}
%\hline
% Parameter & Value & Reference \\ \hline
% $\Rb$       & (21.68 $\pm$ 0.07)\% & \cite{EWWG}  \\
% $\fd$       & (39.5  $^{+1.3}_{-1.4}$)\% & \ref{sec:results} \\
%$\tau(\Bdb)$& (1.56   $\pm$ 0.04) ps & \ref{sec:taubd} \\
%$<x_E>$      & 0.702  $\pm$ 0.008 & \cite{EWWG} \\
%  BR($\Dstarp\rightarrow\Do\pi^+$) & (68.3 $\pm$ 1.4) \% & \cite{PDG00} \\ \hline
%  BR($\Do\rightarrow K^-\pi^+$)&                   &\cite{PDG00} & \\
%  BR($\Do\rightarrow K^-3\pi$) &                   &\cite{PDG00} & \\
%  BR($\Do\rightarrow K^-\pi^+\pi^0$) &             &\cite{PDG00} & \\
%  BR($\Do\rightarrow K^0_s\pi^+\pi^-$) &           &\cite{PDG00} & \\ \hline
%  BR($X_c\rightarrow \ell^-X$) & (0.0122$^{+0.0033}_{-0.0027}$)\% & \cite{EWWG} \\
%  BR(\Btau)           & (1.95  $\pm$ 0.30)\% &\cite{EWWG} \\
%  BR(B$^- \rightarrow \Dstarp \pi^- \ell^-\overline{\nu_{\ell}}$) & (1.24$\pm0.19\pm0.04$) \% & 
%             \cite{DssARGUS,DssALEPH}  \\
%BR($\Bdb~ \rightarrow~ \Dstarp \pi^0 \ell^-\overline{\nu_{\ell}})$ & $(0.62\pm0.10\pm0.02)$\% & ** \\
%BR($\Bsb \rightarrow \Dstarp K \ell^-\overline{\nu_{\ell}})$ & $(0.69\pm0.11\pm0.23)$\% & ** \\ \hline
%\end{tabular}
%\end{center}
%\caption{Values of the most relevant parameters affecting 
%the measurement of $\Vcb$.
%The ones flagged by a ``*'' are likely to change in near future. The three 
%$\Dstarstar$ production rates are fully correlated.}
%\label{tab:summary} 
%\et \ec
%\end{table}

\mysubsubsection {Corrections applied to the measurements}

Since the four LEP measurements have been performed using different 
methods and inputs, they must be put on the same footing before being 
averaged. 
%ALEPH and DELPHI measurements have been updated to use 
%the CLN \cite{CLN} extrapolation method (\ref{eq:clnextr}) and the
%Ligeti \cite{ref:ligeti} \btodss\ model.
%Since OPAL uses the Caprini-Neubert 
%extrapolation method and the JETSET \btodss\ model, corrections to the OPAL
%measurement have been estimated using the ALEPH analysis, which is similar. 
%The three experiments also differ by some of the less important input
%parameters listed in Table~\ref{tab:summary}. 
%Corrections for these 
%parameters, listed in Table \ref{t:sys}
% are calculated in a simple and standard way.
Corrections for changing to the standard input parameters, listed in Table \ref{tab:gensys} or evaluated in previous Sections,
have been calculated as for the $\dmd$ measurement 
(see Section \ref{sec:method}).
The central value of each analysis is adjusted according to the 
difference between the used and desired parameter values and the 
associated systematic error. The systematic error 
itself is then scaled to reflect the desired uncertainty on the input
parameter.
Table~\ref{t:correxp} lists the corrected results.
The uncertainty on the variation with $w$ of the  
 \btodss\  form factors is taken to be fully correlated between experiments.
Published results are using theoretically predicted values for the form
factor ratios $R_1(w)$ and $R_2(w)$ defined in Equation (\ref{eq:r12}).
In the present analysis, experimental results on these quantities, obtained
by CLEO \cite{ref:cleoform} have been used and a corresponding 
systematic uncertainty has
been derived.
The dominant systematic uncertainties on \fvcb\ 
are listed in Table~\ref{t:sys}.
The largest comes from the \btodss\ 
contribution.

%Since ALEPH and OPAL analyses use similar exclusive $\Dstarstar$\ reconstruction
%while DELPHI use an inclusive reconstruction method, it was agreed to 
%determine the necessary corrections to the OPAL measurement from the
%ALEPH analysis.
%Performing the ALEPH analysis using the Caprini-Neubert method instead of 
%the Caprini-Lellouch-Neubert method shows that OPAL result should be corrected 
%in the following way:
%\begin{itemize}
%\item \fvcb\ does not need to be corrected (it changes by only 0.03\%)
%\item the slope should be increased by 0.19 (in good agreement with \cite{CLN} 
%       which estimates a difference of 0.21).
%\end{itemize}
%Performing the ALEPH analysis using the Ligeti \btodss\ model, and taking into 
%account that the fraction of $\Dstarp$\ from $\Dstarstar$\ is 24.4\% in OPAL compared 
%to 11.2\% in ALEPH due to a less efficient  $\Dstarstar$\ rejection cut, shows
%that the following correction should be applied to the OPAL result:
%\begin{itemize}
%\item \fvcb\ should be increased by 9.6\%
%\item the slope should be increased by 0.35
%\item the systematic related to \btodss\ shape and rate is 5.9\% on 
%\fvcb\ and 0.20 on the slope
%\end{itemize}

\begin{figure}[t]
\begin{center}
\begin{tabular}{cc}
\mbox{\epsfxsize7.0cm \epsfysize9.0cm\epsffile{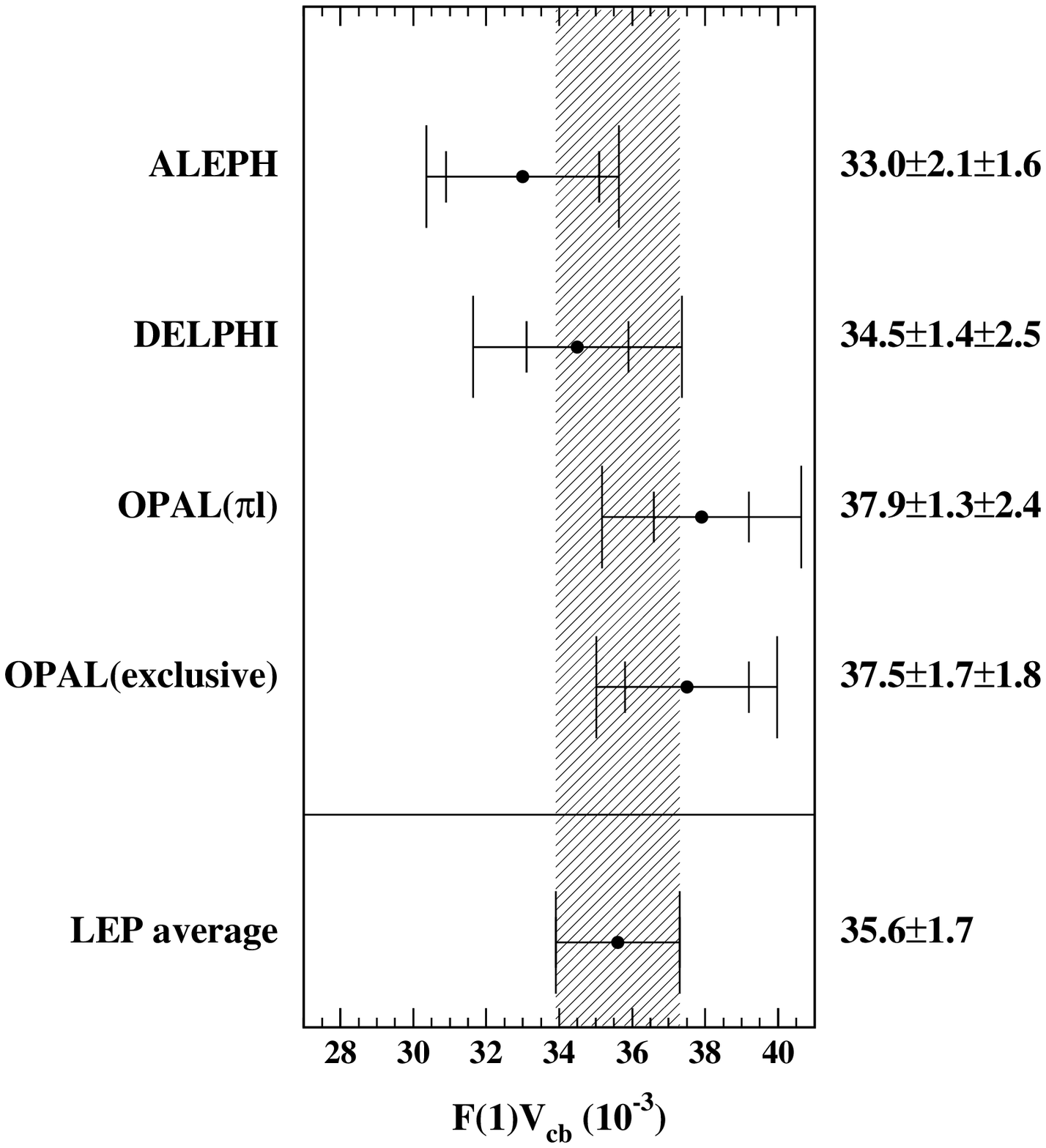}} &
\mbox{\epsfxsize7.0cm \epsfysize9.0cm\epsffile{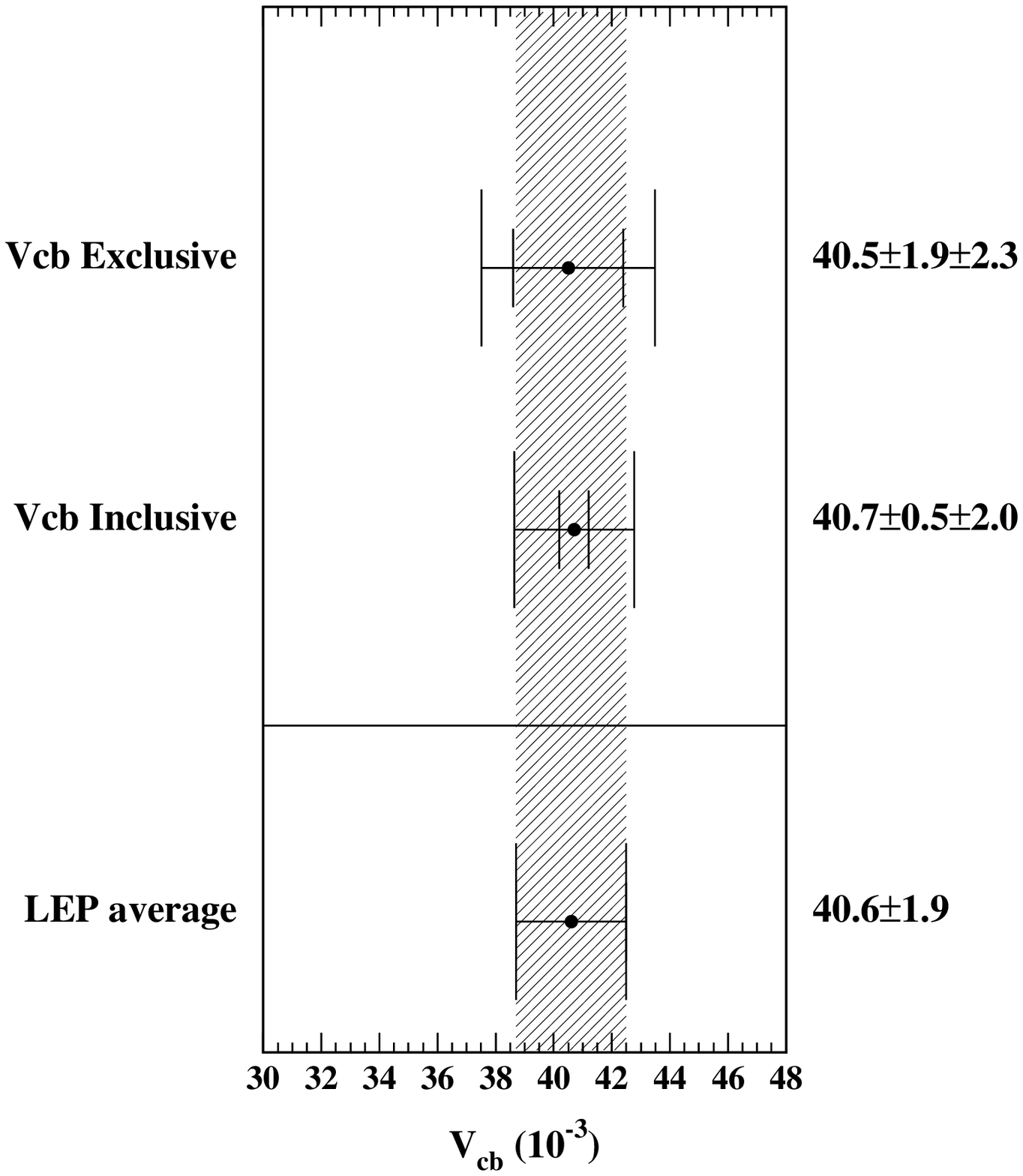}} \\
%\mbox{\epsfxsize7.0cm \epsfysize9.0cm\epsffile{/afs/cern.ch/phys/lep/lepvcb/pdg_rep/vcb_exc_win01_bw.eps}} &
%\mbox{\epsfxsize7.0cm \epsfysize9.0cm\epsffile{/afs/cern.ch/phys/lep/lepvcb/pdg_rep/vcb_all_win01_bw.eps}} \\
\end{tabular}
\caption{{\it Left: Corrected \fvcb\ values and LEP average
 using the exclusive
method.  The values shown in the plot have been adjusted by the 
working group and
are those used to perform the average. The original values
can be found in the experimental papers.
Right: \vcb\ LEP average.}\label{f:vcbav}}

\end{center}
\end{figure}

\mysubsubsection{Exclusive $\Vcb$ average}
The combination method for $\mbox{\fvcb}$ and $\rho_{A_1}^2$ is the same as 
the method used for
$\dmd$
(see Section \ref{sec:method}) generalized to the combination of two or more correlated
parameters.
%After combining the above results 
The LEP average (see Figure \ref{f:vcbav}-right)
gives:
%, using the exclusive method, is:
\begin{eqnarray}
\mbox{\fvcb} & = & \favcb \\
  \rho_{A_1}^2 & = & \rhobb 
\end{eqnarray}
% (which affects mainly the shape of the form factors). 
%It is hoped to reduce it in the future 
%with a better understanding
%of the theoretical uncertainty and with more experimental constraints.
%\begin{table}
%\begin{center}
%\begin{tabular}{|l|c|c|}\hline
%Source &  $\sigma({\cal F}_{D^{*}}(1)|V_{\rm cb}|)/{\cal F}_{D^{*}}(1)|V_{\rm cb}|$& $\sigma(\rho_{A_1}^2)$ \\  \hline
%{\bf    BR's} & & \\
%$\mathrm{BR}(\mathrm D\rightarrow \mathrm K n \pi)$ 
%                        & 1.0 & 0.01  \\
%$\mathrm{BR}(\mathrm D^{*+}\rightarrow\mathrm D^0\pi^+)$  & 1.0 & -\\
%$\Gamma _{{\mathrm b }{\overline{ \mathrm b}}}/\Gamma_{\mathrm{ had}}$
%                        & 0.19 & -\\
%$\fd$          
%                        & 1.2 &  -  \\ \hline
%{\bf         Background} & & \\
%$\rm B^- \rightarrow D^{*+} \pi^- \ell^-\overline{\nu_{\ell}}$  & 4.5 &  0.16 \\
%$\rm \overline{B} \rightarrow D^{*+} X_{\overline{c}}(\rightarrow \ell^- X)$  & 0.20 &  -   \\ 
%$\rm \overline{B} \rightarrow D^{*+} \tau^- \overline{\nu_{\tau}}$ & 0.17 &  -   \\ \hline
%{\bf Detector  }        & 2.1 & 0.10 \\ \hline
%{\bf   Other inputs}    &     &       \\ 
%Fragmentation           & 0.9 &  -    \\ 
% $\Bdb$  lifetime    & 1.2 &  -    \\ \hline
%{\bf Total syst. }      & 5.5 &  0.19  \\  \hline  
%{\bf Statistical }      & 2.7 &  0.09  \\  \hline  
%\end{tabular}                  
%\end{center}
%\caption{{\it Dominant systematic uncertainties on \fvcb~ and $\rho_{A_1}^2$ expressed,
%respectively, as relative and absolute values.}
%{\bf To be updated} \label{t:sys} }
%
%\end{table}

The confidence level of the fit is 12\%. The error ellipses of the 
corrected measurements and of the LEP average are shown on 
Figure~\ref{f:vcbell}.

The theoretical estimate, \fone=$0.88\pm0.05$ (see Appendix 
\ref{appendixC}), is used to determine:
\begin{equation}
\Vcb^{excl.} =  \vcbexc .
\end{equation}

\mysubsection{Overall $\Vcb$ average}
\label{sec:vcbav}

The combined \vcb\ average (see Figure \ref{f:vcbav}-right)
 can be extracted taking into account
correlations between the inclusive and exclusive methods. 
The most important source of correlations comes from theoretical
uncertainties in the evaluation of $\mu_{\pi}$,
the average momentum of the $b$-quark inside the $b$-hadron
 (Appendix \ref{appendixC}). 
In the determination of experimental
systematic uncertainties, theoretical uncertainties in the modelling of 
$b \rightarrow \ell$ decays 
%and the
%exact amount of $b \rightarrow \Dstarstar$ decays 
are taken as fully correlated. 
%If these correlations were neglected, the central value for the 
%$\Vcb$ average would change by only 0.3\%.
Uncertainties from lepton identification and background
also contribute, but to a much lesser extent.
All other sources provide negligible contributions to the
correlated error.
%The various contributions
%to the uncertainty are categorised in Table \ref{corr}. 
The combined value is:
\begin{equation}
\Vcb = \vcbavg
\end{equation}
where, within the total error of $1.9\times 10^{-3}$, 
$1.5\times 10^{-3}$ comes from uncorrelated sources and
$1.0\times 10^{-3}$ from correlated sources. 

%\begin{table}[ht!]
%\begin{center}
%\begin{tabular}{|c||c|c||c|c|} \hline
%source & \multicolumn{2}{|c|} {correlated} & \multicolumn{2}{|c|} {uncorrelated} \\\hline
%         & $\Vcb$ incl. &  $\Vcb$ excl.     & $\Vcb$ incl. &  $\Vcb$ excl. \\\hline
%theory               & 2.4\%     &   2.7\%        & 4.4\%     &   5.0\% \\ 
%exp. syst.           & 0.6\%     &   4.6\%        & 0.7\%     &   3.0\% \\ 
%stat.                &           &                & 0.4\%     &   2.7\% \\ \hline
%total                & 2.5\%     &   5.3\%        & 4.5\%     &   6.4\% \\ \hline
%\end{tabular}   
%\parbox{15cm}{\caption {\it {Contributions to correlated and
%    uncorrelated errors on $\Vcb^{incl.}$ and $\Vcb^{excl.}$,
%expressed as relative errors.
%\label{corr}} {\bf To be updated}}}
%\end{center}
%\end{table}

\mysection{Average of LEP $\Vub$ measurements}
\label{sec:vub}

The LEP combined determinations of 
BR($b \rightarrow \ell^-\overline{\nu_{\ell}}{\rm X}_u$)($\ell^-= e^-$ or $\mu^-$) and the derivation of 
$\Vub$ have been obtained by combining the results reported 
by the 
ALEPH~\cite{aleph_vub},  
DELPHI~\cite{delphi_vub} and L3~\cite{l3_vub} 
Collaborations\footnote{The present members of the $\Vub$ working group are:
D. Abbaneo, M. Battaglia, P. Henrard, S. Mele, E. Piotto and Ph. Rosnet.}. 
The three analyses rely on different techniques to measure the inclusive 
yield of $b \rightarrow u$ transitions in semileptonic $b$-hadron decays. 
The experimental details can be found in the original publications.
All three experiments reported evidence for the 
$b \rightarrow  \ell^-\overline{\nu_{\ell}}{\rm X}_u$ transition and measured 
its rate
BR($b \rightarrow  \ell^-\overline{\nu_{\ell}}{\rm X}_u$). 
DELPHI fitted
$\Vub/\Vcb$ directly 
to the fraction of candidate $b \rightarrow \ell^-\overline{\nu_{\ell}}{\rm X}_u$ decays in the 
selected data sample. For this averaging, the corresponding value of
BR($b \rightarrow \ell^-\overline{\nu_{\ell}}{\rm X}_u$) has been derived by using
the value of $\Vcb$ obtained in Section \ref{sec:vcbav}
%$|V_{cb}| = 0.0395 \pm 0.0017$~\cite{PDG00} 
and the relationship between 
$\Vub$ and  BR($b \rightarrow \ell^-\overline{\nu_{\ell}}{\rm X}_u$) 
given below.

In order to average these results, the sources of systematic uncertainties 
have been divided into two categories. The first contains (a) uncorrelated
systematics due to experimental systematics, such as lepton 
identification, $b$-tagging, vertexing efficiency and energy 
resolution, and (b) uncorrelated systematics from signal modelling and 
background description. 
The second contains correlated systematic uncertainties deriving 
from the simulation of signal $b \rightarrow u$ and background 
$b \rightarrow c$ transitions.
%that have been assumed to be fully 
%correlated for the different analyses. 
The contributions from the statistical,
experimental, and uncorrelated and correlated modelling uncertainties are 
summarised in Table~\ref{tab:brslu}. 

\begin{table}[ht!]
\begin{center}
\begin{tabular}{|l|c c c c c|}
\hline
Experiment & BR & (stat.) & (exp.) & 
(uncorrelated) & (correlated) \\
\hline 
%& & & & &\\
ALEPH~\cite{aleph_vub} & 1.73 & $\pm$ 0.48 & $\pm$ 0.29 &
$\pm$ 0.29 ($^{\pm 0.29~b\rightarrow c}_{\pm 0.04~b\rightarrow u}$) & 
$\pm$ 0.47 ($^{\pm 0.43~b\rightarrow c}_{\pm 0.19~b\rightarrow u}$) \\
 & & & & &\\ 
DELPHI~\cite{delphi_vub} & 1.69 & $\pm$ 0.37 & $\pm$ 0.39 &
$\pm$ 0.18 ($^{\pm 0.13~b\rightarrow c}_{\pm 0.13~b\rightarrow u}$) 
& $\pm$ 0.42 ($^{\pm 0.34~b\rightarrow c}_{\pm 0.20~b\rightarrow u}$) \\
& & & & &\\
L3~\cite{l3_vub} & 3.30 & $\pm$ 1.00 & $\pm$ 0.80 & 
$\pm$ 0.68 ($^{\pm 0.68~b\rightarrow c}_{~~~-~~~b\rightarrow u}$)
& $\pm$ 1.40 ($^{\pm 1.29~b\rightarrow c}_{\pm 0.54~b\rightarrow u}$) \\
\hline
\end{tabular}
\end{center}
\caption[]{{\it The results for 
10$^3 \times${\rm BR}($b \rightarrow  \ell^-\overline{\nu_{\ell}}{\rm X}_u$) from the 
LEP experiments with the statistical, experimental, 
model uncorrelated, and model correlated uncertainties.} \label{tab:brslu}} 
\end{table}

The correlated systematics 
%belong to both the description of  
%background $b \rightarrow c$ and to the modelling of signal 
%$b \rightarrow u$ transitions 
are summarised in Table~\ref{tab:ebrslu}.
Differences in the analysis techniques adopted by the three 
experiments are reflected by the different sizes of the systematics 
uncertainties estimated from each common source. Important common
systematics arise from the D topological branching fractions and the 
rate of ${\rm D} \rightarrow {\rm K}^0$ decays. 
${\rm D}$ decays 
represent a potential source of background for $b \rightarrow u$
decays because both are characterized 
by a small hadronic mass and a low charged 
multiplicity. The sensitivity to the topological branching fractions is 
reduced in the DELPHI analysis by applying a rescaling of the mass 
${\rm M_X}$ of the hadronic system, based on the reconstructed 
$\ell^- \overline{ \nu_{\ell}} {\rm X}$ 
mass, and by the use of identified kaons for separating signal from background 
events. This explains the different sensitivity of the ALEPH and 
DELPHI analyses to these two important sources of systematic uncertainty.
ALEPH and L3 are sensitive to the uncertainties in the $b$ fragmentation
function due to the kinematical variables used for discriminating 
$b \rightarrow \ell^- \overline{\nu_{\ell}} {\rm X}_u$ from 
$b \rightarrow \ell^- \overline{\nu_{\ell}} {\rm X}_c$ 
decays. 
The DELPHI result is sensitive to the assumed production rates of
$b$ hadron species due to the use of kaon 
anti-tagging to reject $b \rightarrow c$, thus rejecting also $\Bs$ and 
$\Lb$ decays; and it is sensitive
to the contribution of ${\rm D}^{(*)} \pi$ and 
${\rm D}^{**}$ 
states in semileptonic decays because the 
resulting difference in the vertex topology is also used for discriminating 
$b \rightarrow u$ from $b \rightarrow c$ decays. 

%The differences in the analysis techniques adopted by the three 
%experiments are reflected by differences in the size of systematic 
%uncertainties estimated from each common source. An important common
%systematic is due to the charm topological branching fractions and to the 
%rate of ${\rm D} \rightarrow {\rm K}^0$ decays. ALEPH and L3 are sensitive to the 
%uncertainties in the $b$ fragmentation function due to the use of kinematical
%variables for discriminating $b \rightarrow \ell^-\overline{\nu_{\ell}}{\rm X}_u$ 
%from $b \rightarrow
%\ell^-\overline{\nu_{\ell}}{\rm X}_c$ decays. 
%The DELPHI result depends on the assumed
%composition in $b$-hadron species due to the use of kaon anti-tagging to 
%reject $b \rightarrow c$, thus rejecting also $\Bs$ and $\Lb$ decays. 

\begin{table}[hb!]
\begin{center}
\begin{tabular}{|l|c|c|c|}
\hline
Source & ALEPH & DELPHI & L3 \\
\hline 
$b$  species                                 & 0.01  & 0.12  &  - \\
$b$ fragmentation                            & 0.22  & 0.03 & 0.32  \\
$b \rightarrow \ell$ model                   & 0.11 & 0.08 & 1.24\\
$c \rightarrow \ell$ model                   & 0.14 & 0.13 & 0.12\\
D topological BR's                           & 0.31 & 0.06 & -\\
BR(D$ \rightarrow {\rm K}^0$)                & 0.08 & 0.19  & -\\
$\Dstarstar$, ${\rm D}^{(*)} \pi$ production & 0.04 & 0.19 & -\\
\hline
$b \rightarrow u$ inclusive model            & 0.18  & 0.08 & 0.25\\
$b \rightarrow u$ exclusive model            & 0.05  & 0.18  & 0.20\\
$\Lambda_b$ decay model                      & 0.04 & -    & 0.44\\
\hline
\end{tabular}
\end{center}
\caption[]{{\it Correlated sources of systematic uncertainties (in units of
10$^{-3}$) entering in the measurement of 
{\rm BR}($b \rightarrow \ell^-\overline{\nu_{\ell}}{\rm X}_u$).} \label{tab:ebrslu}}
\end{table}

The  systematic uncertainties on the $b \rightarrow u$ signal have been 
grouped into {\sl inclusive model} and {\sl exclusive model} classes, 
which are assumed to be fully
correlated. The first corresponds to the uncertainty in the modelling of
the kinematics of the $b$-quark in the heavy hadron. It has been 
estimated from 
the spreads of the results obtained with 
%{\it What difference? half the total difference? or other ways ... ?} 
the ACCMM~\cite{accmm}, the 
Dikeman-Shifman-Uraltsev~\cite{dsu}, and the parton~\cite{parton} models 
in the ALEPH and DELPHI analyses. In the case of the L3 analysis,
the uncertainties in the single pion and in the lepton energy spectra
were evaluated  
from the discrepancies between the model of Ref~\cite{burdman} and the 
ISGW~\cite{isgw} model respectively to the JETSET~7.4~\cite{jetset} 
prediction. 
The {\sl exclusive model} uncertainty corresponds to the modelling of the 
hadronic final state in the $b \rightarrow \ell^-\overline{\nu_{\ell}}
{\rm X}_u$ 
decay.
These uncertainties have been estimated by replacing the parton shower 
fragmentation model in JETSET~\cite{jetset} with the fully exclusive 
ISGW2~\cite{isgw2} 
model by ALEPH and DELPHI, and by propagating a 100\% uncertainty 
on the ${\rm B} \rightarrow \pi \ell^-\overline{\nu_{\ell}}$ rate by L3. 
ALEPH and L3 have also taken into account the uncertainty from the 
modelling of the charmless semi-leptonic decay of $b$-baryons. This has not
been considered by DELPHI as they remove decays containing identified
protons and kaons, thus suppressing the contribution of $b$-baryons as 
mentioned above. In addition to these sources, 
ALEPH has estimated a $b \rightarrow u$ uncertainty from the 
energy cut-off value $\Lambda$ for the hybrid model adopted~\cite{hybrid}; 
DELPHI allowed for the $b$-quark pole mass $m_b$ and the 
expectation value of the kinetic energy operator $<p^2_b>$ uncertainties
which have been assumed to be uncorrelated.

\begin{figure}[th!] 
\begin{center}
\begin{tabular}{cc}
\mbox{\epsfxsize8.0cm \epsfysize10.0cm\epsffile{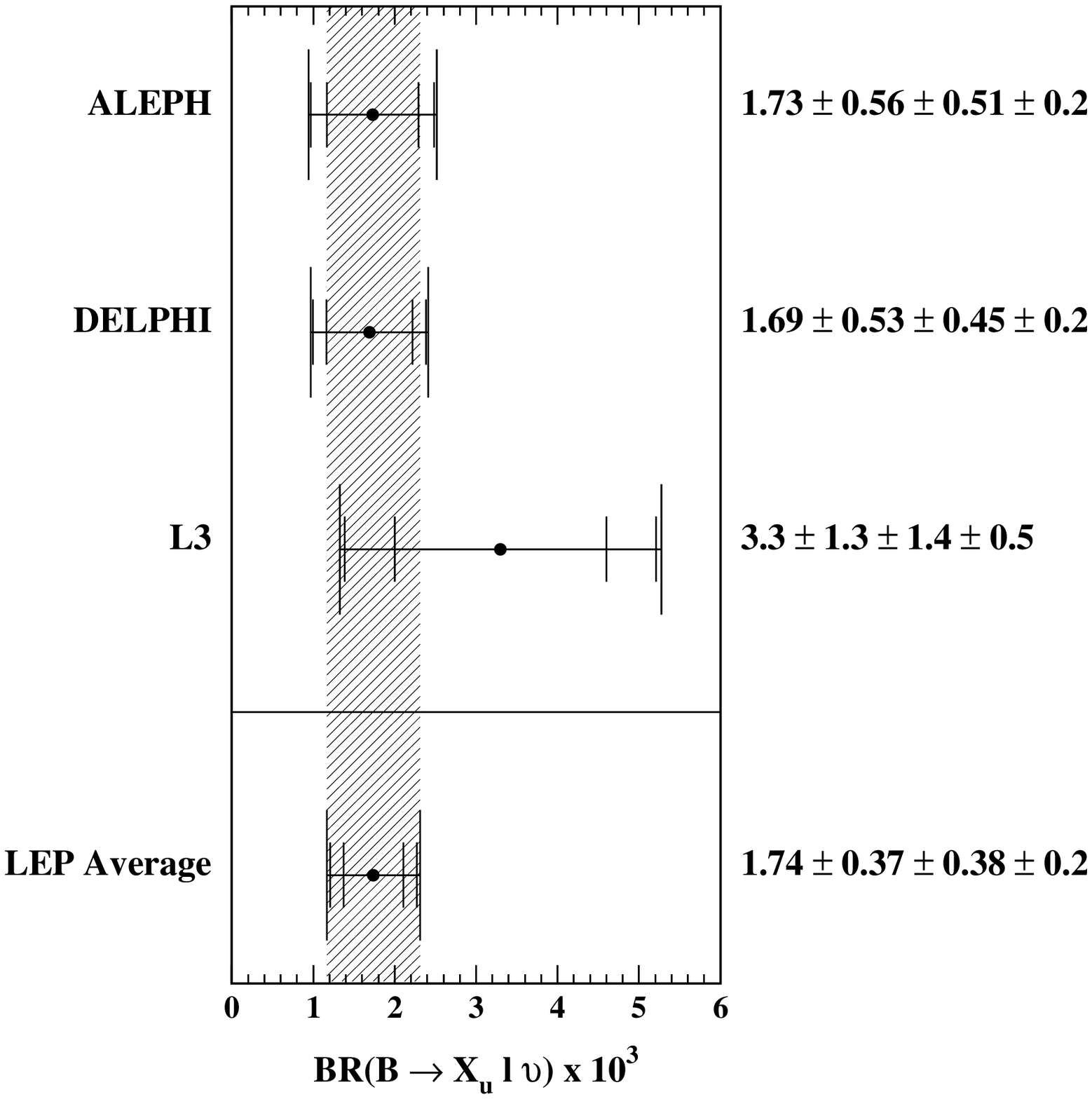}} &
\mbox{\epsfxsize8.0cm \epsfysize10.0cm\epsffile{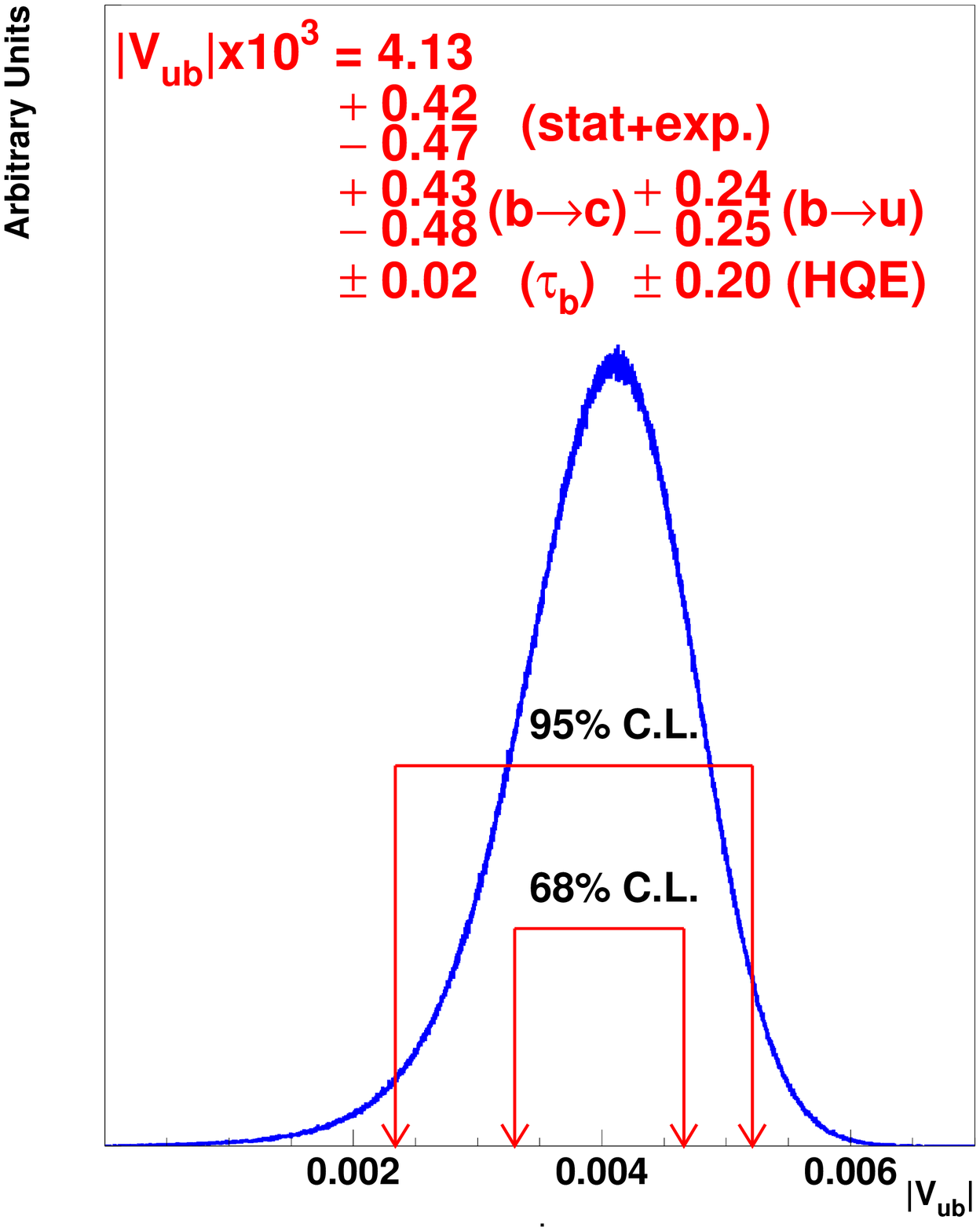}} \\
%\mbox{\epsfxsize8.0cm \epsfysize10.0cm\epsffile{/afs/cern.ch/user/r/roudeau/www/lep_bu_2000_2.eps}} &
%\mbox{\epsfxsize8.0cm \epsfysize10.0cm\epsffile{/afs/cern.ch/user/b/battagl/www/vub/VubPDF2000.eps}} \\
\end{tabular}
\caption{{\it Left: the determinations of 
${\rm BR}(b \rightarrow \ell^-\overline{\nu_{\ell}}{\rm X}_u $) by 
ALEPH, DELPHI and L3 and the resulting LEP average.
Right: the probability density function for $\Vub$ corresponding to 
the LEP average value of ${\rm BR}(b \rightarrow \ell^-\overline{\nu_{\ell}}{\rm X}_u$)
with the value of 
the median and two confidence intervals indicated. 
Unphysical negative entries have been
discarded and the probability density function renormalised accordingly.}
\label{fig:brslu}}
\end{center}
\end{figure}

The three measurements of BR($b \rightarrow  \ell^-\overline{\nu_{\ell}}{\rm X}_u$) 
have been averaged 
using the Best Linear Unbiased Estimate (B.L.U.E.) technique~\cite{blue}.
Using the inputs from Table~\ref{tab:brslu} and Table~\ref{tab:ebrslu}, the 
LEP average value for BR($b \rightarrow \ell^-\nu {\rm X}_u$) was found to be:
\begin{eqnarray} 
{\rm BR}(b \rightarrow \ell^-\overline{\nu_{\ell}}{\rm X}_u)& = &(1.74 \pm 0.37 {\rm (stat.+exp.)} \pm 0.38 (b \rightarrow c) 
\pm 0.21 (b \rightarrow u)) \times 10^{-3} \nonumber \\
 & =& (1.74 \pm 0.57) \times 10^{-3} 
\end{eqnarray} 
with a confidence level for the combination of 0.70 
(see Figure~\ref{fig:brslu}).
% in which the separated and combined values are presented).

The magnitude of the matrix element ${\rm V_{ub}}$
 has been extracted using the 
following relationship derived in the context of (OPE)~\cite{uraltsev},
\cite{hoang}:
\begin{eqnarray}
\Vub & =& 0.00445~ \sqrt{\frac{ \mathrm{BR}(b \rightarrow {\rm X}_u \ell^-\overline{\nu_{\ell}})}
{0.002} \frac{1.55 \mathrm{ps}}{\tau_b}} \times \nonumber \\
 & & (1 \pm 0.010 \mathrm{(pert)} \pm 0.030 (1/m_b^3) \pm 0.035 
(m_b))
\end{eqnarray}
by assuming $m_b$ = (4.58 $\pm$ 0.06)~$\GeV/c^2~$ (Appendix \ref{appendixC}).

From the LEP average of  
${\rm BR}(b \rightarrow \ell^-\overline{\nu_{\ell}}{\rm X}_u $)
obtained above, 
the $b$-lifetime value, $\tau_b$, obtained in Section \ref{sec:Atau},
% of $\tau_b$ = 1.564 $\pm$ 0.014~ps~\cite{PDG00}
and the quoted $\Vub$ uncertainty coming from OPE ($\pm$4.7 \%
relative), a probability density function for $\Vub$ has been calculated.
The resulting distribution is shown in Figure~\ref{fig:brslu}-right,
where all errors are convoluted together
 assuming that they are Gaussian in 
${\rm BR}(b \rightarrow \ell^-\overline{\nu_{\ell}}{\rm X}_u$), 
with the exception of the OPE error assumed 
to be Gaussian in $\Vub$.
% where all the 
%errors are convoluted together  
The small part of 
this function in the negative, unphysical region, corresponding
to only 0.14\%, has been discarded and the 
probability density function renormalised accordingly.
The median of this function has been chosen as the best estimate of
$\Vub$, and $\pm 34.135\%$ and $\pm 47.725\%$ of the integral of 
the probability density function around this value have been used to define
the 1$\sigma$ and 2$\sigma$ confidence regions, denoted hereafter as the 
68\% and 95\% confidence levels, obtaining:
\begin{equation}
\Vub = (4.13^{+0.63}_{-0.75}) \times 10^{-3}~{\rm at~ the~ 68\%~ C.L.}
\end{equation}
and
\begin{equation}
\Vub = (4.13^{+1.18}_{-1.71}) \times 10^{-3}~{\rm at~ the~95\%~ C.L.}
\end{equation}

All the uncertainties have been included in these estimates. 
The application of the above procedure for each error source separately 
yields the following detailed result for the 68\% confidence level:

\begin{eqnarray}
\Vub &= &(4.13 ^{+0.42}_{-0.47}~{\rm (stat.+det.)} 
      ^{+0.43}_{-0.48}~(b \rightarrow c~{\rm syst.)} 
      ^{+0.24}_{-0.25}~(b \rightarrow u~{\rm syst.)}  \nonumber \\
    & &  \pm 0.02~(\tau_b) 
      \pm 0.20~{\rm (OPE)}) \times 10^{-3}.
\end{eqnarray}

%\newpage
%\clearpage

\mysection{Results on CP violation in the $\Bd-\Bdb$ system}
\label{sec:sintwobeta}
The decay $\Bd \rightarrow {\rm J}/\psi \Kos$ is known as the gold-plated mode
to establish the existence of CP violation in $b$-hadron decays, due to
its clean experimental signature and low theoretical uncertainty.
The final state is a CP eigenstate, to which both $\Bd$ and $\Bdb$ can
decay. The interference between their direct and indirect decays via
$\Bd-\Bdb$ mixing leads to a time-dependent CP asymmetry given by:
\begin{equation}
A(t)=\frac{\Gamma(\Bd \rightarrow{\rm J}/\psi \Kos)
-\Gamma(\Bdb \rightarrow{\rm J}/\psi \Kos) }
{\Gamma(\Bd \rightarrow{\rm J}/\psi \Kos)
+\Gamma(\Bdb \rightarrow{\rm J}/\psi \Kos)}=
-\sin{(2 \beta)} \sin{(\dmd t)}.
\end{equation}

Here $\Gamma(\Bd \rightarrow{\rm J}/\psi \Kos)$ represents the rate of 
particles that were produced as $\Bd$ decaying to ${\rm J}/\psi \Kos$
at proper time $t$.

The first published attempt at a direct measurement was made by OPAL
\cite{ref:opalpsi}. They selected 24 candidates with an estimated purity
of 60$\%$ and reported a value of
$\sin{2 \beta}=3.2^{+1.8}_{-2.0}\pm0.5$. CDF published an analysis
based on 395 candidates with a purity of about 40$\%$ \cite{ref:cdfpsi}.
They measured $\sin{2 \beta}=0.79^{+0.41}_{-0.44}$, statistical and
systematic errors combined. Finally ALEPH \cite{ref:alephpsi}
selected 23 candidates
with an estimated purity of 71$\%$ and measured
$\sin{2 \beta}=0.84^{+0.84}_{-1.05}$, where the uncertainty is dominated
by the statistics.
The three log-likelihoods of these measurements have been summed to give 
a combined result, i.e. neglecting any correlation between the 
systematic errors of the different experiments (expected to be small).
It corresponds to \cite{ref:alephpsi}:
\begin{equation}
\sin{2 \beta}=0.88^{+0.36}_{-0.39}.
\end{equation}

A second way to search for CP-violating effects uses inclusive semileptonic
or fully inclusive decays. In the $\rm B^0_d$-$\rm\bar{B}^0_d$ system,
the mass eigenstates $\rm |B_1\rangle$ and $\rm |B_2\rangle$ 
can be described in  terms of the weak eigenstates 
$\rm |B^0_d\rangle$ and $\rm |\bar{B}^0_d\rangle$ by:
\begin{eqnarray}
|{\rm B_1}\rangle & = & \frac{(1+\epsilon_B+\delta_B)|{\rm B^0_d}\rangle
+(1-\epsilon_B-\delta_B)|{\rm\bar{B}^0_d}\rangle}
{\sqrt{2 (1+|\epsilon_B+\delta_B|^2)}}  \\
|{\rm B_2}\rangle & = & \frac{(1+\epsilon_B-\delta_B)|{\rm B^0_d}\rangle
-(1-\epsilon_B+\delta_B)|{\rm\bar{B}^0_d}\rangle}
{\sqrt{2 (1+|\epsilon_B-\delta_B|^2)}} \nonumber
\end{eqnarray}
where the complex parameters $\epsilon_B$ and $\delta_B$ parameterise 
indirect CP and CPT violation \cite{cpttheo1}.
Predictions for $\rm Re(\epsilon_B)$
are of the order of $10^{-3}$ in the Standard Model and up to an 
order of magnitude larger in superweak models \cite{cpttheo2}.
For $\rm B^0_s$ mesons, 
the effects are expected to be at least an order of magnitude smaller,
because of the relative sizes of $\rm Im(V_{td})$ and $\rm Im(V_{ts})$.

The value of $\rm Re (\epsilon_B)$ has been measured by ALEPH, DELPHI and OPAL
by studying time dependent charge asymmetries in semileptonic and fully
inclusive $b$-hadron decays \cite{lepcpt}. The methods are similar to those 
used for
measuring $\rm B^0_d$ oscillations. CDF has also obtained a measurement
using dimuon events \cite{cdfcpt}. 
Assuming $\rm Re(\epsilon_B)=0$ for $\rm B^0_s$ mesons, and that CPT violation
is negligible ({\em i.e.} $\rm Im(\delta_B)=0$), the
LEP and CDF results have been averaged to obtain:
\begin{equation}
{\rm Re (\epsilon_B)/(1+|\epsilon_B|^2)} = -0.0024\pm 0.0039 .
\end{equation}
This value compares well with a recent CLEO result using inclusive lepton
and dilepton events, 
${\rm Re (\epsilon_B)/(1+|\epsilon_B|^2)}=0.0035\pm 0.0103\pm 0.0015$
\cite{cleocpt}.

The measurements of the CPT-violation parameter $\rm Im (\delta_B)$ obtained
by DELPHI and OPAL \cite{lepcpt} have also been averaged to give
\begin{equation}
{\rm Im (\delta_B)} = -0.016\pm 0.012 .
\end{equation}

\mysection{Determination of the parameters of the unitarity triangle}
\label{sec:triangle}
%{\bf Present values for $\rho$ and $\eta$ do not correspond exactly
%to the values of the different parameters quoted in previous sections.
%Updated results will be given once these numbers have been frozen.}
As, in the present report, measurements of $\Vcb$, $\Vub$, $\dmd$,
and limits on $\dms$ and on CP violation in the $\Bd-\Bdb$ system 
have been collected from results made available by CDF, LEP and SLD
Collaborations, it has been considered appropriate to give also an updated
overview
of the constraints provided by these measurements, and by additional
results already published by other collaborations, on the CKM parameters
within the Standard Model framework.
In the Standard Model, weak interactions among quarks are codified
in a $3 \times 3$ unitary matrix: the CKM matrix. 
\begin{equation}
\begin{array}{cccc}
V_{CKM} =
&
\left ( \begin{array}{ccc} 
V_{ud} ~~ V_{us} ~~ V_{ub} \\
V_{cd} ~~ V_{cs} ~~ V_{cb} \\
V_{td} ~~ V_{ts} ~~ V_{tb}
\end{array} \right )
\end{array}
\label{eq:ckmmatrix}
\end{equation}

This matrix can be parametrized using three real quantities and one phase
which cannot be removed by redefining the quark field phases. This
phase leads to the violation of the CP symmetry. 
In the Wolfenstein parametrization \cite{ref:wolf}, these four parameters
are labelled as $\lambda$, $A$, $\rhobar$ and $\etabar$
\footnote{The parameters $\rhobar$ and $\etabar$ are related to the 
original $\rho$ and $\eta$, introduced by Wolfenstein, using the expressions
$\rhobar = \rho \times (1 -\frac{\lambda^2}{2})$
and $\etabar = \eta \times (1 -\frac{\lambda^2}{2})$.}.
Experimentally there
is a hierarchy in the magnitude of the elements of the CKM matrix and
the following expression corresponds to an expansion valid up to
${\cal O}(\lambda^6)$ without any prejudice on the possible values of 
the other parameters \cite{ref:zzx}.

\begin{equation}
\begin{array}{cccc}
V_{CKM} =
&
\left ( \begin{array}{cccc}
1 - \frac{\lambda^{2}}{2} - \frac{\lambda^4}{8} ~~~~~~~~~~~~~~~~~~~~~~~~~~~~~~ ~~~~ \lambda ~~~~~~~~~~~~~~~~~~
A \lambda^{3} (\rho - i \eta) \\
   - \lambda +\frac{A^2 \lambda^5}{2}(1-2 \rho) -i A^2 \lambda^5 \eta~~~~~~~~~~  1 - \frac{\lambda^{2}}{2}
   -\lambda^4(\frac{1}{8}+\frac{A^2}{2})
 ~~~~~~~~           A \lambda^{2}       \\
A \lambda^{3} [1 - (1-\frac{\lambda^2}{2})(\rho +i \eta)] ~~~ -A \lambda^{2}(1-\frac{\lambda^2}{2})[1 + \lambda^{2}(\rho +i \eta)]
 ~~~    1-\frac{A^2 \lambda^4}{2}
\end{array} \right )
& 
%+ O(\lambda^{6}).
\end{array}
\label{eq:eq8}
\end{equation}
It has been obtained using the following definitions for the parameters:
\begin{equation}
V_{us} = \lambda,~V_{cb}= A \lambda^2,~V_{ub}=A \lambda^3 (\rho -i \eta).
\end{equation}
The Standard Model allows the amplitudes for all processes
involving weak transitions to be evaluated in terms of the values of 
these parameters.
%; CP violation is one of those.
Measurements of semileptonic decays of strange and $b$-hadrons are the 
main sources of informations on $\lambda$ and $A$, respectively. 
Measurements of $\epsilonk$, $\vubovcb$, $\dmd$ and $\dms$ are expected
to define compatible values for $\rhobar$ and $\etabar$. The latter provide
a set of four constraints which are obtained by comparing measured and 
expected values of the corresponding quantities, in the framework
of the Standard Model or of any other given model. These constraints depend,
in addition, on parameters which have to be taken from theory or from other
measurements. 

%As there are more constraints (four) than quantities to be determined 
%(two, namely $\rhobar$ and $\etabar$) it is possible
%to remove one or two constraints
%or the knowledge of   similar numbers of
%parameters
%and to compare the value of the
%corresponding quantity, as determined
%by the other constraints, with the expectations.

\mysubsection{The unitarity triangle}

The unitarity of the CKM matrix implies that six equations of the type:
\begin{equation}
\sum_{i=1,3} {\rm V}_{ij}^* {\rm V}_{ik}~=~0~{\rm and}~
\sum_{i=1,3} {\rm V}_{ji} {\rm V}_{ki}^*~=~0,
\label{eq:unitar}
\end{equation}
corresponding to the products of two columns or two lines,
have to be satisfied. They define six triangles of the same area.
The triangle, obtained from the product between the third column and the 
complex
conjugate of the first one, is expected to have sides of similar 
sizes.
Using the Wolfenstein parametrization one obtains:
\begin{equation}
{\rm V}_{ud}^* {\rm V}_{ub}= A \lambda^3 (\rhobar - i \etabar),~
{\rm V}_{cd}^* {\rm V}_{cb}= -A \lambda^3,~
{\rm V}_{td}^* {\rm V}_{tb}= A \lambda^3 (1-\rhobar + i \etabar)
\label{eq:titi}  
\end{equation}
The three expressions are proportional to $A \lambda^3$, which can be factored
out, and the geometrical representation of Equation (\ref{eq:titi}),
in the $(\rhobar,\etabar)$ plane, is a triangle with vertices
at C(0,0), B(1,0) and A($\rhobar,\etabar)$.

The lengths of the sides of the triangle can be related to the values of 
the modulus of CKM matrix elements:
$$
\overline{AC} = \frac{1-\frac{\lambda^2}{2}}{\lambda}  \frac{\mid V_{ub} \mid}{\mid V_{cb} \mid} = 
{\sqrt{\overline{\rho}^{2} + \overline{\eta}^{2}}}~{\rm ~~(\Vub ~and~ \Vcb ~measts.)} 
$$
$$
\overline{AB} = \frac{\mid V_{td} \mid}{\lambda \mid V_{cb} \mid} = \sqrt{(1-\overline{\rho})^{2} +\overline{ \eta}^{2}}~{\rm ~~(\dmd ~and ~\Vcb ~measts.)}  
$$
\begin{equation}
\overline{AB}  =  \frac{1-\frac{\lambda^2}{2}}{\lambda} \frac{\mid V_{td} \mid}{ \mid V_{ts} \mid} = 
\sqrt{(1-\overline{\rho})^{2} +\overline{ \eta}^{2}}~{\rm ~~(\dmd ~and~\dms ~measts.)} 
\end{equation}
these expressions are valid up to  ${\cal O}(\lambda^4)$
\footnote{By definition, the length of the side BC is equal to unity.}.
Angles of the triangle, opposite to the sides BC, AC and AB are designed
respectively as $\alpha,~\beta$ and $\gamma$.

\mysubsection{The constraints which depend on $\rhobar$ and $\etabar$}

%Measurements of strange and $b$-hadron semileptonic decay branching 
%fractions give accurate determinations of the parameters $\lambda$ and
%$\Vcb$. 
%In addition to CP violation studied in
%$\Bd \rightarrow J/\psi \Kos$ decays (Section \ref{sec:sintwobeta}),
%there are, at present, 
Four measurements which provide constraints on the
possible range of variation of the $\rhobar$ and $\etabar$ parameters
have been considered.
%Two-body rare B meson decay rates, measured by CLEO, depend also on 
%these parameters and may provide valuable constraints but they are
%not included in the present analysis. 

\mysubsubsection{Charmless $b$-hadron semileptonic decays}
\label{sec:vub2}
The relative rate between charmed and charmless $b$-hadron semileptonic
decays, proportional to $\vubovcb^2$,
 allows to measure the length of the side AC of the triangle.
Measurements of $\Vub$ obtained at LEP 
from inclusive $b \rightarrow u \ell^- \overline{\nu_{\ell}} {\rm X}$
transitions (Section \ref{sec:vub}) 
and by the CLEO Collaboration of the exclusive
$\overline {{\rm B}} \rightarrow \rho \ell^- \overline{\nu_{\ell}}$ decay 
channel \cite{ref:vubcleo} have been used. The CLEO result is:
\begin{equation}
\Vub=(32.5 \pm 2.9 \pm 5.5) \times 10^{-4}
\end{equation}
where the second uncertainty is theoretical and has been obtained using several
models to describe the decay form factors. The p.d.f. for the CLEO
measurement is thus a convolution of Gaussian and flat distributions.
The p.d.f. for the LEP result is given in Section \ref{sec:vub}. Combining
the two distributions, in practice an almost Gaussian p.d.f. is obtained 
corresponding to:
\begin{equation}
\Vub=(35.5 \pm 3.6) \times 10^{-4}
\end{equation}

\mysubsubsection{The $\Bd-\Bdb$ oscillation parameter $\dmd$}
\label{sec:dudule}
In the Standard Model, the mass difference between the mass eigenstates
of the $\Bd-\Bdb$ system can be written:
\begin{equation}
\dmd =\frac{G_F^2}{6\pi^2}m^2_W m_{\Bd}\eta_c S(\frac{m_t^2}{m_W^2}) A^2 \lambda^6 [(1-\rhobar)^2+\etabar^2] f_{B_d}^2 \hat B_{B_d} 
\label{eq:dmdstd}
\end{equation}
The most uncertain parameter in 
Equation (\ref{eq:dmdstd}) is $f_{B_d} \sqrt{\hat B_{B_d}}$.
Its value has been taken from recent lattice QCD results.
It depends on the values of the $\Bd$ decay constant, $f_{B_d}$, and of
the renormalization scale invariant parameter $\hat B_{B_d}$.
%This last quantity is evaluated \cite{ref:bur1} in a way which is 
%consistent with the conventions used to obtain the other parameters
%which depend on QCD corrections in Equation (\ref{eq:dmdstd}):
It should be noted that the final accuracy on $\rhobar$ and $\etabar$ is weakly
dependent on the assumed uncertainty on this parameter \cite{ref:scripto}. 
The value of $\eta_c=0.55 \pm 0.01$ corresponds to a perturbative QCD 
short-distance NLO correction \cite{ref:bur2}. 
$S$ is the Inami-Lim 
function whose expression can be found also in \cite{ref:bur2}.
The value of $m_t$ if the $\overline{\rm MS}$ top mass,
$m_t^{\overline{\rm MS}}(m_t^{\overline{\rm MS}})=(167 \pm 5) \GeV/{\rm c}^2$, 
deduced from measurements of the physical mass 
performed by CDF and D0 \cite{ref:top}.
Measurements of $\dmd$ constrain the length of the side AB
of the triangle.

\mysubsubsection{The ratio between $\Bd-\Bdb$ and $\Bs-\Bsb$
oscillation periods}
In the Standard Model, the mass difference between the mass eigenstates
of the $\Bs-\Bsb$ system is independent of the values of the 
parameters $\rhobar$ and $\etabar$. 
As a consequence, a measurement of this quantity can be used
as a constraint on the effects of strong interactions appearing in the
parameters $f_{B_q} \sqrt{B_{B_q}}$. 
%Instead of taking from theory
%the absolute values of these quantities, only their relative variation,
%when going from $\Bd$ to $\Bs$ mesons, has to be used. 
The ratio between the values of the mass difference between 
the mass-eigenstates, measured for the two
systems of neutral $b$-mesons, can be expressed as:
\begin{equation}
\frac{\dmd}{\dms}~=~\frac{m_{B_d}f_{B_d}^2 \hat B_{B_d}}
{m_{B_s}f_{B_s}^2 \hat B_{B_s}}
\frac{\lambda^2}{(1-\frac{\lambda^2}{2})^2}[(1-\rhobar)^2+\etabar^2].
\end{equation}

This expression depends on the ratio of $b$-decay constants and bag parameters
for $\Bs$ and $\Bd$ mesons respectively:
\begin{equation}
\xi~=~\frac{f_{B_s}\sqrt{ \hat B_{B_s}}}{f_{B_d}\sqrt{ \hat B_{B_d}}}
\end{equation}
and is expected to have a smaller relative uncertainty than
the two contributing quantities from $\Bd$ and $\Bs$ mesons.

\mysubsubsection{CP violation in the kaon system}

CP violation in the kaon system can be expressed in terms of the 
quantity $\epsilonk$:
\begin{equation}
\epsilonk=C_{\epsilon}A^2 \lambda^6 \hat \BK \etabar~ [-\eta_1 S(\frac{m_c^2}{m_W^2}
)+ \eta_2 S(\frac{m_t^2}{m_W^2})A^2\lambda^4(1-\rhobar)
+\eta_3 S(\frac{m_c^2}{m_W^2},\frac{m_t^2}{m_W^2})]
\label{eq:epskdef}
\end{equation}
where
\begin{equation}
C_{\epsilon} = \frac{G_F^2 f_K^2 m_K m^2_W}{6 \sqrt{2}\pi^2\Delta m_K}.
\label{eq:epsilon}
\end{equation}
Functions $S$ are again appropriate Inami-Lim functions, including
the next to leading order corrections which enter also in the parameters
$\eta_i$.
The most uncertain parameter,
in Equation (\ref{eq:epskdef}) is $\hat \BK$.
Its central value and corresponding uncertainty have been taken
from lattice QCD calculations.
% Values for the other parameters
%are given in Table \ref{tab:alltri}.

\mysubsection{Probability distributions for $\rhobar$, $\etabar$ and
other quantities}

% There are several ways to determine the region of the $(\rhobar,
%\etabar)$ plane which is favoured by present measurements. In the scanning 
%approach \cite{ref:scan} it is considered that it is not possible
%to define probability density distributions for theooretical parameters
%coming from calculations affected by systematic uncertainties or
%educated guesses. For such parameters, it is assumed that one can 
%only define an interval 

Expressions for $\vubovcb$, $\dmd$, $\dmd / \dms$ and $\epsilonk$
relate the corresponding measurements of these quantities 
($c_j,~j=1,4$) to the CKM
matrix parameters $\rhobar$ and $\etabar$, via the set of ancillary parameters
$(x_1,x_2,...,x_N)$ corresponding to all experimentally determined
or theoretically calculated quantities.
The bi-dimensional probability distribution for $\rhobar$ and $\etabar$
% resulting from the combination of the various measurements
is obtained using the Bayes theorem.

\begin{table}[t]
\begin{center}
\small{
\begin{tabular}{|c|c|c|c|c|}
\hline
                       Parameter                 &  Value &
 Gaussian &  Uniform & Ref. \\
 & &   $\sigma$ & half-width &  \\
\hline
$\lambda$    & $0.2237$ & \multicolumn{2}{|c|}{  $0.0033$\,}   & \cite{ref:harimarti} \\
$\left | V_{cb} \right |$ & $40.6 \times 10^{-3}$
     & \multicolumn{2}{|c|}{ $1.9\times 10^{-3}$\,, see text} &Sect. \ref{sec:vcb} \\

$\left | V_{ub} \right |$  & $ 35.5  \times 10^{-4}$ & $ 3.6 \times 10^{-4}$ & -- &Sect. \ref{sec:vub2}\\
$\epsilonk$   &
$2.280 \times 10^{-3}$&
$0.019
 \times 10^{-3}$  & --
     &~\cite{PDG00} \\
     $\Delta m_d$  & $0.487~\mbox{ps}^{-1}$ & $0.014~\mbox{ps}^{-1}$ & --
     &Sect.~\ref{sec:boscill}  \\
$\Delta m_s$  & $>$ 15.0 ps$^{-1}$ at 95\% C.L.
     & \multicolumn{2}{|c|}{see text} &Sect.~\ref{sec:boscill}  \\
$m_t$ & $167~\GeV$ & $ 5~\GeV$ & --
     &~\cite{ref:top} \\
$m_b$ & $4.23~\GeV$ & $ 0.07~\GeV$ & --
     &~\cite{ref:hispanicus} \\
$m_c$  & $1.3~\GeV $ & $0.1~\GeV$
     & --  &~\cite{PDG00} \\
$\hat \BK$    & $0.87$ & $0.06$ &  $0.13$
     & ~\cite{ref:harimarti} \\
$f_{B_d} \sqrt{\hat B_{B_d}}$ & $230~\MeV$  & $25~\MeV$
     &  $20~\MeV$  & ~\cite{ref:harimarti} \\
$\xi=\frac{ f_{B_s}\sqrt{\hat B_{B_s}}}{ f_{B_d}\sqrt{\hat B_{B_d}}}$
     & $1.14$ & 0.03 & $ 0.05 $& ~\cite{ref:harimarti} \\
$\alpha_s$  &  $0.119$ & $0.03$ & -- &~\cite{ref:harimarti} \\
$\eta_1$  &  $1.38$ & $0.53$ & -- &~\cite{ref:harimarti} \\
$\eta_2$  &  $0.574$ & $ 0.004$ & --
     &~\cite{ref:harimarti} \\
$\eta_3$  &  $0.47$ & $ 0.04$ & -- &~\cite{ref:harimarti} \\
$\eta_c$  &  $0.55$ & $ 0.01$ & --
     &~\cite{ref:harimarti} \\
%$f_K$     & $0.161~\GeV$ & \multicolumn{2}{|c|}{fixed}
%     &~\cite{PDG00} \\
%$\Delta m_K$  & $
%0.5301
%\times 10^{-2} ~\mbox{ps}^{-1}$
%     & \multicolumn{2}{|c|}{fixed} &~\cite{PDG00} \\
% $G_F $   & $
% 1.16639
%\times 10^{-5} \GeV^{-2}$
%     & \multicolumn{2}{|c|}{fixed}&~\cite{PDG00} \\
%$ m_{W}$  & $80.41
% ~\GeV/c^2 $
%     & \multicolumn{2}{|c|}{fixed} & ~\cite{PDG00} \\
%$ m_{B^0_d}$ & $5.2792
%~\GeV/c^2 $
%     & \multicolumn{2}{|c|}{fixed} &~\cite{PDG00} \\
%$ m_{B^0_s}$ & $5.3693
%~\GeV/c^2 $
%     & \multicolumn{2}{|c|}{fixed} &~\cite{PDG00} \\
%$ m_K$ & $0.493677
% ~\GeV/c^2$
%     & \multicolumn{2}{|c|}{fixed} &~\cite{PDG00} \\
\hline
\end{tabular} }
\caption[]{ \it {Values of the quantities entering into the expressions of 
$\epsilonk$, $\vubovcb$, $\dmd$ and $\dms$.
In the third column the Gaussian and the flat part of the error are given 
explicitely. Values for the other parameters have been taken from \cite{PDG00}.
 }\label{tab:alltri} }
\end{center}
\end{table}

\begin{equation}
{\cal P}(\rhobar,\etabar,x_1,...,x_{\rm N}|c_1,...,c_{\rm M})
~\propto~\prod_{j=1,{\rm M}}{\cal P}_j(c_j|\rhobar,\etabar,x_1,...,x_{\rm N})
\times \prod_{i=1,{\rm N}}{\cal P}_i(x_i) \times {\cal P}(\rhobar,\etabar).
\label{eq:bayes}
\end{equation} 
%In practice,
%several approaches 
Values used for all relevant parameters and their corresponding probability 
distributions are given in Table \ref{tab:alltri}.
Central values and uncertainties taken for the parameters
obtained from lattice QCD evaluations have been discussed in 
\cite{ref:harimarti} where more details can be found on the present
determination of the CKM unitarity triangle parameters.. 
Rather similar studies have been done to obtain the favoured values
for $\rhobar$ and $\etabar$ and for their corresponding uncertainties
\cite{ref:allrhoeta}.
A different approach, named scanning, has been used in \cite{ref:scan}.
It has been verified that, in practice, similar 95$\%$ C.L. 
regions for $\rhobar$ and $\etabar$ are obtained
if similar central values and corresponding uncertainties are used for 
the different 
parameters \cite{ref:harimarti}.

\begin{figure}[t]
\begin{center}
\vskip -1cm
{\epsfig{figure=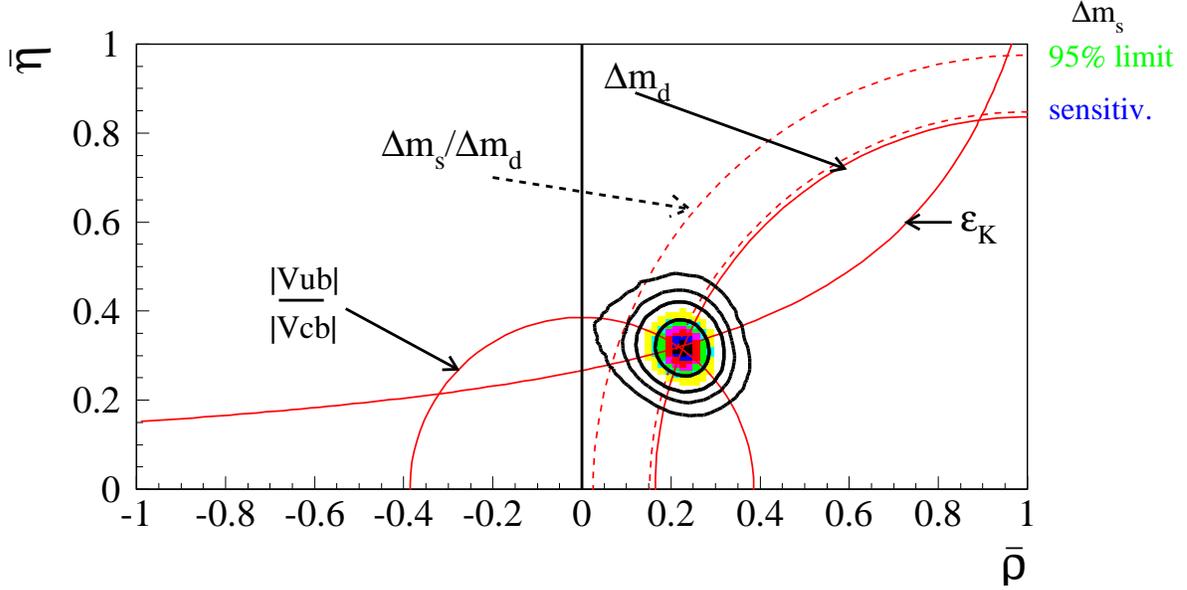,bbllx=30pt,bburx=503pt,bblly=1pt,bbury=270pt,height=8cm}}
\caption{ \it{Allowed regions for $\rhobar$ and $\etabar$ using the parameters listed in Table \ref{tab:alltri}.
The contours at 68 $\%$, 95 $\%$, 99$\%$ and 99.9$\%$ probability are shown. 
The full lines correspond to the central values of the constraints given by
the measurements of  $  \frac{\left | V_{ub} \right |}{\left | V_{cb} \right |} $, $\epsilonk$ and $\dmd$.
The two dotted curves correspond, from left to right respectively to the 95 $\%$ upper limit 
and to the exclusion sensitivity reached in the experimental 
study
 of $\Bs-\Bsb$ oscillations.}
\label{fig:rhoeta_bc99}}
\end{center}
\end{figure}

\mysubsection{The unitarity triangle from all constraints}
The region in the $(\rhobar,~\etabar)$ plane selected by the measurements
of the four constraints is given in Figure \ref{fig:rhoeta_bc99}.
The resulting values of the unitarity triangle
parameters are:
\begin{eqnarray}
\nonumber \rhobar=0.224 \pm 0.038  ,~\etabar=0.317 \pm 0.040 ~~~~~~~~~~~~~~~~~~~~~~~~~~\\
\sin(2\beta)=  0.698 \pm 0.066   ,~\sin(2\alpha)= -0.42  \pm 0.23~{\rm and}~\gamma=(54.8 \pm 6.2)^{\circ}.
\end{eqnarray}

Several comments can be made:
\begin{itemize}
\item 
$\sin(2\beta)$ is determined precisely. This value can be compared
with measurements of this quantity using J/$\psi \Kos$ events
which are analyzed in Section \ref{sec:sintwobeta} and with new results
obtained by BaBar \cite{ref:psiksbabar} and BELLE \cite{ref:psiksbelle}
collaborations. Combining all direct measurements one obtains:
\begin{equation}
\sin(2\beta)=  0.49 \pm 0.16.
\end{equation} 
At present there is no discrepancy with the indirect approach. 
%This comparison
%will be of interest when future direct measurements will have an accuracy
%better than 0.1.

\item
The angle $\gamma$ is known with a $10\%$ accuracy. It may be noted that
the probability that $\gamma$ is greater than 90$^{\circ}$ is only
$0.03\%$. The central value for the angle $\gamma$ is much smaller than 
that obtained in recent fits of rare $b$-meson two-body decays,
based on the factorization hypothesis
\cite{ref:cleotwob}, which found values centered on $\gamma \sim 110^{\circ}$.
Before invoking new physics, the validity of the hypotheses used to
analyze two-body decays of $b$-mesons have to be scrutinized.
\end{itemize}

\mysubsection{The unitarity triangle from $b$-physics measurements
of its sides}

\begin{figure}[t]
\begin{center}
{\epsfig{figure=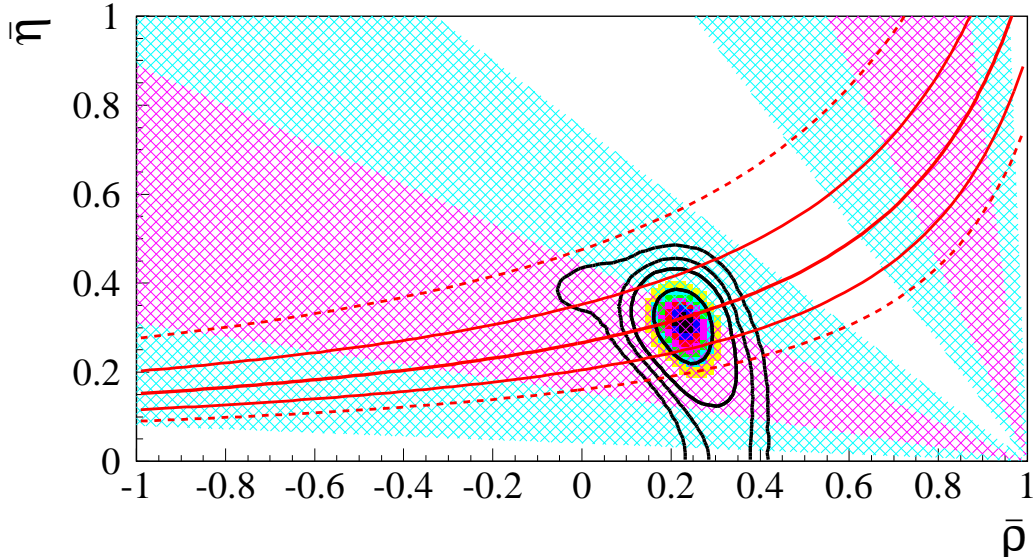,bbllx=30pt,bburx=503pt,bblly=1pt,
bbury=270pt,height=8cm}}
\caption{ \it{ The allowed regions
(at 68$\%$, 95$\%$, 99$\%$ and 99.9$\%$ probability) for $\rhobar$ and $\etabar$ using the constraints given by the measurements of
 $  \frac{\left | V_{ub} \right |}{\left | V_{cb} \right |} $, $\dmd$ and $\dms$. 
The constraint due to $\epsilonk$ is not included. The regions 
(at 68$\%$ and 95$\%$ probability) selected by the measurements of $\epsilonk$
[continuous (1$\sigma$) and dotted (2$\sigma$) curves] and 
$\sin{(2\beta)}$  [darker (1$\sigma$) and clearer (2$\sigma$) zones]
are shown. For $\sin{(2\beta)}$ the two solutions are displayed.
\label{fig:rhoeta_sinepsk_bc99}}}
\end{center}
\end{figure}

Without including the constraint given by the 
measurement of the $\epsilonk$ parameter of CP violation in kaon physics
\footnote{The numerical constraint provided by the measurement of 
$\epsilonk$ has not been applied but the variation of $\etabar$ has been
restricted to positive values only.},
the region of the ($\rhobar$, $\etabar$) plane favoured by semileptonic
decays and oscillations of $b$-hadrons is shown in Figure 
\ref{fig:rhoeta_sinepsk_bc99}. 

The corresponding values of the $\etabar$ and 
$\sin(2\beta)$ parameters defining the unitarity
triangle are:
\begin{eqnarray}
\etabar=0.302^{+0.052}_{-0.061},~\sin(2\beta)= 0.678^{+0.078}_{-0.101} 
\end{eqnarray}

The selected region is well compatible
with the one corresponding to CP violation in kaon physics, displayed
as a band. This comparison is already a test of the compatibility between the 
measurements of the sides of the unitarity triangle and the contraints
coming from the CP violation phase, obtained before the ``B factories era''.
%It was made possible because, mainly, of the impressive progress realized
%in the analyses of $\Bs-\Bsb$ oscillations and lattice QCD.

The value of the $\hat \BK$ parameter given by the constraints from $b$-physics
on $\rhobar$ and $\etabar$ 
and using the Standard Model expression and the measured 
value for $\epsilonk$ is equal to:
\begin{equation}
\hat \BK = 0.90^{+0.30}_{-0.14}.
\end{equation}
It can be compared with the result mentioned before, from lattice QCD:
$\hat \BK=0.87 \pm 0.06 \pm 0.13$ (Table \ref{tab:alltri}). Rather
low values of $\hat \BK$, as obtained from the chiral perturbation
theory \cite{ref:chiralbk} are not favoured by the present analysis.

%The selected region in the $(\rhobar,~\etabar)$ plane
%is also compatible with the one favoured by 
%%present results on CP violation 
%in $b$-physics
%given by direct measurements of $\sin{2 \beta}$ from studies
%of the $J/\psi \Kos$ system (see Section \ref{sec:sintwobeta}).
%the non-B factories measurements of $\sin{(2 \beta)}$
%from the time-dependent CP asymmetry of the $\Bd \rightarrow J/\psi \Kos$
%decay (see Section \ref{sec:sintwobeta}).

\mysubsection{Other consistency checks}

In a similar way as the measurement of $\epsilonk$ which 
has been removed from the 
set of constraints to determine $\rhobar$, $\etabar$
and $\hat \BK$, other constraints can be removed in turn.

\mysubsubsection{Results without including the $\dmd$ measurement}
If the $\dmd$ measurements are removed from the fit, 
the following result is obtained:
\begin{equation}
f_{B_d}\sqrt{ \hat B_{B_d}} = (230 \pm 12)\MeV.
\end{equation}
As the accuracy on this value is better than present theoretical 
uncertainties, from lattice QCD, reported in Table \ref{tab:alltri},
this explains why the characteristics
of the unitarity triangle are weakly dependent on the assumed value
for this quantity.

\mysubsubsection{Results without including the limit on $\dms$}
%The probability distribution for the value of $\dms$, as determined by
%the other constraints, has been obtained and it is expected that
%$\dms$ is in the range $[9.7,~23.2]~ps^{-1}$ at 95$\%$C.L..

Without including measurements on the $\dms$ oscillation amplitude,
obtained in Section \ref{sec:dmsosc}, the probability distribution for
the variable $\dms$ has been obtained, giving the result:
\begin{equation}
\dms =(16.3 \pm 3.4)~\ipsm~{\rm and}~9.7 \leq \dms \leq 23.2~\ipsm~
{\rm at}~95\%~{\rm probability.}
\end{equation}

These values can be compared with recent lattice QCD determinations of $\dms$ 
giving $(15.8 \pm 2.3 \pm 3.3)~\ipsm$ \cite{ref:latticedms1} or
$(18.2 \pm 4.7)~\ipsm$ \cite{ref:latticedms2}. It should be noted that
these predictions are independent of the values of the parameters
$\rhobar$ and $\etabar$. 

Including results of Section \ref{sec:dmsosc}, on searches for
$\Bs-\Bsb$ oscillations, the expected range for $\dms$ gets narrower:
\begin{equation}
\dms =(17.3^{+1.5}_{-0.7})~\ipsm~{\rm and}~15.6 \leq \dms \leq 20.5~\ipsm~
{\rm at}~95\%~{\rm probability.}
\end{equation}

A signal for $\Bs-\Bsb$ oscillations is thus expected at the TeVatron.

%\mysubsubsection{Results without including $\Vub$ measurement}
%The probability distribution for the value of $\vubovcb$ as determined by
%the other constraints is given in Figure {\bf to be prepared}.

\mysection{Summary of all results}
\label{sec:conclusion}
This paper provides precise combined results (see Tables \ref{tab:conclusion}
and \ref{tab:conclusionb}),
from measurements
%available in 1999, 
submitted to the 2000 Summer Conferences,
on parameters which govern
production and decay properties of $b$-hadrons emitted in
high energy $b$-quark jets.
\begin{table}[ht!]
  \begin{center}
    \begin{tabular}{|l| c |}
      \hline
      {\bf {\boldmath $b$}-hadron lifetimes} &  CDF-LEP-SLD, Section \ref{sec:Atau}\\
      \hline
      $\tau(\Bd)$ & $(\taubd)$~ps \\
      $\tau(\Bp)$ & $(\taubp)$~ps \\
      $\tau(\Bs)$ & $(\taubs)$~ps \\
      $\tau(\Lb)$ & $(\taulb)$~ps \\
      $\tau(\Xi_b)$ & $(\tauxib)$~ps \\
      $\tau(b-{\rm baryon})$ & $(\taubbar)$~ps \\
      $\tau_b$ & $(\taubav)$~ps \\
      $\frac{\tau(\Bm)}{\tau(\Bdb)}$ & $ \taubpovertaubd$\\

      \hline
%      \hline
      {\bf {\boldmath $b$}-hadron production rates} & CDF-LEP, Section \ref{sec:boscill}\\
      \hline
      $\fs$ & $\fbs$ \\
      $\fb$ & $\fbar$ \\
      $\fd = \fu$ & $\fbd$ \\
      $\rho(\fs,\fb)$ & $\rhosbar$ \\
      $\rho(\fd,\fs)$ & $\rhosd$ \\
      $\rho(\fs,\fb)$ & $\rhobard$ \\
      \hline
%      \hline
     {\bf {\boldmath ${\rm B}^0 - \overline{{\rm B}^0}$} oscillations} &  ARGUS-CDF-CLEO-LEP-SLD, Section \ref{sec:boscill}\\
      \hline
      $\dmd$ & $(\dmdx)$ \ips \\
      $\chi_d$& $\chix$ \\
      $\dms$ & $> \dmslim$ \ips {\rm at~ the}~95$\%$ C.L. \\
      \hline
%      \hline
    {\bf  Limit on {\boldmath $\Delta \Gamma_{\Bs} $}} &  CDF-LEP-SLD, Section \ref{sec:deltag}\\
      \hline
      $\Delta \Gamma_{\Bs} /\Gamma_{\Bs}  $ & $<0.31  $ at the~95$\%$ C.L.\\
      \hline
    {\bf  {\boldmath $\Lb$} polarization in {\boldmath $\Zz$} decays} &  LEP, Section \ref{sec:systgen}\\
      \hline
      ${\cal P}(\Lb)$ & $ -0.45^{+0.17}_{-0.15} \pm 0.08$ \\
      \hline
%      \hline
     {\bf {\boldmath $\Dstarstar$} properties}  &  ARGUS-CLEO-LEP, Section \ref{sec:systgen}\\
 {\bf in semileptonic B decays}  &  \\
      \hline
${\rm BR}(\Bdb \rightarrow \rm D^{\ast \ast +} \ell^- \overline{\nu_{\ell}})$
& $(3.04 \pm 0.38 ) \% $\\
${\rm BR}(\Bdb \rightarrow \rm D^{\ast \ast +} \ell^- \overline{\nu_{\ell}})
\times {\rm BR}(D^{\ast \ast +} \rightarrow \Dstar X)$ 
& $ (1.82 \pm 0.21 \pm 0.08 ) \% $\\
$\frac{{\rm BR}(\Bdb \rightarrow \Dstarp \ell^- \overline{\nu_{\ell}})+
{\rm BR}(\Bdb \rightarrow \Dstarp \pi \ell^- \overline{\nu_{\ell}})}
{{\rm BR}(\Bdb\rightarrow  \ell^- \overline{\nu_{\ell}} {\rm X})}$ &
$0.50 \pm 0.03 $\\
${\rm BR}({\rm \overline{B}} \rightarrow \rm D_1 \ell^- \overline{\nu_{\ell}})$
& $(0.63 \pm 0.10 ) \% $\\
${\rm BR}({\rm \overline{B}} \rightarrow \rm D_2^* \ell^- \overline{\nu_{\ell}})$
& $<0.4\%~{\rm at~the~95\%~CL.} $\\
\hline
    \end{tabular}
  \end{center}
    \caption{{\it Summary of the results obtained 
on $b$-hadron production rates
and decay properties using data available by mid 2000. More details
on systematic uncertainties can be found in the corresponding sections.}
    \label{tab:conclusion}}
\end{table}

At the $\Zz$ pole, the polarization of $b$-baryons is 
found to be 
significantly different from zero
but it is reduced as compared to the initial $b$-quark polarization of $-0.94$.
Production of $\Sigma_b^{(\ast)}$ baryons has been invoked to explain
this difference.

The accuracy on $\Bdb$ and $\Bm$ lifetimes is better than 2$\%$. The $\Bm$
lifetime is significantly larger than the $\Bdb$ lifetime, in agreement with 
original expectations based on OPE and parton-hadron duality 
\cite{ref:bigilife}. But the clear difference between measured and expected 
lifetimes of $b$-baryons remains to be explained. It may point to a failure
of parton-hadron duality in inclusive B decays or to a failure of quark
models used to evaluate the mean value of operators contributing in 
$b$-baryon decays.
%Is the previous agreement,
%observed for mesons,  purely fortuitous \cite{ref:sacrelife}
%and one may have concern
%about the measurement of $\Vcb$ from inclusive semileptonic decays, 
%or is the discrepancy observed for baryons
%coming from a failure of quark models which need to be used in that case?

The accuracy on $b$-hadron production rates in $b$-quark jets has 
improved by 30$\%$ as compared to published values \cite{PDG00}.

\begin{table}[th!]
  \begin{center}
    \begin{tabular}{|l| c |}
      \hline
     {\bf  Charm counting} &  LEP-SLD, Section \ref{sec:secccbar}\\
      \hline
      $n_c+n_{\overline{c}}$ & $ 1.230 \pm 0.042 $ \\
      ${\rm BR}(b \rightarrow 0c) $ & $(2.0 \pm 1.8)\% $ \\
      ${\rm BR}(b \rightarrow c\overline{c}) $ & $(22.3 \pm 3.2)\% $ \\
      \hline
     {\bf  Measurement of {\boldmath $\Vcb$}} &  LEP, Section \ref{sec:vcb}\\
      \hline
      ${\rm BR}(b \rightarrow \ell X) $ & $(10.58 \pm 0.07 \pm 0.18)\% $ \\
      $\mbox{\fvcb}$ & $\favcb$ \\
      $\rho^2$  & $\rhobb$ \\ 
      $\Vcb^{incl.}$ & $ \vcbinc$ \\
      $\Vcb^{excl.}$ & $ \vcbexc$ \\
      $\Vcb$ & $ \vcbavg$ \\
      \hline
%      \hline
    {\bf  Measurement of {\boldmath $\Vub$}} &  LEP, Section \ref{sec:vub}\\
      \hline
      $\Vub$ & $ (4.04^{+0.62}_{-0.74}) \times 10^{-3}$ \\
\hline
{\bf CP violation in $\Bd$} & CDF-LEP, Section  \ref{sec:sintwobeta}   \\
\hline
$\sin{(2 \beta)}$ & $ 0.88^{+0.36}_{-0.39} $ \\
${\rm Re (\epsilon_B)/(1+|\epsilon_B|^2)}$ & $  -0.0024\pm 0.0039$ \\
${\rm Im (\delta_B)}$ & $ -0.016\pm 0.012 $\\
      \hline
    {\bf  Fitted S.M. observables} & All , Section \ref{sec:triangle}\\
      \hline
      $\rhobar$ & $ 0.224 \pm 0.038 $ \\
      $\etabar$ & $ 0.317 \pm 0.040 $ \\
      $\sin{(2 \beta)}$ & $ 0.698 \pm 0.066 $ \\
      $\sin{(2 \alpha)}$ & $ -0.42 \pm 0.23 $ \\
      $\gamma$ & $ (54.8 \pm 6.2)^{\circ} $ \\
%\hline
%      $\dms$  & $(16.3 \pm 3.4) $ \ips \\
%      $f_{B_d} \sqrt{\hat B_{B_d}}$ & $(230 \pm 12)~MeV $ \\
%      $\hat B_K$ & $0.90^{+0.30}_{-0.14}$\\
      \hline
    \end{tabular}
  \end{center}
    \caption{{\it Summary of the results obtained 
on charm counting and semileptonic $b$-hadron decays.
Constraints on the parameters of the CKM unitarity triangle
are also given. The last free results correspond of the expected values
of the corresponding parameters when they are not included as constraints 
in the
fit. More details
on systematic uncertainties can be found in the corresponding sections.}
    \label{tab:conclusionb}}
\end{table}
In conjunction with the better determination of the $\Bdb$ lifetime,
the improvement in the $\Bd$ production rate has given a more precise
 determination of $\Vcb$
from $\Bdb \rightarrow \Dstarp \ell^- \overline{\nu_{\ell}}$ decays.
The $\Bsb$ fraction in jets is close to 10$\%$ with an uncertainty 
of 12$\%$. Its determination is an important input in
searches for $\Bs$-$\Bsb$ oscillations. 
%The accuracy on the $b$-baryon
%rate has improved by about a factor two because of new measurements
%using spectator baryons.

The mass difference between mass eigenstates of the $\Bd$-$\Bdb$ system
is measured with a relative precision close to 3$\%$.
The corresponding quantity for the $\Bs$-$\Bsb$ system is still
unmeasured, in spite of impressive progress in the 
sensitivity reached by the experiments.
% there
%is still no sign for oscillations of strange B mesons.

A combined result on the decay width difference between 
mass eigenstates of
the $\Bs$-$\Bsb$ system has been obtained which remains far from
theoretical expectations.
%Because of the relatively small fraction of $\Bsb$ mesons produced 
%in $b$-quark jets,
%the limit on the decay widths difference between mass eigenstates of
%the $\Bs$-$\Bsb$ system is still away from theoretical expectations.

Studies of inclusive charm production in $b$-hadron decays have shown that
there is no room left for a large $0c$ component originating from new physics.
Theory can reproduce the values for the semileptonic $b$-hadron branching 
fraction and $n_c + n_{\overline{c}}$ using a standard value for the
charm quark mass and a low value for the scale at which QCD corrections
have to be evaluated. 

Because of the very accurate measurements obtained on the inclusive
$b$-lifetime and semileptonic branching fraction, the accuracy
of $\Vcb$ determined from inclusive semileptonic decays is entirely limited by
theoretical uncertainties. An experimental control of these uncertainties
is thus needed.
%expected to come from studies
%of moments of the lepton momentum
%distribution. The determination of the
%value of the $b$-quark mass, entering into these analyses, 
%appears to be under 
%control from several studies of the $\Upsilon$ system 
%giving similar results.
The measurement of $\Vcb$ from exclusive 
$\Bdb \rightarrow \Dstarp {\rm X} \ell^- \overline{\nu_{\ell}}$
decays is limited by uncertainties related to $\Dstarstar$ 
production and decay mechanisms.
%The measured rate for $\Dstarstar$, the fact that narrow states seem to correspond to only one third of these states and the small rate of the 2$^+$
%are not really expectations from the models. It is thus difficult,
%at present, to quote realistic uncertainties related to the modelling of 
%these states. A conservative attitude has been adopted in the present
%analysis. 
Present experimental results on production characteristics of these states
in $b$-hadron semileptonic decays have been detailed in Section 
\ref{sec:systgen} of the present report.
The fact that the theoretical uncertainties
on the determinations of $\Vcb$ from inclusive and exclusive
measurements are largely uncorrelated has been used in evaluating a 
global average.

The measurement of $\Vub$ from LEP experiments has reached an
accuracy similar to previous determinations 
at the $\Upsilon(4S)$ but with a better sensitivity over a larger
fraction of the phase space than in measurements
using the lepton energy end-point region. 

Because of progress in lattice QCD during the last years,
of improvements in the determination of production and decay properties of
$b$-hadrons given in the previous Sections and, in particular, of 
improvements on the sensitivity of analyses on $\Bs-\Bsb$ oscillations
it is possible, within the framework of the Standard Model, to constrain
the ranges of possible values for the parameters 
%$\rhobar$ and $\etabar$ 
of the CKM 
%matrix. 
%Values for the parameters of the 
unitarity triangle ($\rhobar,~\etabar,
~\sin{2 \alpha},~\sin{2 \beta},~\gamma$). They  have been obtained using
all available information from $b$-physics, CP violation in 
K decays and lattice QCD. Removing, in turn, the corresponding constraints,
expected values for $\hat B_K,~\dms$ and $f_{B_d} \sqrt{\hat B_{B_d}}$
have been also determined which can be compared with future accurate 
direct determinations of these quantities.
%In this way, rather precise evaluations of key parameters
%entering into these analyses have been obtained. 

New measurements from LEP and SLD are still expected in the year 2001 and 
will improve present determinations of $b$-hadron production and decay
properties, which already provide stringent constraints on the shape of the 
CKM unitarity triangle at the start of the era of asymmetric B factories.

\section*{Acknowledgements}
We would like to thank the CERN accelerator divisions for the efficient 
operation of the LEP accelerator, and their close collaboration with the 
four experiments. We would like to thank members of the CDF and SLD 
Collaborations for making results available to us in advance of the 
conferences and for useful discussions concerning their combination.
Useful contacts with members of the CLEO Collaboration are also 
acknowledged. 
%We thank W. Venus for a careful reading of the manuscript
%and for helpful comments.
Results on $\Vcb$ and $\Vub$ have been obtained after discussions
with several theorists to understand the meaning and the importance of 
theoretical uncertainties. Among them we would like to thank in particular:
M. Beneke, I.I. Bigi, G. Buchalla, F. Defazio, A. Hoang, L. Lellouch,
Z. Ligeti
and N. Uraltsev.
Finally, theoretical uncertainties related to quantities evaluated in lattice 
QCD, used in Section \ref{sec:triangle}, have been defined by M. Ciuchini, 
G. D'Agostini, E. Franco, V. Lubicz and G. Martinelli.
Discussions with N. Uraltsev and I.I. Bigi on the evaluation
of $b$- and $c$-quark masses and on different aspects of the determination
of $\Vub$ and $\Vcb$ are acknowledged.

\newpage
\clearpage
\appendix
\mysection{Production rates of the ${\rm D}_1$ and ${\rm D}^*_2$ mesons 
in semileptonic $b$-decays}
\label{appendixA}

\begin{table}[ht!]
\begin{center}
{\small
\begin{tabular}{|l||c|c|c|} \hline
 Experiment    & Channel & Value $\times 10^3$ & Ref. \\   \hline
ALEPH        & ${\rm BR}(b \rightarrow \overline{\rm B}) {\rm BR}({\rm \overline{B}} \rightarrow {\rm D}^{+}_1 \ell^-\overline{\nu_{\ell}}){\rm BR}({\rm D}^{+}_1 \rightarrow \Dstaro \pi^+)$ & $ 2.06^{+0.55~ +0.29}_{-0.51~-0.40} $ & \cite{DssALEPH} \\
ALEPH        & ${\rm BR}(b \rightarrow \overline{\rm B}) {\rm BR}({\rm \overline{B}} \rightarrow {\rm D}^{0}_1 \ell^-\overline{\nu_{\ell}}){\rm BR}({\rm D}^{0}_1 \rightarrow \Dstarp \pi^-)$ & $ 1.68^{+0.40~ +0.28}_{-0.36~-0.29} $ & \cite{DssALEPH} \\
ALEPH        & ${\rm BR}(b \rightarrow \overline{\rm B}) {\rm BR}({\rm \overline{B}} \rightarrow {\rm D}^{0}_1 \ell^-\overline{\nu_{\ell}}){\rm BR}({\rm D}^{0}_1 \rightarrow \Dstarp \pi^-)$ & $ 3.62^{+1.78}_{-1.48}\pm 0.77 $ & \cite{DssALEPH} \\
CLEO     & ${\rm BR}({\rm B^-} \rightarrow {\rm D}^{0}_1 \ell^-\overline{\nu_{\ell}}){\rm BR}({\rm D}^{0}_1 \rightarrow \Dstarp \pi^-)$ & $ 3.73 \pm 0.85 \pm 0.52 \pm 0.24 $ &\cite{ref:cleodss}  \\
DELPHI        & ${\rm BR}(b \rightarrow \overline{\rm B}) {\rm BR}({\rm \overline{B}} \rightarrow {\rm D}^{0}_1 \ell^-\overline{\nu_{\ell}}){\rm BR}({\rm D}^{0}_1 \rightarrow \Dstarp \pi^-)$ & $ 1.8 \pm 0.5\pm 0.3 $ & \cite{Dss2DELPHI} \\
\hline

\end{tabular}}
\caption[]{{\it Measured values of the ${\rm D}_1$ production rate in 
$b$-semileptonic decays.
The third systematic uncertainty, quoted in the CLEO analysis,
comes from the variation of the detection efficiency when changing the 
parameters of the signal form factors in the ISGW2
model \cite{isgw2}.} \label{tab:darate}}

\end{center}
\end{table}

\begin{table}[ht!]
\begin{center}
{\small
\begin{tabular}{|l||c|c|c|} \hline
 Experiment    & Channel & Value $\times 10^4$ & Ref. \\   \hline
ALEPH        & ${\rm BR}(b \rightarrow \overline{\rm B}) {\rm BR}({\rm \overline{B}} \rightarrow {\rm D}^{*+}_2 \ell^-\overline{\nu_{\ell}}){\rm BR}({\rm D}^{*+}_2 \rightarrow \Do \pi^+)$ & $ 3.1^{+2.4~ +0.4}_{-2.2~-0.6} $ & \cite{DssALEPH} \\
ALEPH        & ${\rm BR}(b \rightarrow \overline{\rm B}) {\rm BR}({\rm \overline{B}} \rightarrow {\rm D}^{*0}_2 \ell^-\overline{\nu_{\ell}}){\rm BR}({\rm D}^{*0}_2 \rightarrow \Dstarp \pi^-)$ & $ 5.1^{+3.0~ +0.8}_{-2.6~-0.9} $ & \cite{DssALEPH} \\
ALEPH        & ${\rm BR}(b \rightarrow \overline{\rm B}) {\rm BR}({\rm \overline{B}} \rightarrow {\rm D}^{*0}_2 \ell^-\overline{\nu_{\ell}}){\rm BR}({\rm D}^{*0}_2 \rightarrow \Dp \pi^-)$ & $ 3.8^{+2.4~ +0.8}_{-1.9~-0.8} $ & \cite{DssALEPH} \\
CLEO     & ${\rm BR}({\rm B^-} \rightarrow {\rm D}^{*0}_2 \ell^-\overline{\nu_{\ell}}){\rm BR}({\rm D}^{*0}_2 \rightarrow \Dstarp \pi^-)$ & $ 5.9 \pm 6.6 \pm 1.0 \pm 0.4 $ &\cite{ref:cleodss}  \\
DELPHI        & ${\rm BR}(b \rightarrow \overline{\rm B}) {\rm BR}({\rm \overline{B}} \rightarrow {\rm D}^{*0}_2 \ell^-\overline{\nu_{\ell}}){\rm BR}({\rm D}^{*0}_2 \rightarrow \Dp \pi^-,~\Dstarp \pi^-)$ & $ 11.6 \pm 5.6 \pm 3.2 $ & \cite{Dss2DELPHI} \\
\hline

\end{tabular}}
\caption[]{{\it Measured values of the ${\rm D}^{*}_2$ production rate in 
$b$-semileptonic decays. In the original publication from ALEPH, 
as these values differ from zero only by one
or two standard deviations, only upper limits
were quoted. The third systematic uncertainty, quoted in the CLEO analysis,
comes from the variation of the detection efficiency when changing the 
parameters of the signal form factors in the ISGW2
model \cite{isgw2}.} \label{tab:dbstar}}

\end{center}
\end{table}

\mysection{$\Lb$ polarization measurements}
\label{appendixAb}
\begin{table}[ht!]
\begin{center}
\begin{tabular}{|l||c|c|} \hline
 Experiment    & Value & Ref. \\   \hline
ALEPH        & $ -0.23^{+0.24~ +0.08}_{-0.20~-0.07} $ & \cite{ref:alpol} \\
DELPHI       & $ -0.49^{+0.32~ +0.17}_{-0.30~-0.17} $ & \cite{ref:delpol} \\
OPAL         & $ -0.56^{+0.20~ +0.09}_{-0.13~-0.09} $ & \cite{ref:oppol} \\

\hline
\end{tabular}
\caption[]{{\it Measured values of the $\Lb$ polarization in 
$\Zz$ decays.}\label{tab:polar} }
\end{center}
\end{table}

\newpage
\clearpage

\mysection{Measurements used in the evaluation of 
$b$-hadron lifetimes}
\label{appendixB}
In the following Tables, preliminary numbers are labelled with ``(p)''.

\begin{table}[ht!]
\begin{center}
\begin{tabular}{|l||c|c|c|c|} \hline
 Experiment    & Method                & Data set        & $\tau_{B^0}$ (ps)                & Reference \\   \hline
ALEPH         & D$^{(*)} \ell$         &  91--95        & 1.518$\pm 0.053 \pm 0.034 $ & \cite{ALEB01}  \\
ALEPH          &  Excl. rec.            &  91--94        & 1.25$^{+0.15}_{-0.13} \pm 0.05$ & \cite{ALEB0}  \\
ALEPH          &Partial rec. $\pi^+ \pi^-$&  91--94       & 1.49$^{+0.17+0.08}_{-0.15-0.06}$ & \cite{ALEB0} \\
CDF            & D$^{(*)} \ell$        & 92--95         & 1.474$\pm 0.039^{+0.52}_{-0.51}$  & \cite{CDFB02}  \\
CDF            & Excl. (${\rm J}/\psi K$)    & 92--95         & 1.58$\pm$ 0.09 $\pm$ 0.02        & \cite{CDFB01}  \\
DELPHI         & D$^{(*)} \ell$        &  91--93        & 1.61$^{+0.14}_{-0.13} \pm 0.08$  & \cite{DELB01}  \\
DELPHI         & Charge sec. vtx.      &  91--93        & 1.63 $\pm$ 0.14  $\pm$ 0.13      & \cite{DELB02} \\
DELPHI         & Inclusive D$^* \ell$  &  91--93       & 1.532$\pm$  0.041 $\pm$0.040      & \cite{DELB03} \\
DELPHI (p)        &  Charge sec. vtx.  &  94       & 1.520$\pm$  0.021 $\pm$0.041      & \cite{DELB031} \\
L3             & Charge sec. vtx.      &  94--95        & 1.52$\pm$  0.06  $\pm$ 0.04      & \cite{L3B01}  \\ 
L3 (p)            & Inclusive D$^* \ell$  &  94           & 1.74$\pm$  0.12  $\pm$ 0.04      & \cite{L3B02}  \\ 
OPAL           & D$^{(*)} \ell$        &  91--93        & 1.53$\pm$  0.12  $\pm$ 0.08      & \cite{OPAB0}  \\ 
OPAL           & Charge sec. vtx.      &  93--95        & 1.523$\pm$ 0.057  $\pm$ 0.053    & \cite{OPAB1}  \\ 
OPAL           &  Inclusive D$^* \ell$    &  91--00        & 1.541$\pm$ 0.028  $\pm$ 0.023    & \cite{OPAB11}  \\ 
SLD            & Charge sec. vtx.$\ell$&  93--95        & 1.56$^{+0.14}_{-0.13} \pm$ 0.10 & \cite{SLDB01}  \\ 
SLD  (p)          & Charge sec. vtx.      &  93--98        & 1.565$\pm$ 0.021 $\pm$ 0.043    & \cite{SLDB02}  \\ \hline
Average        &                       &               & $\taubd$                 &    \\   \hline

\end{tabular}
\caption[]{{\it Measurements of the $\Bd$ lifetime.}}
%            a) The combined ALEPH result quoted in \cite{ALEB0} is 1.55 $\pm$ 0.06 $\pm$ 0.03 ps.}
%            b) The combined DELPHI result quoted in \cite{DELB02} is 1.62 $\pm$ 0.12.}\label{tab:bo} }
\end{center}
\end{table}
 
\begin{table}[ht!]
\begin{center}
\begin{tabular}{|l||c|c|c|c|} \hline
 Experiment    & Method            & Data set      & $\tau_{B^+}$(ps)                 & Reference \\   \hline
ALEPH          & D$^{(*)} \ell$    &  91-95        & 1.648$\pm 0.049 \pm 0.035 $ & \cite{ALEB01}  \\
ALEPH          & Excl. rec.        &  91-94        & 1.58$^{+0.21+0.04}_{-0.18-0.03}$ & \cite{ALEB0} \\
CDF            & D$^{(*)} \ell$    & 92-95         & 1.637$\pm 0.058^{+0.45}_{-0.43}$  & \cite{CDFB02}  \\
CDF            & Excl. (${\rm J}/\psi K$)& 92-95         & 1.68 $\pm$ 0.07 $\pm$ 0.02       & \cite{CDFB01}  \\
DELPHI         & D$^{(*)} \ell$    &  91-93        & 1.61 $\pm$ 0.16 $\pm$ 0.12       & \cite{DELB01}$^a$  \\
DELPHI         & Charge sec. vtx.  &  91-93        & 1.72 $\pm$ 0.08 $\pm$ 0.06       & \cite{DELB02}$^a$ \\
DELPHI (p)        & Charge sec. vtx.  &  94        & 1.619 $\pm$ 0.018 $\pm$ 0.035       & \cite{DELB031}$^a$ \\
L3             & Charge sec. vtx.  &  94-95        & 1.66$\pm$  0.06  $\pm$ 0.03      & \cite{L3B01}  \\ 
OPAL           & D$^{(*)} \ell$    &  91-93        & 1.52 $\pm$ 0.14 $\pm 0.09$       & \cite{OPAB0}  \\ 
OPAL           & Charge sec. vtx.  &  93-95        & 1.643$\pm$ 0.037  $\pm$ 0.025    & \cite{OPAB1}  \\ 
SLD            & Charge sec. vtx. $\ell$&  93-95   & 1.61$^{+0.13}_{-0.12} \pm$ 0.07 & \cite{SLDB01}  \\
SLD  (p)          & Charge sec. vtx. &  93-98         & 1.623$\pm$ 0.020 $\pm$ 0.034    & \cite{SLDB02}  \\ \hline
Average        &                   &               & $\taubp$                 &    \\   \hline

\end{tabular}
\caption[]{{\it Measurements of the $\Bp$ lifetime. \\
%           a) The combined ALEPH result quoted in \cite{ALEB0} is 1.58 $\pm$ 0.09 $\pm$ 0.03 ps.\\
           a) The combined DELPHI result quoted in \cite{DELB02} is (1.70 $\pm$ 0.09) ps.} \label{tab:bp}}
\end{center}
\end{table}

\begin{table}[ht!]
\begin{center}
\begin{tabular}{|l||c|c|c|c|} \hline
 Experiment    & Method              & Data set     & $\tau_{B_s}$ (ps)                 & Reference \\   \hline
ALEPH          & D$_s \ell$          & 91-95        & 1.54$^{+0.14}_{-0.13} \pm$ 0.04   & \cite{ALEBS1}  \\
ALEPH          & D$_s h$             & 91-95        & 1.47$\pm$ 0.14 $\pm$ 0.08         & \cite{ALEBS2}  \\
CDF            & D$_s \ell$          & 92-96        & 1.36$ \pm 0.09 ^{+0.06}_{-0.05} $ & \cite{CDFBS}  \\
CDF            & Excl. ${\rm J}/\psi \phi$ & 92-95        & 1.34$^{+0.23}_{-0.19} \pm$ 0.05   & \cite{CDFB01}  \\
DELPHI         & D$_s \ell$          & 91-95        & 1.42$^{+0.14}_{-0.13} \pm 0.03$   & \cite{DELBS0} \\
DELPHI         & D$_s h$             & 91-95        & 1.53$^{+0.16}_{-0.15}$ $\pm 0.07$   & \cite{DELBS1}\\
%DELPHI         & $\phi \ell$         & 91-94        & 1.33$\pm 0.25^{+0.23}_{-0.10}$    & \cite{DELBS1} \\
DELPHI         & D$_s$ inclus.       & 91-94        & 1.60$\pm 0.26^{+0.13}_{-0.15}$    & \cite{DELBS2}\\
OPAL           & D$_s \ell$          & 90-95        & 1.50$^{+0.16}_{-0.15} \pm$ 0.04   & \cite{OPABS1}  \\ 
OPAL           & D$_s$ inclus.       & 90-95        & 1.72$^{+0.20+0.18}_{-0.19-0.17}$  & \cite{OPABS2}  \\ \hline
Average        &                     &              & $\taubs$                  &    \\   \hline

\end{tabular}
\caption[]{{\it Measurements of the $\Bs$ lifetime.} \label{tab:bs} } 
%To extract the $B_s$ lifetime,
%            a mean $B$ hadron lifetime of 1.4 ps was assumed in the papers flagged with an (a). 
%            In this case the numbers quoted in the Table have been scaled 
%            to the current average $B$ hadron lifetime of 1.55 ps.
%           a) The combined DELPHI result quoted 
%           in \cite{DELBS2} is 1.67 $\pm$ 0.14 ps.}
\end{center}
\end{table}
\begin{table}[bth]
\begin{center}
\begin{tabular}{|l||c|c|c|c|} \hline
Experiment&Method               &Data set&$\tau_{\Lambda_{\rm{b}}}$ (ps)    & Reference \\\hline
ALEPH     &$\Lambda   \ell$        &  91-95 & 1.20$^{+0.08}_{-0.08} \pm$ 0.06  & \cite{ALELAM}\\
DELPHI    &$\Lambda \ell \pi$ vtx  &  91-94 & 1.16$\pm 0.20 \pm 0.08$          &  \cite{DELLAM0}$^a$\\
DELPHI    &$\Lambda   \mu$ i.p.    &  91-94 & 1.10$^{+0.19}_{-0.17} \pm 0.09 $ &  \cite{DELLAM1}$^a$ \\
DELPHI    &p$\ell$                 &  91-94 & 1.19$ \pm 0.14 \pm$ 0.07      &  \cite{DELLAM0}$^a$\\
OPAL      &$\Lambda   \ell$ i.p.&  90-94 & 1.21$^{+0.15}_{-0.13} \pm$ 0.10 &  \cite{OPALAM1}$^b$  \\ 
OPAL      &$\Lambda   \ell$ vtx.&  90-94 & 1.15$\pm 0.12 \pm$ 0.06          & \cite{OPALAM1}$^b$ \\ 
\hline
Avg. above 6   &                   &     & $1.170^{+0.066}_{-0.054}$     &      \\   \hline
ALEPH     &$\Lambda_c \ell$        &  91-95 & 1.18$^{+0.13}_{-0.12} \pm$ 0.03  &  \cite{ALELAM}\\
ALEPH     &$\Lambda \ell^- \ell^+$ &  91-95 & 1.30$^{+0.26}_{-0.21} \pm$ 0.04  &  \cite{ALELAM}\\
CDF       &$\Lambda_c \ell$        &  91-95 & 1.32$\pm 0.15        \pm$ 0.06   &  \cite{CDFLAM}\\
DELPHI    &$\Lambda_c   \ell$      &  91-94 & 1.11$^{+0.19}_{-0.18} \pm 0.05$  &  \cite{DELLAM0}$^a$\\
OPAL      &$\Lambda_c \ell \& \Lambda \ell^- \ell^+$ &  90-95 & 1.29$^{+0.24}_{-0.22} \pm$ 0.06  &  \cite{OPALAM2}\\ \hline
Avg. above 5   & $ \tau_{\Lambda_b}$  &        & $\taulb$                             &      \\   \hline
Avg. above 11   &                   &     & $1.208 \pm 0.051$     &      \\   \hline
ALEPH     &$\Xi \ell$              &  90-95 & 1.35$^{+0.37+0.15}_{-0.28-0.17}$ &  \cite{ALELAM1}\\
DELPHI    &$\Xi \ell$           &  91-93 & 1.5 $^{+0.7}_{-0.4} \pm$ 0.3     &  \cite{DELLAM2} \\
\hline
Avg. above 2   &   $ \tau_{\Xi_b}$ &     & $\tauxib$     &      \\   \hline
\end{tabular}
\caption[]{{\it Measurements of the $b$-baryon lifetime.\\
           a) The combined DELPHI result quoted in \cite{DELLAM0} is (1.14$\pm 0.08 $ $\pm$ 0.04) ps. \\
           b) The combined OPAL result quoted in \cite{OPALAM1} is (1.16 $\pm$ 0.11 $\pm$ 0.06) ps. }\\
 \label{tab:lam}} 
\end{center}
\end{table}

\begin{table}[ht!]
\begin{center}
\begin{tabular}{|l||c|c|c|c|} \hline
Experiment & Method            &Data set& $\tau_{B}$ (ps)                 & Reference \\   \hline
ALEPH      & Lepton i.p. (3D)  & 91-93  & 1.533 $\pm$ 0.013 $\pm$      0.022 & \cite{ALEIN1}  \\
L3         & Lepton i.p. (2D)  & 91-94  & 1.544 $\pm$ 0.016 $\pm$      0.021 & \cite{L3IN1}$^b$ \\
OPAL       & Lepton i.p. (2D)  & 90-91  & 1.523 $\pm$ 0.034 $\pm$      0.038 & \cite{OPAIN1}  \\ \hline
Average set 1&                 &        & 1.537 $\pm$ 0.020                  &    \\   \hline
ALEPH      & Dipole            &   91   & 1.511 $\pm$ 0.022 $\pm$      0.078 & \cite{ALEIN2}  \\
ALEPH (p)     & Sec. vert.        &  91-95 & 1.601 $\pm$ 0.004 $\pm$      0.032 & \cite{ALEIN22}  \\
DELPHI     & All track i.p.(2D)& 91-92  & 1.542 $\pm$ 0.021 $\pm$      0.045 & \cite{DELIN0}$^a$ \\
DELPHI     & Sec. vert.        & 91-93  & 1.582 $\pm$ 0.011 $\pm$      0.027 & \cite{DELIN}$^a$ \\
L3         & Sec. vert. + i.p. & 91-94  & 1.556 $\pm$ 0.010 $\pm$      0.017 & \cite{L3IN1}$^b$ \\
OPAL       & Sec. vert.        & 91-94  & 1.611 $\pm$ 0.010 $\pm$      0.027 & \cite{OPAIN2}  \\ 
SLD        & Sec. vert.        &  93    & 1.564 $\pm$ 0.030 $\pm$      0.036 & \cite{SLDIN}  \\ \hline
Average set 2 &                 &         & 1.577 $\pm$ 0.016                  &    \\   \hline
Average sets 1-2   &                   &        & 1.564 $\pm$ 0.014                       &    \\   \hline
CDF        & ${\rm J}/\psi$ vert.    &  92-95 & 1.533 $\pm$ 0.015 $^{+0.035}_{-0.031}$  & \cite{CDFB01}  \\ \hline
\end{tabular}
\caption[]{{\it Measurements of the average $b$-hadron lifetime.\\
            a) The combined DELPHI result quoted in \cite{DELIN} is (1.575 $\pm$ 0.010 $\pm$ 0.026) ps.\\
            b) The combined L3 result quoted in \cite{L3IN1} is (1.549 $\pm$ 0.009 $\pm$ 0.015) ps.} \label{tab:bhad} }

\end{center}
\end{table}

\begin{table}[ht!]
\begin{center}
\begin{tabular}{|l||c|c|c|c|} \hline
 Experiment    & Method            & Data set      & Ratio $\tau_+ /\tau_0$           & Reference \\   \hline
ALEPH          & D$^{(*)} \ell$    &  91-95        & 1.085$\pm 0.059 \pm 0.018$        & \cite{ALEB01}  \\
ALEPH          & Excl. rec.       &  91-94         & 1.27$^{+0.23+0.03}_{-0.19-0.02}$  & \cite{ALEB0}  \\
CDF            & D$^{(*)} \ell$   & 92-95         & 1.110$\pm 0.056^{+0.033}_{-0.030}$  & \cite{CDFB02}  \\
CDF            & Excl. (${\rm J}/\psi K$)& 92-95         & 1.06 $\pm$ 0.07 $\pm$ 0.02          & \cite{CDFB01}  \\
DELPHI         & D$^{(*)} \ell$    &  91-93        & 1.00$^{+0.17}_{-0.15} \pm$ 0.10     & \cite{DELB01} \\
DELPHI         & Charge sec. vtx.  &  91-93        & 1.06$^{+0.13}_{-0.11} \pm 0.10$     & \cite{DELB02} \\
DELPHI         & Charge sec. vtx.  &  94        & 1.065$ \pm$0.022 $\pm$0.033     & \cite{DELB031} \\
L3             & Charge sec. vtx.  &  94-95        & 1.09$\pm$  0.07  $\pm$ 0.03         & \cite{L3B01}  \\ 
OPAL           & D$^{(*)} \ell$    &  91-93        & 0.99$ \pm 0.14^{+0.05}_{-0.04}$     & \cite{OPAB0}  \\ 
OPAL          & Charge sec. vtx.  &  93-95        & 1.079$\pm$ 0.064  $\pm$ 0.041       & \cite{OPAB1}  \\ 
SLD            & Charge sec. vtx. $\ell$&  93-95   & 1.03$^{+0.16}_{-0.14} \pm$ 0.09     & \cite{SLDB01}  \\ 
SLD (p)         & Charge sec. vtx.  &  93-98        & 1.037$^{+0.025}_{-0.024} \pm$ 0.024 & \cite{SLDB02}  \\ \hline
Average        &                   &               & $\taubpovertaubd$                     &    \\   \hline

\end{tabular}
\caption[]{{\it Measurements of the ratio  
$\tau_{\Bp} /\tau_{\Bd}$.} \label{tab:rat}  }
%           a) The combined ALEPH result quoted 
%           in \cite{ALEB0} is 1.03 $\pm$ 0.08 $\pm$ 0.02 ps.}
\end{center}
\end{table}

\newpage
\clearpage

\mysection{Measurements of $b$-hadron production rates}
\label{appendixAc}

\mysubsection{$\Bs$ production rate}
\begin{table}[ht!]
\begin{center}
\begin{tabular}{|l|r@{\,$\pm$\,}l|c|}
\hline
    \multicolumn{1}{|c|}{Quantity}      &  
    \multicolumn{2}{|c|}{Value}         &   
    Ref. \\
\hline
    {\hspace{1mm}}
$\fs~ {\rm BR}(\Bs \rightarrow \Dsm \ell^+ \nu_{\ell} {\rm X})~{\rm BR}(\Dsm \rightarrow \phi \pi^-)  $    {\hspace{1mm}} 
    &   (2.87  &  $ 0.32 \pm {}^{0.25}_{0.41})$ \,$\times 10^{-4}$  &  \cite{A:tbs} \\ %%OK
    {\hspace{1mm}}
$\fs~ {\rm BR}(\Bs \rightarrow \Dsm \ell^+ \nu_{\ell} {\rm X})~{\rm BR}(\Dsm \rightarrow \phi \pi^-) $     {\hspace{1mm}} 
    &   (4.2  &  $ 1.9 )$ \,$\times 10^{-4}$  &  \cite{A:tbs2} \\ %%OK
    {\hspace{1mm}}
$\fs~ {\rm BR}(\Bs \rightarrow \Dsm \ell^+ \nu_{\ell} {\rm X})~{\rm BR}(\Dsm \rightarrow \phi \pi^-)$      {\hspace{1mm}} 
    &   (3.9  &  $ 1.1 \pm 0.8)$ \,$\times 10^{-4}$  &  \cite{A:tbs3} \\ %%OK
\hline
Average of the three measurements   {\hspace{1mm}} 
    &   (3.00  &  $ ^{0.38}_{0.41})$ \,$\times 10^{-4}$  &   \\ %%OK
\hline
    {\hspace{1mm}} $\fs/(\fu+\fd)~{\rm BR}(\Dsm \rightarrow \phi \pi^-)$       {\hspace{1mm}} 
    &   (7.7  &  $ 1.5)$\,$\times 10^{-3}$   &  \cite{A:flamcdf} \\ %%OK
    {\hspace{1mm}} $\fs/(\fu+\fd)~{\rm BR}(\Dsm \rightarrow \phi \pi^-)$       {\hspace{1mm}} 
    &   (7.6  &  $ 1.8)$\,$\times 10^{-3}$   &  \cite{ref:cdfrates} \\ %%OK
\hline
\end{tabular}
\caption []{{\it Inputs used in the calculation of
the $\Bs$ production rate.} \label{tab:sfrac} } 
\end{center}

\end{table}

All published results have been multiplied by the branching fraction
for the decay $\Dsm \rightarrow \phi \pi^-$ to be independent of the assumed
central value and uncertainty of this quantity; quoted uncertainties
have been reevaluated accordingly.

The ALEPH measurement \cite{ref:anotused}
of $\fs~ {\rm BR}(\Bs \rightarrow \Dsm {\rm X})~
 {\rm BR}(\Dsm \rightarrow \phi \pi^-)=(3.1 \pm 0.7 \pm 0.6)\times 10^{-3}$
has not been used to obtain an additional measurement on $\fs$ because of the model dependence attached to the evaluation of 
${\rm BR}(\Bs \rightarrow \Dsm {\rm X})$. Instead, the value of $\fs$, quoted
in Table \ref{tab:rates}, can be used to extract, from this measurement, the
inclusive branching fraction for $\Dsm$ production in $\Bs$ decays:
\begin{equation}
{\rm BR}(\Bs \rightarrow \Dsm {\rm X})~{\rm BR}(\Dsm \rightarrow \phi \pi^-)=
(3.1 \pm 0.7 \pm 0.7)\times 10^{-2}
\end{equation}
and, using the value for ${\rm BR}(\Dsm \rightarrow \phi \pi^-)$
 given in Table \ref{tab:gensys}:
\begin{equation}
{\rm BR}(\Bs \rightarrow \Dsm {\rm X})=0.86 \pm 0.19 \pm 0.29.
\end{equation}

\mysubsection{$b$-baryon production rate}
\begin{table}[ht!]
{\small
\begin{center}
\begin{tabular}{|l|r@{\,$\pm$\,}l|c|}
\hline
    \multicolumn{1}{|c|}{Quantity}      &  
    \multicolumn{2}{|c|}{Value}         &   
    Ref. \\
\hline
    {\hspace{1mm}}
     $\prodlb~{\rm BR}(\Lc \rightarrow {\rm p} {\rm K}^- \pi^+)$     {\hspace{1mm}}
    &   (3.78  & $ 0.31 \pm 0.23)$ \, $\times 10^{-4}$  &  \cite{AD:prodlamb} \\ %%OK
    {\hspace{1mm}}
     $\prodlb~{\rm BR}(\Lc \rightarrow {\rm p} {\rm K}^- \pi^+)$     {\hspace{1mm}}
    &   (5.19  & $ 1.14 \pm {}^{1.19}_{0.66})$ \, $\times 10^{-4}$  &  \cite{AD:prodlamb2} \\ %%OK
\hline
Average of the two measurements   {\hspace{1mm}} 
    &   (3.90  &  $ 0.42)$ \,$\times 10^{-4}$  &   \\ %%OK
\hline
    {\hspace{1mm}} $\fb/(\fu+\fd)~{\rm BR}(\Lc \rightarrow {\rm p} {\rm K}^- \pi^+)$      {\hspace{1mm}} 
    &   (5.9  &  $ 1.4)\times 10^{-3}$   &  \cite{A:flamcdf} \\ %%OK
\hline
    {\hspace{1mm}}     $\prodxb$     {\hspace{1mm}} 
    &  ( 5.4   & $ 1.1 \pm 0.8 )$ $\,\times \ 10^{-4}$  &  \cite{AD:cascb} \\ %%OK
    {\hspace{1mm}}     $\prodxb$     {\hspace{1mm}} 
    &  ( 5.9   & $ 2.1 \pm 1.0 )$ $\,\times \ 10^{-4}$  &  \cite{AD:cascb2} \\ %%OK
\hline
Average of the two measurements   {\hspace{1mm}} 
    &   (5.5  &  $ 1.2)$ \,$\times 10^{-4}$  &   \\ %%OK
\hline
    {\hspace{1mm}} $\fb$      {\hspace{1mm}} 
    &   0.102  &  $ 0.007 \pm 0.027 $   &  \cite{A:flamdir} \\ %%OK
\hline
\end{tabular}
\caption []{{\it Inputs used in the calculation of
the $b$-baryon production rate.} \label{tab:bfrac}} 
\end{center}
}
\end{table}

All published results which are using the $\Lc$ baryon
have been multiplied by the branching fraction
for the decay $\Lc \rightarrow {\rm p} {\rm K}^- \pi^+$ 
to be independent of the assumed
central value and uncertainty of this quantity; quoted uncertainties
have been reevaluated accordingly.

\mysubsection{$\Bp$ production rate}

\begin{table}[ht!]
\begin{center}
\begin{tabular}{|l|r@{\,$\pm$\,}l|c|}
\hline
    \multicolumn{1}{|c|}{Quantity}      &  
    \multicolumn{2}{|c|}{Value}         &   
    Ref. \\
\hline
    {\hspace{1mm}} $\fu$      {\hspace{1mm}} 
    &   0.414  &  $ 0.016 $   &  \cite{ref:delphibplus} \\ %%OK
\hline
\end{tabular}
\caption [] {{\it Direct measurement of the $\Bp$ 
production rate.} \label{tab:bpfrac}} 
\end{center}

\end{table}

This value is obtained from the measurement of the production rate
of charged weakly decaying $b$-hadrons. A small correction has been applied
to account for $\Xi_b^-$ production, as given in 
Table \ref{tab:bfrac}.
\newpage
\clearpage

\mysection{Measurements of $c$ and $\overline{c}$ production rates in
$b$-hadron decays}
\label{appendixAe}

\begin{table}[ht!]
\begin{center}
\begin{tabular}{|l|c|} \hline
Experiment   & ${\rm P}(b \rightarrow \Dn~{\rm or}~\Dnb) \times 
{\rm BR}(\Dn \rightarrow \Km \pi^+)(\%)$\\
 \hline
 ALEPH ~~\cite{ref:alinc} &  $2.32 \pm 0.090 \pm 0.048 \pm 0.040(0.035)$\\
 DELPHI~\cite{ref:delinc}      &   $2.308 \pm 0.075 \pm0.045 \pm0.134(0.121)$\\
 OPAL~~~\cite{ref:opainc}  &       $2.099 \pm0.106 \pm 0.076 \pm0.094(0.053)$\\
 \hline
 CLEO~~~\cite{ref:cleinc}  &       $2.487 \pm 0.055 \pm0.074$ \\
\hline
\end{tabular}
\end{center}
\caption{{\it Experimental results on $\Do$ production rates
in $b$-hadron decays.} \label{tab:bxcbr1}}

\end{table}
\begin{table}[ht!]
\begin{center}
\begin{tabular}{|l|c|} \hline
Experiment   & ${\rm P}(b \rightarrow \Dp+\Dm) \times 
{\rm BR}(\Dp \rightarrow \Km \pi^+ \pi^+)(\%)$ \\
 \hline
 ALEPH ~~\cite{ref:alinc}      & 
                         $2.13 \pm 0.120 \pm 0.052 \pm0.073(0.029)$ \\
 DELPHI~\cite{ref:delinc}      & 
                         $2.092 \pm 0.094 \pm 0.057 \pm 0.088(0.060)$\\
 OPAL~~~\cite{ref:opainc}  &  
                         $1.752 \pm 0.143 \pm 0.074 \pm 0.098 \pm 0.065$ \\
 \hline
 CLEO~~~\cite{ref:cleinc}  &  
                         $2.160 \pm 0.083 \pm 0.083$ \\
\hline

\end{tabular}
\end{center}
\caption{{\it Experimental results on $\Dp$ production rates
in $b$-hadron decays.} \label{tab:bxcbr2}}

\end{table}

\begin{table}[ht!]
\begin{center}
\begin{tabular}{|l|c|} \hline
Experiment   & ${\rm P}(b \rightarrow \Ds+\Dsb) \times 
{\rm BR}(\Ds \rightarrow \phi \pi^+)(\%)$\\
 \hline
 ALEPH ~~\cite{ref:alinc}      &  $0.652 \pm 0.061 \pm 0.022 \pm 0.026(0.018)$\\
 DELPHI~\cite{ref:delinc}      &   $0.574 \pm 0.078 \pm0.070 \pm0.036(0.021)$\\
 OPAL~~~\cite{ref:opainc}  &       $0.768 \pm0.083 \pm 0.048 \pm0.057(0.027)$\\
 \hline
 CLEO~~~\cite{ref:cleinc2}  &       $0.424 \pm 0.014 \pm0.031$ \\
\hline
\end{tabular}
\end{center}
\caption{{\it Experimental results on $\Ds$ production rates
in $b$-hadron decays.} \label{tab:bxcbr3}}

\end{table}

\begin{table}[ht!]
\begin{center}
\begin{tabular}{|l|c|} \hline
Experiment   & ${\rm P}(b \rightarrow \Lc+\Lcb) \times 
{\rm BR}(\Lc \rightarrow p \Km \pi^+)(\%)$ \\
 \hline
 ALEPH ~~\cite{ref:alinc}      &
                         $0.48 \pm 0.06 \pm 0.02 \pm0.019(0.011)$ \\
 DELPHI~\cite{ref:delinc}      & 
                         $0.445 \pm 0.086 \pm 0.026 \pm 0.028(0.022)$\\
 OPAL~~~\cite{ref:opainc}  &  
                         $0.564 \pm 0.106 \pm 0.032 \pm 0.034 \pm 0.019$ \\
 \hline
 CLEO~~~\cite{ref:cleinc3}  & 
                         $0.273 \pm 0.051 \pm 0.039$ \\
\hline
\end{tabular}
\end{center}
\caption{{\it Experimental results on $\Lc$ production rates
in $b$-hadron decays.} \label{tab:bxcbr4}}

\end{table}
\clearpage

\begin{table}[ht!]
\begin{center}
\begin{tabular}{|l|c|c|} \hline
Experiment   & ${\rm BR}(b \rightarrow 0{\rm D~X})(\%)$ & 
${\rm BR}(b \rightarrow {\rm D}~\overline{\rm D}~{\rm X})(\%)$\\
 \hline
 DELPHI~\cite{ref:incldel}  &   $3.3 \pm 1.8 \pm 1.0$ & $13.6 \pm 3.0 \pm 3.0$\\
 SLD~~~\cite{ref:inclsld}  & $5.6 \pm 1.1 \pm 2.0$ & $24.6 \pm 1.4 \pm 4.0$\\
\hline
\end{tabular}
\end{center}
\caption{{\it Experimental results on no-open charm and 
two charm hadron branching fractions. The first uncertainty corresponds 
to statistical errors and un-correlated systematics whereas
the second number is for correlated systematic uncertainties}
\label{tab:inclbr}}
\end{table}

\begin{table}[ht!]
\begin{center}
\begin{tabular}{|l|c|} \hline
Decay channel   & BR ($\%$)\\
 \hline
$b \rightarrow \Do \Dsb {\rm X}$    &  $9.1^{+2.0~+1.3+3.1}_{-1.8~-1.2~-1.9}$ \\
$b \rightarrow \Dp \Dsb {\rm X} $   &  $4.0^{+1.7}_{-1.4}\pm0.7^{+1.4}_{-0.9}$ \\
\hline
$b \rightarrow \Do \Dob {\rm X} $   &  $5.1^{+1.6~+1.2}_{-1.4~-1.1}\pm0.3$ \\
$b \rightarrow \Do \Dm,~\Dp \Dob {\rm X} $   &  $2.7^{+1.5~+1.0}_{-1.3~-0.9}\pm0.2$ \\
$b \rightarrow \Dp \Dm {\rm X}$    &  $<0.9$ at 90$\%$C.L.\\
\hline
$b \rightarrow \Dstarp \Dsb {\rm X} $   &  $3.3^{+1.0}_{-0.9}\pm0.6^{+1.1}_{-0.7}$ \\
$b \rightarrow \Dstarp \Dob,~\Do \Dstarm {\rm X}$    &  $3.0^{+0.9~+0.7}_{-0.8~-0.5}\pm0.2$ \\
$b \rightarrow \Dstarp \Dm,~\Dp \Dstarm {\rm X}  $  &  $2.5^{+1.0~+0.6}_{-0.9~-0.5}\pm0.2$ \\
\hline
$b \rightarrow \Dstarp \Dstarm {\rm X} $   &  $1.2^{+0.4}_{-0.3} \pm 0.2 \pm 0.1$ \\
\hline
\end{tabular}
\end{center}
\caption{{\it Measurements of different double charm final states by the 
ALEPH Collaboration \cite{ref:alephdoubled}. 
The first error is statistical, the second is the sum of
all systematic errors except those from D branching fractions  
which determine the third error.} \label{tab:alephcc}}
\end{table}

Only measured branching fractions into a D or a $\overline{{\rm D}}$ meson
have been
used in the determination of the double charm rates given in Section 
\ref{sec:alephm}.
Decay channels with one or two ${\rm D}^*$ can be used to evaluate the
branching channel for $b \rightarrow \Dp \Dm {\rm X}$ which is unmeasured.
Two models have been considered. In the first model 
\footnote{R. Barate, private communication.},
two parameters are introduced: the total double charm rate
with no $\Ds$ in the final state and the relative rate, $r$, for producing a 
$\Dstar$ instead of a D meson, in the decay of the $b$-hadron. 
The value $r=3$ allows to reproduce the different measured decay rates
and gives BR$(b \rightarrow \Dp \Dm {\rm X})=0.5\%$. In another approach,
only non-strange final states have been considered and
three parameters have been used corresponding to the respective 
production rates for D$\overline{{\rm D}}$,  D$\overline{{\rm D}}^*$ and
 ${\rm D}^* \overline{{\rm D}}^*$ assumed to be equal in all possible
charge combinations. These parameters have been fitted using measured
decay rates for the different decay channels. The fit prefers no prompt
production of $b \rightarrow {\rm D} \overline{{\rm D}} {\rm X}$ and,
fitting only the two remaining parameters, it gives:
BR$(b \rightarrow \Dp \Dm {\rm X})=0.35\%$. 

These two results are compatible
with the quoted limit for this channel and the importance of $\Dstar$
production in double-charm $b$-hadron decays can explain the smallness
of the $\Dp \Dm$ contribution.

\clearpage

\mysection{Theoretical uncertainties relevant to the measurements of
$\Vub$ and $\Vcb$}
\label{appendixC}

%At beginning of June 1999 a workshop entitled
%``Informal Workshop on the Derivation of $\Vcb$
%and $\Vub$: Experimental Status and 
%Theory Uncertainties''\footnote{Participants at this workshop were:
%D. Abbaneo, P. Ball, E. Barberio, M. Battaglia, M. Beneke, 
%I.I. Bigi, 
%G. Buchalla, M. Calvi, O. Cooke, F. Defazio, L. di Ciaccio, R. Fleischer,
% P. Gagnon, P. Henrard, A. Hoang, L. Lellouch, J. Lu, S. Mele,
%E. Piotto, Ph. Rosnet, P. Roudeau, D. Rousseau, Ch. Schwick and F. Simonetto.}
%was held at CERN.
%
%As, at that time, there was not a general 
%consensus on the values for theoretical 
%errors to be used to evaluate $\Vcb$ and $\Vub$~\cite{Babar}, 
%the main aim of this workshop has been to scrutinize
%the uncertainties attached to the different parameters and to define
%a common set of values to be adopted for the derivation of $\Vub$ and $\Vcb$.
%%In the following note, 

The determination of the values of $\Vcb$ and $\Vub$ from present measurements
depends on the knowledge of the values of other parameters and on the 
use of theoretical expressions of limited accuracy.

Some of these parameters, such as $\alpha_S$ (the strong coupling constant),
are measured and it can be assumed
that the corresponding error has a Gaussian distribution. Other quantities
contributing to the error (parameters, series truncated at finite orders, ..)
are taken from theory. For a parameter, the quoted uncertainty is expected to 
correspond to a range which contains its exact value. 
If the uncertainty is related to a limitation of the theoretical approach
in evaluating higher order contributions, the quoted uncertainty is expected
to have been deduced from the evaluation of an upper limit on the contribution
from neglected terms. To ensure that the combined uncertainty corresponds to
intervals that contain the exact values of 
$\Vcb$ or $\Vub$, uncertainties of purely theoretical origin have to be added
linearly. Other type of uncertainties, assumed to be Gaussian distributed, can be convoluted with the previous one to get the final error distribution.

In practice, there are no complete analytical expressions allowing
to update the values of the different parameters and to evaluate correlations.
%it is not easy to collect the needed information from existing
%publications because there are correlations between the different
%contributing components and 
In addition, uncertainties on some parameters have already
been obtained by summing in quadrature individual errors. As a result,
in the present analysis, systematic uncertainties have been added in quadrature
and an interval of $\pm1\sigma$ centered on the quoted value
for $\Vcb$ or $\Vub$ is supposed to contain the exact value of these
parameters with 68$\%$ probability (only).
%It is thus important that, not only the values for the uncertainties are given
%but also that the way they have been obtained be made available, At 
%present such detailed studies do not exist and, in the present analysis 
%the various contributing systematic uncertainties have been 
%added in quadrature. 
%This procedure is justified when several independent 
%sources of uncertainties contribute, independently of their exact
%distributions, unless a single source clearly dominates. A linear sum
%can be justified if uncertainties
%are fully and positively correlated, which is not really the case of the
%presently considered quantities.
%In addition, theoretical errors correspond usually to 
%a confidence level larger than 68$\%$, which is the confidence level assumed
%when these uncertainties are taken as standard deviations of Gaussian
%distributions.
%
%It is clear that the present procedure is not entirely satisfactory;
%it is simply pragmatic in the absence, at present, of a direct experimental 
%control,
%or of other evaluations with different theoretical techniques, of the 
%main sources of systematic theoretical uncertainties.
%Additional details can be found in \cite{ref:bigi0}. 

\mysubsection{The $b$ and $c$-quark masses}
\label{sec:aaa}
The $b$-quark mass used to determine $\Vub$ and $\Vcb$ must be running and 
evaluated at a low scale to improve
the convergence of the perturbative QCD series used to compute semileptonic 
widths of $b$-hadrons \cite{ref:bs11}. In this way, concerns \cite{ref:bs12}
originating
from the importance of QCD corrections affecting the $b$-quark mass,
getting amplified by the power (about five) at which it is raised in the 
theoretical expression of $\Gamma_{sl}(b)$, are controlled.
Typical scales have to be around 1 $\GeV$. 

Values for the $b$-quark mass, using the definition given in \cite{ref:bs11},
have been obtained at NNLO from analyses 
~\cite{ref:mass1,ref:mass2} of sum rules in the $\Upsilon$
system:
\begin{equation}
m_b^{kin.}(1~\GeV)=(4.56 \pm 0.06)~\GeV/c^2,~m_b^{kin.}(1~\GeV)=(4.57 \pm 0.04)~\GeV/c^2.
\end{equation}

Whereas the pole or the $\overline{{\rm MS}}$ $b$-quark masses are not adapted
to extract $\Vub$ or $\Vcb$, there exist other mass definitions which
have similar properties as the kinetic mass mentioned above. These are
the 1S ($m_b^{1S}$) and the potential subtracted 
($m_b^{PS}$) masses \cite{ref:mass2,ref:mass3}. All these values of the $b$-quark mass have been obtained using sum rules which relate the masses and the electronic decay widths
of $\Upsilon$ mesons to moments of the vacuum polarization function, as 
proposed initially in \cite{ref:bs13}.

In the following the value:
\begin{equation}
m_b^{kin.}(1~\GeV)=(4.58 \pm 0.06)~\GeV/c^2
\end{equation}
has been used where the quoted uncertainty is the quadratic sum of 
several contributions.

The inclusive and exclusive determinations of $\Vcb$ need also a value for
the $c$-quark mass. At present, it is considered \cite{ref:bs11}
that the mass difference between $b$ and $c$ quarks is under better 
control than the absolute determination of the $c$-quark mass itself.
Using the definition of the kinetic mass:
\begin{equation}
\overline{M_{H_Q}}= m_Q^{kin.}(\mu) + \overline{\Lambda}(\mu)
+\frac{\mu_{\pi}^2(\mu)}{2 m_Q(\mu)}
+\frac{\rho_D^3(\mu)-\overline{\rho}^3(\mu)}{4 m_Q^2(\mu)}
+{\cal O}\left ( \frac{1}{m_Q^3}\right )
\label{eq:mkin}
\end{equation}
in which, $\overline{M_{H_Q}}$ is the spin-averaged heavy hadron mass, 
$\overline{\Lambda}(\mu)$ is a function which is independent of 
the heavy quark,
$\mu_{\pi}^2(\mu)$ is the average of the square of the heavy quark momentum
inside the hadron, $\rho_D^3(\mu)$ is the expectation value of the Darwin term
and $\overline{\rho}^3(\mu)$ is the sum of positive non-local correlators
\cite{ref:bs15}.

Taking the difference of the previous expression for $b$ and $c$ quarks
one obtains:
\begin{equation}
\begin{array}{cc}
m_b^{kin.}(\mu)-m_c^{kin.}(\mu) =
& \overline{M_B}-\overline{M_D}
-\frac{\mu_{\pi}^2(\mu)}{2} 
\left ( \frac{1}{m_b(\mu)} -\frac{1}{m_c(\mu)} \right )\\
 & -\frac{\rho_D^3(\mu)-\overline{\rho}^3(\mu)}{4}
\left ( \frac{1}{m_b^2(\mu)} -\frac{1}{m_c^2(\mu)} \right )
+{\cal O}\left ( \frac{1}{m_{b,c}^3} \right ).\\
\end{array}
\end{equation}
As the $\rho^3$ terms are expected to be of order $0.1~\GeV^3$ for 
$\mu\sim 1 ~\GeV$, the ${\cal O}\left ( \frac{1}{m_{b,c}^2}\right )$
correction has already a small value. Numerically this gives:
\begin{equation}
m_b^{kin.}(1~\GeV)-m_c^{kin.}(1~\GeV) =
(~3.50~+~0.040 ~\frac{\mu_{\pi}^2-0.5}{0.1}~+~\Delta M_2~ )~(\GeV/c^2)
\end{equation}
where $\left |\Delta M_2 \right| \leq 0.015 ~\GeV/c^2$.

It can be noted that $m_c^{kin.}(1 ~\GeV) \sim 1.08~\GeV/c^2$ is obtained 
in this way. This may seem to be a rather low value but it is
in agreement with the independent determination of
this quantity obtained using the value of the $\overline{{\rm MS}}$
charm quark mass, 
$\overline{m_c}(\overline{m_c}) = (1.23 \pm 0.09)~\GeV/c^2$ 
\cite{ref:mc} 
%mentioned previously when using the expression,
and the theoretical expression
given in \cite{ref:cza}, which relates the kinetic and the 
$\overline{{\rm MS}}$ heavy quark mass definitions.

Even if, numerically there is no evidence 
(by chance?) for such problems,
it may be argued that the $1/m_c$ is less under control than the $1/m_b$
expansion as $m_c\sim 1~\GeV$. As a result it has been decided to double
the theoretical error attached to the inclusive determination of $\Vcb$.

\mysubsection{Measurement of $\Vub$ using the decay 
$b \rightarrow  \ell^-\overline{\nu_{\ell}}{\rm X}_{u}$}
\label{sec:A}

Based on studies developed independently by two 
groups \cite{uraltsev},\cite{hoang}, 
a relative  theoretical uncertainty of 5\% has been evaluated for the 
extraction of $\Vub$ from the measured inclusive charmless semileptonic rate 
${\rm BR}(b\rightarrow \ell^-\overline{\nu_{\ell}}{\rm X}_{u})$. The central values
of the two analyses also 
agree\footnote{In practice the initial central value quoted in \cite{uraltsev}
has been corrected and the value obtained 
in \cite{hoang} is
$\Vub~=~0.00443 \left ( \frac{{\rm BR}(b \rightarrow  \ell
\overline{\nu_{\ell}}{\rm X}_u)}{0.002}  \right )^{1/2}
\left ( \frac{1.55~{\rm ps}}{\tau_b} \right )^{1/2}
 \times \left (1 \pm 0.020_{pert.} \pm 0.030_{m_b} \right ) $.},
and the following relationship 
%(from Ref.~\cite{ref:vub0}) 
has been adopted in the extraction of $\Vub$:

\begin{equation}
\begin{array}{cc}
\Vub=
& 0.00445 \left ( \frac{{\rm BR}(b \rightarrow \ell^-
\overline{\nu_{\ell}} {\rm X}_u)}{0.002}  \right )^{1/2}
\left ( \frac{1.55~{\rm ps}}{\tau_B} \right )^{1/2}\\
  & \times \left (1 \pm 0.010_{pert.} \pm 0.030_{1/m_b^3} \pm 0.035_{m_b} \right ) \\
\end{array}
\label{eq:vub}
\end{equation}

The measurements performed at LEP reported in Section \ref{sec:vub}, 
based on the inclusive selection
of $b \rightarrow \ell^-\overline{\nu_{\ell}}{\rm X}_u $ decays, are sensitive to
a large fraction of the mass distribution of the hadronic system
and to the full range of the lepton energy spectrum.
This mass distribution $M_X$ has been studied by theorists \cite{ref:mass} for
the $b \rightarrow \ell^-\overline{\nu_{\ell}}{\rm X}_u $ transition.
The conclusion was\footnote{In practice experimental 
analyses take into account the expected
mass distribution of the hadronic system, and the variation of the
detection efficiencies as a function of this mass, to evaluate the 
real importance
of this systematic uncertainty.}
that the additional theoretical uncertainty on $\Vub$ 
from $m_b$ and $\mu_{\pi}^2$ is less than 10\% if the experimental inclusive
technique is sensitive up to at least 
$M_X \simeq 1.5$~$\GeV/c^2$.

%It has thus been concluded that the theoretical errors attached to the 
%determination of
%$\Vub$ using the present experimental technique are under control.

\mysubsection{Measurement of $\Vcb$ using the decay 
$\Bdb \rightarrow \Dstarp \ell^-\overline{\nu_{\ell}}$}
\label{sec:B}

It has been proposed to use the parametrization given in
\cite{ref:vcb1} to account for the dependence in $w$ and to 
extract the value at $w=1$ of the differential decay rate.
The quantity ${\cal F}_{D^{\ast}}(1) \Vcb$ is then obtained.

In order to measure $\Vcb$,
the value at $w=1$ of the form factor is taken from theory by evaluating
different corrections \cite{ref:bigi1} which have to be applied to the naive 
expectation of  ${\cal F}_{D^{\ast}}(1)=1$:
\begin{equation}
\left | {\cal F}_{D^{\ast}}(1) \right |^2~=~\xi_A(\mu) -\Delta^A_{1/m^2} 
 -\Delta^A_{1/m^3} -\sum_{0<\epsilon_i<\mu} \left | {\cal F}_i \right |^2
\end{equation}

%The central value of ${\cal F}_{D^{\ast}}(1)$
%has been recently lowered by 2$\%$
%by evaluating higher order perturbative corrections \cite{ref:ural1}.
%\begin{equation}
%F_{D^{\ast}}(1)~=~0.89
%\end{equation}

The central value of the $b$ quark mass and its uncertainty are the
same as given in Section \ref{sec:aaa}.
%The charm quark mass is obtained from the mass difference 
%$m_b-m_c$, inferred from the measured values of the 
%spin-averaged beauty and charm meson masses:
%\begin{equation}
%\begin{array}{cc}
%m_b~-~m_c~=~
%&  \left < M_B \right > -\left < M_C \right >~+~
%\mu_{\pi}^2 \left ( \frac{1}{2m_c}- \frac{1}{2m_b} \right )~+~
%{\cal O} \left ( 1/m^2_{c,b} \right ) \\
% ~~~~~~~~~~=~ &  (~3.50~+~0.040 \frac{\mu_{\pi}^2-0.5}{0.1}~+~\Delta M_2~ )~(\GeV/c^2)\\
%\end{array}
%\label{eq:mc}
%\end{equation}
The main difference between the theoretical analyses described in
\cite{ref:bigi1} and \cite{ref:neu1} comes from the evaluation of the 
 uncertainties on non-perturbative corrections. A different
approach can be found in \cite{ref:grin}.

The detailed balance of uncertainties given in \cite{ref:bigi1}, 
revised 
%by L. Lellouch and P. Roudeau at this workshop 
for consistency
with the values adopted for the different parameters, is
the following:
\begin{equation}
{\cal F}_{D^{\ast}}(1)~=~0.880 -0.024 \frac{\mu^2_{\pi}-0.5~ \GeV^2}{0.1~\GeV^2}
\pm 0.035_{excit.} \pm 0.010_{pert.} \pm 0.025_{1/m^3}
\label{eq:fdstar}
\end{equation}

%In the following, are reviewed 
The hypotheses or results on which
the contributions of the different terms were based
are reviewed in the following.

\mysubsubsection{$\mu^2_{\pi}$} 
 
The value $\mu^2_{\pi}=(0.5 \pm 0.1)~\GeV^2$ has been used and it gives
$\pm0.024$ uncertainty on ${\cal F}_{D^{\ast}}(1)$.

From QCD sum rules, the following value has been obtained~\cite{ref:ball1}:
\begin{equation}
\mu^2_{\pi}=(0.5 \pm 0.1)~\GeV^2
\end{equation}
and a model-independent lower bound has been established~\cite{ref:bigi2}:
\begin{equation}
\mu^2_{\pi}>\mu^2_G \simeq \frac{4}{3}(M^2_{B^\ast}-M^2_B)
\approx 0.4 ~\GeV^2
\end{equation}

\mysubsubsection{High mass excitations}

In QCD sum rules, used in
\cite{ref:bigi1}, \cite{ref:neu1}, the effect expected
from high mass hadronic states has been introduced in a rather
arbitrary way. In \cite{ref:bigi1}, it has been assumed that their effect,
on the correction terms which behave as $m_Q^{-2}$, can vary between
0 and 100$\%$ of $\Delta^A_{1/m^2}$. Such a variation corresponds
to $\pm 0.035$ on ${\cal F}_{D^{\ast}}(1)$.

\mysubsubsection{Higher order non-perturbative corrections}

The value of $\pm 0.025$ uncertainty related to the contribution from
terms of order at least $m_Q^{-3}$ is considered as being optimistic
from the authors of \cite{ref:bigi1}.

\mysubsubsection{Adopted value}
Combining in quadrature the uncertainties quoted in Equation
(\ref{eq:fdstar})
gives:
\begin{equation}
{\cal F}_{D^{\ast}}(1)~=~0.88 ~\pm 0.05.
\label{eq:fok}
\end{equation}
%Individual uncertainties quoted in \cite{ref:bigi1},
%which correspond essentially to those quoted in Equation (\ref{eq:fdstar}),
% are larger as compared to 
%those mentioned in \cite{ref:neu1} mainly for those related to the 
%contributions of excited states and to the remaining $1/m^3$ terms.
%From \cite{ref:neu1}, a linear (in quadrature) combination of uncertainties
%gives $\pm 0.042~(\pm 0.027)$.
%
%A linear sum of the different sources of theoretical errors can be justified
%if all components appear to be positively correlated which does not correspond
%to the present situation as for instance concerning the two components
%mentioned before and which are dominant.
%
%It is thus proposed to use:
%
%\begin{equation}
%F_{D^{\ast}}(1)~=~0.890 \pm 0.046
%\end{equation}
%
%The quoted error is slightly larger than the value obtained by summing in 
%quadrature the different contributions from \cite{ref:bigi1}. It corresponds 
%also to
%the use of theoretical errors added linearly but considering that the
%values mentionned in (\ref{eq:fdstar}) correspond to a 90$\%$C.L..

It exists several other determinations of ${\cal F}_{D^{\ast}}(1)$,
 collected in Table \ref{tab:fb}.
\begin{table}[h]
\begin{center}
\begin{tabular}{|c|c|} \hline
\fone\ & ref \\ \hline
$0.89  \pm 0.08$ & \cite{ref:bigi0} \\
$0.913 \pm 0.042$ & \cite{Babar} \\
$0.935 \pm 0.035$ & (lattice)\cite{simone}\\ \hline
\end{tabular}
\end{center}
\caption{Different values for ${\cal F}_{D^{\ast}}(1)$.}
\label{tab:fb}
\end{table}
The first value is obtained from the same expressions as used in the present
analysis but uncertainties have been combined linearly. The $\pm 0.08$
interval corresponds to a range inside which the exact value of 
${\cal F}_{D^{\ast}}(1)$ is expected. This uncertainty is thus rather similar
to the $\pm 0.1$ interval which corresponds to the 95$\%$ probability region
when using Equation (\ref{eq:fok}).
The quoted uncertainty for the second value in Table \ref{tab:fb} corresponds
also to a range and is two times smaller as compared with the first value.
The third result has been obtained using lattice QCD and uncertainties have
been combined in quadrature, the uncertainty originating from the 
quenching approximation remains to be evaluated.
From these comparisons it results that the uncertainty adopted on 
${\cal F}_{D^{\ast}}(1)$ agrees with the most conservative estimate.

%\mysubsubsection{Expected improvements in the control of theoretical errors}
%
% The uncertainty attached to ``excited states'' is expected to be decreased
%if better measurements of 
%$\Dstarstar$
% production rates and
%hadronic mass distributions are 
%obtained\footnote{$\Dstarstar$ denotes all decay modes which are not
%$\overline{{\rm B}} \rightarrow {\rm D} \ell^-\overline{\nu_{\ell}}$ and
%$\overline{{\rm B}} \rightarrow \Dstar \ell^-\overline{\nu_{\ell}}$.}.
%
% The uncertainty related to $\mu^2_{\pi}$ 
%is expected to decrease by measuring
%moments of the lepton momentum in the B rest frame. 
%Is this feasible
%at LEP? Can we only use the transverse lepton momentum? Do we have
%enough control on the B energy reconstruction? What theoretical
%expressions can we exactly use?

%\mysubsubsection{Form factors for $\Dstarstar$ production near to  $w=1$}
%
%It has been realized that the $w$-dependence of the form factors used
%to describe $\Dstarstar$ production in $b$-hadron semileptonic decays is
%important for evaluating the systematic uncertainty on $\Vcb$
%related to the fraction of $\Dstar$ mesons coming from $\Dstarstar$ decays.
%%The present model used by DELPHI, based on HQET parametrizations, valid
%%in the limit of an infinite quark mass value, is too optimistic.
%It has been proposed to use the model described in \cite{ref:ligeti},
%which predicts a rate for the ratio of 2$^+$ over 1$^+$ narrow states
%more in agreement with the experimental results. 
%Additional information on $\Dstarstar$ production which may bring 
%constraints 
%on such models would be welcome.

\mysubsection{Measurement of $\Vcb$ using the inclusive semileptonic decay 
$b \rightarrow \ell^-X$ rate}
\label{sec:C}

The expression relating $\Vcb$ to the inclusive semileptonic
branching fraction can be found in \cite{ref:bigi1}:

\begin{equation}
\begin{array}{cc}
\Vcb=
& 0.0411 \left ( \frac{{\rm BR}(b \rightarrow \ell^-
\overline{\nu_{\ell}}{\rm X}_c)}{0.105}  \right )^{1/2}
\left ( \frac{1.55~{\rm ps}}{\tau_b} \right )^{1/2}\\
  & \times \left ( 1 -0.012 \frac{\mu_{\pi}^2-0.5 \GeV^2}{0.1 \GeV^2} \right )\\
  & \times \left (1 \pm 0.015_{pert.} \pm 0.010_{m_b} \pm 0.012_{1/m_Q^3}\right ) \\
\end{array}
\label{eq:huit}
\end{equation}
The central value of reference \cite{ref:bigi1} has been lowered by 2$\%$ 
to account for a different choice of the $b$-quark mass. 
A very similar result for the central value and the  uncertainties 
can be found in \cite{hoang}
\footnote{Using the same values for the lifetime and the semileptonic 
branching fraction of $b$-hadrons it gives:
$\Vcb = 0.0426 \times (1 \pm 0.019 \pm 0.017 \pm 0.012)$.}.

\mysubsubsection{Uncertainties related to quark masses}
It has been assumed that $m_b(\mu)$ can be determined with an uncertainty of
$\pm$60~$\MeV/c^2$, as discussed in Section~\ref{sec:aaa}, and that the difference 
between the 
$b$- and $c$-quark masses
is known with an uncertainty of $\pm$40~$\MeV/c^2$, related to the 
error on $\mu_{\pi}^2$. These variations induce $\pm 0.010$ 
and $\pm 0.012$ uncertainties
on $\Vcb$, respectively.

% Quoted uncertainties on quark masses appear to be realistic.
%The analysis done by \cite{ref:hoang} fully supports quoted uncertainties
%and indicates that an even smaller uncertainty on the value of
%the $b$ quark mass could have been used.

\mysubsubsection{Adopted value}

Adding in quadrature the quoted uncertainties
in Equation (\ref{eq:huit}) leads to a $\pm$2.5$\%$ relative 
uncertainty on $\Vcb$. 
Since these errors have not been cross-checked by other theoretical
approaches or experimental measurements,
%This is considered, at present, without a control
%by other theoretical approaches or experimental measurements, as being
%not safe enough (see reference \cite{ref:bigi0} for a more detailed 
%discussion) and, 
as mentioned in Section \ref{sec:aaa}, it has been decided
to inflate the total error by an arbitrary 
factor of two, leading to 
a theoretical uncertainty of $\pm$5$\%$ (see reference \cite{ref:bigi0} for 
a more detailed discussion).
This implies, in practice, that uncertainties of
$\pm 120~\MeV/c^2$, $\pm 80~\MeV/c^2$, and $\pm 0.2~\GeV^2$ have been
attached to the values of $m_b$, $m_b-m_c$,
and $\mu_{\pi}^2$ respectively.
%in a rather arbitrary way, uncertainties have been 
%inflated by a factor of two.
% in the present evaluation.

% A value of 5$\%$ relative uncertainty on $\Vcb$ has been adopted.
%It corresponds to the linear sum of quoted theoretical errors to account
%for possible positive correlations between the different sources of errors.
%This differs from the exclusive determination of $\Vcb$
%in which the largest contributing uncertainties were found to be
%of independent origins.

% If progresses in the experimental determination of $\mu^2_{\pi}$ are obtained,
%it can be envisaged to decrease the present values of theoretical 
%uncertainties.

\mysubsection{Common sources of theoretical errors for the two determinations of
$\Vcb$}

Theoretical uncertainties attached to the exclusive and inclusive
 measurements of $\Vcb$
are largely uncorrelated. When evaluating
the average, only the uncertainties related
to $\mu_{\pi}^2$ and $m_b$ have been considered fully correlated.

\mysubsection{Sources of theoretical errors entering into the measurement
of the ratio $\frac{\Vub}{\Vcb}$}

In the evaluation of the ratio $\frac{\Vub}{\Vcb}$, several common experimental
and model systematics have a reduced effect.
% Therefore, it has been agreed to 
%proceed to the extraction of $\frac{|V_{ub}|_{incl}}{|V_{cb}|_{incl}}$ and 
%$\frac{|V_{ub}|_{incl}}{|V_{cb}|_{excl}}$ separately and to combine 
%subsequently into the average $\frac{|V_{ub}|}{|V_{cb}|}$.
 
For the  ratio $\frac{|{\rm V}_{ub}|_{incl}}{|{\rm V}_{cb}|_{incl}}$, uncertainties
originating from the determination of
 $m_b$ have to be considered as
fully correlated. Leading effects of order ${\cal O}(1/m^3)$ are expected
to be uncorrelated between $\Vub$ and $\Vcb$ determinations 
\cite{ref:bigi0}. This is also expected for perturbative uncertainties.

With the conventions used in the previous sections, the relative errors
on the ratio $\frac{|{\rm V}_{ub}|_{incl}}{|{\rm V}_{cb}|_{incl}}$ are:

\begin{equation}
\pm 0.015(m_b)~\pm 0.032(pert.)~\pm 0.024(\mu_{\pi}^2) \pm 0.038(1/m^3)
~=~\pm 0.06
\label{eq:rincl}
\end{equation}

For the ratio $\frac{|{\rm V}_{ub}|_{incl}}{|{\rm V}_{cb}|_{excl}}$, all uncertainties
can be considered as uncorrelated giving:
\begin{equation}
\pm 0.035(m_b)~\pm 0.014(pert.)~\pm 0.024(\mu_{\pi}^2) \pm 0.039(1/m^3)
\pm 0.035(excit.)~=~\pm 0.07
\label{eq:rexcl}
\end{equation}

In practice, the ratio $\frac{|{\rm V}_{ub}|}{|{\rm V}_{cb}|}$ will be obtained using the average
of the two measurements of $\Vcb$ and taking into account common systematics
with $\Vub$.

\mysubsection{Conclusions and summary}
%The conclusions of the workshop which are given in this Appendix
%have left aside
%%This note is mainly a working document and
% several subtleties concerning
%the exact meaning of the parameters entering into the different expressions
%for $\Vub(incl.)$, $\Vcb(incl.)$ and $\Vcb(excl.)$.
%% are not properly described.
%The interested reader has therefore to consult the documents quoted in 
%the references
%for more information.

Two groups (at least) have obtained consistent results on central values 
and uncertainties for $\Vub(incl.)$ and $\Vcb(incl.)$.

Uncertainties on $\Vcb(incl.)$ have been enlarged, in a rather arbitrary way,
by a factor two to have some margin because of possible additional 
contributions (reliability of $1/m_c$ expansion, need for other techniques
to evaluate $m_b$,...) to the error. 
%As an example, this factor two can account
%for a
%twice larger uncertainty on the $b$-quark mass.

Dedicated experimental studies on $\Dstarstar$ production and on the 
distributions of moments of the lepton momentum in the B rest frame
are needed to evaluate with  better accuracy and greater confidence the most
important 
sources of systematic errors contributing in the theoretical expressions.

As said in the introduction, this proposal is a first conservative attempt, 
and it is hoped that in future quoted values for the uncertainties
can be reduced and/or corrected for possible mistakes.

The following central values and theoretical uncertainties have been used 
in the measurements of $\Vub$ and $\Vcb$ obtained from combined
LEP analyses available for Summer 2000:
\begin{itemize}
\item{inclusive measurement of $\Vub$:}
\begin{equation}
\begin{array}{cc}
\Vub=
& 0.00445 \left ( \frac{{\rm BR}(b \rightarrow \ell^-
\overline{\nu_{\ell}}{\rm X}_u)}{0.002}  \right )^{1/2}
\left ( \frac{1.55~{\rm ps}}{\tau_b} \right )^{1/2}\\
  & \times \left (1 \pm 0.010_{pert.} \pm 0.030_{1/m_b^3} \pm 0.035_{m_b} 
\right ) \\
\end{array}
\end{equation}

\item{inclusive measurement of $\Vcb$:}
\begin{equation}
\begin{array}{cc}
\Vcb=
& 0.0411 \left ( \frac{{\rm BR}(b \rightarrow \ell^-
\overline{\nu_{\ell}}{\rm X}_c)}{0.105}  \right )^{1/2}
\left ( \frac{1.55~{\rm ps}}{\tau_b} \right )^{1/2}\\
  & \times \left ( 1 -0.024 \frac{\mu_{\pi}^2-0.5 \GeV^2}{0.2 \GeV^2} \right )\\
  & \times \left (1 \pm 0.030_{pert.} \pm 0.020_{m_b} \pm 0.024_{1/m_Q^3}
\right ) \\
\end{array}
\end{equation}

\item{exclusive measurement of $\Vcb$:}
\begin{equation}
{\cal F}_{D^{\ast}}(1)~=~0.880 -0.024 \frac{\mu^2_{\pi}-0.5~ \GeV^2}{0.1~\GeV^2}
\pm 0.035_{excit.} \pm 0.010_{pert.} \pm 0.025_{1/m^3}
\end{equation}
\end{itemize}

All uncertainties have been added in quadrature.

The correlated uncertainty between $\Vcb(incl.)$ and $\Vcb(excl.)$ measurements
stem from the effect of $\mu^2_{\pi}$.

The correlated uncertainty between $\Vub(incl.)$ and $\Vcb(incl.)$ measurements
is limited to the effect of $m_b$.
Hence no correlated uncertainty
between $\Vub(incl.)$ and $\Vcb(excl.)$ measurements is assumed.

\end{document}